\documentclass[11pt,letterpaper,english]{article}
\usepackage{microtype}
\usepackage{float}
\usepackage{amsmath} 
\usepackage{amsthm}
\usepackage{mathrsfs}
\usepackage{graphicx} 
\usepackage{caption}
\usepackage{setspace}
\usepackage{amssymb}
\usepackage{multirow}
\usepackage{rotating}
\usepackage{array}
\usepackage{booktabs}
\usepackage{multibib}
\usepackage{xr}
\usepackage{algorithm}
\usepackage{algpseudocode}
\usepackage{color}
\usepackage[normalem]{ulem} % Section 8.3 revision markup
\usepackage[small,compact]{titlesec} 
\DeclareMathAlphabet{\mathcalligra}{T1}{calligra}{m}{n}
\usepackage[sectionbib,round,authoryear]{natbib}
\usepackage{makecell}

\usepackage{times}
\usepackage{abstract}

\usepackage[hang,flushmargin]{footmisc}
\usepackage{fullpage} 
\usepackage{hyperref}
\usepackage[compact]{titlesec}
\usepackage[section]{placeins}
\usepackage{footnote}
\usepackage{epstopdf}
\usepackage{placeins}
\usepackage[toc,page]{appendix}
\usepackage{graphicx}
\usepackage{subcaption}
\usepackage{amsmath} 
\usepackage{amsthm}
\usepackage{graphicx} 
\usepackage{blkarray}
\usepackage{stackengine}
\usepackage{accents}
\usepackage{enumitem}
\usepackage[mathscr]{euscript}

\usepackage{pdflscape}
\usepackage{lipsum}    % For placeholder text, if needed

\bibpunct[, ]{(}{)}{;}{a}{}{,}

\hypersetup{
  colorlinks = false,
  urlcolor = darkblue
}

\definecolor{darkblue}{rgb}{0.0, 0.0, 0.55}
\theoremstyle{definition}

\theoremstyle{definition}

\usepackage[utf8]{inputenc}
\usepackage{longtable}
\usepackage{graphicx}
\usepackage{epstopdf}
\usepackage{bbm}
\usepackage{dsfont}
\usepackage{comment}
\usepackage{pgfplots}
\usepackage{xfrac}
\usepackage[flushleft]{threeparttable}

\usepackage{pifont}% http://ctan.org/pkg/pifont

\usepackage{caption}
\theoremstyle{plain}

\newcommand{\squishlist}{
   \begin{list}{$\bullet$}
    { \setlength{\itemsep}{0pt} \setlength{\parsep}{1pt}
      \setlength{\topsep}{1pt} \setlength{\partopsep}{1pt}
      \setlength{\leftmargin}{1.5em} \setlength{\labelwidth}{1em}
      \setlength{\labelsep}{0.5em} } }

\newcommand{\squishlisttwo}{
   \begin{list}{$\bullet$}
    { \setlength{\itemsep}{0pt} \setlength{\parsep}{0pt}
      \setlength{\topsep}{0pt} \setlength{\partopsep}{0pt}
      \setlength{\leftmargin}{1em} \setlength{\labelwidth}{1.5em}
      \setlength{\labelsep}{0.5em} } }

\newcommand{\squishend}{
    \end{list}  }

\newtheoremstyle{mycustom}
  {3pt}{3pt}
 {\normalfont}{}
 {\itshape}{.}{0.5em}{}
\theoremstyle{mycustom}
\newtheorem{defnitalic}{Definition}
\makeatletter
\let\NAT@force@numbers\relax
\makeatother

\begin{document}
%\begin{bibunit}

\title{How does online shopping affect offline price sensitivity?}
%\date{}
\author{Shirsho Biswas, University of Washington\thanks{Authors contributed equally and names are listed in alphabetical order. E-mails: shirsho@uw.edu, hemay@uw.edu, hzhang96@uw.edu. We thank an anonymous pet supplies retailer for providing the data and for thoughtful discussions on many aspects of the paper. The opinions expressed in this paper are the authors' and do not reflect those of the data providers.} \\ Hema Yoganarasimhan, University of Washington\\ Haonan Zhang, University of Washington}

\maketitle

\begin{abstract} 

\noindent \singlespacing 

The rapid rise of e-commerce has transformed consumer behavior, prompting questions about how online adoption influences offline shopping. We examine whether consumers who adopt a retailer's online shopping channels become more price-sensitive in their subsequent \emph{offline} purchases with that retailer. Using transaction-level data from a large Brazilian pet supplies retailer operating both online and offline, we compare “adopters”—customers who began shopping online after a period of offline-only purchasing—with “non-adopters” who remained offline-only. We estimate a discrete choice logit model with individual-level heterogeneity, based on an algorithm that can handle both high-dimensional fixed effects and price endogeneity. We then apply a staggered difference-in-differences approach to the estimated price elasticities and obtain the Average Treatment Effect on the Treated (ATT). We find that offline price sensitivity increases significantly after online adoption in three out of four product categories. Our results underscore the importance of recognizing cross-channel effects in consumer behavior and contribute to the literature on pricing and multichannel retailing by identifying online adoption as a key driver of offline price sensitivity.

\end{abstract}


\clearpage
\section{Introduction}
\label{sec:intro}

Over the past two decades, online shopping has grown rapidly, driven by the widespread adoption of computers and smartphones. U.S. e-commerce sales more than doubled from \$571 billion in 2019 to approximately \$1.19 trillion in 2024, and the share of e-commerce in total US retail sales has grown from 15.4\% to 22.7\% over the same period \citep{digitalcommerce2024}.  Online grocery continues to grow faster than the overall grocery market, as consumers increasingly prioritize the convenience of online shopping \citep{mckinsey2026grocery}. This shift has prompted traditional brick-and-mortar retailers such as Walmart, Target, and Costco to increase investments into their e-commerce channels as a major driver of growth.  Moreover, e-commerce participation differs substantially across countries and product categories.\footnote{ OECD (2019), for example, reports that travel services were purchased by 79\% of online buyers in Norway and 23\% in Latvia; within the EU28, clothing, footwear, and sporting goods were purchased by 64\% of online buyers, compared with 53\% for travel products. See OECD, \emph{Unpacking E-commerce: Business Models, Trends and Policies} (2019), Chapter 2: \href{https://doi.org/10.1787/23561431-en}{report}. These figures describe purchases among online buyers, rather than online shares of retail sales.} As retailers expand online in these markets, understanding the implications for their existing offline business becomes increasingly important.

Given the rise of online shopping and investments in e-commerce, a natural question that arises for retailers is whether a consumer's adoption of a given retailer's online channels influences her shopping behavior at that retailer's physical stores, and if so, how? In this paper, we focus on one important aspect of consumer behavior -- their price sensitivity. Specifically, we seek to answer the following research question: When an offline-only consumer starts using the online shopping channel(s) of a given retailer, does it affect her offline price sensitivity (i.e., price sensitivity at that retailer's physical stores)? If so, does offline price sensitivity increase or decrease post-online adoption? There are two main reasons why answering this question is important from both researchers' and managers' perspectives. First, from a substantive perspective, understanding the determinants of price elasticity is important -- doing so gives us insights into what factors make consumers more or less price sensitive, and how. Second, from a managerial perspective, quantifying the spillover effects of online shopping on offline price sensitivity can have significant implications for retailers' pricing strategies and profits. Given the migration of consumers from offline stores to multi-channel shopping, quantifying the existence and impact of cross-channel externalities can help firms optimize and coordinate pricing and promotion across channels. More broadly, there is a growing interest among managers in understanding how interactions and spillovers between online and offline shopping channels shape consumer behavior, given the growth of omnichannel retailing across the globe \citep{zhang_etal_2019, biswasexpress}.

While there is a rich literature studying the determinants of price sensitivity, this literature largely focuses on determinants of price elasticities \textit{within} a channel (see $\S$\ref{sec:literature} for a detailed discussion). In contrast, our focus is on the externalities or spillovers \emph{across} channels. From a conceptual perspective, there are reasons why we may expect offline price sensitivity to increase post-online adoption as well as reasons that suggest the opposite. On the one hand, online shopping may make consumers less price sensitive and more brand loyal because online recommendation systems promote easy repeat purchases and lead to habit formation by highlighting prior purchases, e.g., ``Buy again" or ``Frequently purchased items" \citep{danaher2003comparison}. This increase in brand loyalty in a retailer's online channels could spill over to offline shopping occasions and may make consumers more brand loyal and less price-sensitive in that retailer's physical stores too (after adopting online shopping). On the other hand, online shopping makes it easy for consumers to engage in price comparisons (due to features like sorting by price and lower search costs of browsing through many options on an app or website). This habit of checking and comparing prices on a retailer's online channels can potentially carry over to the retailer's offline channels, and consumers may become more aware of price differences in the retailer's physical stores as well. Another potential reason that may make consumers more price-sensitive after adopting online shopping is that, in many contexts, prices are lower online than offline \citep{brynjolfsson2000frictionless, cooper2006prices}. Being exposed to lower prices for the same products online can make in-store prices seem inflated by comparison, prompting greater scrutiny of the prices of different brands. Finally, recent research suggests that when consumers are exposed to a wider variety of prices (which is likely to occur when they adopt online shopping), they become more price-sensitive \citep{aparicio2024algorithmic}. Therefore, the question of whether the adoption of a retailer's online channels affects consumers' offline price sensitivities at that retailer's physical stores, and if so, in which direction, is an empirical one.

We address this question using data from a large pet supplies retailer in Brazil that sells through both online and offline channels. The dataset includes detailed transaction-level information for all customers across both channels from January 1, 2019, to June 30, 2021. In Brazil, the share of Internet users purchasing goods or services online in the preceding twelve months rose from 39\% in 2019 to 46\% in 2021 \citep{cetic2022households2021}. This growth during our observation period makes Brazil a useful setting for examining how adding online shopping to an existing offline relationship changes subsequent offline behavior. In our data, we observe some consumers who were initially offline-only and then made their first online purchase with the retailer within our data window, while others remained offline-only throughout. This allows us to track the offline purchases of online shopping adopters, both pre- and post-adoption, and also all purchases of those consumers who remained offline only. We focus on four major product categories: dog food, cat food, dog hygiene, and cat hygiene, which together account for more than 50\% of the retailer's total revenue. These categories are also relatively frequently purchased and are characterized by a good amount of price variation both across brands and across time, which allows us to measure individual-level price elasticities at the monthly level. 

From an empirical perspective, a key question relates to the estimand of interest and how to measure it. Note that adopting online shopping is a consumer-level decision, i.e., firms cannot randomly assign consumers to adopt online shopping. Therefore, we cannot quantify the Average Treatment Effect (ATE) of adopting online shopping on offline price sensitivity. In addition, even if obtaining the ATE was possible, firms cannot act on it because they cannot force consumers to adopt online shopping. Therefore, obtaining the ATE is not only impractical but also lacks actionable implications. Hence, we focus on identifying the Average Treatment Effect on the Treated (ATT), which captures the impact of treatment on those who adopted online shopping. 

Our empirical strategy consists of two steps. In the first stage, we specify and estimate a discrete choice logit model for each category, controlling for unobserved customer heterogeneity in brand preferences. In our model, we allow consumers' offline price sensitivity to be a function of the extent to which a given consumer adopts online shopping at a point in time, measured by the cumulative fraction of their online spending with the retailer up to that point in time.\footnote{We also include an alternate specification with a binary online adoption variable, as a robustness check.} This modeling approach allows us to measure the effect of interest and examine how the price elasticity of consumers who have adopted online shopping diverges over time from that of those who have not.  We encounter two challenges in estimating this demand model. First, maximum likelihood estimation of a non-linear model, such as the logit model, is quite challenging in the presence of high-dimensional fixed effects (customer-brand fixed effects in our case). To accommodate these high-dimensional fixed effects, we build on the Minorization-Maximization algorithm proposed by \cite{chen2022mm} and allow for covariates that vary by individual, by alternative, and also over time. We then apply this algorithm to our setting for estimation. To the best of our knowledge, this is the first empirical paper in marketing to efficiently estimate high-dimensional customer-brand fixed effects in a demand model using the MM algorithm.  Second, we must address potential endogeneity in prices using our observational data \citep{nevo2001measuring}. To that end, we use costs as an instrument \citep{berry1994estimating} in conjunction with a control function approach \citep{villas1999endogeneity, petrin2010control}. We then use the estimates from this model to compute customer-month-level price elasticities.

In the second stage, we use the estimated customer-month-level offline price elasticities of adopters and non-adopters from the first stage to estimate our ATT of interest, which is the effect of adopting online shopping on consumers' offline price elasticities. Again, our observational data pose some challenges for the identification of the ATT. First, consumers in our sample are either treated or untreated in each time period, and we do not observe their counterfactual outcomes in the unobserved condition (i.e., their behaviors if they had not adopted online shopping). Moreover, the staggered timing of online adoption among different consumers requires careful computation and aggregation of treatment effects \citep{goodman2021difference, callaway2021difference, baker2022much}. To deal with this, we employ a staggered difference-in-differences (DID) framework, as proposed by \cite{callaway2021difference}, utilizing the variation in staggered adoption timing of online shopping. We identify different cohorts of consumers who adopt online shopping (i.e., adopters) in each month of our data and compare changes in their behavior to those of a control group of customers who shop exclusively offline throughout the observation period (i.e., non-adopters). 

Our main finding is that consumers become \textit{more} price sensitive in their offline shopping trips after adopting online shopping for three out of four categories. The relative increase in offline price elasticity, defined as the ATT relative to the pre-adoption average price elasticity of the treated group is heterogeneous across categories. We find a large effect in the dog and cat hygiene categories (where price elasticity increases by 72.7\% and 34.7\%, respectively). We also observe a significant but smaller increase in the dog food category (8.4\%) and no effect for cat food.  The larger effects for hygiene products may reflect greater flexibility to time purchases around promotions.

We also examine the behavioral mechanisms underlying this increase in offline price sensitivity. Using weekly SKU--store price histories, we find that, conditional on an offline purchase in a focal category, adopters become 6.2--10.9 percentage points more likely to purchase during promotional-price periods. This pattern holds after accounting for customer, year-month, and SKU fixed effects, linking adoption to more favorable purchase timing for the same products. In contrast, brand-share analyses show no systematic shift toward cheaper brands. A separate decomposition shows that total purchasing with the retailer increases as activity shifts from offline to online channels, consistent with continued retailer engagement. Together, these descriptive findings are consistent with online shopping increasing price attention and encouraging more strategic timing of subsequent offline purchases.

We also provide a series of tests and alternative specifications that establish the robustness of our results and provide additional empirical support for the generalizability of our findings. First, we show that using a binary online adoption variable in our demand model does not change our conclusions. Second, we examine pre-adoption price elasticities to rule out pre-adoption trends in the price elasticities of consumers who chose to adopt online shopping, and we also use Inverse Propensity Weighting (IPW) approaches to control for self-selection of the online adoption decision. Third, we consider an alternative estimator where later adopter groups serve as a control group for earlier adopters. Fourth, we consider two alternate definitions of purchase occasions to also include online shopping visits and to include months with no purchases online or offline. Fifth, we consider alternate approaches to address price endogeneity and alternate product market definitions. Our conclusions and substantive findings remain unchanged across all of these robustness checks. More broadly, we also present some evidence in favor of portability to other settings. For instance, our data period includes COVID-19, an exogenous macro-environmental shock that drove many consumers to adopt online shopping. Interestingly, we do not find differences in the adoption effect between customers who adopted online shopping during COVID-19 vs. those who adopted organically before that. We also find similar results for consumers who adopted online shopping during the retailer's online-only Black Friday promotion. These results indicate that the effect is not sensitive to the reasons/drivers of a consumer's decision to adopt online shopping.

Finally, counterfactual pricing simulations show that accounting for the online adoption effect increases simulated monthly offline profit by 1.17\% under uniform pricing and 4.1\% under personalized pricing, relative to ignoring the effect within each pricing regime. These results highlight the value of incorporating customers' online channel usage into offline pricing decisions alongside their brand preferences and purchase histories.

%Based on our empirical results, we consider counterfactual pricing scenarios and document the value of correctly accounting for the online adoption effect on offline prices. We first consider a uniform pricing strategy, where all customers pay the same price, and the firm correctly accounts for the online adoption effect when optimizing prices. We demonstrate that this approach can increase the firm's offline profit (summed across the three categories where there is a significant online adoption effect) by 1.17\%, compared to a scenario where the manager does not account for the online adoption effect in their demand model and attempts to set optimal offline prices. Next, we perform a similar comparison under a personalized pricing scenario, where the firm can personalize prices (similar to \cite{rossi1996value,smith2023optimal, jain2024effective}). We show that incorporating the customer's online spend in the behavioral features in the demand model and optimizing prices based on this demand model increase profits by 4.1\% (compared to a setting where the firm does not include the online adoption effect in the demand model but still personalizes prices). Together, these counterfactual results suggest correctly accounting for online spend in offline demand models can help retailers improve their pricing and personalization strategies. 

Our paper contributes to the broad literature on online-offline retailing and pricing. From a substantive perspective, we contribute to the literature on price sensitivity and multichannel retailing by identifying a new determinant of offline price sensitivity: the adoption of online shopping. We demonstrate that consumers become more price-sensitive in a retailer's physical stores after adopting that retailer's online channels for shopping. This finding is important and distinct from previous research, which has mainly focused on cross-sectional or within-household comparisons of price sensitivity between online and offline settings. From a managerial perspective, our findings have implications for firms' pricing strategies and profitability, emphasizing the need for multichannel retailers to account for the online adoption effect (i.e., heightened offline price sensitivity when consumers adopt a retailer's online channels) in their pricing decisions. 

%our counterfactual simulations suggest that incorporating the online adoption effects into price elasticity measurements and subsequent price optimization can significantly improve retailer profits. Overall, these findings highlight the importance of correctly accounting for cross-channel spillovers in retailers' marketing decisions.

The rest of the paper is organized as follows. In \S\ref{sec:literature}, we discuss the related literature, and in \S\ref{sec:set_data}, we discuss the setting and data. In \S\ref{sec:descriptive.analysis}, we document descriptive evidence, and in \S\ref{sec:firms_problem}, we specify the firm's problem and provide an overview of the empirical strategy. \S\ref{sec:model} presents the demand model, and \S\ref{sec:att} presents the ATT estimates of online adoption on offline price sensitivity. \S\ref{sec:robust} discusses the generalizability and robustness checks of our main results. Finally, in \S\ref{sec:conclusion}, we summarize our findings and their implications, and also discuss avenues for future research.

\section{Related Literature}
\label{sec:literature}

Our study relates to multiple streams of literature on the determinants of price elasticity, digital platforms, and multi-channel retailing.

%(1) the determinants of consumer price sensitivity, (2) channel-specific price sensitivity, and (3) multichannel retailing.

First, our work relates to the large literature in marketing and economics that has focused on documenting the drivers of consumers' price sensitivity. Prior research has identified various determinants of price sensitivity, particularly in grocery settings. These include (1) macroeconomic conditions, such as GDP growth and inflation rates \citep{gordon2013does, tellis1988price, bijmolt2005new}; (2) marketing interventions, such as advertising \citep{kanetkar1992price, kaul1995empirical, kalra1998impact, jedidi1999managing}, promotions \cite{mela1997long, jedidi1999managing, kopalle1999dynamic} and category expansion \citep{yu2021dark}; (3) brand strength \citep{guadagni1983logit, kamakura1989probabilistic, guadagni2008logit}; (4) consumer loyalty \citep{danaher2003comparison}; (5) time between purchase and consumption \citep{joo2020temporal}. Early research has also examined the relationship between customer demographic characteristics -- age, education, income, and household size -- and price sensitivity using household-level scanner panel data \citep{rossi1993bayesian, narasimhan1984price, bawa1987coupon, bawa1989analyzing}. However, later studies have concluded these relationships are generally weak, suggesting that demographics do not explain a large portion of the variation in household-level price sensitivity \citep{hoch1995determinants, rossi1996value}.

We consider the adoption of a given retailer's online channels as a novel factor that may affect consumers' offline price sensitivity at that retailer's physical stores. Given the rise of digitization and online shopping, this factor can be particularly important for multi-channel retailers (who are increasingly investing in online channels). Existing literature provides some insights into how consumer behavior differs between online and offline channels. Early studies focusing on cross-sectional comparisons find that consumers exhibit lower price sensitivity and less brand switching when shopping online compared to offline \citep{degeratu2000consumer, andrews2004behavioural}. \citet{danaher2003comparison} shows that high market share brands exhibit greater brand loyalty online, whereas low market share brands exhibit greater brand loyalty in offline stores. However, these studies are ``cross-sectional" in nature, i.e., they compare the set of consumers shopping offline vs. those shopping online, who may be very different from each other. More recent research has shifted toward within-household analysis. For example, \cite{chu2008research} compare price sensitivity in online and offline grocery shopping within the same household and find that consumers are generally less price-sensitive online. Further, \citet{chu2010empirical} investigate the role of household characteristics in moderating price sensitivity across channels. However, none of this research looks into {\it if} and {\it how} the {\it adoption} of online shopping affects consumers' offline price sensitivity, which is the focus of our paper.

More broadly, our study contributes to the broader multichannel retailing literature; see \cite{Cui_etal_2021} for a comprehensive overview. One key stream of research highlights that online and offline channels can function as complements \citep{avery2012adding, wang2017can, bell2018offline} or substitutes \citep{forman2009competition,li2021books,shriver2022demand}, depending on the context. Other studies have examined how digitization affects consumer shopping behavior across channels, focusing on spend patterns and purchase volume \citep{wang2015go,narang2019mobile,biswasexpress}, brand exploration \citep{pozzi2012shopping}, and purchase variety \citep{chintala2023browsing}. Building on this body of work, our study provides new insights into multichannel strategy \citep{neslin2022omnichannel} by investigating changes in offline price sensitivity when consumers adopt a retailer's online channels for shopping.

\section{Setting and Data}
\label{sec:set_data}

\subsection{Setting}
\label{ssec:setting}

We use data from a leading pet supplies retailer in Brazil. This firm operates in a multichannel environment, with over 200 brick-and-mortar stores nationwide and a digital presence through both a mobile app and an e-commerce website. The retailer offers a broad selection of pet-related products, including food, medicines, hygiene and grooming items, toys, and accessories. To track consumer purchases across online and offline channels, the retailer uses Brazil’s National Identification Number, which provides a unique identifier for each customer. 
Online purchases - whether through the app or website - require customers to log into their accounts, making all such transactions traceable to individual IDs. For brick-and-mortar purchases, customers provide their ID at checkout, although it is not mandatory. According to the retailer, around a quarter of offline transactions are not associated with a customer ID, and this rate has remained stable throughout the study period.\footnote{The fact that this rate has remained stable despite the fact that there was a spike in online shopping adoption during COVID-19 reassures us that the adoption of online shopping does not affect it.} 

The dataset spans 30 months, from January 1, 2019, to June 30, 2021, and covers all transactions (online and offline) linked to identifiable customers. Each transaction includes a unique order ID and detailed product-level information, such as price, quantity, brand, category (e.g., dry food or hygiene), and purchase channel. The retailer also supplies the wholesale procurement costs recorded for the purchased items in its internal transaction database. These records provide the underlying cost data for our instruments. We explain the procurement setting and identifying assumption in $\S$\ref{ssec:endogeneity} and the construction of the cost indices in Web Appendix $\S$\ref{appsssec:price}.

Based on our discussions with the retailer, pricing and product assortment decisions are managed separately across the online and offline channels, without a coordinated joint strategy.\footnote{We also verify in our data that price promotions are not highly coordinated across channels. Details in Web Appendix \ref{appssec:cross.channel.price.differences}.} Our focal product markets comprise brands and product types available in both channels; categories offered exclusively online are outside these markets. These institutional features help distinguish changes in consumers' offline price sensitivity from coordinated channel policies. 

\subsection{Dataset Construction}
\label{ssec:construction}
We now discuss how we construct the dataset used for our empirical analysis.

\noindent \textbf{Selection of Product Categories:} 
We focus on the four top-selling non-medical categories: Dry Dog Food, Dry Cat Food, Dog Hygiene (i.e., Hygienic Mats), and Cat Hygiene (i.e., Sands). These categories account for more than 50\% of the focal retailer's revenue (see Table \ref{2_Category_By_Sales}). Products within these four major categories are generally measured in comparable units, allowing us to construct brand-level prices per unit. 

%Further, they tend to be sold in non-comparable units/packs, making price measurement challenging. 

We exclude certain categories from our analysis. First, we exclude medical products such as prescription drugs and anti-flea medication (second category shown in Table \ref{2_Category_By_Sales}) because many products in this category require a doctor's prescription and purchase does not depend solely on consumers' preferences. Further, medical products can be quite differentiated and target specific ailments, and as such, do not serve as natural substitutes for each other. Second, we exclude all small categories that account for less than 3\% of the total sales revenue during the observation period, e.g., snacks and toys. Products in these categories are often unique, serve diverse functions, and are not necessarily natural substitutes for each other.

%% comment out table separating anti-flea and medical
\begin{comment}
\begin{table}[htp!]
\caption{Top Categories Ranked By Total Sales (Offline and Online) During the Observation Period.}
\label{2_Category_By_Sales}
\centering
\footnotesize{
\begin{tabular}{lrr}
\toprule
Category & Share of sales(\%) & Cumulative share (\%) \\
\midrule
Dry Dog Food & 32.62 & 32.62 \\
Dry Cat Food & 9.41 & 42.03 \\
Anti-Flea & 6.27 & 48.30 \\
Medical & 5.91 & 54.21 \\
Dog Hygiene & 5.48 & 59.68 \\
Cat Hygiene & 3.62 & 63.30 \\
%Snacks & 2.58 & 65.88 \\
\bottomrule
\end{tabular}
}
\end{table}
\end{comment}

\begin{table}[htp!]
\caption{Top Categories Ranked By Total Sales (Offline and Online) During the Observation Period.}
\label{2_Category_By_Sales}
\centering
\footnotesize{
\begin{tabular}{lrr}
\toprule
Category & Share of sales(\%) & Cumulative share (\%) \\
\midrule
Dry Dog Food & 32.62 & 32.62 \\
Medical & 12.18 & 44.80 \\
Dry Cat Food & 9.41 & 54.21 \\
Dog Hygiene & 5.48 & 59.68 \\
Cat Hygiene & 3.62 & 63.30 \\
\bottomrule
\end{tabular}
}
\end{table}

\noindent \textbf{Selection of Brands and Stock Keeping Units (SKUs):} Each category contains a large number of pack sizes and SKUs per brand. First, within each category, we focus on specific pack size(s) to ensure that the price variation for each brand is captured accurately -- this avoids the need to aggregate price movements across a number of different pack sizes. Unit-price normalization makes prices comparable in units of measurement, but does not remove differences in pack size or product function that can affect substitution. This restriction helps us compare brand choices among products of similar size and use \citep{allenby1998marketing,kim2002modeling,chintagunta2005beyond,dube2010state}. We assess sensitivity of our results to this restriction by using a broader market definition in $\S$\ref{sssec:product_market_definition}. The pack sizes we select for each category are as follows: 15KG for dry dog food, 10 KG for dry cat food, 80cm $\times$ 60cm 30-piece mats for dog hygiene, and silica cat sand (1.6KG, 1.8KG, and 2.0KG) for cat hygiene. We provide a detailed discussion of the reasons for selecting these pack sizes in Web Appendix $\S$\ref{appsssec:brand_sku}.

Next, we aggregate SKUs to the brand level and estimate consumer preferences and price elasticities at the brand level for each product category, similar to \cite{hitsch2021prices,gordon2013does}.  Specifically, we follow a two-step process for each category. First, for each category, we select the top-selling brands and SKUs that are comparable in usage and are substitutes for each other. Next, we aggregate prices and costs (our instrumental variable to deal with price endogeneity) to the brand level from the SKU level, following the approach described in  \cite{hitsch2021prices}. We provide details on both these steps in Web Appendix $\S$\ref{appssec:dataset_const}.

\noindent \textbf{Selection of Customers:} For each category, we retain customers who purchase only the selected pack size(s), so that their observed within-category brand substitutions remain within the modeled product set \citep{kim2002modeling,chintagunta2005beyond,dube2010state}. The expanded-market check in $\S$\ref{sssec:product_market_definition} also includes customers who switch pack sizes. Next, given that our goal is to investigate the effect of consumers' adoption of online shopping at a given retailer\footnote{``Adoption of online shopping'' refers to a customer making his/her first online purchase with the focal retailer in \textit{any} category, and not just one of the four focal categories.} on their \textit{offline} price sensitivity at that retailer's physical stores, we need to ensure that we have sufficiently long pre- and post-adoption periods for each adopter in our sample to effectively measure price elasticities pre- and post-adoption. Therefore, we focus on customers who adopt online shopping with the retailer in the middle of our observation period, i.e., July 1st, 2019 -- June 30th, 2020. So our 30-month data period is divided into three phases: a pre-adoption period (January 1st, 2019 -- June 30th, 2019) when all adopters are in their pre-adoption phase, an adoption period (July 1st, 2019 -- June 30th, 2020) during which time the adopters in our sample make their first online purchase, and a post-adoption period (July 1st, 2020 - June 30th, 2021), when all the adopters in our sample are in their post-adoption phase. To clarify, a given adopter's pre- and post-adoption phases are defined as the months before and after they make their first online purchase. Non-adopters only shop offline through the entire data period.\footnote{Note that we also verified with the retailer that the identified adopters and non-adopters in our data did not adopt online shopping before our data period begins.} 

\begin{figure}[htp!]
    \centering
    \caption{Sample Sizes of Adopters by Cohorts and Product Category. Corresponding sample sizes for each cohort are shown in Table \ref{tab:2_append_adopter_samplesize} in Web Appendix \ref{appsssec:sample.customer}.}
    \label{fig:2_adopter_samplesize}
    \begin{subfigure}{0.49\textwidth}
        \centering
        \caption{Dry Dog Food}
        \includegraphics[scale=0.25]{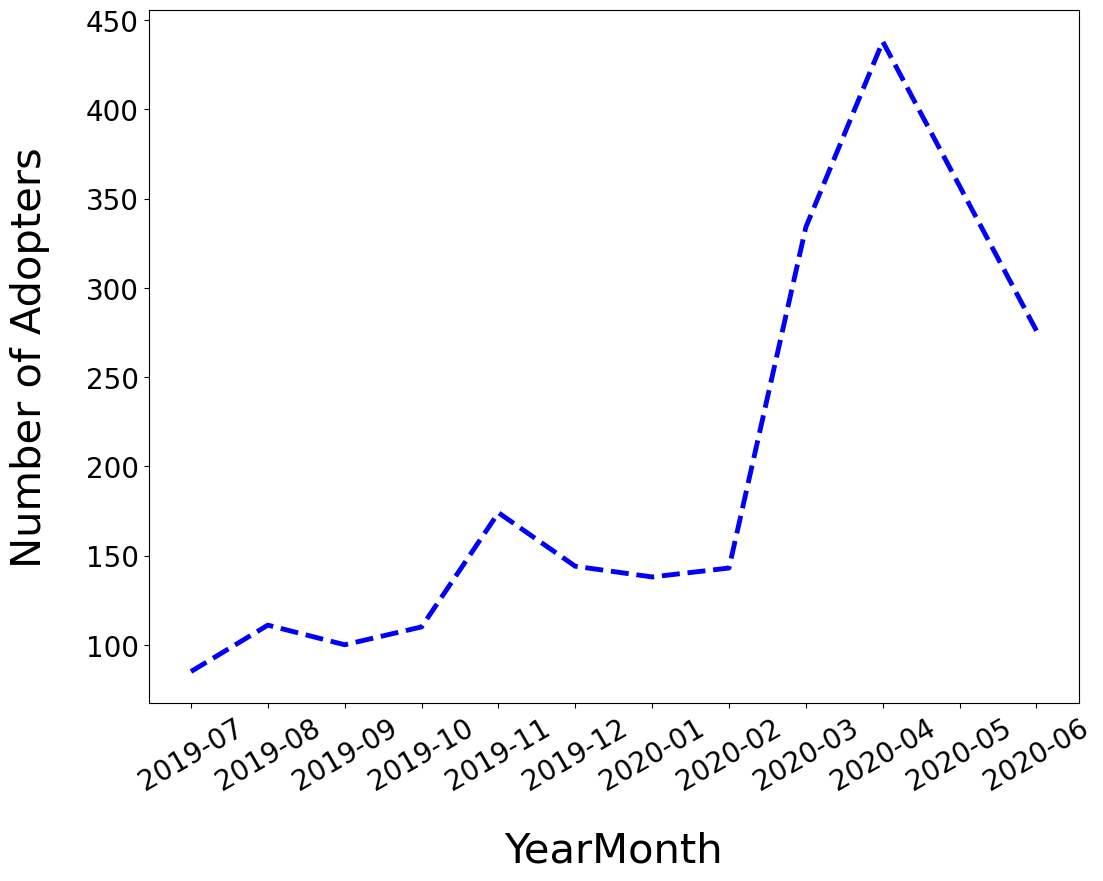}
    \end{subfigure}
    ~ %add desired spacing between the subfigures, if needed
    \begin{subfigure}{0.49\textwidth}
        \centering
        \caption{Dry Cat Food}
        \includegraphics[scale=0.25]{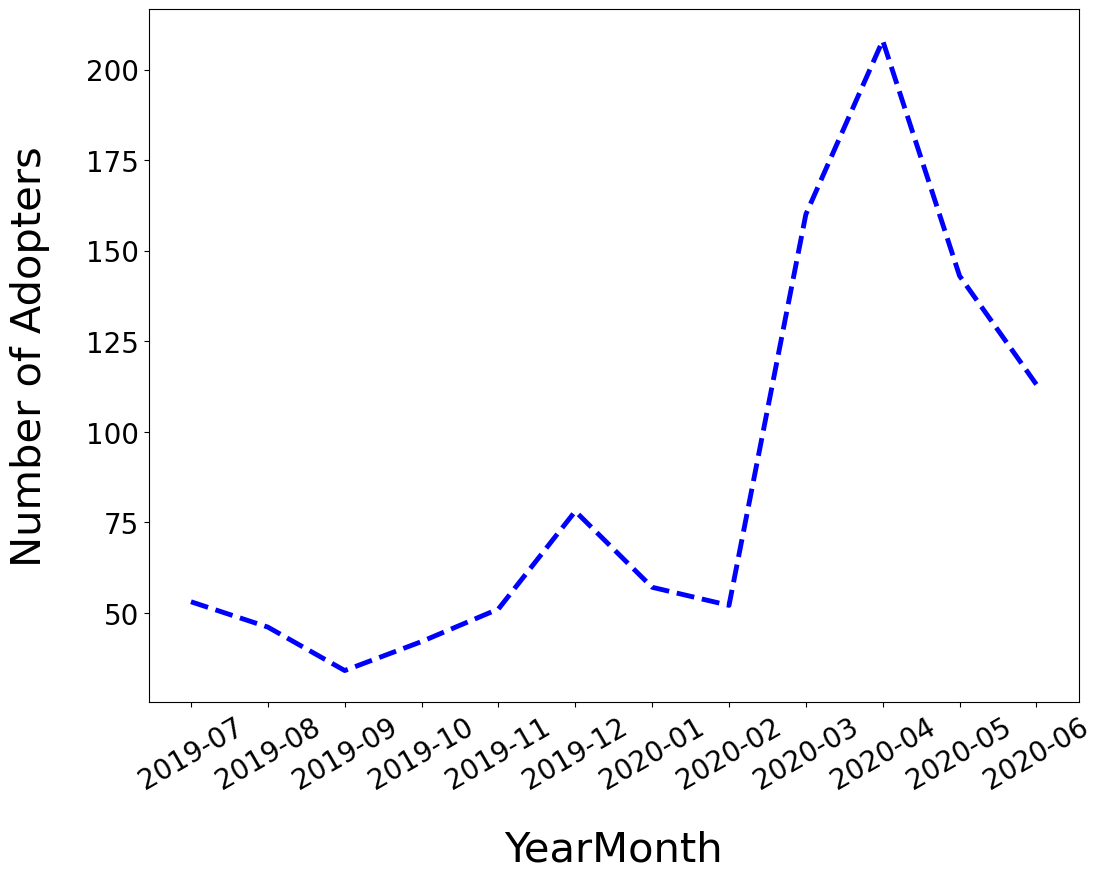}
    \end{subfigure}

    \vspace{1ex} % add some vertical space between the rows

    \begin{subfigure}{0.49\textwidth}
        \centering
        \caption{Dog Hygiene}
        \includegraphics[scale=0.25]{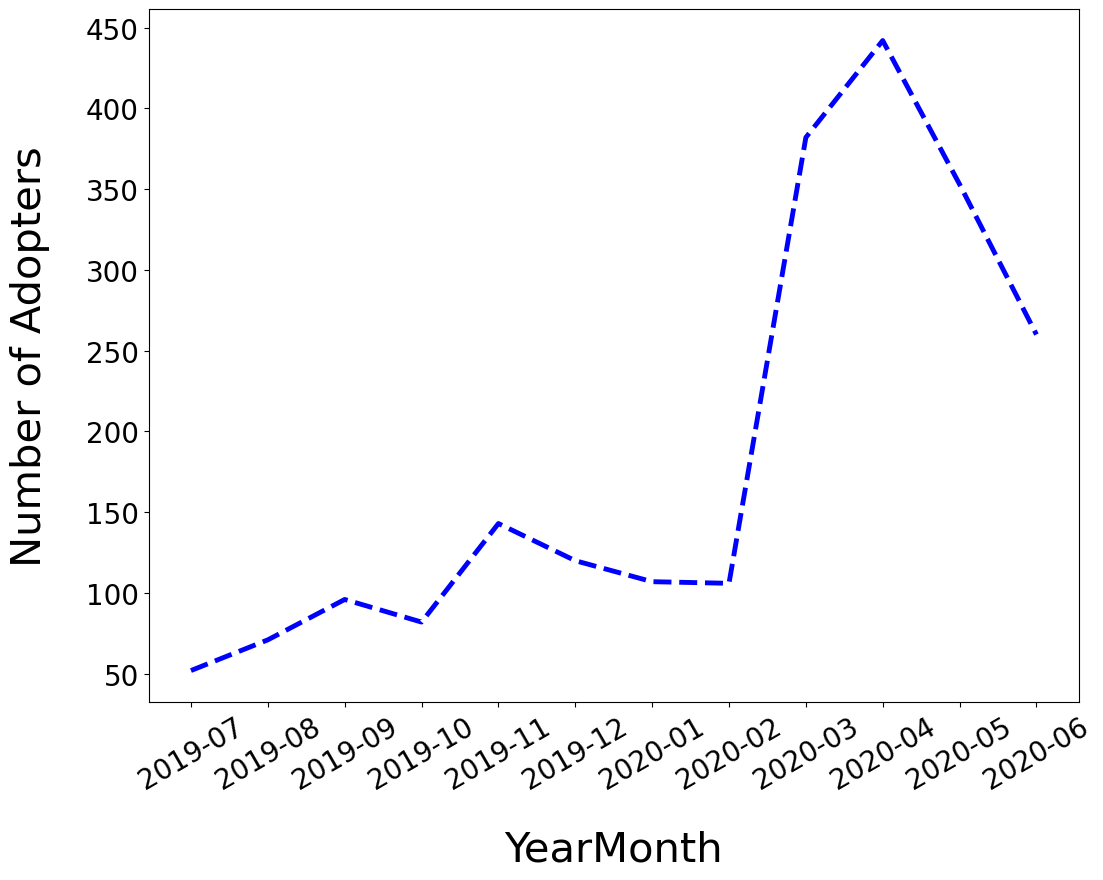}
    \end{subfigure}
    ~ %add desired spacing between the subfigures, if needed
    \begin{subfigure}{0.49\textwidth}
        \centering
        \caption{Cat Hygiene}
        \includegraphics[scale=0.25]{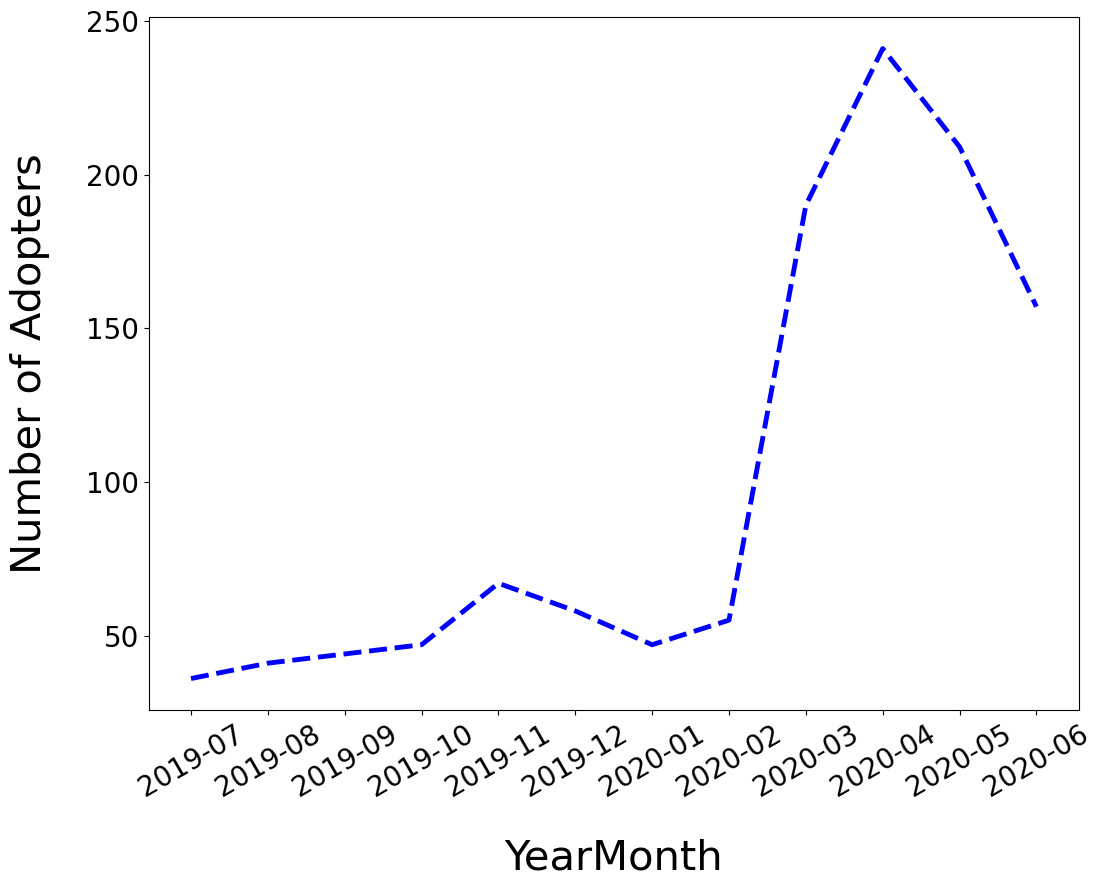}
    \end{subfigure}
\end{figure}

We now outline the criteria used to select adopters and non-adopters for each product category below. 
\noindent {\it Adopters} in a given category must satisfy the following criteria:
\squishlist
\item  Made their first-ever online purchase (in any product category, not just the focal category) between July 1st, 2019, and June 30th, 2020. 
\item Made at least one offline purchase in the focal category both before and after online adoption. 
\squishend
Overall, this gives us 12 cohorts of adopters, where each cohort represents the month in which the user adopted online shopping. For example, the first cohort of adopters consists of users who adopted online shopping during July 2019 while the twelfth cohort consists of users who adopted online shopping during June 2020. Figure \ref{fig:adopter_def} in Web Appendix \ref{appsssec:sample.customer} shows stylistic examples of users in different cohorts. Figure \ref{fig:2_adopter_samplesize} shows the number of adopters across the different periods in each category. Notice that there is a spike in the number of adopters during March--April 2020, which is likely driven by the onset of the COVID-19 pandemic. Our analysis is agnostic to the reasons for adopting online shopping, and in our empirical analysis, we will allow for the possibility that these cohorts may be different from earlier/later cohorts.  

\noindent {\it Non-adopters} in a given category must meet the following criteria:
\squishlist    
\item Did not make any online purchases at any point during the entire observation period.
\item Made at least one offline purchase in the category during both the pre-adoption (i.e., January 1st, 2019 -- June 30th, 2019) and the post-adoption periods (i.e., July 1st, 2020 -- June 30th, 2021). 
\squishend

\noindent \textbf{Estimation Sample Construction:} 
For each customer in a given category (both adopters and non-adopters), we organize the data into a panel that captures purchase occasions and brand choices for individual customers over time. Specifically, we define purchase occasions at the customer-month level. If a customer visits a physical store of the retailer during a month and purchases in any category (not limited to the categories we focus on in this study), that customer-month is included as a purchase occasion for all four categories. If a customer makes no purchases in any category during the month, that customer-month is not included as a purchase occasion. In other words, a purchase occasion represents any customer-month during which we observe any offline interaction with the retailer (similar to week-store visit in \cite{gordon2013does}).\footnote{ We combine a customer's offline visits within a month into a single purchase occasion, even if the customer visits more than once. More than 96\% of these customer-months involve purchases at only one store. For the remaining occasions involving multiple stores, we assign a single store for each category: the store where the customer spent the most in that category during the month. This assignment links each purchase occasion to the corresponding store's prices and costs (Web Appendix $\S$\ref{appsssec:price}).}

We use observed purchases in any category to identify months with offline retailer activity and the associated store, while retaining non-purchase of the focal category as an outcome. This defines our main analysis as brand choice and category purchase incidence conditional on observed offline activity \citep{chib2004model,chintagunta2005beyond,dube2010state,gordon2013does}. Channel choice itself can depend on transportation costs, distance, and expected prices; modeling it jointly would require additional assumptions about the sequence of channel, store, and brand decisions. To assess sensitivity to conditioning on offline activity, $\S$\ref{sssec:alternative.def.purchase.occasions} considers months with activity in either channel and also a complete customer-month panel including months without retailer purchases. Both yield findings consistent with the main specification (Web Appendices $\S$\ref{appssec:alternative.def.po}--$\S$\ref{appssec:customer.month.full.panel}). 

%\hy{Shirsho -- can you check?} \sbis{I think we should move footnote 4 back to the main text because the comment on why we do not model channel choice has come up consistently every time we have presented this paper} %In this literature, the timing and incidence of store visits are assumed to be exogenous, and we assume this too. 

For each purchase occasion, we identify the brand chosen by the consumer as the brand that has positive spend.\footnote{We rarely observe customer-month purchase occasions that involve a purchase of more than one brand. All product categories have fewer than 0.86\% of total purchase occasions (e.g., including those with multiple visits to a single store in a month, visits to multiple stores and a single visit) with multiple brand purchases. For these few purchase occasions with multiple brand purchases, we assign the brand with the largest spend for that customer during the month, similar to previous brand choice literature, e.g. \cite{erdem1996decision}. Some researchers have handled this issue differently, for instance, by having each brand chosen count as an independent single-brand purchase occasion \citep{krishnamurthi1988model,chib2004model}. Since the fraction of purchases is so small, we do not think this decision makes a material impact on our results.} If the customer does not have positive spend on any brand during a purchase occasion, we label it as choosing the outside option (i.e., ``no-purchase").

\subsection{Summary Statistics}
\label{ssec:sum_statistics}
We now present some summary statistics on the estimation sample.

First, we find that there are systematic differences in the demographics (age, income, and gender) of adopters and non-adopters across all four categories. Across all categories, women tend to adopt online shopping more than men. Additionally, online adopters tend to be younger and from higher-income households. This makes intuitive sense because younger and higher-income consumers are often more familiar with online technology and seek time-saving and convenient shopping options. Details of these demographic characteristics are shown in Web Appendix $\S$\ref{appssec:demographic}.

Next, Table \ref{tab:2_Summary_Statistics} summarizes the category-level statistics, including the number of brands, sample sizes for adopters and non-adopters (as defined above), the total number of purchase occasions, and the number of purchase occasions that contain a brand purchase within each category. 

\begin{table}[htp!]
    \centering
  \begin{threeparttable}
\caption{Summary Statistics By Category}
\label{tab:2_Summary_Statistics}
\footnotesize{
\begin{tabular}{lllll}
\toprule
& Dry Dog Food & Dry Cat Food & Dog Hygiene & Cat Hygiene \\ \midrule
Number of Brands & 8 & 4 & 4 & 4 \\
Number of Adopters & 2,410 & 1,037 & 2,214 & 1,192 \\
Number of Non-Adopters & 7,938 & 2,297 & 3,582 & 2,725 \\
Number of Purchase Occasions & 136,411 & 45,443 & 78,253 & 55,410 \\
Purchase Occasions with a Category Purchase & 99,233 & 27,623 & 40,183 & 42,596 \\ %\hline
\bottomrule
\end{tabular}
}
  \end{threeparttable}
\end{table}

\begin{table}[htp!]
    \centering
  \begin{threeparttable}
\caption{Summary Statistics of Prices and Costs}
\label{tab:2_Summary_Statistics_Estimation_Sample}
\footnotesize{
\begin{tabular}{lllll}
\toprule
  & Dry Dog Food & Dry Cat Food & Dog Hygiene & Cat Hygiene \\ \midrule
%Estimation Sample Observations & 1,227,699	 & 227,215 &	391,265	& 277,050 \\ 
Mean Price & 10.17 & 11.89 & 2.28 & 12.25 \\
SD Price & 2.82 & 2.40 & 0.28 & 2.46 \\
Mean Cost & 6.29 & 8.04 & 0.96 & 4.95 \\
SD Cost & 2.32 & 3.39 & 0.29 & 1.54 \\
%Number of Observations & 1,091,288 & 181,772 & 313,012 & 221,640 \\
%Total Observations & 1,227,699	 & 227,215 &	391,265	& 277,050 \\ 
\bottomrule
\end{tabular}
}
\end{threeparttable}
\end{table}

Finally, in Table \ref{tab:2_Summary_Statistics_Estimation_Sample}, we show some summary statistics of the prices and costs for each category. Specifically, we show the mean and standard deviation (SD) of the brand price (or cost) per pack size unit (e.g, KG or piece) for each category, for the estimation sample. These values are calculated by averaging across all purchase occasions and brands. Specifically, the mean price for category $c$ is calculated as: $\textrm{Mean Price}_{c} = \frac{1}{N_{c}}\frac{1}{J_{c}}\sum_{t=1}^{N_{c}}\sum_{j=1}^{J}\textrm{Price}_{jt}$ where $\textrm{Price}_{jt}$ is the price of brand $j$ on the purchase occasion $t$. $N_{c}$ is the total number of purchase occasions in category $c$, and $J_{c}$ is the number of brands of category $c$. The standard deviation is computed as $\textrm{SD Price}_{c} = \sqrt{\frac{1}{N_{c}}\frac{1}{J_{c}}\sum_{t=1}^{N_{c}}\sum_{j=1}^{J}(\textrm{Price}_{jt} - \textrm{Mean Price}_{c})^2}$. The mean cost and SD cost are defined similarly.\footnote{All price and cost figures are inflation-adjusted. Brazil exhibited significant inflation during our study period, and adjusting for inflation is crucial to reflect the real price changes. The adjustment is relative to the first observation period, January 2019.} Overall, we see that the data contain significant price and cost variation for each category. This variation is important and helps us estimate price elasticities. 

\section{Descriptive Analysis}
\label{sec:descriptive.analysis}

We now present some descriptive analysis that examines if/how the adoption of online shopping with the retailer changes consumer behavior during their offline shopping trips at the retailer's physical stores in the post-adoption period. Specifically, we focus on two variables that can capture changes in \textit{within-customer} offline price sensitivity:
\squishlist
    \item Price Paid: Conditional on making an offline purchase in the focal category, do adopters shift toward lower-priced after online adoption (compared to non-adopters)? 
    \item Category Purchase Frequency: Conditional on an offline purchase occasion (as defined earlier), do adopters purchase the focal category less frequently after they adopt online shopping, compared to non-adopters?
\squishend

\subsection{Price Paid}
\label{ssec:price.paid}

\begin{table}[htp!]
   \caption{Two-Way Fixed Effects Model Estimates -- ln(Price\_Paid)}
   \label{tab:twfe_price_paid}
   \centering
   \footnotesize{
 \begin{tabular}{lcccc}
      \tabularnewline \midrule \midrule
      Dependent Variable: & \multicolumn{4}{c}{ln(Price\_Paid)}\\
                         & Dry Dog Food  & Dry Cat Food   & Dog Hygiene & Cat Hygiene \\   
      Model:             & (1)           & (2)            & (3)         & (4)\\  
      \midrule
      \emph{Variables}\\
      $\mathcal{I}\{\textrm{Adopter}_{i}\} \times \mathcal{I}\{ t \geq T_{i}^{\textrm{adopt}}\}$ & -0.0102$^{*}$ & -0.0106$^{**}$ & -0.0046     & -0.0160$^{*}$\\   
                         & (0.0040)      & (0.0038)       & (0.0030)    & (0.0063)\\   
      \midrule
      \emph{Fixed-effects}\\
      Customer           & Yes           & Yes            & Yes         & Yes\\  
      YearMonth          & Yes           & Yes            & Yes         & Yes\\  
      \midrule
      \emph{Fit statistics}\\
      Observations       & 99,233        & 27,623         & 40,183      & 42,596\\  
      R$^2$              & 0.89566       & 0.70460        & 0.62699     & 0.66927\\  
      Within R$^2$       & 0.00030       & 0.00065        & 0.00012     & 0.00074\\  
      \midrule \midrule
      \multicolumn{5}{l}{\emph{Clustered (Customer \& YearMonth) standard-errors in parentheses}}\\
      \multicolumn{5}{l}{\emph{Signif. Codes: ***: 0.001, **: 0.01, *: 0.05}}\\
   \end{tabular}
   }
\end{table}

We examine how the prices paid by adopters change after online adoption (compared to non-adopters). To that end, we focus on all offline purchase orders (made by adopters and non-adopters) that contain a focal category purchase and estimate the following two-way fixed-effects regression for each category at the customer-order level: 
\begin{equation}
    \textrm{ln(Price\_Paid)}_{it}  = \delta \times \mathcal{I}\{\textrm{Adopter}_{i}\} \times \mathcal{I}\{ t \geq T_{i}^{\textrm{adopt}}\} +
     \textrm{Customer}_{i} + \textrm{YearMonth}_{t} + \epsilon_{it},
\end{equation}
where  $\mathcal{I}\{\textrm{Adopter}_{i}\}$ is an indicator variable which is 1 for adopters and 0 for non-adopters. $T_{i}^{\textrm{adopt}}$ is the adoption time of customer $i$, and $\mathcal{I}\{ t \geq T_{i}^{\textrm{adopt}}\}$ is an indicator which is 1 if the order $t$ is during customer $i$'s post-adoption period. We also include customer-fixed effects, $\textrm{Customer}_{i}$, to measure the within-customer change, and time-fixed effects, $\textrm{YearMonth}_{t}$, to control for time-specific unobserved shocks. A negative $\delta$ indicates that adopters, on average, pay lower prices or shift towards the lower end of the price distribution after adopting online shopping, relative to non-adopters. A positive $\delta$ would indicate the opposite. 

%Standard errors are two-way clustered at the customer and date levels.

Table \ref{tab:twfe_price_paid} presents the within-customer changes in the prices paid by adopters relative to non-adopters, controlling for customer and date fixed-effects. Across all product categories, we see that adopters pay lower prices offline after they adopt shopping online in the retailer, compared to non-adopters. Further, three of the categories show a statistically significant effect at the 5\% level of significance. In Web Appendix \ref{appsec:descriptive} Table \ref{tab:twfe_percentile_price}, we repeat this analysis with the percentile of price paid as the dependent variable (instead of ln(Price\_Paid)) and find consistent results --  customers shift towards lower-priced products during their offline purchases after adopting online shopping. Together, these results suggest that, conditional on purchasing, consumers are paying lower prices, possibly because they are more price-sensitive. 

%\footnote{Dog hygiene is significant at the 10\% level of significance.}

\subsection{Category Purchase Frequency} 
\label{ssec:purchase.frequency}

\begin{table}[htp!]
   \caption{Two-Way Fixed Effects Model Estimates - BuyCategory}
   \label{tab:twfe_buyCategory}
   \centering
   \footnotesize{
   \begin{tabular}{lcccc}
      \tabularnewline \midrule \midrule
      Dependent Variable: & \multicolumn{4}{c}{BuyCategory}\\
                         & Dry Dog Food        & Dry Cat Food        & Dog Hygiene     & Cat Hygiene \\   
      Model:             & (1)             & (2)             & (3)             & (4)\\  
      \midrule
      \emph{Variables}\\
      $\mathcal{I}\{\textrm{Adopter}_{i}\} \times \mathcal{I}\{ t \geq T_{i}^{\textrm{adopt}}\}$ & -0.0751$^{***}$ & -0.0497$^{***}$ & -0.0617$^{***}$ & -0.0654$^{***}$\\   
                         & (0.0077)        & (0.0109)        & (0.0075)        & (0.0106)\\   
      \midrule
      \emph{Fixed-effects}\\
      Customer       & Yes             & Yes             & Yes             & Yes\\  
      YearMonth         & Yes             & Yes             & Yes             & Yes\\  
      \midrule
      \emph{Fit statistics}\\
      Observations       & 136,411         & 45,443          & 78,253          & 55,410\\  
      R$^2$              & 0.30479         & 0.25032         & 0.23015         & 0.26662\\  
      Within R$^2$       & 0.00147         & 0.00066         & 0.00113         & 0.00121\\  
      \midrule \midrule
      \multicolumn{5}{l}{\emph{Clustered (Customer) standard-errors in parentheses}}\\
      \multicolumn{5}{l}{\emph{Signif. Codes: ***: 0.001, **: 0.01, *: 0.05}}\\
   \end{tabular}
   }
\end{table}

Next, we examine how the frequency of adopters' purchase of a given category in the offline channel changes compares to non-adopters. Specifically, during each offline purchase occasion, we construct a binary dependent variable, $\mathcal{I}\{\textrm{BuyCategory}_{it}\}$, which equals 1 if customer $i$ buys the focal category in purchase occasion $t$, and 0 otherwise. We then estimate the following two-way fixed effects regression at the customer-month purchase occasion level: 
\begin{equation}
\begin{aligned}
    \mathcal{I}\{\textrm{BuyCategory}_{it}\} & = \delta \times \mathcal{I}\{\textrm{Adopter}_{i}\} \times \mathcal{I}\{ t \geq T_{i}^{\textrm{adopt}}\} +
     \textrm{Customer}_{i} + \textrm{YearMonth}_{t} + \epsilon_{it} 
\end{aligned}
\end{equation}
As before, we include customer and year-month fixed effects. The coefficient $\delta$ captures the change in the probability of buying the category for adopters relative to non-adopters after adoption.

Table \ref{tab:twfe_buyCategory} shows the results on this estimation. We find that, conditional on a purchase occasion, adopters purchase dog food and cat food 7.5\% and 5.0\% less frequently than non-adopters in the offline channel after they begin using the retailer's online channel. Similarly, adopters buy dog and cat hygiene 6.2\% and 6.5\% less frequently than non-adopters, respectively. These findings suggest that, compared to non-adopters, adopters are more likely to choose the ``no-purchase" option after they start shopping online.

In summary, both the descriptive analyses suggests that consumers are less likely to buy the focal category offline and when they do buy, they pay lower prices, on average. Together these findings provide some suggestive evidence for potentially higher offline price sensitivity after online adoption. In the next few sections, we build on these findings and estimate customer-level price elasticities as a function of their online adoption status and then quantify the effect of online adoption on offline price elasticities. To interpret these patterns, $\S$\ref{ssec:mechanisms} examines purchases during promotional-price periods, changes in brand composition, and reallocation between the retailer's offline and online channels. These complementary analyses connect the descriptive results to greater price attention and strategic purchase timing, with detailed evidence in Web Appendix $\S$\ref{appssec:purchase_behavior_mechanisms}.

\section{Firm's Problem and Overview of Empirical Strategy}
\label{sec:firms_problem}

The primary objective of the firm is to quantify the causal effect of consumers' adoption of its online channels on their offline price sensitivity at its physical stores, compared to consumers who shop exclusively offline. However, quantifying this impact presents both methodological challenges and limited managerial relevance for several reasons. 

First, adopting the online channel is a consumer-level decision, and firms cannot randomly assign consumers to adopt the online channel or remain offline-only through field experiments. This makes it impossible to directly measure the Average Treatment Effect (ATE) of online adoption on consumers' offline price sensitivity. Second, although firms could implement an intent-to-treat (ITT) design - such as randomly offering incentives to some offline-only consumers to encourage their online adoption - this approach still does not give us ATE given that some consumers would never adopt online \footnote{We do not obtain ATE under the ITT design, since the causal quantity obtained from 2SLS estimators represents the Average Causal Response, which is a weighted sum of individual treatment effects where weights are compliance scores; For more details, see \cite{angrist1995two, syrgkanis2019machine,chen2024new,mummalaneni2026content}.}.Third, even if a firm could somehow identify a significant effect of adopting online shopping on consumers' offline price sensitivity, it cannot force consumers to choose a specific channel. Therefore, obtaining the ATE is not only impractical but also lacks actionable implications. Instead, the main quantity of interest for the firm is the change in offline price sensitivity among consumers who adopt online shopping compared to those who continue to shop exclusively offline i.e. the Average Treatment Effect on the Treated (ATT). Measuring this would help firms adjust their offline pricing strategies in response to their customers adopting online shopping. Thus, our goal is to identify the ATT of online adoption on offline price sensitivity.

We adopt a two-step empirical strategy to solve this problem:
\squishlist
\item First, we measure category-level own-price elasticities at the customer-month level: In $\S$\ref {sec:model}, we use a logit brand choice model where each consumer's price sensitivity coefficient varies with the extent of their online spending and include customer-brand fixed effects. To address price endogeneity, we use the control function approach \citep{petrin2010control}, using wholesale cost as an instrument. To efficiently estimate the model given a large number of customer-brand fixed effects, we implement a Minorization-Maximization (MM) algorithm for the logit model \citep{chen2022mm}.
\item Next, we use the estimated elasticities in the first step to measure the change in offline price elasticities due to online adoption. In $\S$\ref {sec:att}, we use a staggered difference-in-differences approach \citep{callaway2021difference} to estimate the ATT of online adoption on offline price elasticities after accounting for differences in the timing of adoption between different adopter cohorts.
\squishend

\section{Demand Model}
\label{sec:model}

\subsection{Choice Model and Utility Specification} 
\label{ssec:choice_model}
For each category, we use a logit model of brand choice \citep{guadagni1983logit,chintagunta1991investigating} to estimate offline price elasticities at customer-brand-month-level.\footnote{As explained in $\S$\ref{ssec:construction}, the model captures brand choice and category purchase incidence conditional on observed offline retailer activity. The alternative purchase-occasion definitions in $\S$\ref{sssec:alternative.def.purchase.occasions} include months with activity in either channel and months without retailer purchases; both yield consistent findings (Web Appendices $\S$\ref{appssec:alternative.def.po}--$\S$\ref{appssec:customer.month.full.panel}).} Suppose that for a given category, there are $J$ brands indexed by $j = 1, 2,\ldots, J$ plus an outside option indexed by 0. The utility that consumer $i$ obtains from choosing brand $j$ during offline purchase occasion $t$ is given by \footnote{Here, we use a single index $t$ to denote a purchase occasion, which contains both the time (YearMonth) and store information associated with the purchase occasion.}:
\begin{equation}
    \begin{aligned}
    U_{ijt} & = \alpha_{0} \textrm{Price}_{jt} +
    \alpha_{1} \textrm{cumPercOnlineSpend}_{it} + 
    \alpha_2  \textrm{Price}_{jt}\times \textrm{cumPercOnlineSpend}_{it} \\
& + \alpha_{3} \mathcal{I}\{\textrm{BuyPrevOcca}_{ijt}\} + \delta_{ij} + \epsilon_{ijt} ~~ \forall j \in \{1, \ldots, J\} \\
    \end{aligned}
\label{eq:util_choice_M1}
\end{equation}
In this utility specification, $\textrm{Price}_{jt}$ is the price of brand $j$ during purchase occasion $t$. We allow a consumer's price sensitivity coefficient to depend on their cumulative percentage of online spending, denoted as $\textrm{cumPercOnlineSpend}_{it}$. This variable represents the proportion of the customer's total spending with the retailer that has occurred online up to the purchase occasion $t$ since their first transaction with the retailer. We use this continuous variable to allow for variation in the online adoption effect based on the \textit{extent} of a consumer's online spend. We also consider an alternate specification where we use a binary online adoption indicator in \S\ref{sssec:alt_demand}. The coefficient $\alpha_0$ represents the baseline price sensitivity for purely offline customers. The interaction coefficient $\alpha_2$ reflects how price sensitivity changes as consumers spend more online. A negative $\alpha_2$ would suggest that higher online spending share leads to greater offline price sensitivity.\footnote{Interaction terms in logit and other nonlinear models are best interpreted via marginal effects or elasticities \citep{ai2003interaction}.} The variable $\mathcal{I}\{\textrm{BuyPrevOcca}_{ijt}\}$ is an indicator that equals 1 if customer $i$ purchased brand $j$ at the previous purchase occasion $t-1$, and 0 otherwise, capturing state dependence via $\alpha_3$. The customer-brand fixed effects $\delta_{ij}$ capture consumer $i$'s time-invariant preference for brand $j$. These fixed effects capture systematic time-invariant consumer-level heterogeneity in brand preferences. However, including these high-dimensional fixed effects introduces some estimation challenges which we discuss in \S\ref{ssec:estimation}. The term $\epsilon_{ijt}$ is an idiosyncratic error term. Finally, the utility of choosing the outside option is normalized as $U_{i0t} = 0 + \epsilon_{i0t}$ where $\epsilon_{i0t}$ also follows an i.i.d. Type I extreme value distribution.

\subsection{Price Endogeneity}
\label{ssec:endogeneity}
A common challenge in estimating price elasticities with observational choice data is price endogeneity. This issue arises when the error term, $\epsilon_{ijt}$, contains some time-varying unobserved demand shocks that may be correlated with prices; e.g., if promotions or advertising are correlated with prices. To address this, we apply the control function approach for two endogenous variables, $\textrm{Price}_{jt}$ and $\textrm{Price}_{jt}\times \textrm{cumPercOnlineSpend}_{it}$ \citep{petrin2010control, wooldridge2015control}. We use the weighted wholesale cost indices for each brand $j$ (constructed as described in $\S$\ref{ssec:construction}), $\textrm{Cost}_{jt}$, as the instrument. 

In this setting, temporary promotions and store-level pricing decisions may respond to local demand, inventory, or competitive conditions observed by the retailer but unobserved to us. These conditions can also affect brand choice, so customer--brand fixed effects need not eliminate price endogeneity. Our cost data are actual wholesale procurement costs supplied directly by the retailer from its transaction records, rather than costs inferred from retail prices or assumed markups. According to the retailer, procurement costs are determined through manufacturer contracts and adjust less frequently than retail prices. This separation between procurement and local promotional decisions motivates our identifying assumption: conditional on the model controls, wholesale costs affect brand choice through retail prices and are unrelated to the remaining unobserved demand shocks \citep{berry1993automobile}. Web Appendix $\S$\ref{appsssec:price} describes the construction of the cost indices. We summarize additional first-stage diagnostics, price-and-cost plots, price-variation decompositions, and checks without control functions or with a single price control function in $\S$\ref{sssec:control_function_price_endogeneity}, with detailed results in Web Appendices $\S$\ref{appssec:alternative.imple.no.control.function}--$\S$\ref{appssec:price_variation_control_function}.

To obtain control functions for two endogenous variables, we run the following two regressions  and obtain the residuals; see \citep{aparicio2024algorithmic} for a similar use case of this approach.\footnote{Alternatively, some earlier work has suggested that a more parsimonious way to deal with the interaction term between an endogenous regressor and an exogenous variable is to estimate first-stage regression for price in Equation \eqref{eq:Price_FirstStage} and add the fitted residuals as an additional regressor in the model equation, without the need to add another first stage regression deal with the interaction term between price and online share \citep{ebbes2016dealing}. We also consider this alternative approach and find that the results remain consistent with our main findings. Please see Web Appendix $\S$\ref{appssec:alternative.imple.control.function} for details.}
\begin{equation}
\label{eq:Price_FirstStage}
    \textrm{Price}_{jt} = \gamma_{1}\textrm{Cost}_{jt} + \textrm{Brand}_{j} + \textrm{Store}_{t} + \textrm{YearMonth}_{t} + \xi_{jt}^{(1)}
\end{equation}
\begin{equation}
\label{eq:PriceInteraction_FirstStage}
    \textrm{Price}_{jt} \times \textrm{cumPercOnlineSpend}_{it}=\gamma_{2} \textrm{Cost}_{jt} + \gamma_{3} \textrm{Cost}_{jt}\times \textrm{cumPercOnlineSpend}_{it} + \boldsymbol{\gamma}_{4}'\textrm{Control}_{ijt} + \xi_{ijt}^{(2)}
\end{equation}
Here, $\textrm{Control}_{ijt}$ is a vector of control variables (e.g., $\textrm{cumPercOnlineSpend}_{it}, \mathcal{I}\{\textrm{BuyPrevOcca}_{ijt}\}, \textrm{Customer}_{i},$
$\textrm{Brand}_{j}, \textrm{YearMonth}_{t}$). Next, following \cite{petrin2010control}, the original error term in the utility specification, $\epsilon_{ijt}$, can be decomposed as follows:
\begin{equation}
    \epsilon_{ijt} = \lambda_1 \textrm{CF\_Price}_{jt} + \lambda_2 \textrm{CF\_Price\_cumPercOnlineSpend}_{ijt} + \tilde{\epsilon}_{ijt}
\end{equation}
where $\tilde{\epsilon}_{ijt}$ is assumed to follow i.i.d. Type I extreme value distribution. We then include two control functions into the utility equation as additional regressors, which is now given by: 
\begin{equation}
    \begin{aligned}
    U_{ijt} & = \alpha_{0} \textrm{Price}_{jt} +
    \alpha_{1} \textrm{cumPercOnlineSpend}_{it} + 
    \alpha_2  \textrm{Price}_{jt}\times \textrm{cumPercOnlineSpend}_{it} \\
    & + \alpha_{3} \mathcal{I}\{\textrm{BuyPrevOcca}_{ijt}\} + \delta_{ij} +\lambda_1 \textrm{CF\_Price}_{jt} + \lambda_2 \textrm{CF\_Price\_cumPercOnlineSpend}_{ijt} + \tilde{\epsilon}_{ijt} \\
    & = \psi_{ijt} + \tilde{\epsilon}_{ijt}   ~~ \forall j \in \{1, \ldots, J\}, \\
    \end{aligned}
\label{eq:util_choice_M2}
\end{equation}
where $\psi_{ijt}$ is the deterministic part of the utility term and $\psi_{i0t} = 0$ for the outside option. Under the Type I extreme value assumption for $\tilde{\epsilon}_{ijt}$, the probability that consumer $i$ chooses brand $j$ at shopping occasion $t$ is given by the following formula:
\begin{equation}
    Pr_{ijt} = \frac{\exp(\psi_{ijt})}{\sum_{j'=0}^{J} \exp(\psi_{ij't})} 
\end{equation}
Let $\boldsymbol{\psi}_{it} = (\psi_{i0t}, \psi_{i1t}, \psi_{i2t}, \ldots, \psi_{iJt})\in \mathbb{R}^{J+1}$ be the vector of nominal utilities and $\boldsymbol{y}_{it} = (y_{i0t}, y_{i1t},\ldots, y_{iJt}) \in \mathbb{R}^{J+1}$ be the binary choice outcomes for consumer $i$ at shopping occasion $t$, with $y_{ijt} = 1$ if consumer $i$ chooses brand j at shopping occasion $t$ and 0 otherwise. The likelihood of consumer $i$ at purchase occasion $t$ is given by: 
\begin{equation}
    l(\boldsymbol{\psi}_{it}; \boldsymbol{y}_{it}) = \prod_{j=0}^{J} (Pr_{ijt}(\boldsymbol{\psi}_{it}))^{y_{ijt}}
\end{equation}
The log-likelihood function over all consumers and purchase occasions can be written as
\begin{equation}
\label{eq:log_likelihood}
    \mathcal{L}(\boldsymbol{\theta}) = \sum_{i=1}^{N} \sum_{t\in T_{i}} \log l(\boldsymbol{\psi}_{it}; \boldsymbol{y}_{it})
\end{equation}
where $T_{i}$ is the set of purchase occasions associated with customer $i$. The parameter vector of interest is $\boldsymbol{\theta} = \{ \alpha_0, \alpha_1, \alpha_2, \alpha_3, \lambda_1, \lambda_2, \{\delta_{ij}\}_{i=1,\ldots, N; j=1,\ldots,J} \}$. Note that this is a high-dimensional parameter vector -- in addition to population parameters such as $\alpha$s and $\lambda$s, we also have customer-brand fixed-effects ($\delta_{ij}$s), which scale with the sample size. 

\subsection{Estimation} 
\label{ssec:estimation}

Our main estimation strategy
follows the standard two-step approach used in control function approaches; see  \citet{petrin2010control} for details. However, the key additional challenge in our setting arises from the presence of high-dimensional customer-brand fixed effects. While high-dimensional fixed effects generally allow for more flexible substitution patterns \citep{jiang2025high}, doing so complicates the estimation procedure. We now discuss both the overview of our estimation strategy and also provide a summary of the solution concept that we adopt to address the high-dimensional fixed-effects. 

First, we run OLS regressions for $\textrm{Price}_{jt}$ in Equation \eqref{eq:Price_FirstStage} and $\textrm{Price}_{jt}\times \textrm{cumPercOnlineSpend}_{it}$ in Equation \eqref{eq:PriceInteraction_FirstStage}, respectively. We then obtain the residuals, $\textrm{CF\_Price}_{jt}$ and $\textrm{CF\_Price\_cumPercOnlineSpend}_{ijt}$, which serve as control functions. Second, we include these residuals as additional regressors in the logit model (i.e., Equation \eqref{eq:util_choice_M2}). In a standard setting, estimation can proceed as usual at this stage. However, since we have a large number of customer-brand fixed effects $(\delta_{ij})$, standard maximum likelihood estimation becomes computationally infeasible. To overcome this, we adopt a Minorization-Maximization (MM) algorithm for the logit model with fixed effects, an efficient estimation method proposed by \citet{chen2022mm}. The MM algorithm transforms the non-linear optimization problem (i.e., MLE estimation) into a sequence of simpler problems by iteratively constructing and maximizing a surrogate transfer minorization function $S(\cdot)$ that minorizes the log-likelihood \citep{james2017mm}. The transfer minorization function $S(\cdot)$ is less than the log-likelihood function everywhere except at the current best guess of the parameters, where it is equal. \cite{chen2022mm} derives a transfer minorization function for the standard multinomial logit model in which there are no alternative-specific time-varying covariates in the utility function. In the theorem below, we extend their proof to cover a fully flexible logit model with covariates that may vary by individual, by alternative (brand), and over time.\footnote{While our paper was the first to present this extension and proof for this specific case, in a recently posted update, \citet{chen2025mm} present a similar extension.}  We then derive the corresponding transfer minorization function $S(\cdot)$ for the log-likelihood $\mathcal{L}$. 

\noindent \textbf{Theorem}. Let $\mathcal{L}(\boldsymbol{\theta})$ be the log-likelihood of the logit model, defined as in Equation \eqref{eq:log_likelihood}. Let $S(\boldsymbol{\theta}; \boldsymbol{\theta}^{(k)})$ be defined as: 
\begin{equation}
\label{eq:LL_SO_Taylor}
    S(\boldsymbol{\theta}; \boldsymbol{\theta}^{(k)}) = \mathcal{L}(\boldsymbol{\theta}^{(k)}) + \frac{1}{2}\sum_{i,j,t} h_{j}(\boldsymbol{\psi}_{it}^{(k)}; \boldsymbol{y}_{it})^2 - \frac{1}{2}\sum_{i,j,t} [\psi_{ijt}^{(k)} + h_{j}(\boldsymbol{\psi}_{it}^{(k)}; \boldsymbol{y}_{it}) - \psi_{ijt}]^2,
\end{equation}
where 
\begin{equation}
    h_{j}(\boldsymbol{\psi}_{it}^{(k)}; \boldsymbol{y}_{it}) = \frac{\partial \log l(\boldsymbol{\psi}_{it}^{(k)}; \boldsymbol{y}_{it})}{\partial \psi_{ijt}^{(k)}} = y_{ijt} - Pr_{ijt}(\boldsymbol{\psi}_{it}^{(k)})
\end{equation}
\begin{equation}
   \boldsymbol{\psi}_{it}^{(k)} = (\psi_{i0t}^{(k)}, \psi_{i1t}^{(k)}, \psi_{i2t}^{(k)}, \ldots, \psi_{iJt}^{(k)}) 
\end{equation}
Here, $h_{j}(\boldsymbol{\psi}_{it}^{(k)};\boldsymbol{y}_{it})$ is the $j_{th}$ entry of the gradient vector of log-likelihood with respect to the nominal utility of alternative $j$, $\psi_{ijt}^{(k)}$. $\boldsymbol{\psi}_{it}^{(k)}$ denotes the vector of nominal utilities for consumer $i$ at iteration $k$. 
Then, the function $S(\boldsymbol{\theta}; \boldsymbol{\theta}^{(k)})$ is a transfer minorization of $\mathcal{L}(\boldsymbol{\theta})$.

\noindent \textit{Proof.} See Web Appendix \ref{appssec:mm.algorithm}. 

Based on the theorem above and proof from \cite{chen2022mm}, the iterative sequence $\boldsymbol{\theta}^{(k+1)} = \arg\max_{\theta}S(\boldsymbol{\theta}; \boldsymbol{\theta}^{(k)})$, $k=0,1,\ldots,$ converges to a local, if not global, maximum of $\mathcal{L}(\boldsymbol{\theta})$.\footnote{For more details, please refer to the proof of Theorem 1 in \cite{chen2022mm}, which shows the convergence of this iterative estimator to the true parameters.} 

Equation \eqref{eq:LL_SO_Taylor} is a second-order Taylor expansion of the log-likelihood function at $\boldsymbol{\theta}^{(k)}$. The model parameters we are searching over in the iterative MM algorithm $(\boldsymbol{\theta}$, rather than $\boldsymbol{\theta}^{(k)})$, appear only in the third term. Because the only term in $S(\boldsymbol{\theta}; \boldsymbol{\theta}^{(k)})$ that depends on $\boldsymbol{\theta}$ is the sum of squared residuals, maximizing $S(\boldsymbol{\theta}; \boldsymbol{\theta}^{(k)})$ is equivalent to minimizing this least squares objective, and the coefficients can be obtained by regressing $\psi_{ijt}^{(k)} + h_{j}(\boldsymbol{\psi}_{it}^{(k)};\boldsymbol{y}_{it})$ on the vector of explanatory variables $\boldsymbol{x}_{ijt} = \{\textrm{Price}_{jt}, \textrm{cumPercOnlineSpend}_{it}, \textrm{Price}_{jt}\times \textrm{cumPercOnlineSpend}_{it},\mathcal{I}\{\textrm{BuyPrevOcca}_{ijt}\},  \textrm{CF\_Price}_{jt},
\textrm{CF\_Price\_cumPercOnlineSpend}_{ijt},\delta_{ij}\}$ in the utility function. This linearization enables us to apply standard panel data methods despite the large number of fixed effects $\delta_{ij}$. Moreover, another advantage of using the MM algorithm to estimate the logit model is that for customer-brand pairs with few or no purchases, maximum likelihood estimation formally drives $\delta_{ij}$ toward $-\infty$, so that the predicted choice probability is exactly zero. In practice, this boundary solution is numerically unstable. The MM algorithm we employ circumvents this by iteratively solving a series of linearized regressions: never-chosen customer-brand pairs converge to large negative but finite estimates, which approximate the boundary solution while preserving numerical stability. Thus, the algorithm can include all $(i,j)$ pairs without manual dropping, effectively regularizing the fixed effects associated with rare or absent choices. 

In summary, the MM-based estimation algorithm that we employ is as follows: 
\squishlist
\item Initialization: Set the initial parameter estimates $\boldsymbol{\theta}^{(0)}$ including all customer-brand fixed effects as zero.

\item Iterative Steps:  
    \begin{enumerate}
    \item Minorization step: Given $\boldsymbol{\theta}^{(k)}$, calculate $h_{j}(\boldsymbol{\psi}_{it}^{(k)};\boldsymbol{y}_{it})$ and obtain $\psi_{ijt}^{(k)} + h_{j}(\boldsymbol{\psi}_{it}^{(k)};\boldsymbol{y}_{it})$.

    \item Maximization step: Regress $\psi_{ijt}^{(k)} + h_{j}(\boldsymbol{\psi}_{it}^{(k)};\boldsymbol{y}_{it})$ on the vector of variables $\boldsymbol{x}_{ijt}$ in the utility equation \eqref{eq:util_choice_M2}. 

    \item Repeat until convergence, that is, $\Vert \boldsymbol{\theta}^{(k+1)} -\boldsymbol{\theta}^{(k)}\Vert_{\infty} < 10^{-3}$.

    \end{enumerate}
\squishend
Since the estimated residuals from the first stage enter the estimation in the second stage, the standard errors in the second step estimation tend to be downward biased, although the parameter estimates themselves are consistent \citep{wooldridge2010econometric}. Therefore, we use block bootstrapping with 500 replications, which treats individual customers as blocks. In each replication, we resample the same number of customers as in the original sample with replacement, re-estimate the demand model, and then obtain the bootstrapped standard errors for the demand parameters and price elasticities \citep{cameron2005microeconometrics}. 

\subsection{Demand Model Parameter Estimates}
\label{ssec:parameter.estimates}

\begin{table}[htp!]
\caption{Coefficient Estimates of Demand Model - Main Results}
\label{append.main.demand.model}
\centering
\footnotesize{
\begin{tabular}{lcccc}
\toprule
 & Dry Dog Food & Dry Cat Food & Dog Hygiene & Cat Hygiene \\
\textit{Variables} &  &  &  &  \\
\midrule
Price & \makecell{-0.3159$^{***}$ \\(0.0235)} & \makecell{-0.1603$^{***}$ \\(0.0188)} & \makecell{-1.2464$^{***}$ \\(0.123)} & \makecell{-0.221$^{***}$ \\(0.0209)} \\
cumPercOnlineSpend & \makecell{-0.3357\\(0.4601)} & \makecell{-1.4046\\(1.0888)} & \makecell{8.7512$^{***}$ \\(1.8675)} & \makecell{3.4655$^{**}$ \\(1.0615)} \\
Price * cumPercOnlineSpend & \makecell{-0.1115$^{**}$ \\(0.0421)} & \makecell{0.018\\(0.091)} & \makecell{-4.1601$^{***}$ \\(0.8197)} & \makecell{-0.284$^{***}$ \\(0.0713)} \\
BuyPrevOcca & \makecell{0.1512$^{***}$ \\(0.0273)} & \makecell{-0.5037$^{***}$ \\(0.0451)} & \makecell{-0.1821$^{***}$ \\(0.0284)} & \makecell{0.3247$^{***}$ \\(0.0295)} \\
CF\_Price & \makecell{0.0515\\(0.0293)} & \makecell{0.0468\\(0.0258)} & \makecell{0.1911\\(0.1815)} & \makecell{0.2805$^{***}$ \\(0.0279)} \\
CF\_Price\_cumPercOnlineSpend & \makecell{0.0227\\(0.0803)} & \makecell{-0.0604\\(0.1232)} & \makecell{3.6335$^{***}$ \\(1.0007)} & \makecell{0.3006$^{**}$ \\(0.0936)} \\
Customer\_Brand\_FE & Yes & Yes & Yes & Yes\\ \midrule
Observations       & 136,411         & 45,443          & 78,253         & 55,410\\ 
Log-likelihood & -84210.21 & -30698.052& -55444.644& -37327.537 \\ 
\bottomrule
\multicolumn{5}{l}{\emph{Bootstrapped standard-errors in parentheses via 500 replications}}\\
\multicolumn{5}{l}{\emph{Signif. Codes: ***: 0.001, **: 0.01, *: 0.05}}\\
\end{tabular}
}
\end{table}

Table \ref{append.main.demand.model} presents the estimation results of the demand model based on Equation \eqref{eq:util_choice_M2}. 
%\footnote{\textcolor{red}{We also present a version of demand estimation results based on Equation \eqref{eq:util_choice_M1} without any control functions for each of the four category. We obtain qualitatively similar coefficient estimates for price and price interaction terms. We notice that control functions effectively correct for price endogeneity for cat hygiene with control functions being significant. See Web Appendix \ref{appssec:alternative.imple.no.control.function}}} 
The standard errors of the parameter estimates are derived using block bootstrap with 500 replications. All the price coefficients are negative and statistically significant. The price interaction terms are negative and significant for three out of the four categories we study: dry dog food and for dog and cat hygiene categories, indicating that the adoption (and extent of adoption) of a retailer's channels makes consumers \emph{more} price sensitive during their subsequent offline shopping trips at that retailer's physical stores. The coefficient is insignificant for dry cat food. We discuss the potential sources for this heterogeneity in $\S$\ref{ssec:elasticity_est}.

The magnitude of the coefficient of the interaction term in the logit model is difficult to interpret directly. Instead, we can interpret the magnitude of the effect we find by computing the marginal effects and price elasticities \citep{ai2003interaction, karaca2012interaction}. To that end, we calculate price elasticities using the statistically significant coefficients, as specified in Equation \eqref{eq:elas_main_model} and discuss their implications in the next section.

\section{ATT of Online Adoption on Offline Price Elasticity}
\label{sec:att}

Now that we have the demand parameter estimates and we see a negative and significant effect on the interaction term of interest, we measure the category-level own-price elasticities and formalize the ATT of the adoption of a retailer's online channels on offline price elasticities. In $\S$\ref{ssec:elasticity_est}, we present the elasticity estimates from the estimation procedure. Next, in $\S$\ref{ssec:within.customer.analysis}, we perform a within-customer comparison of adopters’ offline price elasticity before and after the adoption of online shopping. In $\S$\ref{ssec:did}, we formalize the changes in offline price elasticity due to the adoption of online shopping using a staggered difference-in-differences (DID) model.

\subsection{Elasticity Estimates}
\label{ssec:elasticity_est}

Given the demand model parameters, we can then derive a customer $i$'s own-price elasticity for brand $j$ in a purchase occasion $t$ as shown below:
\begin{equation} \hspace{-0.5in}
\label{eq:elas_main_model}
\begin{aligned}
\quad  \eta_{ijt} =   \frac{\partial Pr_{ijt}}{\partial \textrm{Price}_{jt}}\cdot \frac{\textrm{Price}_{jt}}{ Pr_{ijt}} = (\alpha_0 + \alpha_2 \times \textrm{cumPercOnlineSpend}_{it}) \times \textrm{Price}_{jt} \times (1 - Pr_{ijt})
\end{aligned}
\end{equation}

To estimate category-level own-price elasticity, we proceed in two steps. First, for each customer $i$, brand $j$ in the customer-month purchase occasion $t$, we calculate the own-price elasticity $\eta_{ijt}$ using Equation \eqref{eq:elas_main_model}. We then average across the $J$ brands to obtain a customer-month level own-price elasticity for each category, denoted by $\tilde{\eta}_{it} = \frac{1}{J}\sum_{j'}\eta_{ij't}$. Hereafter, we let $t$ index the calendar time (i.e., year-month) rather than a purchase occasion. $\tilde{\eta}_{it}$ serves as the dependent variable in our subsequent regression analysis.

\begin{figure}[htp!]
    \centering
    \caption{Average price elasticity between adopters (shown in twelve dotted lines) and non-adopters (shown in one dashed line) for each category.}
    \label{fig:5_Main_Elas_Prediction}

    \begin{subfigure}{0.49\textwidth}
        \centering
        \caption{Dry Dog Food}
        \includegraphics[scale=0.26]{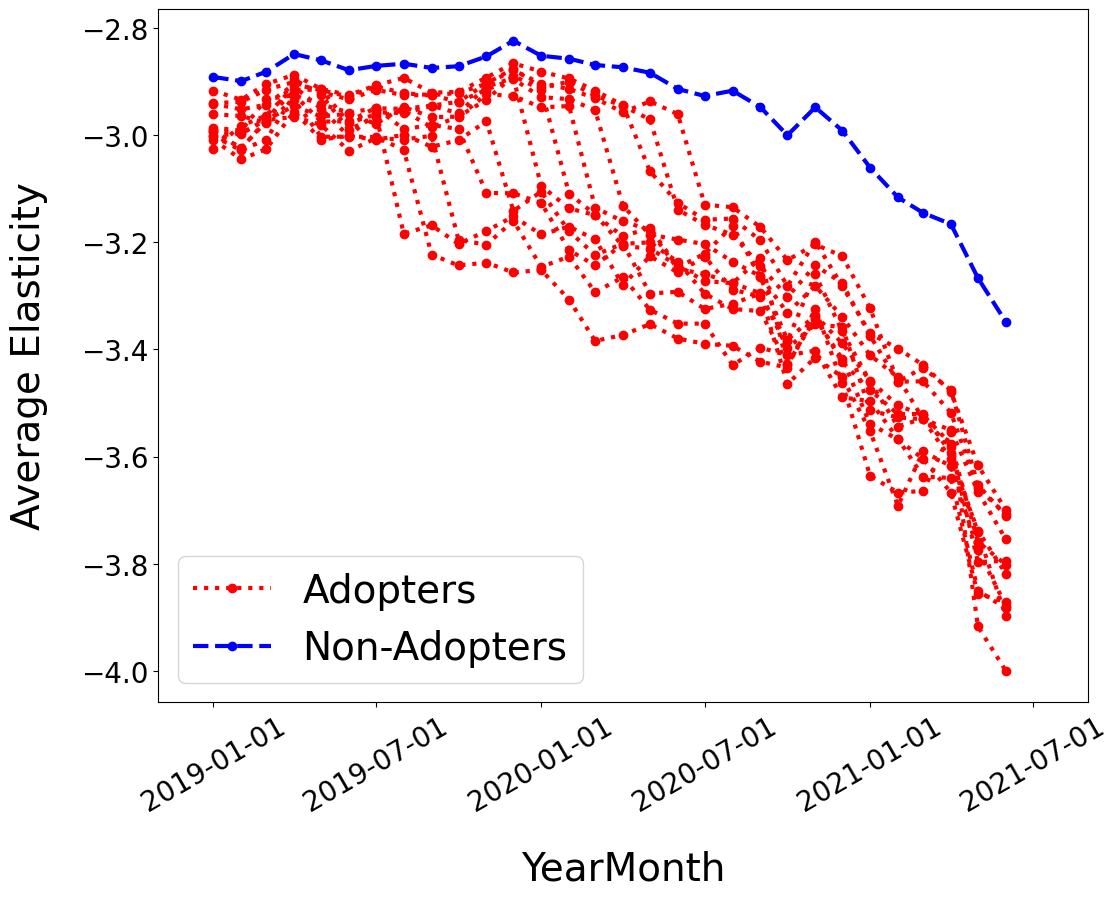}
    \end{subfigure}%
    ~ %add desired spacing between the subfigures, if needed
    \begin{subfigure}{0.49\textwidth}
        \centering
        \caption{Dry Cat Food}
        \includegraphics[scale=0.26]{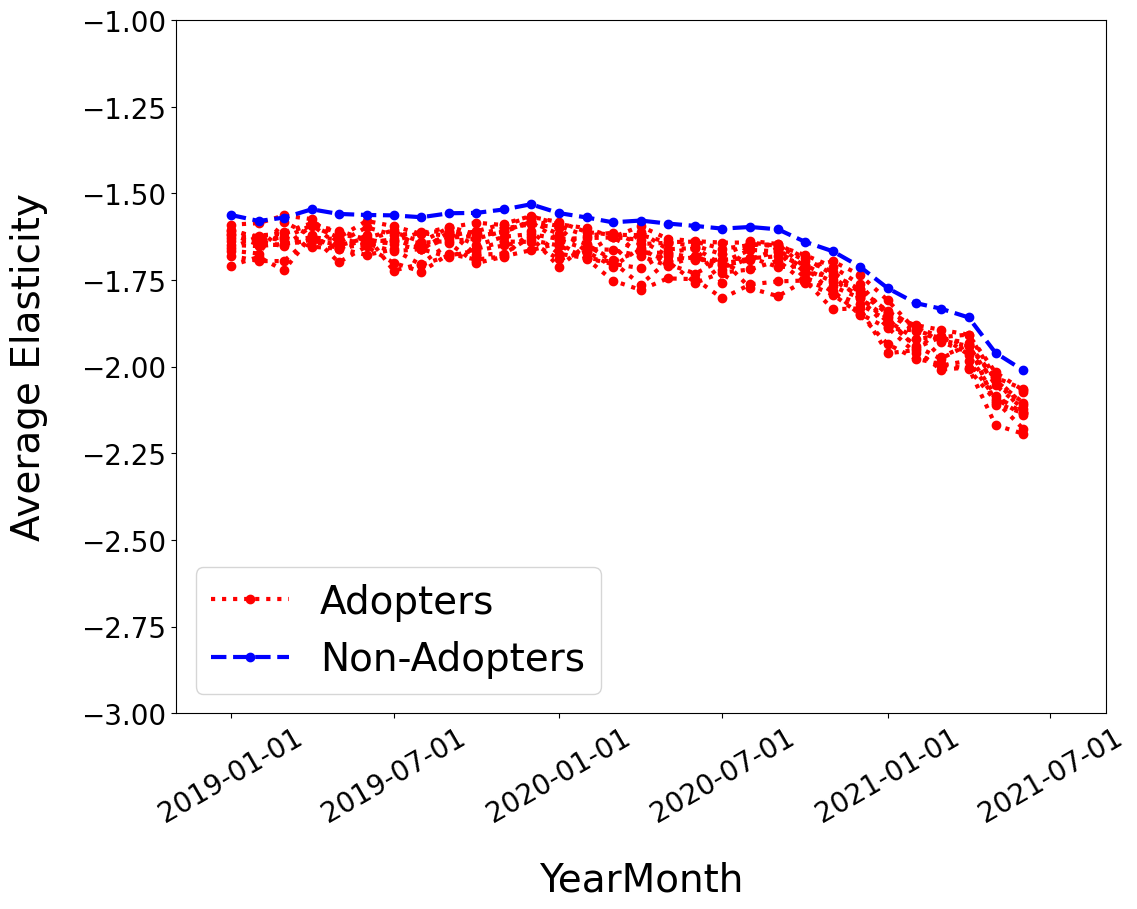}
    \end{subfigure}

    \vspace{1ex} % add some vertical space between the rows

    \begin{subfigure}{0.49\textwidth}
        \centering
        \caption{Dog Hygiene}
        \includegraphics[scale=0.26]{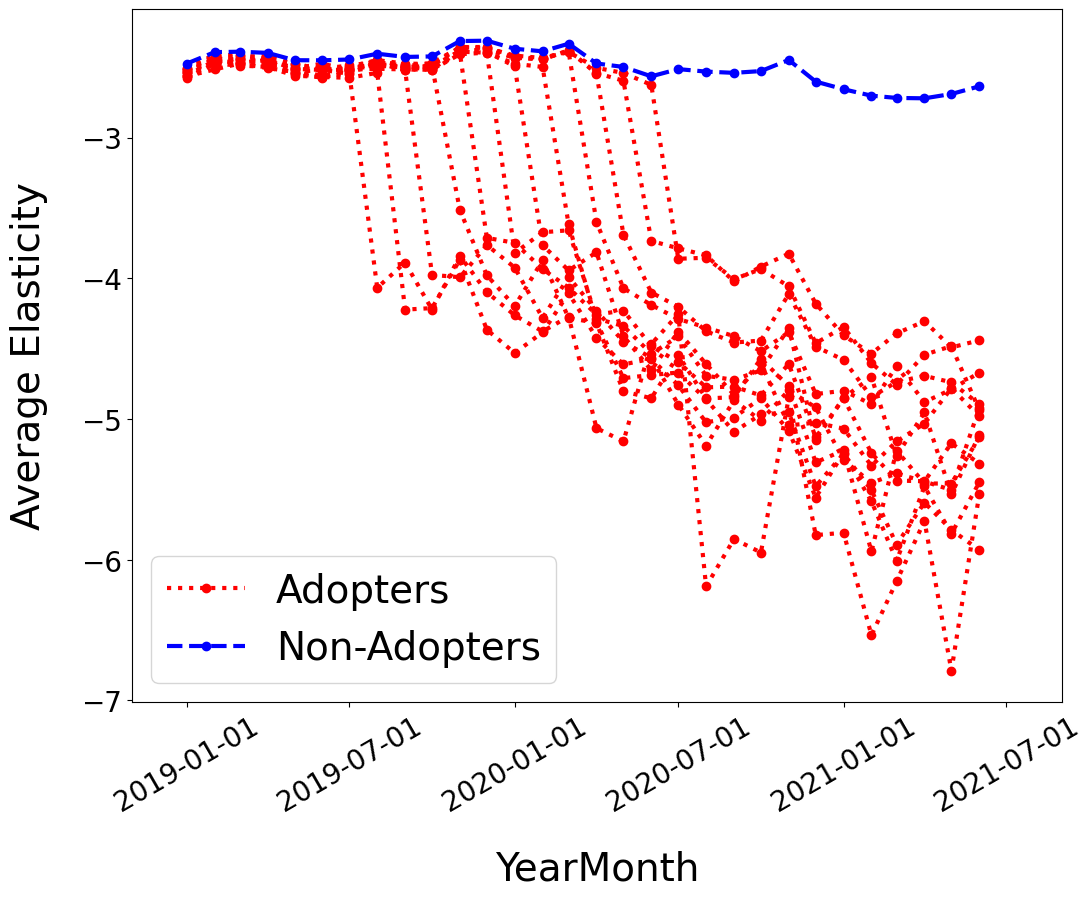}
    \end{subfigure}%
    ~ %add desired spacing between the subfigures, if needed
    \begin{subfigure}{0.49\textwidth}
        \centering
        \caption{Cat Hygiene}
        \includegraphics[scale=0.26]{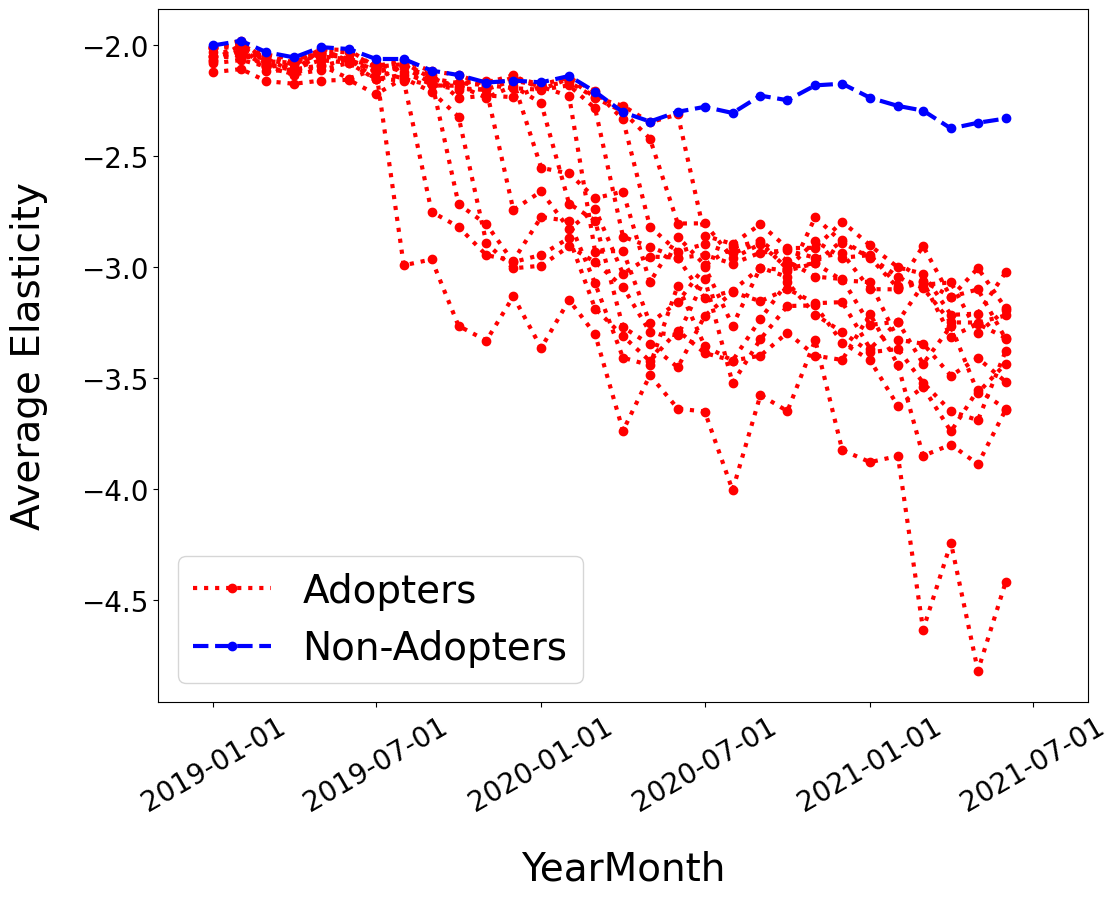}
    \end{subfigure}
\end{figure}

For easier interpretation, we further calculate the monthly average elasticity for each adopter and non-adopter cohort by taking the cross-sectional mean across all customers' own-price elasticities within that cohort. Figure \ref{fig:5_Main_Elas_Prediction} shows these monthly average offline price elasticities for both adopters and non-adopters of these product categories. The dashed line represents the baseline for non-adopters, while the twelve dotted lines represent adopter cohorts, each defined by the month in which customers first purchased online between July 2019 and June 2020.

%Specifically, for each consumer $i$, brand $j$ in the purchase occasion at month $t$, we compute the brand-level own-price elasticity $\eta_{ijt}$, given by Equation \eqref{eq:elas_main_model}. We then average across the $J$ brands to obtain a customer-month level elasticity for each category, denoted by $\tilde{\eta}_{it} = \frac{1}{J}\sum_{j'}\eta_{ij't}$, which is our dependent variable. 

%% main effect figure for all categories
%\input{0_Table_Figure_Summary/6_Main_Result_Elas_Prediction}

Before online adoption, consumers' offline price elasticities closely mirror those of non-adopters, suggesting comparable pre-trends. However, after adopting online shopping, adopters consistently become more price sensitive in their offline shopping. The magnitude of the increase in offline price elasticity is most pronounced for dog hygiene and cat hygiene, moderate for dry dog food, and barely visible for dry cat food. One possible explanation for the larger changes in hygiene categories is greater flexibility in purchase timing. An owner with hygiene mats still at home may wait for a promotion before buying the next pack. When a good deal appears, buying extra packs can help avoid paying regular prices later. Food purchases may be harder to rearrange: if supplies run low, feeding a pet cannot wait until its usual food goes on sale. Although dry food can also be bought ahead, freshness concerns may discourage owners from keeping several extra bags. Spare hygiene mats or litter do not raise the same concerns about freshness. Since the main coefficient of interest in Table \ref{append.main.demand.model} for cat food is not statistically significant at the 5\% level, and the elasticity figure also shows no effect for cat food, we drop this category in our subsequent analyses.

Online shopping may help consumers recognize good deals, but these differences affect how readily they act on that information. A shopper might become more aware of prices in both food and hygiene categories while adjusting hygiene purchases more substantially, consistent with research showing that promotions affect purchase timing and quantities differently across categories \citep{bell1999decomposition}. Our mechanism analyses in $\S$\ref{ssec:mechanisms} show that adopters become more likely to purchase the same SKU offline during promotional-price periods, without a systematic shift toward cheaper brands.  Thus, these differences in effects across categories likely arise due to greater flexibility purchase timing and home inventory management.

%\hy{Is this really true. Check with some cat person} 
%\sbis{This paper doesn't say that explicitly. I did find another reference saying that cats are very picky eaters and have especially strong food preferences in terms of shape, smell and texture (https://www.vetmed.wisc.edu/expert-qa-understanding-pet-food/). I changed the text to say that. Hema - this looks fine, right?}
%Second, the online channel makes price comparison nearly frictionless, especially for hygiene products with standard and simple attributes such as the pad size. Once customers learn how easy it is to find lower prices online, their heightened price awareness likely spills over to their offline purchases, accelerating brand switching in-store. These factors raise the hurdle for brand switching for food categories and may blunt the impact of online adoption. 

In sum, Figure \ref{fig:5_Main_Elas_Prediction} provides clear visual evidence supporting the idea that adoption of online shopping heightens consumers' offline price elasticity. Furthermore, this pattern holds true across all cohorts, as well as for the dry dog food and dog and cat hygiene categories. For cat food, we find a null effect; hence, we drop the category from our subsequent analyses. A drawback of the plots above is that we don't control for common time-specific shocks and/or unobserved customer heterogeneity. In the next few sections, we address these confounds using within-customer and staggered difference-in-differences (DID) analysis.

\subsection{Within-Customer Analysis}
\label{ssec:within.customer.analysis}

We now quantify how adopters' offline price elasticity changes after they start shopping online at the focal retailer. We estimate two model specifications on the sample of adopters for each category:
\begin{equation}
    \begin{aligned}
        \tilde{\eta}_{it} & = \phi_{0} + \phi_{1} \times \mathcal{I}\{t \geq T_{i}^{\textrm{adopt}}\} + e_{it}
    \end{aligned}
    \label{eq:within_customer_base}
\end{equation}
\begin{equation}
    \begin{aligned}
        \tilde{\eta}_{it} & = \textrm{Customer}_{i} + \textrm{YearMonth}_{t} + \phi_{2} \times \mathcal{I}\{t \geq T_{i}^{\textrm{adopt}}\} + e_{it},
    \end{aligned}
    \label{eq:within_customer_FE}
\end{equation}
where the binary indicator $\mathcal{I}\{t \geq T_{i}^{\textrm{adopt}}\}$ indicates the post-online-adoption period, which equals 1 if customer $i$ has adopted online shopping at time $t$, and 0 otherwise. $\textrm{Customer}_{i}$ and $\textrm{YearMonth}_{t}$ are customer- and year-month fixed effects, which control for individual-specific time-invariant unobserved heterogeneity and time-specific unobserved shocks, respectively. Finally, because the outcome variable is an estimate (rather than data), we bootstrap standard errors. Specifically, we employ a block bootstrap with 500 replications, treating individual customers as blocks (as described in \S\ref{sec:model}) to obtain standard errors. 

% The $\textit{Constant}$ term gives the average pre-adoption offline price elasticity.

Table \ref{tab:5_within_analysis} presents the parameter estimates (and bootstrapped standard errors) for these categories. We find that the coefficient on $\mathcal{I}\{t \geq T_{i}^{\textrm{adopt}}\}$ is consistently negative across all specifications and categories. This suggests that, on average, adopters become more sensitive in offline channels in these product categories after adopting online shopping (even after controlling for customer and time fixed effects).

\begin{table}[htp!]
   \caption{Within-Customer Analysis of Offline Price Elasticity - Main Results}
   \centering
   \label{tab:5_within_analysis}
   \footnotesize{
\begin{tabular}{lcccccc}
      \tabularnewline \midrule \midrule
      Dependent Variable: & \multicolumn{6}{c}{Offline Price Elasticity}\\
       & \multicolumn{2}{c}{Dry Dog Food}  & \multicolumn{2}{c}{Dog Hygiene} & \multicolumn{2}{c}{Cat Hygiene} \\ 
      Model:             & (1)             & (2)             & (3)             & (4)             & (5)            & (6)             \\  
      \midrule
      \emph{Variables}\\
      Constant           & -2.935$^{***}$  &                 & -2.464$^{***}$ &                 & -2.125$^{***}$  &   \\   
                         & (0.2142)        &                 & (0.2429)       &                 & (0.1998)        &   \\   
      $\mathcal{I}\{t \geq T_{i}^{\textrm{adopt}}\}$ & -0.3870$^{***}$ & -0.1173$^{*}$  & -1.916$^{***}$ & -0.7239$^{**}$ & -0.8860$^{***}$ & -0.3347$^{**}$\\   
                         & (0.0759)        & (0.0543)              & (0.3563)       & (0.2361)        & (0.1727)        & (0.1250)\\   
      \midrule
      \emph{Fixed-effects}\\
      Customer       &                 & Yes                   &                & Yes             &                 & Yes\\  
      YearMonth         &                 & Yes                   &                & Yes             &                 & Yes\\  
      \midrule
      \emph{Fit statistics}\\
      Observations       & 29,416          & 29,416                & 28,375         & 28,375          & 15,081          & 15,081\\  
      R$^2$              & 0.31325         & 0.86657               & 0.34866        & 0.72137         & 0.37526         & 0.75980\\  
      Within R$^2$       &                 & 0.04011               &                & 0.03352         &                 & 0.03960\\  
      \midrule \midrule
        \multicolumn{5}{l}{\emph{Bootstrapped standard-errors in parentheses via 500 replications}}\\
      \multicolumn{7}{l}{\emph{Signif. Codes: ***: 0.001, **: 0.01, *: 0.05}}\\
   \end{tabular}
   }
\end{table}

Although this pre/post comparison isolates within-customer changes, it remains a before-after contrast confined to adopters. It does not control for common time-varying or macro shocks that could have affected the trend of price elasticities of both adopters and non-adopters. To effectively address this issue, we need to construct a counterfactual or control group of non-adopters, who were also offline customers of the firm in the pre-adoption period and continue to be offline only. To address this issue, we outline a difference-in-differences strategy in the next section.

\subsection{Difference-In-Differences (DID) Analysis}
\label{ssec:did}

To estimate the causal effect of interest -- the average treatment effect on the treated (ATT), we use a difference-in-differences (DID) approach. This method compares adopters with non-adopters before and after online adoption. As a baseline specification, we estimate the following two-way fixed effects (TWFE) model: 
\begin{equation}
    \begin{aligned}
        \tilde{\eta}_{it} = \textrm{Customer}_{i} + \textrm{YearMonth}_{t} + \phi_{TWFE}  \times \mathcal{I}\{\textrm{Adopter}_{i}\} \times \mathcal{I}\{ t \geq T_{i}^{\textrm{adopt}}\} + e_{it}.
    \end{aligned}
    \label{eq:DID_TWFE}
\end{equation}
Here, $\phi_{TWFE}$ is our parameter of interest. $\mathcal{I}\{\textrm{Adopter}_{i}\}$ is a binary indicator that is 1 if customer $i$ is an adopter and 0 otherwise. $\mathcal{I}\{ t \geq T_{i}^{\textrm{adopt}}\}$ is a binary indicator equal to 1 if the year-month $t$ is in the post-online adoption phase of customer $i$. We present the estimation results of the two-way fixed effects estimator in Table \ref{tab:5_Main_Result_TWFE} in Web Appendix \S\ref{appsec:did}. The estimates show that there is a significant increase in price elasticity for adopters post-adoption, as indicated by the negative and significant coefficient on $\phi_{TWFE}$ for these categories.

Although TWFE regression is widely used in empirical studies involving staggered adoption, it only provides consistent estimates of ATT under the assumption of homogeneous treatment effects, a condition that is rarely met in practice; see \citet{de2020two, callaway2021difference, sun2021estimating, goodman2021difference} for detailed discussions of this issue. Specifically, \cite{goodman2021difference} shows that the TWFE estimator $(\phi_{TWFE})$ effectively represents a weighted average of all possible 2-by-2 combinations of pairwise difference-in-differences comparisons between groups of units treated at different times. 

To address this issue, we adopt the staggered difference-in-differences (DID) framework proposed by \cite{callaway2021difference}, hereafter referred to as the CS estimator as our main specification. The CS estimator addresses the biases associated with TWFE estimators by eliminating problematic comparisons between late-treated and already-treated units and has been applied recently in the marketing and economics literature \citep{ang2021effects,braghieri2022social, bekkerman2023effect}. This method is particularly suited to our context, because we have customers adopting online shopping over a one year period, in 12 cohorts (one cohort per month). As discussed earlier, because there were exogenous events during this time period (e.g., the onset of COVID-19), it is possible that the ATT is different across cohorts. Therefore, we use this estimator in our main analysis and then supplement the analysis with TWFE estimates as a robustness check (which provide similar results in terms of direction and magnitude).  
We refer the reader to the original paper for a discussion of the technical details of how the CS estimator works. The main idea is to identify the Average Treatment Effect on the Treated (ATT) for each post-treatment period, for consumers who are treated at different times, and then calculate the overall treatment effect as a weighted average of these ATTs. In our context, the ATT in month $t$ for customers who adopt online shopping in month $g$ can be defined as: 
\begin{equation}
    ATT(g,t) = \mathbb{E}[\tilde{\eta}_{it}(g) - \tilde{\eta}_{it}(0) | G_{ig} = 1],
\label{eq:ATT_CS}
\end{equation}
where $\tilde{\eta}_{it}(g)$ is the observed elasticity of customer $i$ at month $t$ who adopts online shopping in month $g$. $G_{ig}$ is a binary indicator which is set to one if the customer $i$ adopts online shopping in month $g$. Since the adoption window in our study spans from July 2019 to June 2020, $g$ ranges from 7 to 18. $\tilde{\eta}_{it}(0)$ represents the potential outcome (i.e. counterfactual elasticity) of customer $i$ at month $t$, if they did not adopt online shopping. Under the ``no anticipation of treatment'' assumption from CS, $\tilde{\eta}_{it}(0) =  \tilde{\eta}_{it}(g)$ for all $t < g$. In other words, the offline price elasticity of any customer $i$ during the pre-treatment period does not depend on whether and when customer $i$ adopts online shopping, and this represents the counterfactual elasticity if they had not adopted online shopping.

As discussed by \cite{callaway2021difference}, this effect, $ATT(g,t)$, can be identified provided that parallel pre-trends hold. We provide a test of parallel pre-trends in Figure \ref{fig:7_Lead_Lag_Model_CS} in Web Appendix \S\ref{appsec:parallel_pretrend}. 

The overall treatment effect ($\theta_o$) is then calculated as a weighted average of $ATT(g,t)$ for all periods post-adoption across all groups: 
\begin{equation}
    \theta_o = \frac{1}{C_{o}} \sum_g \sum_{t > g} w_g  ATT(g,t)
\end{equation}
where $w_g$ is the weight assigned to units in group $g$, which is the proportion of the number of online adopters in treatment group $g$ relative to the total number of adopters. $C_o$ normalizes the weights to sum to one.

\begin{table}[htp!]
   \caption{ATT of Online Adoption on Offline Price Elasticity Over All Post-Adoption Period (CS Estimator)}
   \centering
   \label{tab:5_CS_Result}
   \footnotesize{
   \begin{tabular}{lccc}
      \tabularnewline \midrule \midrule
      Dependent Variable: & \multicolumn{3}{c}{Offline Price Elasticity}\\
       & \multicolumn{1}{c}{\makecell{Dry Dog Food}}  
       & \multicolumn{1}{c}{\makecell{Dog Hygiene}}
       & \multicolumn{1}{c}{\makecell{Cat Hygiene}}  \\ 
      Model:             & (1)                         & (2)             & (3)             \\  
      \midrule
%      \emph{Variables}\\
      ATT               & -0.2482$^{**}$  & -1.7922$^{***}$ & -0.7381$^{***}$\\   
                         & (0.0945)       & (0.3678)        & (0.1794)       \\   \midrule
                                         \makecell{ATT Relative to Adopters' \\ Pre-Adoption Elasticity}      & 8.4\%  & 72.7\% & 34.7\%     \\   \midrule
      \midrule
 %     \emph{Fit statistics}\\
     Observations       & 136,411              & 78,253         & 55,410\\  
      \midrule %\midrule
        \multicolumn{4}{l}{\emph{Bootstrapped standard-errors in parentheses via 500 replications}}\\
      \multicolumn{4}{l}{\emph{Signif. Codes: ***: 0.001, **: 0.01, *: 0.05}}\\
   \end{tabular}
   }
\end{table}

We present the overall treatment effect estimates $(\theta_{o})$ in Table \ref{tab:5_CS_Result}. Since the customer-month level price elasticities are themselves estimates from our demand model with confidence intervals around them (and not directly observed from data), we compute standard errors for our ATT estimates using customer-level block bootstraps with 500 replications. Figure \ref{fig:5_Boot_Est} in Web Appendix \S\ref{appsec:did} shows the bootstrapped distribution of these ATT estimates. The results in Table \ref{tab:5_CS_Result} suggest that consumers become more price-sensitive in the offline channel after they adopt online shopping, and this finding is consistent across these categories. The effect is stronger for the dog and cat hygiene categories, consistent with the elasticity visualization in Figure \ref{fig:5_Main_Elas_Prediction}. To help with interpretation, we calculate the relative magnitude of the elasticity increase in percentage terms by dividing the ATT by the average pre-adoption price elasticity of adopters and show these percentage changes in the second row of Table \ref{tab:5_CS_Result}. 

The most substantial effects are observed for dog hygiene and cat hygiene categories; offline price elasticity increases by 72.7\% (i.e., -1.792 / -2.464) for dog hygiene and 34.7\% (i.e., -0.738 / -2.125) for cat hygiene. The increase in offline price elasticity is lower for dry dog food, at 8.4\% (i.e., -0.248 / -2.935). As discussed in \S\ref{ssec:elasticity_est}, these differences may reflect greater flexibility to time hygiene purchases around discounts. In sum, consistent with the descriptive evidence earlier, we find that consumers become more price-sensitive in their offline shopping trips when they start shopping online.

\section{Generalizability, Potential Mechanisms, and Robustness Checks}
\label{sec:robust}

We now discuss the generalizability of our findings, provide evidence on potential mechanisms, and present a series of tests and alternative specifications that establish the robustness of our results.

\subsection{Generalizability}
\label{ssec:generalizability}

We first discuss the generalizability of our findings and provide evidence supporting their portability to other settings. 

Consumers may adopt online shopping for different reasons, and a natural question is whether our results are specific to consumers who adopt online shopping for certain reasons but not others. To examine this question, we leverage the fact that our observation period starts before the onset of the COVID-19 pandemic and continues through it. The cohorts that adopted from July 2019 to February 2020 are likely to have done so organically, while those who adopted from March 2020 are likely to have done so for exogenous reasons, i.e., the COVID-19 pandemic. Indeed, there is a big surge in online adoptions starting March 2020 (see Figure \ref{fig:2_adopter_samplesize}). We use this exogenous shock in the reason for adoption across cohorts to examine this issue. We also observe a spike in adoption around the retailer's Black Friday online-only promotion event on November 29, 2019. We separately examine customers who adopted during this promotion, whom we refer to as Black Friday adopters. Overall, we find that the reason why a consumer switches to online shopping does not appear to have a significant impact on post-adoption price sensitivity: the ATT of cohorts that switched during the COVID-19 pandemic is similar to that of cohorts that switched organically before the onset of COVID-19 (see Figure \ref{fig:5_Main_Elas_Prediction} and cohort-specific ATT estimates in Web Appendix $\S$\ref{appsec:covid_att}). We also find similarly sized effects for Black Friday and COVID-19 adopters (see Table \ref{tab:att_price_elasticity_cohort_BF_Covid} in Web Appendix $\S$\ref{appsec:covid_att}). These results suggest that the online adoption effect is not sensitive to the reasons that drive a consumer's decision to adopt online shopping.

\subsection{Potential Mechanisms} 
\label{ssec:mechanisms}

We next discuss potential mechanisms through which online adoption may increase consumers' offline price sensitivity. Conceptually, online adoption may affect offline price sensitivity through at least two related processes. First, the online channel may lower search costs and make price comparisons easier. For example, consumers can sort products by price, compare brands side by side, and observe a wider set of alternatives more easily than they can in a physical store. They can also track price changes over time more easily. This may make consumers more attentive to price differences and price changes over time, and this greater price awareness may carry over to their subsequent offline shopping occasions. Second, if online prices are lower or more dispersed than offline prices, exposure to the online channel may make price differences more salient and increase consumers' responsiveness to price changes more generally. Both mechanisms imply that online adoption can change how consumers process prices even when they later shop offline.

We first examine the extent to which our results are likely to be driven by the fact that online prices are slightly lower than offline prices in our setting. To that end, we examine the magnitude of the price differences between online and offline channels in our study (see details in Web Appendix $\S$\ref{appssec:cross.channel.price.differences}). We find that online prices are only 3--6\% lower, on average, than offline in our data.\footnote{The small average price difference does not reflect a persistent difference in base prices. Instead, it appears to arise
primarily from more frequent promotions in the online channel as detailed on Web Appendix $\S$\ref{appssec:cross.channel.price.differences}} Furthermore, the variation in prices explained by channel is significantly smaller than the variation in prices explained by SKU-level differences -- across all categories, only less than 1\% of the variation in prices is explained by channels, whereas when we include SKU fixed effects, this increases to more than 84\%. Overall, these data patterns suggest that the differences between online and offline prices are relatively small and are unlikely to be the primary driver of our results.

We then examine how the increase in offline price sensitivity is reflected in consumers' subsequent offline purchases. First, in $\S$\ref{sec:descriptive.analysis}, we show that conditional on making an offline store visit, consumers become less likely to purchase the focal category offline after online adoption, thus their responses to price changes are heightened.\footnote{One concern is that the decline in offline category purchases may reflect customers leaving the focal retailer rather than changes in offline price sensitivity. To assess this, we decompose total, offline, and online purchasing at the customer-month level. We find that for each focal category, total purchase incidence, quantity, and spending increase after online adoption, while offline purchases decline and online purchases increase. This pattern suggests that the decline in offline category purchasing primarily reflects within-retailer channel reallocation rather than customer attrition. See Web Appendix $\S$\ref{appssec:purchase_behavior_mechanisms}.} Next, we provide descriptive evidence that adopters become more likely to time their offline purchases around promotional prices after online adoption. Specifically, conditional on making an offline purchase in a focal category, adopters become significantly more likely to purchase the same SKU during promotional-price periods after adopting the online channel. \footnote{We define promotional price period as those weeks that a SKU's sales price is 10\% off than the regular price within a store.} The estimated increase ranges from 6.2 to 10.9 percentage points across the four product categories (see Table \ref{tab:purchase_with_promotion_price} in Web Appendix $\S$\ref{appssec:purchase_behavior_mechanisms}). This pattern is consistent with consumers becoming more price attentive and tracking price changes over time after online adoption, and then applying this awareness making purchase decisions offline as well. 

Importantly, the evidence does not suggest that the main effect is driven by consumers switching to cheaper brands. We examine brand-level purchase shares, classify brands into high- and low-price groups based on their average prices, and estimate whether adopters become less likely to purchase high-price brands after online adoption. The estimates are small and mostly statistically insignificant (see Table \ref{tab:2_BuyHighPriceBrand} in Web Appendix $\S$\ref{appssec:purchase_behavior_mechanisms}). We also do not observe systematic changes in consumer-level brand variety during offline store visits. These results suggest that the increase in offline price sensitivity is better explained by more strategic timing of purchases and greater attention to price variation within products, rather than by substitution toward lower-priced brands. We also rule out the potential mechanism that consumers only buy a product offline, post-online adoption, if it is available at a lower price than online. We find that among weeks when a SKU is purchased offline, it is available at the same or lower price online more than 65\% of the time. (See details in Web Appendix $\S$\ref{appssec:cross.channel.price.differences}.)

Taken together, the descriptive evidence points to a mechanism in which online adoption increases consumers' price awareness and induces more strategic offline purchase timing. Other potential mechanisms, such as consumers becoming more accustomed to comparing prices across brands or their exposure to a wider variety of prices, may be at play as well (as discussed in $\S$\ref{sec:intro}). Thus, we expect that our results are likely to generalize to other empirical settings where online prices are the same or higher than offline prices.\footnote{In practice, there is considerable heterogeneity in whether online prices are lower, the same, or higher than offline prices across retailers, countries, and product categories \citep{cavallo2017online}.} 

Nevertheless, we note that we are unable to tease out the exact source of these effects because we lack granular data on consumers' search and shopping behavior in the online channel. We do not observe online click-stream data, so we cannot comment on price search behavior online or the variety of prices seen online. We also do not observe any search behavior in offline stores. Further, while there are some differences in offline prices across physical stores at any given point in time, we do not have a sufficiently large proportion of stores that are consistently higher or lower priced than others throughout our entire time period. Therefore, we cannot conduct a clean test of whether the effect we observe is driven more by price differences across the offline and online channels, the variety of prices that the consumer sees, or changes in search behavior that occur after adopting online shopping. We hope that future research can further disentangle which of these mechanisms are at play and to what extent. Further studies of such spillover effects in other countries, product categories, and retailers can also provide a more comprehensive understanding of the extent to which these effects manifest in different settings.

% Replacement for Section 8.3 in MS_Feb2026.tex.
% Add to the MAIN DOCUMENT PREAMBLE if not already loaded:
%   \usepackage[normalem]{ulem}
% The manuscript already loads color. Blue denotes new text relative to the
% supplied draft; prior red author markup is treated as baseline text.
% Paragraph-level redline: removed paragraphs are struck through, followed by
% their replacements. Deleted headings are unnumbered to preserve numbering.
% The displayed pre-adoption demand equation and its label are retained.
% The new selection-on-observables and market-definition subsections add two
% subsection numbers. Existing labels are preserved; external hard-coded
% section numbers should be updated separately.

\begingroup
\sloppy % Allow long struck-through formulas to wrap without margin overflow.
\subsection{Robustness Checks}
\label{ssec:robustness_checks}

We assess the sensitivity of our findings to the demand specification, selection into online adoption, and the construction of the estimation sample. The checks below are organized by the concern they address. Unless otherwise noted, the findings remain qualitatively consistent with the main analysis. We report the detailed specifications and estimates in the corresponding Web Appendices.

\subsubsection{Alternative Demand Specifications}
\label{sssec:alt_demand}

First, we examine whether the estimated increase in offline price sensitivity depends on how we measure online adoption or on changes in brand loyalty and recent channel use. We estimate three alternative demand specifications, retaining customer-brand fixed effects and the corresponding control-function terms.

Binary online adoption variable: In our main demand model, we use the continuous variable \mbox{$\textrm{cumPercOnlineSpend}_{it}$} to capture how consumers' offline price sensitivity changes after adopting online shopping. We now consider an alternative specification where we treat online adoption as a binary treatment using the binary indicator, \mbox{$\textrm{Post\_OnlineAdoption}_{it}$}, which takes value one if customer \mbox{$i$} has adopted online shopping by time \mbox{$t$}, and zero otherwise.  The specification and estimates for this check are reported in Web Appendix $\S$\ref{appsec:alternative.demand.binary}.

Accounting for a potential change in state dependence or brand loyalty due to online adoption: In our main model, we include customer-brand fixed effects to capture time-invariant preferences and use \mbox{$\mathcal{I}\{\textrm{BuyPrevOcca}_{ijt}\}$} to capture the state dependence. We now consider a specification with an additional control variable, \mbox{$\mathcal{I}\{\textrm{BuyPrevOcca}_{ijt}\} \times \textrm{cumPercOnlineSpend}_{it}$}, which captures the interaction between state dependence (e.g., repeated purchase of a brand) and the share of online spending. This can account for any effects that online adoption may have on brand loyalty. The specification and estimates for this check are reported in Web Appendix $\S$\ref{appsec:alternative.demand.state}.

Accounting for the recent online use: We also estimate an additional specification that includes \mbox{$\mathcal{I}\{\textrm{lastMonthOnline}_{it}\}$}, an indicator equal to one if customer \mbox{$i$} made an online purchase in any category during the month preceding the focal purchase occasion. We also include its interaction with price, \mbox{$\textrm{Price}_{jt}\times \mathcal{I}\{\textrm{lastMonthOnline}_{it}\}$}, to capture whether recent online channel use is associated with greater offline price sensitivity. The specification and estimates for this check are reported in Web Appendix $\S$\ref{appsec:alternative.demand.recent}.

The utility specifications and the estimation results from these exercises are shown in Web Appendix \mbox{$\S$}\mbox{\ref{appsec:alternative.demand.model}}. Across all of these alternate specifications, we find that the results are consistent with our main specification. We also do not find evidence of significant changes to offline brand loyalty or a significant effect of recent online channel usage on offline price sensitivity after consumers adopt online shopping.

\subsubsection{Pre-Adoption Trends and Adoption Timing}
\label{sssec:self_selection}

One potential concern with the adoption design is that adopters may already be becoming more price sensitive before they first purchase online. Therefore, we examine pre-adoption patterns in two ways. First, we confirm that the event-study estimates based on our main demand model show no statistically significant pre-treatment effects (see Web Appendix $\S$\ref{appsec:parallel_pretrend}). 

Second, we estimate a demand specification that allows the price coefficient itself to vary before adoption, rather than depend only on cumulative online spending share. For each of the twelve adoption cohorts, we estimate a demand model using the focal adopters and their corresponding not-yet-adopted comparison group. We allow adopters' price sensitivity to differ in each of the five months, from six to two months before adoption, relative to the month immediately before adoption, and separately after adoption. We retain customer-brand fixed effects, lagged brand purchase, and the corresponding control-function terms. We then use these demand estimates to construct cohort-specific elasticity paths and test for differential linear pre-trends in a panel that stacks the six pre-adoption months across all twelve cohorts. We see that the estimated differential slopes are not statistically distinguishable from zero in the three focal categories. See Web Appendix $\S$\ref{appsec:adoption_timing_parallel_pretrend}  for the detailed specification and results.

\subsubsection{Selection on Observable Characteristics}
\label{sssec:selection_observables}

Consumers self-select into adopting a retailer's online channel, and as shown in Web Appendix \mbox{$\S$}\mbox{\ref{appssec:demographic}}, there are some systematic differences between adopters and non-adopters in terms of observable characteristics like demographics. In general, such differences are not an issue as long as the parallel pre-trends assumption holds (see Chapter 5 of \mbox{\citet{angrist2008mostly}} and \mbox{\citet{roth2023s}} for discussions). 

%For instance, there may be some factor that affects younger customers who are more likely to adopt online shopping, and only shows up in the later part of our data period. 

Nevertheless, differences in characteristics between groups may still potentially lead to differences in price elasticities over time in some way that is not showing up in the first few months of data or is not being accounted for in our model. To provide additional credibility to our findings and to ensure that they are robust to any time-varying factors correlated with observable characteristics, we use propensity score-based methods like Inverse Propensity Weighting (IPW) and Propensity Score Matching (PSM) to account for differences in observable characteristics between the groups, which improves the comparability between the groups. These methods improve covariate balance, discard poorly matched customers, and make our results more robust to model misspecification with respect to any time-varying factors that are correlated with observable characteristics.

We perform the propensity score estimation and matching for each adoption cohort separately. Specifically, we define the pre-adoption period separately for each cohort and measure pre-adoption variables (e.g., monthly average spend) using all purchase data up to the month of adoption \mbox{\citep{gu2021dark}}. For example, when identifying matched non-adopters for customers who adopt in July 2019, we measure the pre-adoption variables for both adopters and non-adopters using the data up to June 30, 2019, and estimate the propensity score model for the adoption cohort in July 2019. We repeat this process for each cohort and estimate twelve different propensity score models corresponding to each monthly adoption cohort. We compute propensity scores with a logistic regression model and the variables included are demographics (e.g., gender, age, and household income) and pre-adoption purchase variables (i.e. average monthly spend, number of unique orders, unique brands, and unique categories). With the estimated propensity scores, we use our CS estimator to estimate the ATT using both IPW and PSM methods. The propensity score model specification and detailed results from the IPW and PSM methods are shown in Web Appendix \S\mbox{\ref{appsec:selection.psm}}. We find that our conclusions remain the same.

\subsubsection{Alternative Control Groups}
\label{sssec:alternative.control}

Next, we assess whether our findings are sensitive to the choice of comparison group. We consider two concerns: differences between customers who eventually adopt online shopping and those who never adopt, and asymmetric purchase requirements during the adoption window.

\textbf{Not-yet-treated adopters.} Even after accounting for observable characteristics, adopters and never-adopters may respond differently to time-varying unobserved factors. Following \cite{narang2019mobile} and \cite{manchanda2015social}, we therefore use later adoption cohorts as comparison groups for earlier cohorts, restricting their use as controls to periods before they adopt. Because all customers in this analysis eventually adopt online shopping, this comparison minimizes differences associated with the propensity to adopt. We report the resulting CS estimates in Web Appendix $\S$\ref{appssec:alternative_control_not_yet_treated}.

\textbf{Purchase requirements during the adoption window.} In our main sample, described in $\S$\ref{ssec:construction}, both adopters and non-adopters must make focal-category purchases during the common pre- and post-adoption periods. Adopters also necessarily transact during the twelve-month adoption window through their first online purchase, whereas non-adopters face no corresponding requirement. To make these observation requirements more symmetric, we restrict the comparison group to non-adopters who make at least one offline purchase during the adoption window, leaving the adopter definition and all other sample criteria unchanged. We then re-estimate the demand model and ATT using this comparison group. Detailed results are reported in Web Appendix $\S$\ref{appssec:alternative.filter.control.group}.

Overall, both checks yield conclusions consistent with our main specification: consumers become more offline price-sensitive after adopting online shopping, with negative and statistically significant ATT estimates in the three focal categories.

\subsubsection{Alternative Definition of Purchase Occasions}
\label{sssec:alternative.def.purchase.occasions}

In our main analysis, we define a purchase occasion (in \S\mbox{\ref{ssec:construction}}) as any customer-month during which we observe any offline interaction with the retailer (similar to a week-store visit in \mbox{\cite{gordon2013does}}). We consider two alternative definitions.

Under the first alternative definition, a customer-month is included as a purchase occasion if a customer visits the retailer (i.e., either in-store or online) and purchases in any category during the month (not limited to the four categories we focus on in this study). If a customer makes no purchases in any category during the month, that customer-month is not included as a purchase occasion. In other words, a purchase occasion represents any customer-month during which we observe any interaction with the retailer, either offline or online. Under the second alternative definition, we construct a complete monthly panel for each customer, beginning in the month of the customer's first observed purchase and ending in the last month in which the customer purchases any category offered by the retailer. The detailed demand model estimates and ATT estimates are presented in Web Appendix \S\mbox{\ref{appssec:alternative.def.po}} and \S\mbox{\ref{appssec:customer.month.full.panel}}. In both cases, the conclusions remain consistent with those from the main specification.

\subsubsection{Alternative Specifications to Address Price Endogeneity}
\label{sssec:control_function_price_endogeneity}

Our preferred specification addresses potential price endogeneity by instrumenting retail prices with wholesale costs and including the corresponding first-stage residuals in the utility function. However, the extent of residual price endogeneity may differ across categories, and the estimated control-function terms are not statistically significant in every category. We therefore conduct several additional analyses to assess whether our substantive conclusions depend on the particular control-function specification.\footnote{\label{fn:cf_bias_sensitivity}\citet{degrange2024estimating} highlight potential bias in alternative-specific constants under the control-function approach, even when the slope coefficients on endogenous variables are well estimated. However, this is not a primary concern for us since our parameters of interest---the coefficients on price and its interaction with cumulative online spending share---are slope coefficients that govern price sensitivity and its variation with online engagement. Thus intercept-bias concern is  not central to our analysis.}

First, in Web Appendix \mbox{$\S$}\mbox{\ref{appssec:alternative.imple.no.control.function}}, we re-estimate the demand model without either control-function term. The resulting specification includes price, cumulative online spending share, their interaction, lagged brand purchase, and customer-brand fixed effects, but excludes \mbox{$\textrm{CF\_Price}_{jt}$} and \mbox{$\textrm{CF\_Price\_cumPercOnlineSpend}_{ijt}$}. We use the resulting demand estimates to recompute customer-month-level offline price elasticities and then re-estimate the ATT using the same staggered difference-in-differences procedure. The demand estimates, elasticity trajectories, and ATT estimates remain qualitatively consistent with our main results.

%This exercise provides a benchmark showing that our substantive conclusions are not impacted by the inclusion of the control-function terms.

Second, in Web Appendix \mbox{$\S$}\mbox{\ref{appssec:alternative.imple.control.function}}, we implement a more parsimonious control-function specification following \mbox{\cite{ebbes2016dealing}}. Rather than constructing separate residuals for price and the interaction between price and cumulative online spending share, we include only the residual from the first-stage price equation. This alternative addresses the endogeneity of price while treating the interaction through the same underlying price correction. The estimated demand parameters and ATT estimates again remain consistent with those from our preferred two-control-function specification.

Third, in Web Appendix \mbox{$\S$}\mbox{\ref{appssec:price_variation_control_function}}, we provide additional evidence on the relevance of the cost instruments and the sources of identifying price variation. The first-stage estimates show that wholesale costs significantly predict offline retail prices after controlling for store, brand, and year-month fixed effects, with conditional F-stats greater than 10 for each category. Similarly, the interaction between cost and cumulative online spending share strongly predicts the endogenous interaction between price and cumulative online spending share. We also decompose retail-price variation and document meaningful variation across stores and over time after accounting for persistent product and time components. Together with the evidence that systematic online--offline price differences are modest relative to product-level price variation, these analyses indicate that our estimates are supported by cost-related supply-side variation rather than being mechanically driven by large channel-specific price differences.

Taken together, these exercises show that the estimated increase in offline price sensitivity after online adoption is robust to omitting the control-function correction, using a more parsimonious single-control-function specification, and examining the relevance and underlying variation of the cost-based instruments. We therefore retain the two-control-function model as our preferred specification because it directly addresses the potential endogeneity of both price and its interaction with cumulative online spending share. See Web Appendix \S\mbox{\ref{appsec:price.endogeneity.control.function}} for detailed estimation results.

\par\endgroup

\subsubsection{Alternative Product Market Definition}
\label{sssec:product_market_definition}

The main analysis focuses on selected pack sizes and excludes customers who switch to other sizes within a category. This restriction may omit a margin of substitution used by price-sensitive consumers. We therefore expand each product market to include pack sizes accounting for at least 80\% of category sales and include customers who switch among these sizes. Re-estimating the demand model and ATT yields negative and significant adoption effects in the three focal categories. Web Appendix $\S$\ref{appsec:alternative.product.category.definition} describes the expanded markets and reports the estimates.

\section{Managerial Implications and Conclusions}
\label{sec:conclusion}

The rapid growth of e-commerce has fundamentally reshaped consumer behavior, but its spillover effects on offline shopping dynamics remain understudied. In this paper, we investigate how the adoption of online shopping using a retailer's e-commerce channels influences consumers' offline price sensitivity when shopping at that retailer's physical stores, utilizing transaction-level data from a multichannel pet supplies retailer in Brazil. We find that consumers who adopt online shopping become more price-sensitive in their offline purchases. Further descriptive analysis of the data suggests that this effect is likely driven by strategic timing of offline purchases around promotions rather than switching to cheaper brands.  

From a substantive standpoint, our results contribute to the literature on price elasticity by identifying online adoption as a novel determinant of offline price sensitivity. This factor has so far been ignored in the demand estimation literature, which usually only accounts for factors like consumer demographics and macro-economic shocks to affect consumer price sensitivity. From a managerial standpoint, our findings have implications for retailer offline pricing and profitability, emphasizing the need for multichannel retailers to account for the online adoption effect (i.e., heightened offline price sensitivity when consumers adopt their online channels) in their pricing decisions. Further, our approach can be used to personalize prices depending on customers' channel adoption status. 

\par  From a managerial perspective, our findings can be used to improve the firm's pricing and targeting strategies. To illustrate how differences in online adoption can inform offline pricing, we conduct counterfactual simulations comparing uniform and personalized prices under demand models that account for or ignore the online adoption effect. Our model allows each customer's offline price coefficient to vary with their cumulative online spending share, so customers with different levels of online engagement can face different personalized prices, implemented, for example, through targeted coupons. We calculate profit-maximizing prices under each model and evaluate the resulting offline profits using the demand model that incorporates the adoption effect. Accounting for this effect increases simulated monthly offline profit by 1.17\% under uniform pricing and 4.1\% under personalized pricing, relative to ignoring the effect within each pricing regime. These findings illustrate the value of incorporating customers' channel usage into offline pricing decisions alongside their brand preferences and purchase histories. Please see Web Appendix $\S$\ref{appsec:counterfactual.implication} for the detailed analysis and results. While existing literature has largely focused on pricing and personalization based on customer demographics and prior purchase behavior \citep{rossi1993bayesian,chen2001individual,jiang2021consumer, jain2024effective}, our work suggests that incorporating cross-channel effects into measures of price elasticity and pricing strategy can further enhance profits.

Finally, our paper opens many avenues for future research. First, future research could explore whether these effects generalize to other markets (e.g., non-discretionary goods or regions with lower e-commerce penetration) and investigate mediating factors, such as the role of mobile price-checking apps or subscription models. Further, while our paper documents one aspect of consumer behavior (price sensitivity), multi-channel shopping can induce other spillover effects from one channel to another. For instance, product categories purchased offline may change after adopting online shopping \citep{huyghe2017clicks}, and there may be cross-media effects between channels \citep{joo2014television, du2019immediate}. Future research can investigate other such spillover effects and their effect on cross-channel purchases. Ultimately, as retail continues to evolve toward omnichannel integration, understanding these interdependencies will be critical for both academics and firms aiming to thrive in a digitally transformed landscape.

\section*{Funding and Competing Interests}
The authors were not paid by the data provider. A preliminary version of this manuscript was reviewed by the data provider to ensure that data was accurately depicted, but they did not review/comment on the content of the manuscript.

%\bibliography{main_bib.bib}{}
%\bibliographystyle{apalike}

%\putbib

%\end{bibunit}

\newpage

\begin{appendices}

\setcounter{table}{0}
\setcounter{figure}{0}
\setcounter{equation}{0}
\setcounter{page}{0}
\renewcommand{\thetable}{A\arabic{table}}
\renewcommand{\thefigure}{A\arabic{figure}}
\renewcommand{\theequation}{A\arabic{equation}}
\renewcommand{\thepage}{\roman{page}}
\pagenumbering{roman}
%\begin{bibunit}

\section{Appendix to Setting and Data}
\label{appsec:data}

\subsection{Details of Dataset Construction}
\label{appssec:dataset_const}

\subsubsection{Selection of Brands and Stock Keeping Units (SKUs)} 

\label{appsssec:brand_sku}

For each product category (i.e., dry dog food, dry cat food, dog hygiene and cat hygiene), we focus on a single pack size that accounts for the largest sales revenue and offers sufficient brand representation. Our goal is to study consumer brand choice and price elasticity, so consistent with previous brand choice literature \citep{allenby1998marketing,kim2002modeling,chintagunta2005beyond,dube2010state} we focus on a market (i.e., single most popular pack size within each category) that initially (as of January 1st, 2019 - June 30th, 2019) features at least three brands. We exclude any market with fewer than three brands or with one brand contributing more than 60\% of total sales. Applying these criteria yields the top one selling  pack sizes for each category during the observation period: 15KG adult daily dry dog food, 10.1KG adult daily dry cat food, and an 80cm $\times$ 60cm dog hygiene mat containing 30 pieces.

However, we observe high market concentration for certain cat hygiene pack sizes. The 12KG, 4KG, and 3KG sizes are among the top-selling segments (e.g., ranked 1st, 3rd, and 4th in offline sales). However, each has less than three brands, and a single brand contributes over 60\% of sales revenue (85\%, 63\%, and 88\%, respectively), making these pack sizes unsuitable for brand choice analysis. Furthermore, the excluded pack sizes differ significantly in sand types, compositions, and functional characteristics, limiting their substitutability in the consumer's mind. For example, 3KG pack sizes SKUs are primarily made from cereal- and vegetable-based materials that are eco-friendly, bio-degenerate, lightweight, and prone to scattering, whereas 4KG and 12KG pack sizes consist of granulated smectite clay, which is heavy, non-biodegradable and derived from mine materials. Hence, these pack sizes cater to distinct functional needs and consumer preferences. \footnote{For a similar reason, we exclude the 10KG pack size, which ranks 6th and sells sands with pine and wood.} The differences in product attributes and target consumer preferences complicate defining a cohesive market that includes these pack sizes. In addition, even if 4KG and 12KG pack sizes are grouped into a single market, the top brand still accounts for 77\% of the sales, leaving little room for brand choice analysis. 

Consequently, we focus on silica cat sand in the 1.6KG, 1.8KG, and 2.0KG, which together account for 29\% of the store's cat sand sales. Among these, the 1.8KG size is the second largest in the offline cat sand sales. Despite slight variations in pack sizes across brands, all three pack sizes offer the same sand category (i.e., silica), and we observe significant overlap in consumer purchases. Specifically, among customers who purchased the 1.8KG size at least once during the sample period, 28\% also purchased the 1.6KG pack size, and 20\% also purchased the 2.0KG pack size during the observation period, indicating a clear overlap of consumer preferences in this segment. 

After defining the market with the top one selling pack size for each category, we rank brands by offline sales in descending order and include those that collectively account for at least 80\% of the sales, following earlier literature \citep{gordon2013does}. The remaining brands are grouped into a single `Other Brand' for each product category, ensuring broad market coverage.

\subsubsection{Price and Cost Index Calculations} 
\label{appsssec:price}

We construct brand-level price indices for each category by aggregating SKU-level price data from {\it offline} transactions \citep{gordon2013does,hitsch2021prices}. First, we construct a store-SKU-month level price matrix from the available transactions data.
To begin with, this price matrix could contain both outliers and missing values. Any observed price that is more than four times higher than or less than one-quarter of the median price for each SKU (across all store-month observations) is treated as an outlier and labeled as missing, following \cite{hitsch2021prices}. This process filters out unreasonable prices before imputing any missing values of store-SKU-month price (when there is no sale or there is an outlier). For these missing prices, we use the highest observed price from the previous three months for the same SKU at the same store to impute the missing value \citep{gordon2013does}. Finally, we exclude SKUs for a particular store and week if we have never observed them being sold in that store previously.

Next, we aggregate SKUs to create brand-level price indices by converting all SKU prices to a per-unit basis for comparable units. All product categories (e.g., dry food and cat sands) are measured by price per kilogram (KG) except for hygienic mats, which are measured by price per piece. Following the approach of \cite{hitsch2021prices}, the brand price for brand $j$ at store $s$ in month $t$, denoted as $\textrm{BrandPrice}_{jst}$, is calculated as the weighted average price per unit across all SKUs in that brand in the defined market. The weights are based on each SKU's share of offline sales within the brand at that store over the full 30-month data period. 
\begin{equation}
    \textrm{Brand\_Price}_{jst} = \sum_{k\in C_{js}} w_{jsk} \times \textrm{SKU\_Price}_{jstk},
\end{equation}
where the weight of SKU $k$ is $w_{jsk} = \frac{\textrm{Store\_Sales}_{jsk}}{\sum_{k'\in C_{js}} \textrm{Store\_Sales}_{jsk'}}$, 
and $\textrm{Store\_Sales}_{jsk}$ is the total offline sales of SKU $k$ for brand $j$ at store $s$ from January 1, 2019 to June 30, 2021. $C_{js}$ represents the set of SKUs for brand $j$ available at store $s$, and $\textrm{SKU\_Price}_{jstk}$ is the price of SKU $k$ for brand $j$ sold in store $s$ during month $t$, as recorded in the store-SKU-month price matrix.

The underlying wholesale costs are obtained directly from the retailer's procurement records, rather than inferred from retail prices or an assumed retail margin. The source and institutional rationale for these instruments are discussed in $\S$\ref{ssec:setting} and $\S$\ref{ssec:endogeneity}.

We then apply a similar approach to construct a store-SKU-month level cost matrix based on wholesale cost data, and aggregate SKU-level costs to create brand-level cost indices that correspond to each price index. These cost indices serve as the instrument for addressing price endogeneity, as discussed in Section \S\ref{sec:model}. Finally, we use Brazil’s Extended National Consumer Index (IPCA) to obtain inflation-adjusted prices and costs \citep{ibge2024}, with January 2019, the first month of observation, as the base period. 

%\footnote{Brazil's Extended National Consumer Index (IPCA) is provided by the Brazil Institute of Geography and Statistics (IBGE). The IPCA data is available at \url{https://www.ibge.gov.br/en/statistics/economic/prices-and-costs/}.} 

\newpage

\subsubsection{Supplementary material for Selection of Customers}
\label{appsssec:sample.customer}

\begin{figure}[htp!]
    \centering
        \caption{Staggered Adoption Cohorts of Online Shopping (12 Adoption Cohorts from July 2019 to June 2020)}
\label{fig:adopter_def}\includegraphics[width=0.9\linewidth]{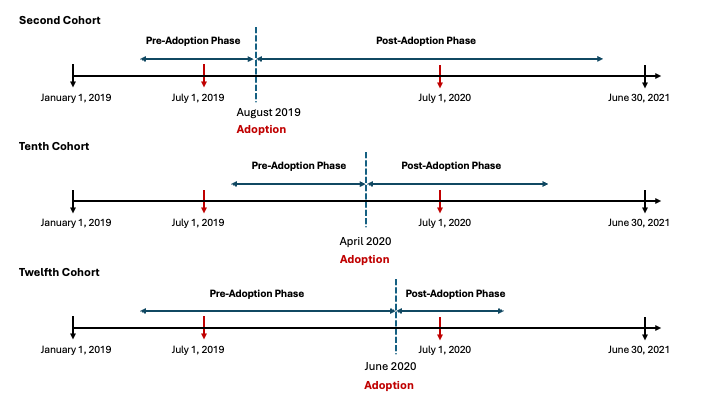}
\end{figure}

Figure \ref{fig:adopter_def} illustrates three examples of adopters defined in \S\ref{ssec:construction}, including adopters whose first online purchase in any category during August 2019, April 2020, and June 2020, respectively. Table \ref{tab:2_append_adopter_samplesize} presents the sample sizes for each adoption cohort and each category.

\begin{table}[htp!]
    \caption{Adopters Sample Size by Adoption Time - By Product Category}
    \label{tab:2_append_adopter_samplesize}
    \centering
    \footnotesize{
\begin{tabular}{lrrrr}
\toprule
Adoption Month & Dry Dog Food & Dry Cat Food & Dog Hygiene & Cat Hygiene \\
\midrule
2019-07 & 85 & 53 & 52 & 36 \\
2019-08 & 111 & 46 & 71 & 41 \\
2019-09 & 100 & 34 & 96 & 44 \\
2019-10 & 110 & 42 & 82 & 47 \\
2019-11 & 174 & 51 & 143 & 67 \\
2019-12 & 144 & 78 & 120 & 58 \\
2020-01 & 138 & 57 & 107 & 47 \\
2020-02 & 143 & 52 & 106 & 55 \\
2020-03 & 334 & 160 & 382 & 190 \\
2020-04 & 438 & 208 & 442 & 241 \\
2020-05 & 357 & 143 & 353 & 209 \\
2020-06 & 276 & 113 & 260 & 157 \\ \midrule
Total & 2410 & 1037 & 2214 & 1192 \\
\bottomrule
\end{tabular}
}
\end{table}

\clearpage
\subsection{Summary Statistics for Customer Demographics}
\label{appssec:demographic}
\begin{figure}[htp!]
    \centering
    \caption{Customer demographics of adopters and non-adopters of product category - age and monthly household income}
    \label{fig:2_demographic_all_category}
    \begin{subfigure}{0.49\textwidth}
        \centering
        \caption{Dry Dog Food - Age}
        \includegraphics[scale=0.20]{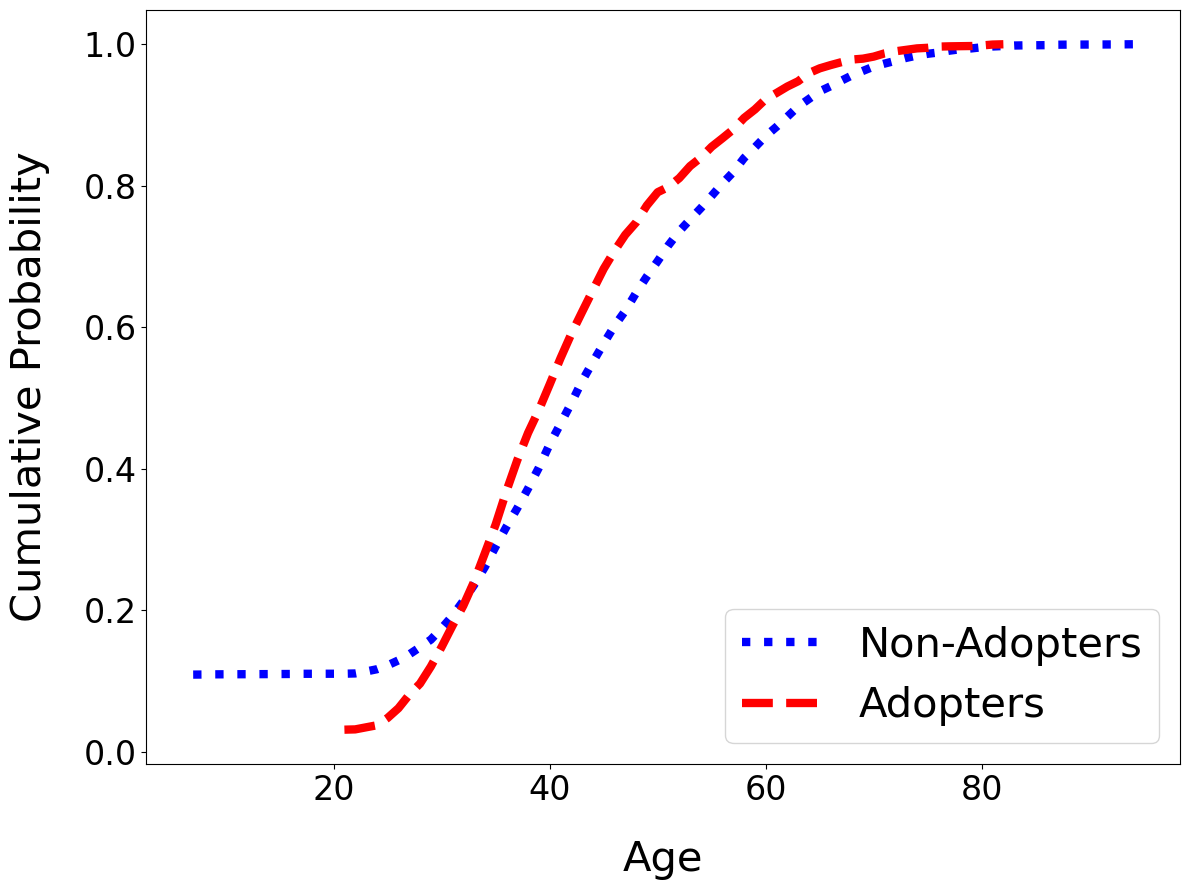}
    \end{subfigure}%
    ~ %add desired spacing between the subfigures, if needed
    \begin{subfigure}{0.49\textwidth}
        \centering
        \caption{Dry Dog Food - Income}
        \includegraphics[scale=0.20]{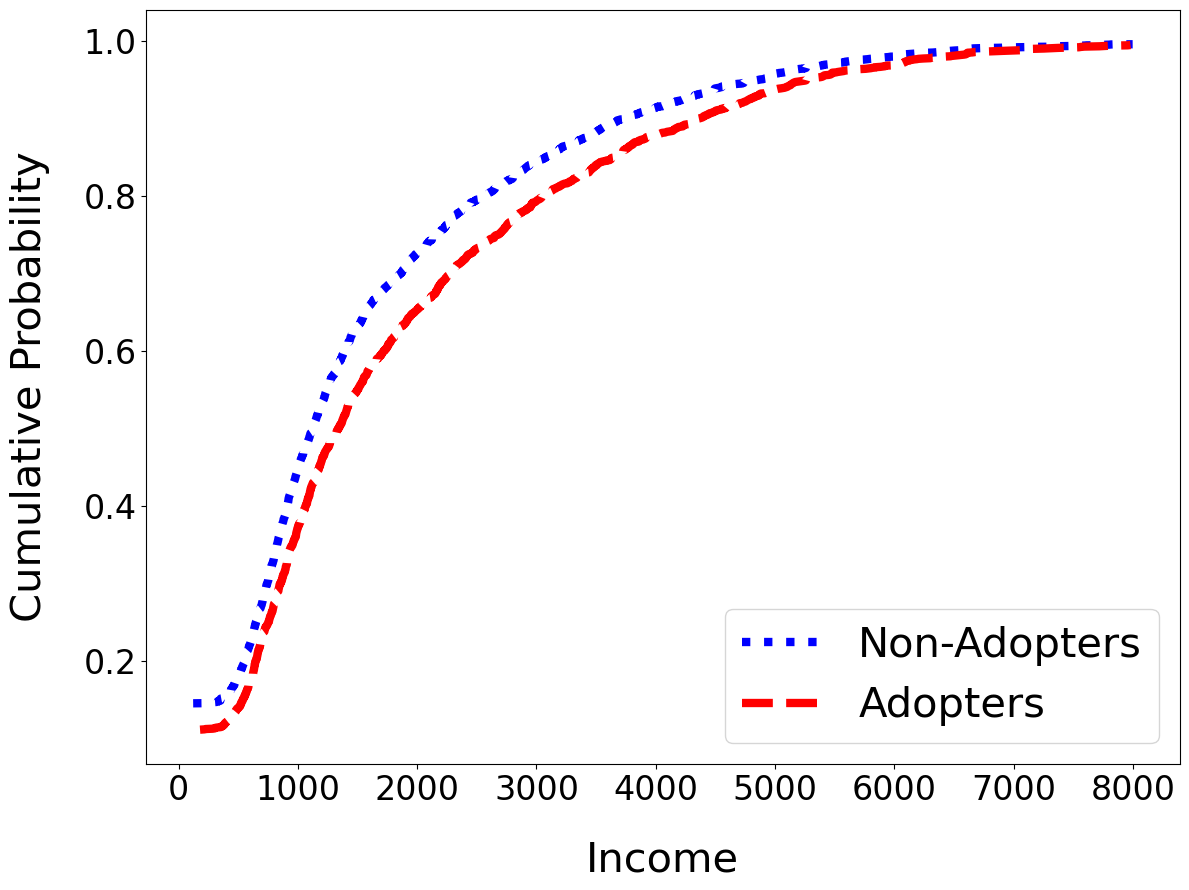}
    \end{subfigure}

    \vspace{1ex} % add some vertical space between the rows
            \begin{subfigure}{0.49\textwidth}
        \centering
        \caption{Dry Cat Food - Age}
        \includegraphics[scale=0.20]{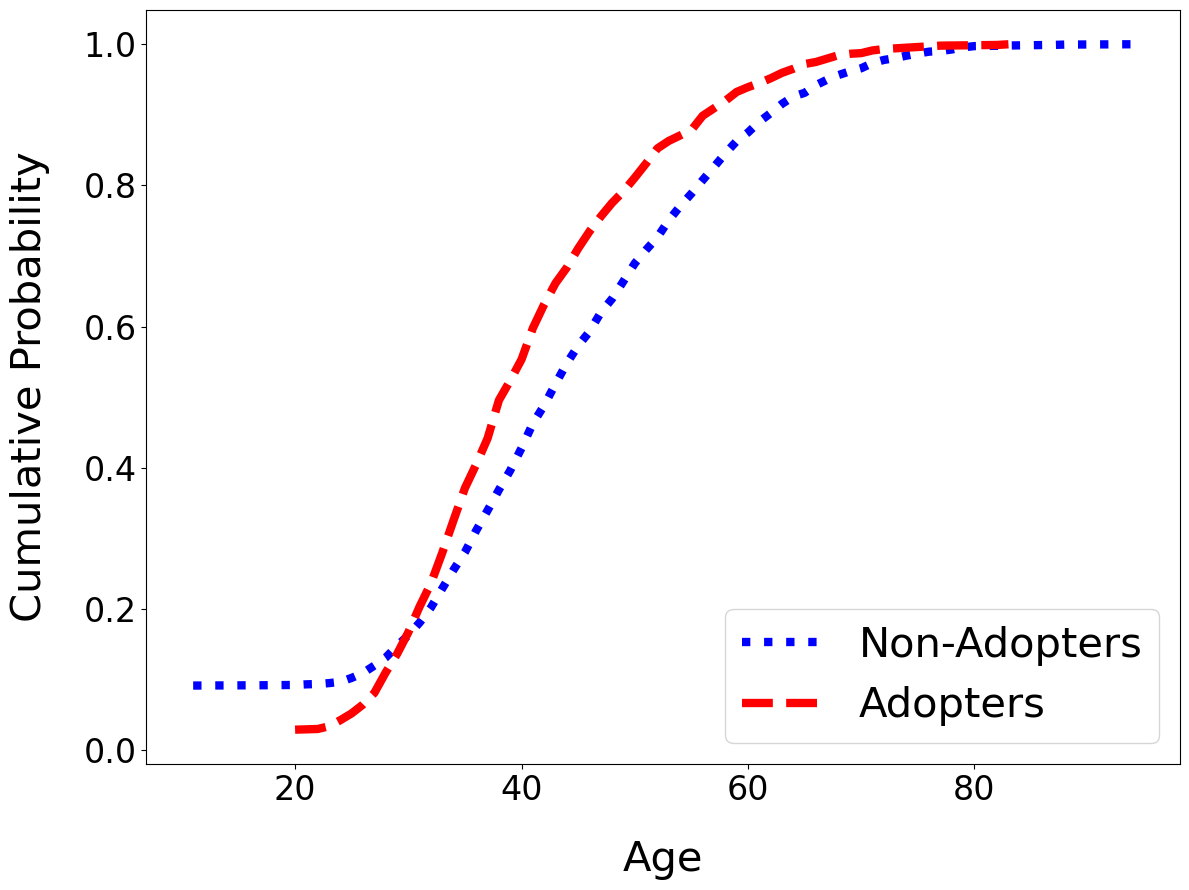}
    \end{subfigure}%
    ~ %add desired spacing between the subfigures, if needed
    \begin{subfigure}{0.49\textwidth}
        \centering
        \caption{Dry Cat Food - Income}
        \includegraphics[scale=0.20]{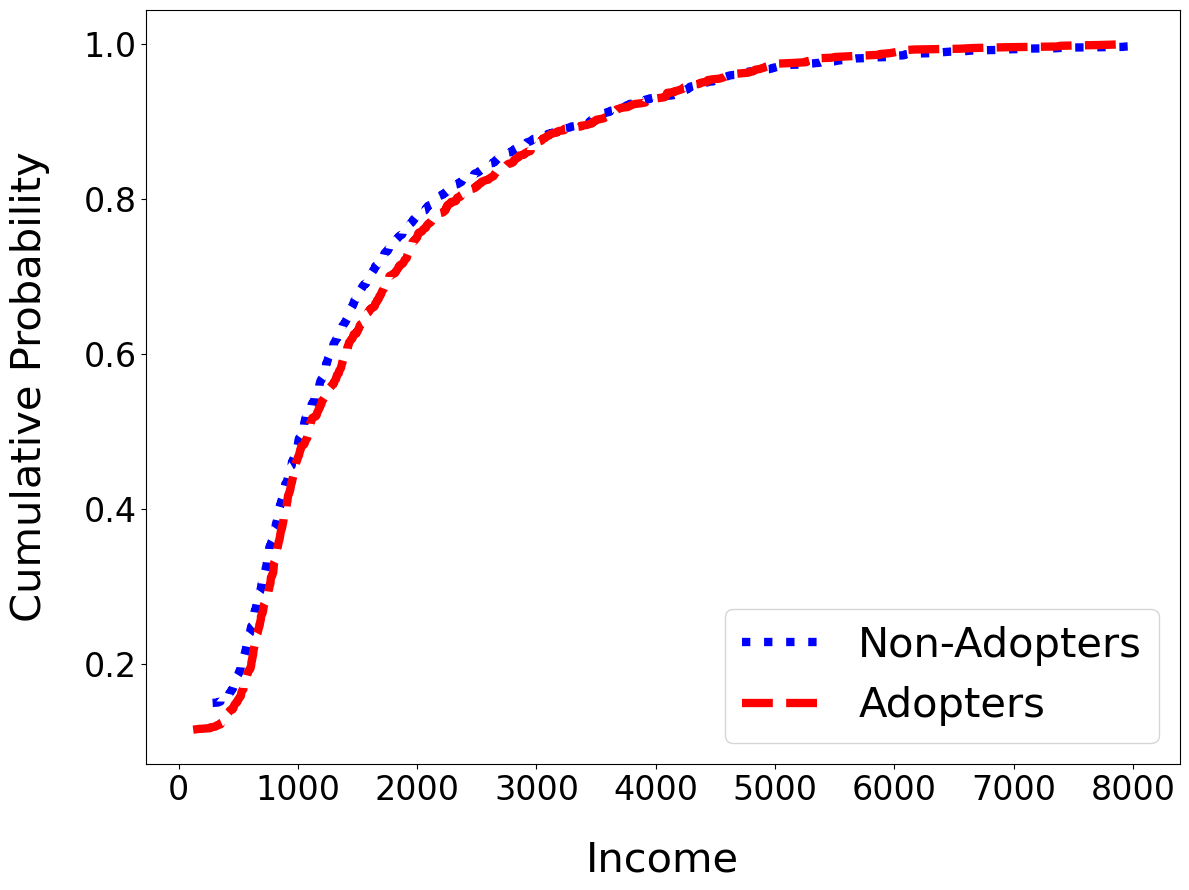}
    \end{subfigure}

    \begin{subfigure}{0.49\textwidth}
        \centering
        \caption{Dog Hygiene - Age}
        \includegraphics[scale=0.20]{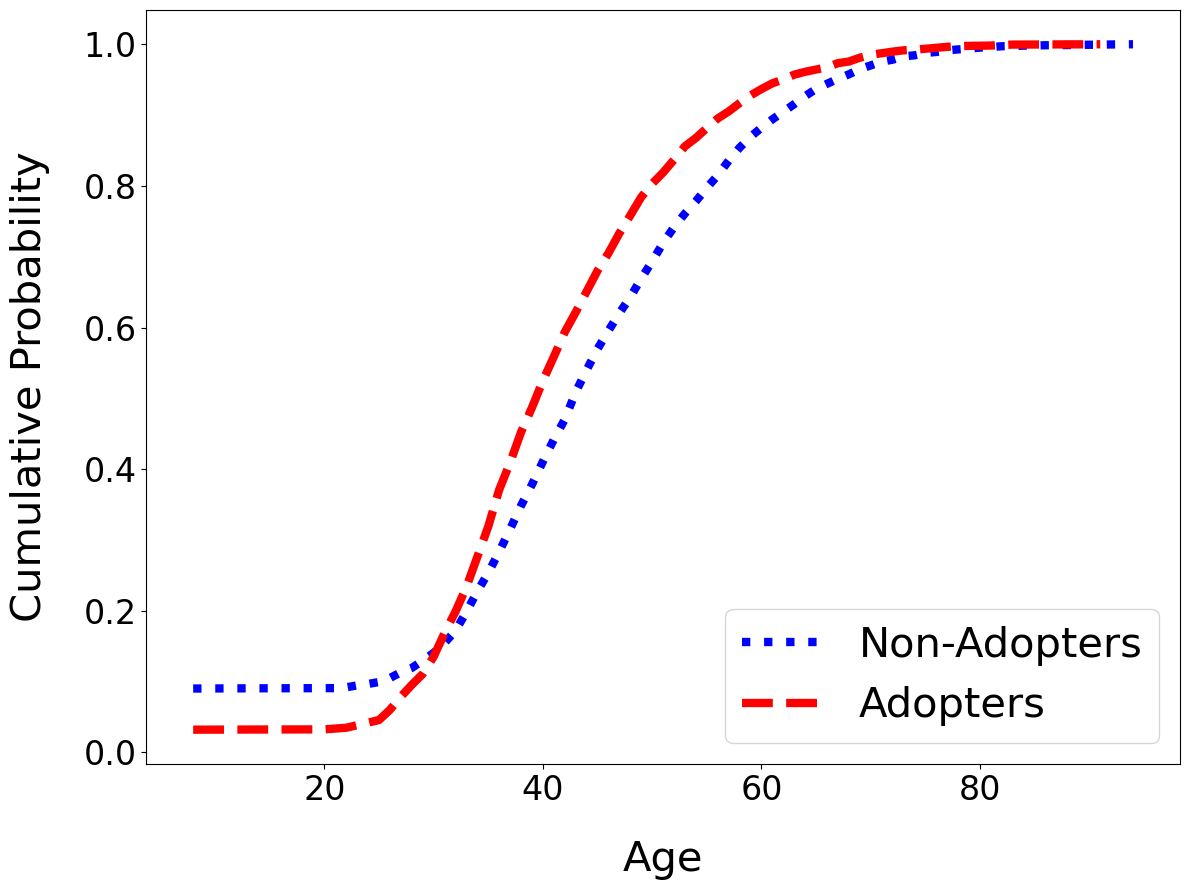}
    \end{subfigure}%
    ~ %add desired spacing between the subfigures, if needed
    \begin{subfigure}{0.49\textwidth}
        \centering
        \caption{Dog Hygiene - Income}
        \includegraphics[scale=0.20]{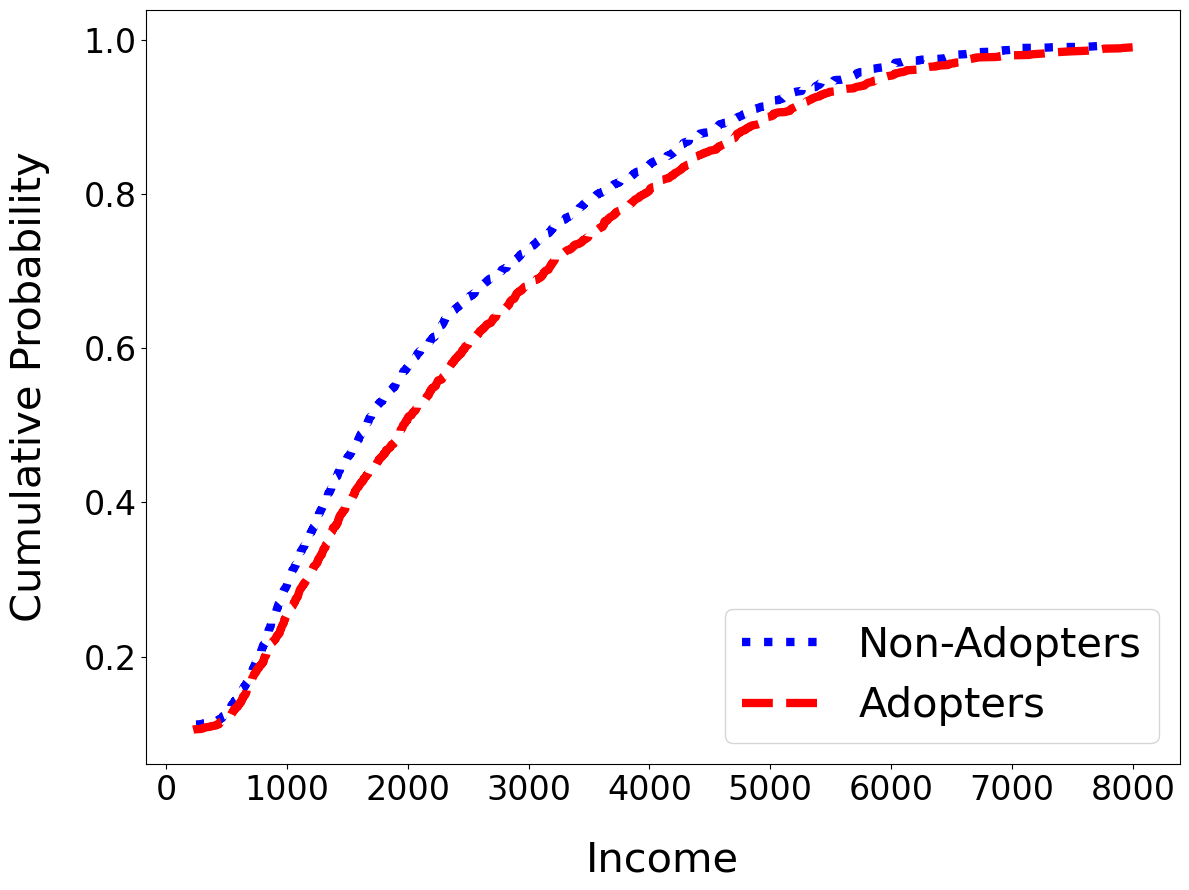}
    \end{subfigure}
    \vspace{1ex} % add some vertical space between the rows

    \begin{subfigure}{0.49\textwidth}
        \centering
        \caption{Cat Hygiene - Age}
        \includegraphics[scale=0.20]{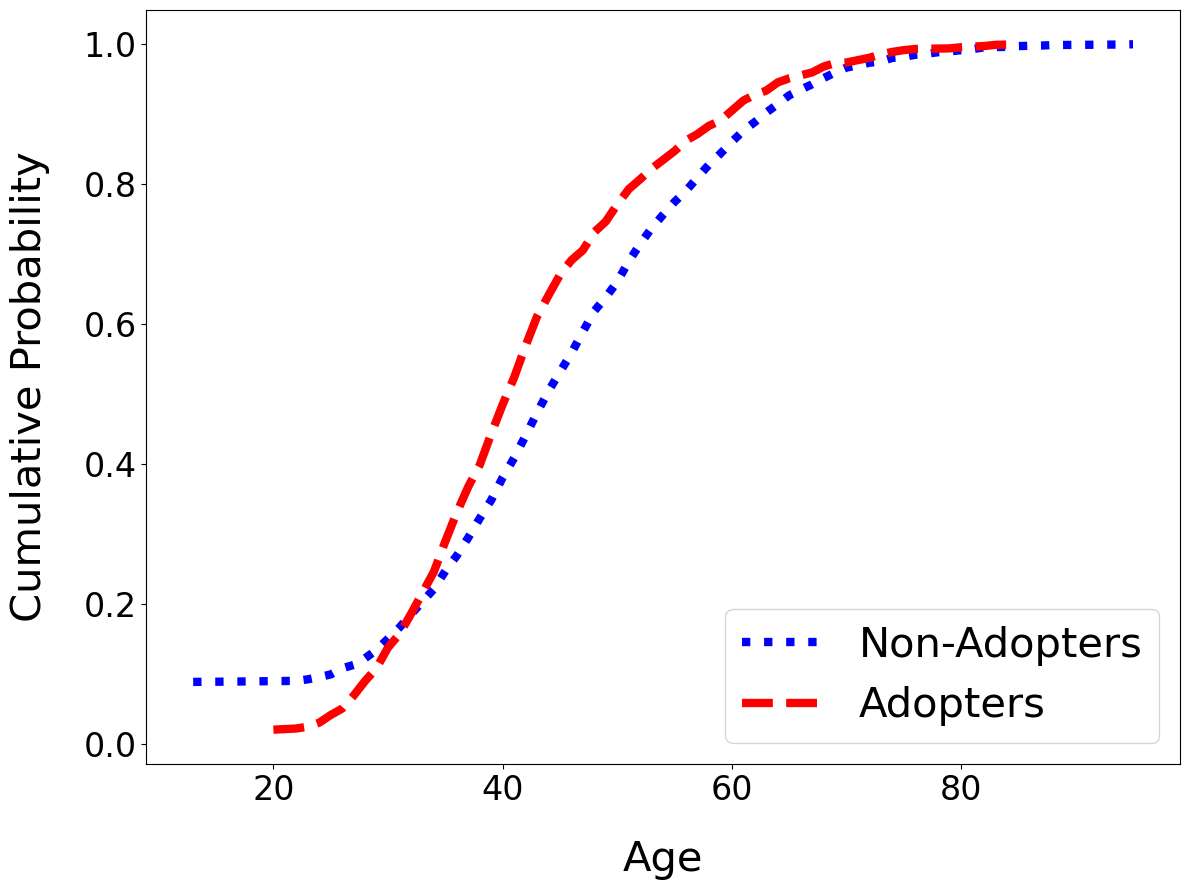}
    \end{subfigure}%
    ~ %add desired spacing between the subfigures, if needed
    \begin{subfigure}{0.49\textwidth}
        \centering
        \caption{Cat Hygiene - Income}
        \includegraphics[scale=0.20]{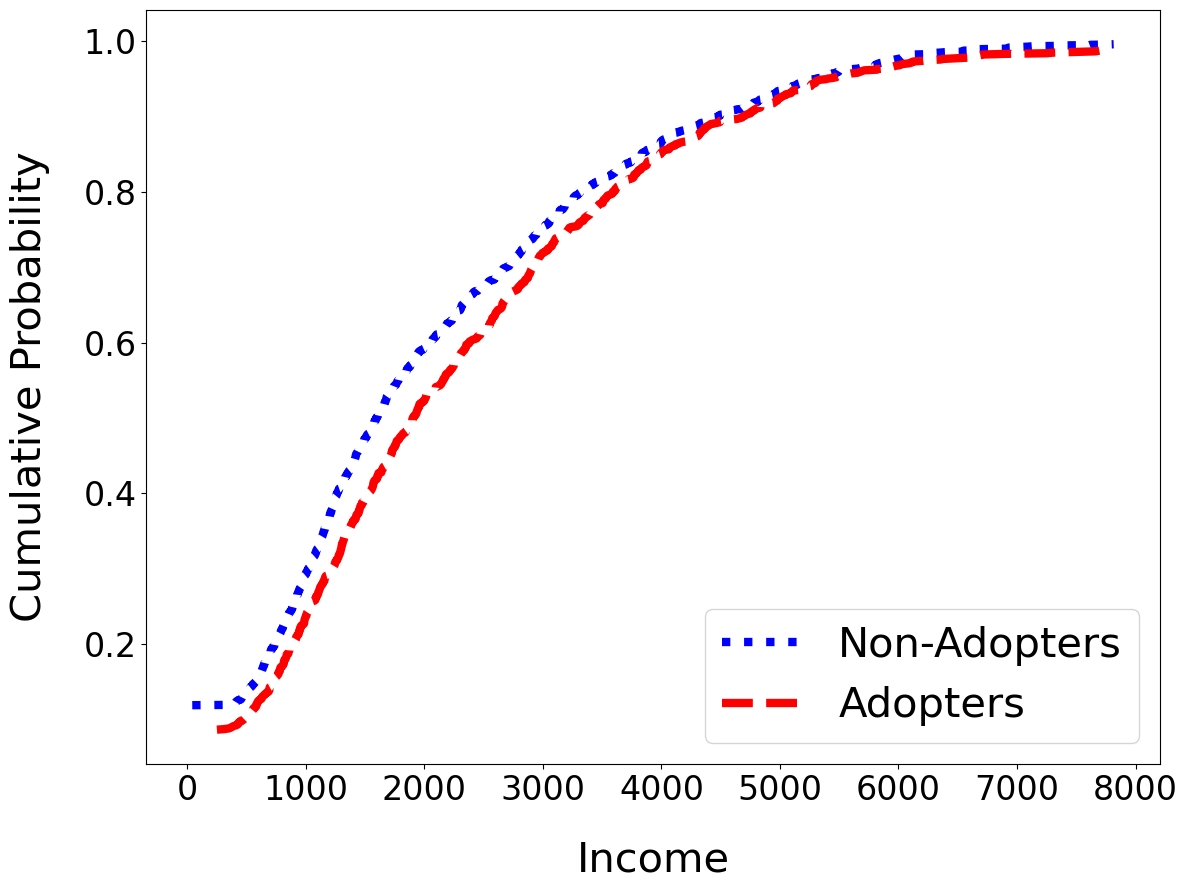}
    \end{subfigure}
\end{figure}

We summarize the customer demographics for each product category. 
Table \ref{2_Gender} presents the gender distribution for the different product categories amongst adopters and non-adopters. In all categories, adopters exhibit a higher proportion of women compared to non-adopters. For example, in the dry dog food category, 48\% of adopters are female, whereas only 32.6\% of non-adopters are female. Figure \ref{fig:2_demographic_all_category} shows the cumulative age and income distributions for adopters and non adopters. The height of the leftmost point on each curve represents the fraction of customers with missing age or income information. Across all categories, adopters tend to be younger compared to non-adopters. In terms of monthly household income, adopters generally exhibit higher incomes relative to compared to non-adopters.

\begin{table}[htp!]
\caption{Gender by Product Categories}
\label{2_Gender}
\centering
\footnotesize{
\begin{tabular}{lrrrrrrrr}
\toprule
\multicolumn{1}{c}{} & \multicolumn{2}{c}{Dry Dog Food} & \multicolumn{2}{c}{Dry Cat Food} & \multicolumn{2}{c}{Dog Hygiene} & \multicolumn{2}{c}{Cat Hygiene} \\ \midrule
Gender & Adopters & Non-Adopters & Adopters & Non-Adopters & Adopters & Non-Adopters & Adopters & Non-Adopters \\
\midrule
Female & 0.480 & 0.326 & 0.570 & 0.418 & 0.601 & 0.470 & 0.602 & 0.462 \\
Male & 0.398 & 0.522 & 0.286 & 0.417 & 0.285 & 0.379 & 0.279 & 0.387 \\
NaN & 0.122 & 0.152 & 0.144 & 0.165 & 0.113 & 0.152 & 0.118 & 0.151 \\ \midrule
Total & 1 & 1 & 1 & 1 & 1 & 1 & 1 & 1\\
\bottomrule
\end{tabular}
}
\end{table}

\clearpage

\section{Appendix to Descriptive Analysis}
\label{appsec:descriptive}

\subsection{Supplementary Result: Percentile Price Paid}
\label{appssec:price}

As a robustness check for the results in \S\ref{sec:descriptive.analysis}, Table \ref{tab:twfe_percentile_price} presents the within-customer changes in the percentile of prices paid by adopters relative to non-adopters, controlling for customer and year-month fixed effects. The percentile of price paid is calculated by first constructing the empirical price distribution for each month, based on all transactions in that month involving SKUs from the focal category. Each transaction price is then ranked relative to this monthly price distribution to obtain its percentile.

%To calculate the percentile of price paid, the price distribution for a given month is constructed as the empirical distribution of prices observed across all transactions occurring during that month that contain purchases of SKUs in the focal category.  
 
\begin{table}[htp!]
   \caption{Two-Way Fixed Effects Model Estimates - Percentile\_Price\_Paid}
   \label{tab:twfe_percentile_price}
   \centering
      \footnotesize{
 \begin{tabular}{lcccc}
      \tabularnewline \midrule \midrule
      Dependent Variable: & \multicolumn{4}{c}{Percentile\_Price\_Paid}\\
                         & Dry Dog Food      & Dry Cat Food      & Dog Hygiene & Cat Hygiene \\   
      Model:             & (1)           & (2)           & (3)         & (4)\\  
      \midrule
      \emph{Variables}\\
       $\mathcal{I}\{\textrm{Adopter}_{i}\} \times \mathcal{I}\{ t \geq T_{i}^{\textrm{adopt}}\}$ & -1.437$^{**}$ & -3.318$^{**}$ & -0.9569     & -4.104$^{***}$\\   
                         & (0.4141)      & (0.9727)      & (0.6351)    & (1.100)\\   
      \midrule
      \emph{Fixed-effects}\\
      Customer           & Yes           & Yes           & Yes         & Yes\\  
      YearMonth          & Yes           & Yes           & Yes         & Yes\\  
      \midrule
      \emph{Fit statistics}\\
      Observations       & 99,233        & 27,623        & 40,183      & 42,596\\  
      R$^2$              & 0.86731       & 0.61190       & 0.50654     & 0.42914\\  
      Within R$^2$       & 0.00057       & 0.00135       & 0.00010     & 0.00126\\  
      \midrule \midrule
      \multicolumn{5}{l}{\emph{Clustered (Customer \& YearMonth) standard-errors in parentheses}}\\
      \multicolumn{5}{l}{\emph{Signif. Codes: ***: 0.001, **: 0.01, *: 0.05}}\\
   \end{tabular}
   }
\end{table}

\newpage
\section{Appendix to Model and Estimation}
\label{appsec:model}

\subsection{Proof for Minorization-Maximization Estimation}
\label{appssec:mm.algorithm}

\noindent \textbf{Proof of Theorem.} 
We show that $S(\boldsymbol{\theta};\boldsymbol{\theta}^{(k)})$ is a minorization of $\mathcal{L}(\boldsymbol{\theta}^{(k)})$ at $\boldsymbol{\theta}^{(k)}$ for maximization by checking the following requirements.

\begin{itemize}
    \item $\mathcal{S}(\boldsymbol{\theta}; \boldsymbol{\theta}^{(k)})\leq \mathcal{L}(\boldsymbol{\theta}^{(k)})$. For a given consumer $i$ at purchase occasion $t$, we denote the choice probability of consumer $i$ toward brand $j$ during purchase occasion $t$ as $Pr_{ijt}$:
    \begin{equation}
        l(\boldsymbol{\psi}_{it}; \boldsymbol{y}_{it}) = \prod_{j} \left(\frac{\exp(\psi_{ijt})}{1 + \sum_{j=1}^{J} \exp(U_{ijt})}\right)^{y_{ijt}}
    \end{equation}

    \begin{equation}
        h_{j}(\boldsymbol{\psi}_{it};\boldsymbol{y}_{it}) = \frac{\partial \log l(\boldsymbol{\psi}_{it}; \boldsymbol{y}_{it})}{\partial \psi_{ijt}} = y_{ijt} - Pr_{ijt}
    \end{equation}

        \begin{equation}
        h_{jk}(\boldsymbol{\psi}_{it};\boldsymbol{y}_{it}) = \frac{\partial^2 \log l(\boldsymbol{\psi}_{it}; \boldsymbol{y}_{it})}{\partial \psi_{ijt} \partial \psi_{ikt}} = \begin{cases} 
- Pr_{ijt}(1 - Pr_{ikt}), & j = k, \\
Pr_{ijt}Pr_{ikt}, & j \neq k.
\end{cases}
    \end{equation}

The Taylor expansion of $\log l(\boldsymbol{\psi}_{it})$ at $\tilde{\boldsymbol{\psi}}_{it}$ is
\begin{equation}
    \log l(\boldsymbol{\psi}_{it}; \boldsymbol{y}_{it}) =  \log l(\tilde{\boldsymbol{\psi}}_{it}; \boldsymbol{y}_{it}) + (\boldsymbol{\psi}_{it} - \tilde{\boldsymbol{\psi}}_{it})' \nabla \log l(\tilde{\boldsymbol{\psi}}_{it}; \boldsymbol{y}_{it}) \nonumber +  \frac{1}{2} (\boldsymbol{\psi}_{it} - \tilde{\boldsymbol{\psi}}_{it})' \nabla^2 \log l(\boldsymbol{\psi}^{*}_{it}; \boldsymbol{y}_{it}) (\boldsymbol{\psi}_{it} - \tilde{\boldsymbol{\psi}}_{it}).
\end{equation}
Where
\begin{equation}
\begin{aligned}
\nabla^2 \log l(\boldsymbol{\psi}_{it}^{*}; \boldsymbol{y}_{it}) &=
\begin{bmatrix}
h_{11}(\boldsymbol{\psi}_{it}^{*}; \boldsymbol{y}_{it}) & h_{12}(\boldsymbol{\psi}_{it}^{*}; \boldsymbol{y}_{it}) & \cdots & h_{1J}(\boldsymbol{\psi}_{it}^{*}; \boldsymbol{y}_{it}) \\
h_{21}(\boldsymbol{\psi}_{it}^{*}; \boldsymbol{y}_{it}) & h_{22}(\boldsymbol{\psi}_{it}^{*}; \boldsymbol{y}_{it}) & \cdots & h_{2J}(\boldsymbol{\psi}_{it}^{*}; \boldsymbol{y}_{it}) \\
\vdots & \vdots & \ddots & \vdots \\
h_{J1}(\boldsymbol{\psi}_{it}^{*}; \boldsymbol{y}_{it}) & h_{J2}(\boldsymbol{\psi}_{it}^{*}; \boldsymbol{y}_{it}) & \cdots & h_{JJ}(\boldsymbol{\psi}_{it}^{*}; \boldsymbol{y}_{it})
\end{bmatrix} \nonumber \\
&=
\begin{bmatrix}
-{Pr}_{i1t}(1 - {Pr}_{i1t}) & {Pr}_{i1t}{Pr}_{i2t} & \cdots & {Pr}_{i1t}{Pr}_{iJt} \\
{Pr}_{i2t}{Pr}_{i1t} & -{Pr}_{i2t}(1 - {Pr}_{i2t}) & \cdots & {Pr}_{i2t}{Pr}_{iJt} \\
\vdots & \vdots & \ddots & \vdots \\
{Pr}_{iJt}{Pr}_{i1t} & {Pr}_{iJt}{Pr}_{i2t} & \cdots & -{Pr}_{iJt}(1 - {Pr}_{iJt})
\end{bmatrix}
\end{aligned}
\end{equation}

We have $\nabla^2 \log l(\boldsymbol{\psi}_{it}^{*}; \boldsymbol{y}_{it}) \geq -I$, i.e., $\nabla^2 \log l(\boldsymbol{\psi}_{it}^{*}; \boldsymbol{y}_{it}) + I$ is semi-positive definite.

Therefore,
\begin{align}
\log l(\boldsymbol{\psi}_{it}; \boldsymbol{y}_{it}) &\geq \log l(\tilde{\boldsymbol{\psi}}_{it}; \boldsymbol{y}_{it}) + (\boldsymbol{\psi}_{it} - \tilde{\boldsymbol{\psi}}_{it})' \nabla \log l(\tilde{\boldsymbol{\psi}}_{it}; \boldsymbol{y}_{it}) \nonumber - \frac{1}{2} (\boldsymbol{\psi}_{it} - \tilde{\boldsymbol{\psi}}_{it})' (\boldsymbol{\psi}_{it} - \tilde{\boldsymbol{\psi}}_{it}) \nonumber \\
&= \log l(\tilde{\boldsymbol{\psi}}_{it}; \boldsymbol{y}_{it}) - (\tilde{\boldsymbol{\psi}}_{it} - \boldsymbol{\psi}_{it})' \nabla \log l(\tilde{\boldsymbol{\psi}}_{it}; \boldsymbol{y}_{it}) - \frac{1}{2} (\tilde{\boldsymbol{\psi}}_{it} - \boldsymbol{\psi}_{it})' (\tilde{\boldsymbol{\psi}}_{it} - \boldsymbol{\psi}_{it}) \nonumber \\
&= \log l(\tilde{\boldsymbol{\psi}}_{it}; \boldsymbol{y}_{it}) - \sum_j h_j(\tilde{\boldsymbol{\psi}}_{it}; \boldsymbol{y}_{it})( \tilde{\psi}_{ijt} - \psi_{ijt}) \nonumber - \frac{1}{2} \sum_j (\tilde{\psi}_{ijt} - \psi_{ijt})^2 \nonumber \\
&= \log l(\tilde{\boldsymbol{\psi}}_{it}; \boldsymbol{y}_{it}) + \frac{1}{2} \sum_j h_j(\tilde{\boldsymbol{\psi}}_{it}; \boldsymbol{y}_{it})^2 - \frac{1}{2} \sum_j (\tilde{\psi}_{ijt} +h_j(\tilde{\boldsymbol{\psi}}_{it}; \boldsymbol{y}_{it})- \psi_{ijt})^2. 
\end{align}

For now, we derive a lower bound for $\log l(\boldsymbol{\psi}_{it};\boldsymbol{y}_{it})$, the log-likelihood for each choice instance (i.e., customer-month purchase occasion). The log-likelihood function is its sum over user $i$ and time $t$. We substitute $\boldsymbol{\psi}_{it}$ and $\tilde{\boldsymbol{\psi}}_{it}$ with the nominal utility $\boldsymbol{x}_{ijt}'\boldsymbol{\alpha} + \delta_{ij}$ and nominal utility evaluated at the current estimates in the $k_{th}$ iteration $\boldsymbol{x}_{ijt}'\boldsymbol{\alpha}^{(k)} + \delta_{ij}^{(k)}$, respectively. Summing over all customers over all periods, the log-likelihood $\mathcal{L}(\boldsymbol{\theta})$ satisfies

\begin{equation}
\begin{aligned}
\mathcal{L}(\boldsymbol{\theta}) & \geq \mathcal{L}(\boldsymbol{\theta}^{(k)}) + \frac{1}{2}\sum_{i,j,t} h_{j}(\tilde{\boldsymbol{\psi}}_{it}^{(k)}; \boldsymbol{y}_{it})^2 - \frac{1}{2}\sum_{i,j,t}(\tilde{\psi}_{ijt}^{(k)} + h_{j}(\tilde{\boldsymbol{\psi}}_{it}^{(k)};\boldsymbol{y}_{it}) - \boldsymbol{x}_{ijt}' \boldsymbol{\alpha} - \delta_{ij})^2 \\
& = \mathcal{L}(\boldsymbol{\theta}^{(k)}) + \frac{1}{2}\sum_{i,j,t} h_{j}(\boldsymbol{x}_{ijt}'\boldsymbol{\alpha}^{(k)} + \delta_{ij}^{(k)}; \boldsymbol{y}_{it})^2 \\
& - \frac{1}{2}\sum_{i,j,t}(\boldsymbol{x}_{ijt}'\boldsymbol{\alpha}^{(k)} + \delta_{ij}^{(k)} + h_{j}(
\boldsymbol{x}_{ijt}'\boldsymbol{\alpha}^{(k)} + \delta_{ij}^{(k)}; \boldsymbol{y}_{it})
- \boldsymbol{x}_{ijt}' \boldsymbol{\alpha} - \delta_{ij})^2 \\
\end{aligned}
\end{equation}
where the RHS of the above inequality is $S(\boldsymbol{\theta}; \boldsymbol{\theta}^{(k)})$.

\item $S(\boldsymbol{\theta}^{(k)}; \boldsymbol{\theta}^{(k)}) = \mathcal{L}(\boldsymbol{\theta}^{(k)})$.

By definition, 
\begin{equation}
\begin{aligned}
    S(\boldsymbol{\theta}^{(k)}; \boldsymbol{\theta}^{(k)}) & = 
    \mathcal{L}(\boldsymbol{\theta}^{(k)}) + \frac{1}{2}\sum_{i,j,t} h_{j}(\tilde{\boldsymbol{\psi}}_{it}^{(k)}; \boldsymbol{y}_{it})^2 - \frac{1}{2}\sum_{i,j,t}(\tilde{\psi}_{ijt}^{(k)} + h_{j}(\tilde{\boldsymbol{\psi}}_{it}^{(k)};\boldsymbol{y}_{it}) - \boldsymbol{x}_{ijt}' \boldsymbol{\alpha}^{(k)} - \delta_{ij}^{(k)})^2 \\
    & =   \mathcal{L}(\boldsymbol{\theta}^{(k)}) + \frac{1}{2}\sum_{i,j,t} h_{j}(\tilde{\boldsymbol{\psi}}_{it}^{(k)}; \boldsymbol{y}_{it})^2  - \frac{1}{2}\sum_{i,j,t} h_{j}(\tilde{\boldsymbol{\psi}}_{it}^{(k)}; \boldsymbol{y}_{it})^2 \\
    & =  \mathcal{L}(\boldsymbol{\theta}^{(k)}) 
\end{aligned}
\end{equation}
\end{itemize}

\newpage
\section{Appendix to Difference-in-Differences (DID) Analysis}
\label{appsec:did}

\subsection{Bootstrapped distribution of ATT estimates}

Since the customer-month level price elasticities are estimates and not directly observed from data, to obtain correct standard errors, we perform a customer-level block bootstrap with 500 replications. This includes (1) resampling customers with replacement; (2) re-estimating the demand model on each bootstrapped dataset; (3) using the estimated coefficients from that replication to compute price elasticities at the customer-month level (e.g., insignificant demand model coefficients evaluated based on the analytical standard error in the current bootstrap iteration is set to 0); (4) estimating the ATT via staggered DID based on these elasticity estimates; (5) plotting and calculating the bootstrapped standard errors for the ATT and demand parameters.

\begin{figure}[htp!]
    \centering
    \caption{Bootstrapped distribution of Overall ATT Estimates based on 500 replications with the point estimate Table \ref{tab:5_CS_Result}.}
    \label{fig:5_Boot_Est}
    \begin{subfigure}{0.32\textwidth}
        \centering
        \caption{Dry Dog Food}
        \includegraphics[width=\linewidth]{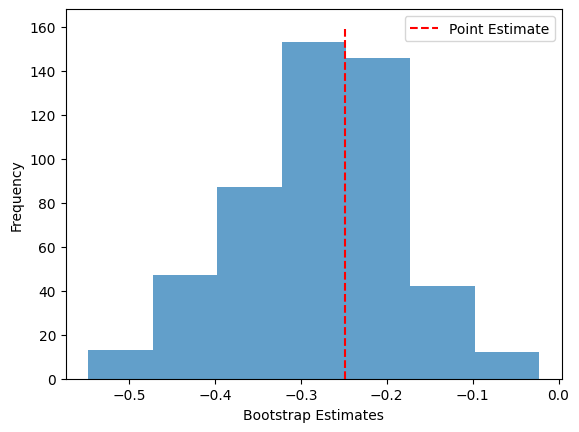}
    \end{subfigure}%
    \hfill
    \begin{subfigure}{0.32\textwidth}
        \centering
        \caption{Dog Hygiene}
        \includegraphics[width=\linewidth]{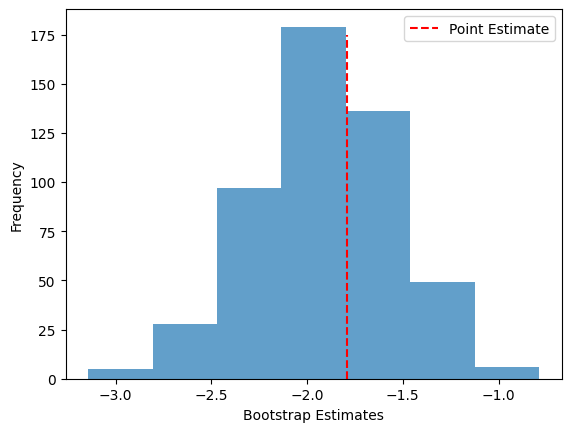}
    \end{subfigure}%
    \hfill
    \begin{subfigure}{0.32\textwidth}
        \centering
        \caption{Cat Hygiene}
        \includegraphics[width=\linewidth]{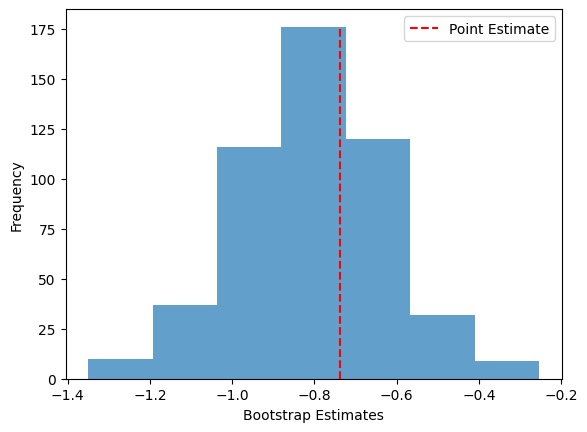}
    \end{subfigure}
\end{figure}

Figure \ref{fig:5_Boot_Est} presents the bootstrapped distribution of ATT estimates — that is, the effect of online adoption on price elasticity, based on 500 replications.

%%% 2 * 2 bootstrap figure
\begin{comment}
\begin{figure}[htp!]
    \centering
    \caption{Bootstrapped distribution of Overall ATT Estimates based on 500 replications with the point estimate Table \ref{tab:5_CS_Result}.}
    \label{fig:5_Boot_Est}
    \begin{subfigure}{0.49\textwidth}
        \centering
        \caption{Dry Dog Food}
        \includegraphics[scale=0.5]{0_Table_Figure_Summary/6_Bootstrap/5_DogSP_Boot_Dist500_CS.png}
    \end{subfigure}%
    ~ %add desired spacing between the subfigures, if needed
    \begin{subfigure}{0.49\textwidth}
        \centering
        \caption{Dry Cat Food}
        \includegraphics[scale=0.5]{0_Table_Figure_Summary/6_Bootstrap/5_CatSP_Boot_Dist500_CS.png}
    \end{subfigure}

    \vspace{1ex} % add some vertical space between the rows

    \begin{subfigure}{0.49\textwidth}
        \centering
        \caption{Dog Hygiene}
        \includegraphics[scale=0.5]{0_Table_Figure_Summary/6_Bootstrap/5_HM_Boot_Dist500_CS.png}
    \end{subfigure}%
    ~ %add desired spacing between the subfigures, if needed
    \begin{subfigure}{0.49\textwidth}
        \centering
        \caption{Cat Hygiene}
        \includegraphics[scale=0.5]{0_Table_Figure_Summary/6_Bootstrap/5_Sand_Boot_Dist500_CS.png}
    \end{subfigure}
\end{figure}
\end{comment}

%%% TWFE including cat
\begin{comment}
\begin{table}[htp!]
   \caption{Two-Way Fixed Effects Estimate - Offline Elasticity}
   \centering
   \label{tab:5_Main_Result_TWFE}
   \footnotesize{
\begin{tabular}{lcccc}
      \tabularnewline \midrule \midrule
      Dependent Variable: & \multicolumn{4}{c}{Offline Price Elasticity}\\
                         & Dry Dog Food    & Dry Cat Food    & Dog Hygiene    & Cat Hygiene \\   
      Model:             & (1)             & (2)             & (3)            & (4)\\  
      \midrule
      \emph{Variables}\\
      PostOnlineAdoption & -0.2145$^{**}$ & -0.0206$^{***}$ & -1.544$^{***}$ & -0.6368$^{***}$\\   
                         & (0.0697)        & (0.0038)        & (0.3308)       & (0.1605)\\   
      \midrule
      \emph{Fixed-effects}\\
      Customer       & Yes             & Yes             & Yes            & Yes\\  
      YearMonth         & Yes             & Yes             & Yes            & Yes\\  
      \midrule
      \emph{Fit statistics}\\
      Observations       & 136,411         & 45,443          & 78,253         & 55,410\\  
      R$^2$              & 0.91126         & 0.95578         & 0.76395        & 0.82317\\  
      Within R$^2$       & 0.24557         & 0.01260         & 0.32091        & 0.32582\\  
      \midrule \midrule
   \multicolumn{5}{l}{\emph{Bootstrapped standard-errors in parentheses via 500 replications}}\\
      \multicolumn{5}{l}{\emph{Signif. Codes: ***: 0.001, **: 0.01, *: 0.05}}\\
   \end{tabular}
   }
\end{table}
\end{comment}

\subsection{Two-Way Fixed Effects (TWFE) Estimates}

Table \ref{tab:5_Main_Result_TWFE} provides TWFE estimates of online adoption on offline price elasticities estimated from the main demand model described in Equation \eqref{eq:util_choice_M2}. 

\begin{table}[htp!]
   \caption{Two-Way Fixed Effects Estimate - Offline Price Elasticity}
   \centering
   \label{tab:5_Main_Result_TWFE}
   \footnotesize{
\begin{tabular}{lccc}
      \tabularnewline \midrule \midrule
      Dependent Variable: & \multicolumn{3}{c}{Offline Elasticity}\\
                         & Dry Dog Food      & Dog Hygiene    & Cat Hygiene \\   
      Model:             & (1)             & (2)             & (3)          \\  
      \midrule
      \emph{Variables}\\
      PostOnlineAdoption & -0.2145$^{**}$  & -1.544$^{***}$ & -0.6368$^{***}$\\   
                         & (0.0697)        & (0.3308)       & (0.1605)\\   
      \midrule
      \emph{Fixed-effects}\\
      CUSTOMER\_ID       & Yes             & Yes             & Yes          \\  
      MONTHSTART         & Yes             & Yes             & Yes            \\  
      \midrule
      \emph{Fit statistics}\\
      Observations       & 136,411                 & 78,253         & 55,410\\  
      R$^2$              & 0.91126                & 0.76395        & 0.82317\\  
      Within R$^2$       & 0.24557              & 0.32091        & 0.32582\\  
      \midrule \midrule
   \multicolumn{4}{l}{\emph{Bootstrapped standard-errors in parentheses via 500 replications}}\\
      \multicolumn{4}{l}{\emph{Signif. Codes: ***: 0.001, **: 0.01, *: 0.05}}\\
   \end{tabular}
   }
\end{table}

\newpage

\clearpage

\section{Appendix to Generalizability and Robustness Checks}
\label{appsec:robustness}

\subsection{Detailed Results for Cohort-Specific, COVID-19, and Black Friday Adopters}
\label{appsec:covid_att}

\begin{figure}[htp!]
    \centering
     \caption{Cohort-Specific ATT Estimates with Bootstrapped 95\% Confidence Intervals}
\label{fig:AE_Cohort_Specific_ATT}   \includegraphics[width=0.5\linewidth]{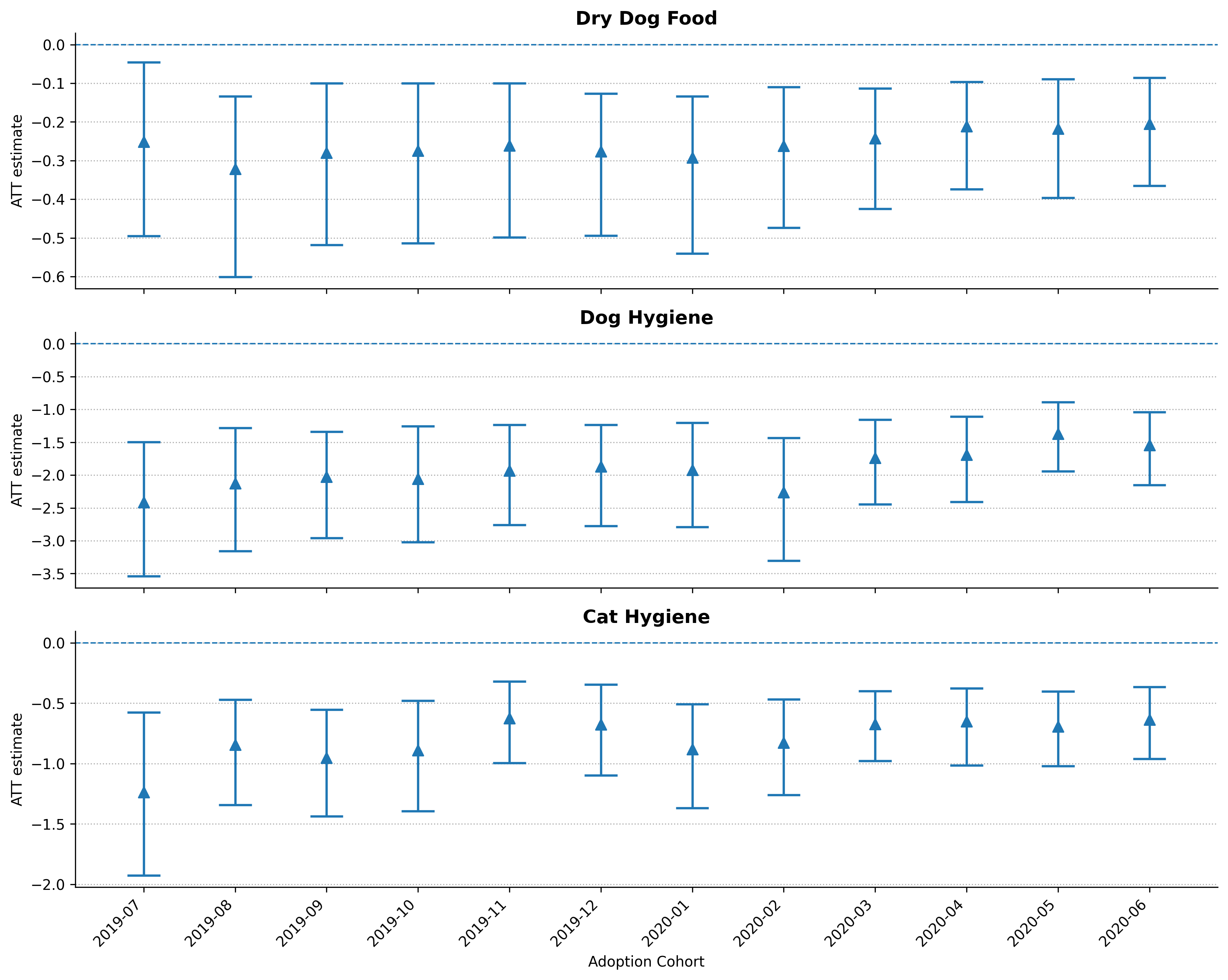}
\end{figure}

\noindent \textbf{Cohort-Specific ATT Estimates.} First, we report cohort-specific ATT estimates using \cite{callaway2021difference} in Table \ref{tab:CS_Result_Cohort_Specific} and the corresponding 95\% bootstrap confidence intervals in Figure~\ref{fig:AE_Cohort_Specific_ATT}. The estimates are consistently negative and statistically significant across all 12 adoption cohorts and three product categories, suggesting that our main findings are not driven by a particular adoption cohort.

%As shown in Table~\ref{tab:att_price_elasticity_cohort_BF_Covid} and Figure~\ref{fig:AE_Cohort_Specific_ATT}, both event-induced groups exhibit the same pattern as in the main analysis: after online adoption, customers become more price elastic in their offline brand choices. The only exception is Cat Hygiene for Black Friday adopters, where the estimate is imprecise due to the very small sample size of Black Friday adopters in this category.
%Overall, these results help address the selection concern by showing that the estimated increase in offline price sensitivity is robust across adoption cohorts and also holds for more plausibly event-induced adopters.

\begin{table}[htp!]
   \caption{ATT of Online Adoption on Offline Price Elasticity Over All Post-Adoption Period (CS Estimator) -- Cohort-Specific Estimates}
   \centering
   \label{tab:CS_Result_Cohort_Specific}
   \footnotesize{
   \begin{tabular}{lccc}
      \midrule \midrule
      Dependent Variable: & \multicolumn{3}{c}{Offline Price Elasticity} \\
                          & \multicolumn{1}{c}{\makecell{Dry Dog Food}}
                          & \multicolumn{1}{c}{\makecell{Dog Hygiene}}
                          & \multicolumn{1}{c}{\makecell{Cat Hygiene}} \\
      Model:              & (1) & (2) & (3) \\
      \midrule
      Adoption Cohort     &     &     &     \\
      2019-07 & \makecell{-0.2516$^{*}$\\(0.1061)}
              & \makecell{-2.4115$^{***}$\\(0.5203)}
              & \makecell{-1.2383$^{***}$\\(0.3556)} \\
      2019-08 & \makecell{-0.3213$^{*}$\\(0.1279)}
              & \makecell{-2.1267$^{***}$\\(0.4816)}
              & \makecell{-0.8431$^{***}$\\(0.2208)} \\
      2019-09 & \makecell{-0.2796$^{*}$\\(0.1103)}
              & \makecell{-2.0271$^{***}$\\(0.4088)}
              & \makecell{-0.9531$^{***}$\\(0.2248)} \\
      2019-10 & \makecell{-0.2742$^{*}$\\(0.1076)}
              & \makecell{-2.0577$^{***}$\\(0.4531)}
              & \makecell{-0.8899$^{***}$\\(0.2278)} \\
      2019-11 & \makecell{-0.2612$^{*}$\\(0.1048)}
              & \makecell{-1.9306$^{***}$\\(0.3950)}
              & \makecell{-0.6237$^{***}$\\(0.1693)} \\
      2019-12 & \makecell{-0.2763$^{**}$\\(0.0964)}
              & \makecell{-1.8661$^{***}$\\(0.4023)}
              & \makecell{-0.6752$^{***}$\\(0.1867)} \\
      2020-01 & \makecell{-0.2921$^{**}$\\(0.1056)}
              & \makecell{-1.9222$^{***}$\\(0.4103)}
              & \makecell{-0.8818$^{***}$\\(0.2250)} \\
      2020-02 & \makecell{-0.2615$^{**}$\\(0.0953)}
              & \makecell{-2.2618$^{***}$\\(0.4894)}
              & \makecell{-0.8275$^{***}$\\(0.1958)} \\
      2020-03 & \makecell{-0.2423$^{**}$\\(0.0834)}
              & \makecell{-1.7370$^{***}$\\(0.3385)}
              & \makecell{-0.6717$^{***}$\\(0.1573)} \\
      2020-04 & \makecell{-0.2108$^{**}$\\(0.0739)}
              & \makecell{-1.6918$^{***}$\\(0.3347)}
              & \makecell{-0.6486$^{***}$\\(0.1569)} \\
      2020-05 & \makecell{-0.2175$^{**}$\\(0.0793)}
              & \makecell{-1.3697$^{***}$\\(0.2701)}
              & \makecell{-0.6916$^{***}$\\(0.1628)} \\
      2020-06 & \makecell{-0.2047$^{**}$\\(0.0713)}
              & \makecell{-1.5432$^{***}$\\(0.3047)}
              & \makecell{-0.6349$^{***}$\\(0.1468)} \\
      \midrule \midrule
      \multicolumn{4}{l}{\emph{Bootstrapped standard errors in parentheses via 500 replications.}} \\
      \multicolumn{4}{l}{\emph{Significance codes: ***: 0.001, **: 0.01, *: 0.05.}} \\
   \end{tabular}
   }
\end{table}

%Specifically, we identify (1) adopters due to online Black Friday sales; (2) covid adopters who adopt online channel between the onset of covid (2020-03-16) and 2020-04-30. We still obtain consistent findings with the main results, showing that irrespective of reasons for online adoption, consumers become more offline price elastic after the online adoption. The sample size for cat hygiene is insignificant due to a small sample size of Black Friday adopters (n = 13). 

\noindent \textbf{ATT Estimates for Event-Induced Adopters} Second, we examine consumers who adopted online shopping during two major events: the retailer's Black Friday online promotion on November 29, 2019, and the onset of COVID-19. Specifically, we identify Black Friday adopters who first purchased online during November 29, 2019, and COVID-19 adopters who first purchased online between March 16, 2020 and April 30, 2020, respectively. These events may have encouraged online adoption for reasons beyond consumers' underlying preferences for online shopping, providing an additional robustness check against potential selection concerns. 

Table~\ref{tab:att_price_elasticity_cohort_BF_Covid} presents the corresponding ATT estimates. Consistent with our main findings, both Black Friday and COVID-19 adopters exhibit increased offline price sensitivity following online adoption. The estimates are negative and statistically significant across all three categories, except for Black Friday adopters in Cat Hygiene, where the estimate is statistically insignificant, potentially due to the small sample size (n = 13). Overall, these findings suggest that the increase in offline price sensitivity is robust across adoption cohorts and also holds among consumers whose adoption may have been induced by external events.

\begin{table}[!htp]
\centering
\caption{ATT of Online Adoption on Offline Price Elasticity Over All Post-Adoption Period (CS Estimator) -- Cohort-Specific Estimates}
\label{tab:att_price_elasticity_cohort_BF_Covid}
\footnotesize{
\begin{tabular}{lccc}
\toprule\toprule
\multicolumn{1}{l}{Dependent Variable:} 
& \multicolumn{3}{c}{Offline Price Elasticity} \\
\cmidrule(lr){2-4}
& Dry Dog Food & Dog Hygiene & Cat Hygiene \\
Model: & (1) & (2) & (3) \\
\midrule
Adoption Cohort & & & \\

Black Friday 
& \makecell{-0.1927$^{*}$ \\ (0.0856)}
& \makecell{-1.8646$^{***}$ \\ (0.4778)}
& \makecell{-0.3389 \\ (0.1926)} \\

COVID-19 (between 2020-03-16 and 2020-03-31)
& \makecell{-0.2155$^{**}$ \\ (0.0730)}
& \makecell{-1.6422$^{***}$ \\ (0.3225)}
& \makecell{-0.5736$^{***}$ \\ (0.1431)} \\

COVID-19 (between 2020-04-01 and 2020-04-30) & \makecell{-0.2108$^{**}$\\(0.0739)}
& \makecell{-1.6918$^{***}$\\(0.3347)}
& \makecell{-0.6486$^{***}$\\(0.1569)} \\

\bottomrule\bottomrule
\multicolumn{4}{l}{\emph{Bootstrapped standard errors in parentheses via 500 replications.}} \\
      \multicolumn{4}{l}{\emph{Significance codes: ***: 0.001, **: 0.01, *: 0.05.}} \\
\end{tabular}
}
\end{table}

\subsection{Detailed Results for Potential Mechanisms}
\label{appsec:potential_mechanisms}

\subsubsection{Detailed Results for the Role of Price Difference Across Channels}
\label{appssec:cross.channel.price.differences}

\begin{figure}[htp!]
    \centering
    \caption{Distribution of Online and Offline Prices (Natural Log Scale) over SKU-Date - PDF}
    \label{fig:A9_Online_Offline_LogPrice}
    \begin{subfigure}{0.32\textwidth}
        \centering
        \caption{Dry Dog Food}
        \includegraphics[width=\linewidth]{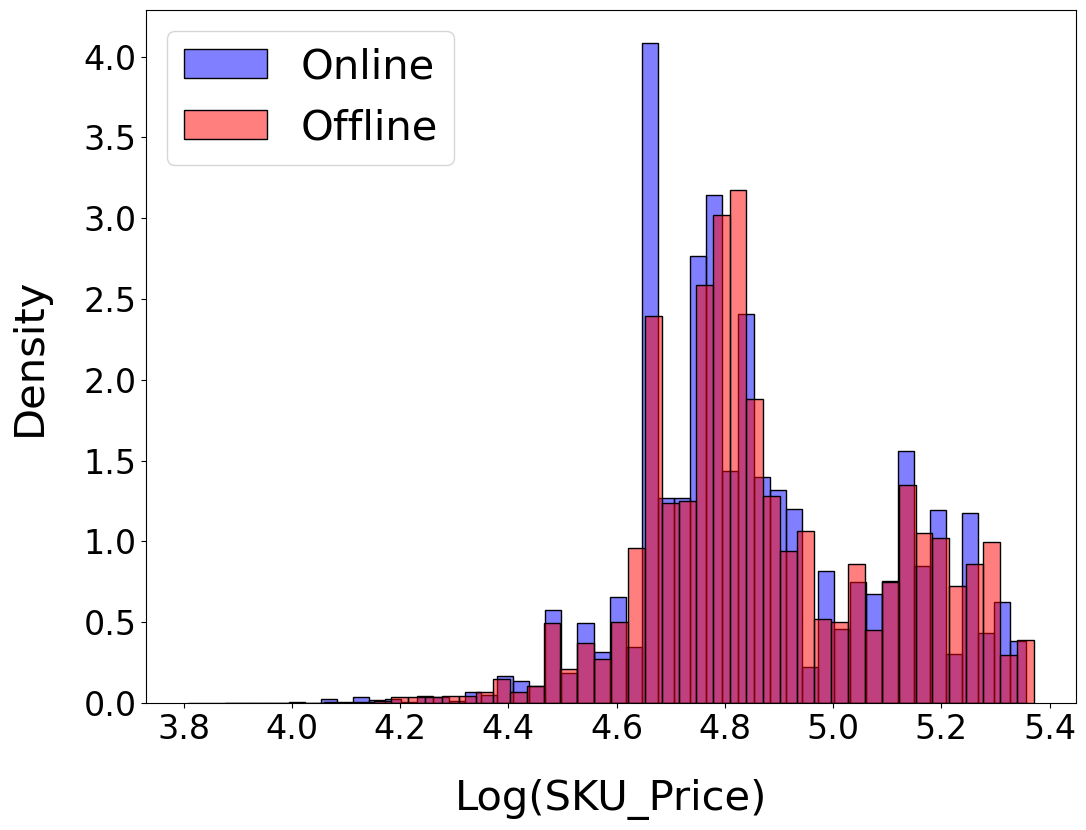}
    \end{subfigure}%
    \hfill
    \begin{subfigure}{0.32\textwidth}
        \centering
        \caption{Dog Hygiene}
        \includegraphics[width=\linewidth]{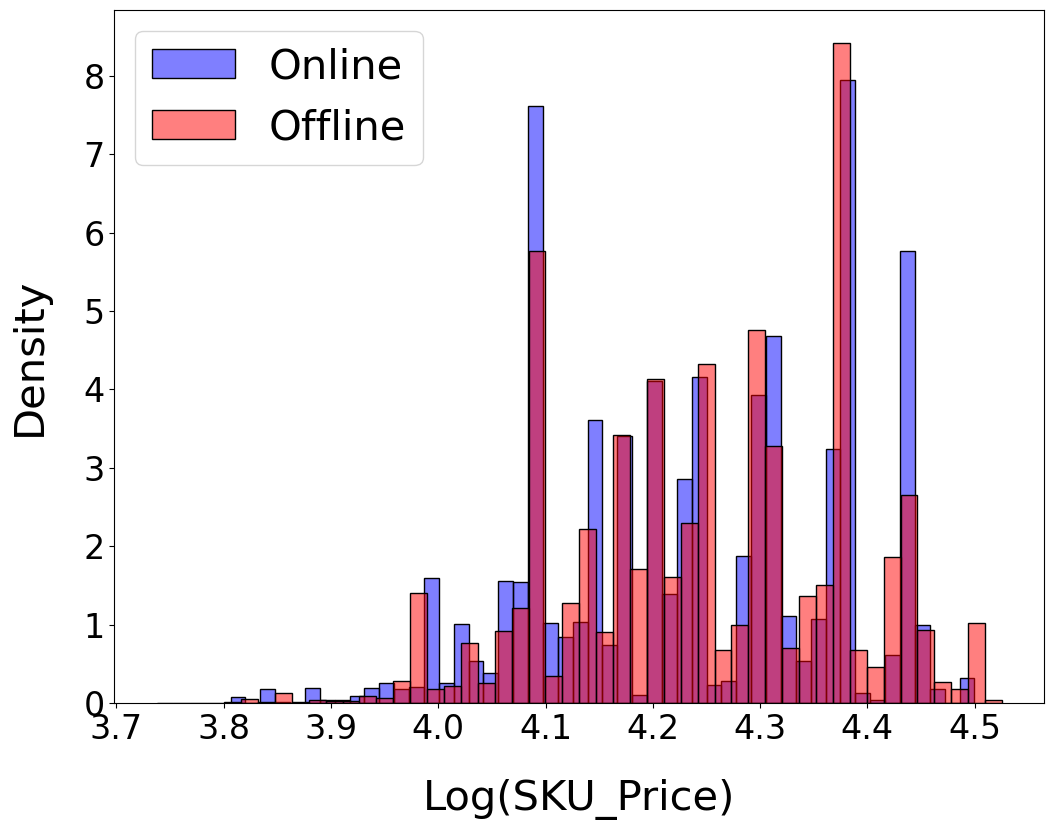}
    \end{subfigure}%
    \hfill
    \begin{subfigure}{0.32\textwidth}
        \centering
        \caption{Cat Hygiene}
        \includegraphics[width=\linewidth]{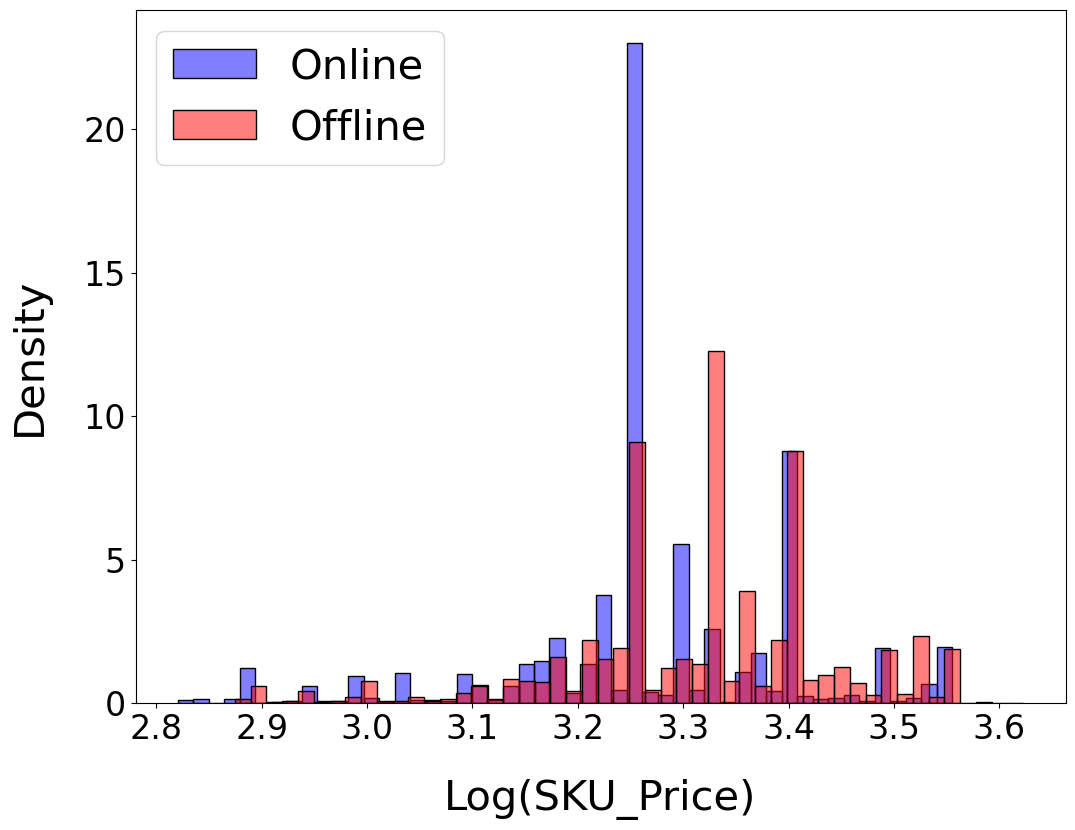}
    \end{subfigure}
\end{figure}

Figure \ref{fig:A9_Online_Offline_LogPrice} presents the probability density function (PDF) of offline and online prices (in natural log scale) for each category at the SKU-Store-Date level. While there is substantial price variation across SKUs, the central tendency, skewness, and dispersion of distributions for online and offline channels are quite similar. The substantial overlap in online and offline price distribution suggests no systematic difference between online and offline prices overall. %Furthermore, we can visually observe that the price differences across SKUs are significantly larger than those across channels. This finding also aligns with \cite{cavallo2017online}, which observes that within a multichannel retailer, online and offline prices are identical approximately 72\% of the time.

\begin{comment}
\begin{figure}[htp!]
    \centering
    \caption{Distribution of Online and Offline Prices (Natural Log Scale) over SKU-Date - PDF}
    \label{fig:A9_Online_Offline_LogPrice}
    \begin{subfigure}{0.49\textwidth}
        \centering
        \caption{Dry Dog Food}
        \includegraphics[scale=0.28]{0_Appendix_Table_Figure_Summary/A9_Online_Offline_Price_Figure/9_Dog_PDF_Log_SKU_Price.png}
    \end{subfigure}%
    ~ %add desired spacing between the subfigures, if needed
    \begin{subfigure}{0.49\textwidth}
        \centering
        \caption{Dry Cat Food}
        \includegraphics[scale=0.28]{0_Appendix_Table_Figure_Summary/A9_Online_Offline_Price_Figure/9_Cat_PDF_Log_SKU_Price.png}
    \end{subfigure}

    \vspace{1ex} % add some vertical space between the rows
    \begin{subfigure}{0.49\textwidth}
        \centering
        \caption{Dog Hygiene}
        \includegraphics[scale=0.28]{0_Appendix_Table_Figure_Summary/A9_Online_Offline_Price_Figure/9_HM_PDF_Log_SKU_Price.png}
    \end{subfigure}%
    ~ %add desired spacing between the subfigures, if needed
    \begin{subfigure}{0.49\textwidth}
        \centering
        \caption{Cat Hygiene}
        \includegraphics[scale=0.28]{0_Appendix_Table_Figure_Summary/A9_Online_Offline_Price_Figure/9_Sand_PDF_Log_SKU_Price.png}
    \end{subfigure}
\end{figure}
\end{comment}

To formally compare offline and online prices, we construct a sample at the SKU-Channel-Store-Date level. Each row represents the price of a specific SKU on a particular channel (online or offline) at a specific store and date. We then estimate the following model:
\begin{equation}
\label{eq:sku_price_regression}
\ln(\textrm{SKU\_Price}_{kstc}) = \beta \times \textrm{isOnline}_{c} + \textrm{SKU}_{k} + \textrm{Date}_{t} + \epsilon_{kstc}   
\end{equation}
where the $\textrm{SKU\_Price}_{kstc}$ refers to the selling price of SKU $k$ of store $s$ at time $t$. The index $c$ refers to the online or offline channel. Here, a negative $\beta$ would indicate that, on average, online prices are lower than offline prices for the same SKU on the same date, and vice versa. Table \ref{tab:A9_price_online_offline_diff} presents the regression result for comparing online and offline prices in Equation \eqref{eq:sku_price_regression} for three categories. Table \ref{tab:A9_price_anova} presents the regression results of decomposing SKU price variation for each category. Column (1) of Table \ref{tab:A9_price_anova} only includes the channel indicator, \textit{isOnline}, and Column (2) includes both channel indicator and SKU fixed effects. Comparing the R-squared values in Table \ref{tab:A9_price_anova}, we see that the price variation across channels is far smaller than the price variation across SKUs. Nevertheless, we see that online prices are lower on average by 3.2\%--6.1\% across categories. 

\begin{table}[htp!]
   \caption{Descriptive Analysis - Online vs. Offline Prices at the SKU-Date Level}
    \label{tab:A9_price_online_offline_diff}
   \centering
   \footnotesize{
   \begin{tabular}{lccc}
      \tabularnewline \midrule \midrule
      Dependent Variable: & \multicolumn{3}{c}{ln(SKU\_Price)}\\
                   & Dry Dog Food    & Dog Hygiene   & Cat Hygiene \\   
      Model:       & (1)             & (2)             & (3)        \\  
      \midrule
      \emph{Variables}\\
      isOnline     & -0.0322$^{***}$  & -0.0361$^{*}$ & -0.0607$^{***}$\\   
                   & (0.0022)              & (0.0135)      & (0.0087)\\   
      \midrule
      \emph{Fixed-effects}\\
      SKU          & Yes                     & Yes           & Yes\\  
      DATE         & Yes                       & Yes           & Yes\\  
      \midrule
      \emph{Fit statistics}\\
      Observations & 554,191                 & 328,323       & 244,723\\  
      R$^2$        & 0.95202               & 0.98493       & 0.91763\\  
      Within R$^2$ & 0.01819              & 0.00912       & 0.02441\\  
      \midrule \midrule
      \multicolumn{4}{l}{\emph{Clustered (SKU \& DATE) standard-errors in parentheses}}\\
      \multicolumn{4}{l}{\emph{Signif. Codes: ***: 0.001, **: 0.01, *: 0.05}}\\
   \end{tabular}
   }
\end{table}

\begin{table}[htp!]
   \caption{Descriptive Analysis - SKU-Level Price Variation Decomposition. Compared to column 1, column 2 controls for SKU fixed effects, indicating that after holding SKU constant, the price variation attributable to the online-offline channel is very small.}
   \label{tab:A9_price_anova}
   \centering
   \footnotesize{
   \begin{tabular}{lcccccc}
      \tabularnewline \midrule \midrule
      Dependent Variable: & \multicolumn{6}{c}{ln(SKU\_Price)}\\
       & \multicolumn{2}{c}{Dry Dog Food}  & \multicolumn{2}{c}{Dog Hygiene} & \multicolumn{2}{c}{Cat Hygiene} \\ 
      Model:       & (1)           & (2)                       & (1)                   & (2)           & (1)           & (2)\\  
      \midrule
      \emph{Variables}\\
      Constant     & 4.941$^{***}$ &                              & 4.043$^{***}$         &               & 3.224$^{***}$ &   \\   
                   & (0.0439)      &                              & (0.1420)              &               & (0.0769)      &   \\   
      isOnline     & 0.0659$^{*}$  & -0.0313$^{***}$  & -0.0123               & -0.0382$^{*}$ & -0.0726$^{*}$ & -0.0594$^{***}$\\   
                   & (0.0287)      & (0.0025)         & (0.0269)              & (0.0135)      & (0.0304)      & (0.0087)\\   
      \midrule
      \emph{Fixed-effects}\\
      SKU          &               & Yes                      &                       & Yes           &               & Yes\\  
      \midrule
      \emph{Fit statistics}\\
      Observations & 554,191       & 554,191               & 328,323               & 328,323       & 244,723       & 244,723\\  
      R$^2$        & 0.00398       & 0.89516                & $1.61\times 10^{-5}$  & 0.95365       & 0.00309       & 0.84228\\  
      Within R$^2$ &               & 0.00797              &                       & 0.00333       &               & 0.01235\\  
      \midrule \midrule
      \multicolumn{7}{l}{\emph{Clustered (SKU \& DATE) standard-errors in parentheses}}\\
      \multicolumn{7}{l}{\emph{Signif. Codes: ***: 0.001, **: 0.01, *: 0.05}}\\
   \end{tabular}
   }
\end{table}

We next examine whether these average price differences are associated with differences in promotional activity across channels. The conclusion is that the small average price difference does not reflect a persistent difference in base prices. For approximately 85\% of store-SKU-day observations, the same SKU has an identical price across channels. Thus, it appears to arise primarily from more frequent promotions in the online channel. To examine this further, we measure the frequency of promotions the fraction of SKU-week observations with a promotion in each channel. Following \cite{hitsch2021prices}, we construct weekly SKU--store price series and classify each observed offline purchase as occurring at either a regular or promotional price (promotion is defined as a 10\% or larger drop from the regular base price). Taking the dry dog food category as an example, we find that 7.3\% of store-SKU-week observations have an offline promotion, compared with 17.5\% of online SKU-week observations with an online promotion. This shows that there are indeed more frequent promotions online. Moreover, only 4.6\% of store-SKU-week observations have concurrent promotions in both channels, indicating limited joint co-ordination of promotions across channels.Conditional on a SKU being promoted in a given week, the average promotion depth is approximately 18\% in both channels, indicating that online and offline promotions are similar in magnitude.

\begin{figure}[htp!]
    \centering
     \caption{CDF of online-offline price difference over each offline purchase of a SKU -- Dry Dog Food}
\label{fig:SE_diff_offline_online_price_paid} 
\includegraphics[width=0.6\linewidth]{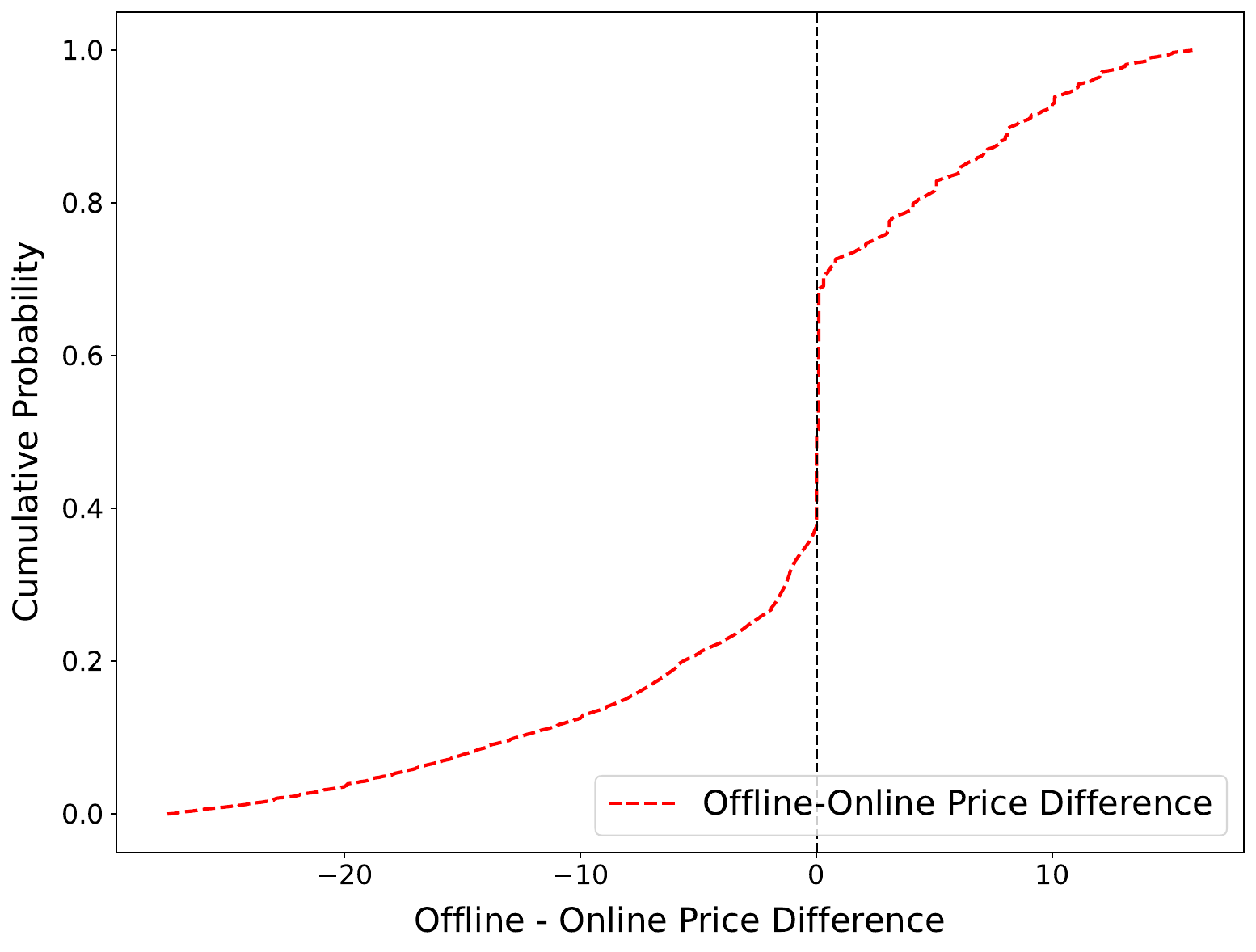}
\end{figure}

Despite lower average online prices and more frequent online promotions, consumers continue to make offline purchases even when online prices are lower. Figure \ref{fig:SE_diff_offline_online_price_paid} presents the cumulative distribution function (CDF) of the offline-online price differences in the week of purchase for each SKu purchased offline, using the dry dog food category as an example. We find that for about 65\% (i.e. a majority) of offline purchase occasions when a SKU was purchased offline, it was available online at the same or lower price. Of these occasions, it was available at a lower price online about 30\% of the time and at the same price online about 35\% of the time. This pattern helps rule out an alternate explanation that consumers who adopt online shopping only purchase a product offline when its price is cheaper than online.

\newpage

\subsubsection{Detailed Results for Price Search and Learning, Switching to Lower Price Brands, Strategic Timing of Purchase during Promotions}
\label{appssec:purchase_behavior_mechanisms}

This section provides supporting descriptive evidence for the potential mechanisms discussed in $\S$\ref{ssec:mechanisms}. The main paper first shows that, conditional on an offline purchase in the focal category, adopters pay lower prices after online adoption (Table~\ref{tab:twfe_price_paid}; see also Web Appendix Table~\ref{tab:twfe_percentile_price} for the corresponding analysis of price percentiles). It also shows that, conditional on an offline purchase occasion, adopters become less likely to purchase the focal category offline (Table~\ref{tab:twfe_buyCategory}). We next examine whether these patterns are associated with (i) more strategic timing of offline purchases around promotions, (ii) substitution toward lower-priced brands, or (iii) reallocation of purchases from the retailer's offline channel to its online channel. Because these analyses are descriptive, we interpret them as evidence consistent with the proposed mechanisms rather than as definitive identification of consumers' underlying search process.

As shown in $\S$~\ref{appssec:cross.channel.price.differences} (Figure~\ref{fig:A9_Online_Offline_LogPrice} and Tables~\ref{tab:A9_price_online_offline_diff} and \ref{tab:A9_price_anova}), online and offline price distributions overlap substantially, and online prices for the same SKU and date are only modestly lower on average. Thus, large or persistent cross-channel price gaps are unlikely to be the primary explanation for the increase in offline price sensitivity. The analyses below instead examine how adopters respond to price variation in their subsequent offline purchases.

\paragraph{Strategic timing of offline purchases.} We first test whether adopters become more likely to purchase offline during promotional-price periods. Following \cite{hitsch2021prices}, we construct weekly SKU--store price series and classify each observed offline purchase as occurring at either a regular or promotional price (promotion is defined as a 10\% or larger drop from the regular base price). We then restrict the sample to offline orders containing a purchase in the focal category and estimate the following customer-order-level regression separately for each category:
\begin{equation}
\label{eq:post_online_adoption}
    \mathcal{I}\{\text{PurchaseWithPromotionPrice}_{it}\} = \text{Customer}_{i} + \text{YearMonth}_{t} + \text{SKU}_{t} + \delta \times \text{PostOnlineAdoption}_{it} + \epsilon_{it}
\end{equation}
Each observation corresponds to an offline order in which customer $i$ purchases a single SKU from the focal category at purchase order $t$. The dependent variable, $\mathcal{I}\{\text{PurchaseWithPromotionPrice}_{it}\}$, equals one if the purchased SKU is offered at a promotional price and zero otherwise.

\begin{table}[htp!]
   \caption{Two-Way Fixed Effects Descriptive Analysis - Purchase with Promotion Price}
   \label{tab:purchase_with_promotion_price}
   \centering
   \footnotesize{
   \begin{tabular}{lccc}
      \tabularnewline \midrule \midrule
      Dependent Variable: & \multicolumn{3}{c}{Purchase with Promotion Price}\\
                         & Dry Dog Food       & Dog Hygiene    & Cat Hygiene \\   
      Model:             & (1)            & (2)            & (3)\\  
      \midrule
      \emph{Variables}\\
      PostOnlineAdoption & 0.0749$^{***}$ & 0.0619$^{***}$ & 0.1086$^{***}$\\   
                         & (0.0114)       & (0.0187)       & (0.0171)\\
      \midrule
      \emph{Fixed-effects}\\
      Customer           & Yes            & Yes            & Yes\\  
      YearMonth          & Yes            & Yes            & Yes\\  
      SKU                & Yes            & Yes            & Yes\\  
      \midrule
      \emph{Fit statistics}\\
      Observations       & 99,233         & 40,183         & 42,596\\  
      R$^2$              & 0.35911        & 0.30840        & 0.28754\\  
      Within R$^2$       & 0.00174        & 0.00111        & 0.00331\\  
      \midrule \midrule
      \multicolumn{4}{l}{\emph{Clustered (Customer) standard-errors in parentheses}}\\
      \multicolumn{4}{l}{\emph{Signif. Codes: ***: 0.001, **: 0.01, *: 0.05}}\\
   \end{tabular}
   }
\end{table}

Table~\ref{tab:purchase_with_promotion_price} shows that online adoption is associated with a 6.2--10.9 percentage-point increase in the probability that an offline category purchase occurs during a promotional-price period. The estimates are positive and statistically significant in all four categories. Holding the purchased SKU fixed, this pattern is consistent with adopters becoming more attentive to price variation and timing their offline purchases more strategically.

\paragraph{Substitution toward lower-priced brands.} We next examine whether the lower prices paid by adopters are instead explained by broad substitution toward cheaper brands. Within each category, we classify brands as high- or low-priced using the median of their average prices over the sample period.\footnote{We do not see a change in price rank for brands over time under each category, and see the price trends for brand prices evolve the same, and lower price brands do not cross the higher price brands during the sample period.} We then estimate
\begin{equation}
\label{eq:high_price_brand_twfe}
\mathcal{I}\{\text{BuyHighPriceBrand}_{it}\}
=
\delta \times \,\text{PostOnlineAdoption}_{it}
+
\textrm{Customer}_i
+
\textrm{YearMonth}_{t}
+
\epsilon_{it}.
\end{equation}

\begin{table}[htp!]
\caption{Two-Way Fixed Effects Descriptive Analysis -- BuyHighPriceBrand}
\centering
\label{tab:2_BuyHighPriceBrand}
\footnotesize{
\begin{tabular}{lccc}
\tabularnewline \midrule \midrule
Dependent Variable: & \multicolumn{3}{c}{BuyHighPriceBrand}\\
& Dry Dog Food & Dog Hygiene & Cat Hygiene \\   
Model:             & (1)      & (2)         & (3)\\  
\midrule
\emph{Variables}\\
PostOnlineAdoption & -0.0080  & -0.0003     & -0.0078\\   
                   & (0.0057) & (0.0098)    & (0.0051)\\   
\midrule
\emph{Fixed-effects}\\
Customer           & Yes      & Yes         & Yes\\  
YearMonth          & Yes      & Yes         & Yes\\  
\midrule
\emph{Fit statistics}\\
Observations       & 99,233   & 40,183      & 42,596\\  
R$^2$              & 0.91257  & 0.71035     & 0.89465\\  
Within R$^2$       & 0.00010  & $8.03\times 10^{-8}$ & 0.00025\\  
\midrule \midrule
\multicolumn{4}{l}{\emph{Clustered (Customer) standard-errors in parentheses}}\\
\multicolumn{4}{l}{\emph{Signif. Codes: ***: 0.001, **: 0.01, *: 0.05}}\\
\end{tabular}
}
\end{table}

Table~\ref{tab:2_BuyHighPriceBrand} provides no evidence that adopters systematically shift away from high-priced brands: none of the three estimates is significantly negative.

\paragraph{Within-retailer channel reallocation.}

The decline in offline category purchasing documented in Table~\ref{tab:twfe_buyCategory} could reflect purchases shifting to the retailer's online channel or customers leaving the focal retailer. To examine these possibilities, we construct a customer-month panel and separately measure store-visit (e.g., whether a customer visits any stores/interacts with the retailer at a given month), purchase incidence for a given category, quantity, and spending across all channels, the offline channel, and the online channel. We present the result of dry dog food and obtain qualitative similar results for other categories. We estimate
\begin{equation}
\label{eq:channel_reallocation_twfe}
y_{it}
=
\delta \times \,\text{PostOnlineAdoption}_{it} +
\text{Customer}_{i} + \text{YearMonth}_{t}
+ \epsilon_{it},
\end{equation}
where $y_{it}$ refers to each of the dependent variable of our interest.

Table~\ref{tab:post_adoption_total_offline_online} shows that total store-visit frequency, category purchase incidence, quantity, and spending increase after online adoption, even as the corresponding offline outcomes decline and online outcomes increase. This pattern is consistent with within-retailer channel reallocation rather than customer attrition. It also clarifies that the lower offline category-purchase frequency in Table~\ref{tab:twfe_buyCategory} does not reflect an overall reduction in customers' engagement with the retailer.

Taken together, the evidence is consistent with online adoption increasing consumers' price awareness and encouraging more strategic timing of subsequent offline purchases. The results are not well explained by large online--offline price gaps or by systematic substitution toward lower-priced brands. Nevertheless, because we do not observe online clickstream data or in-store search behavior, we cannot separately identify whether the effect operates through lower online search costs, exposure to a wider range of prices, or other changes in price attention. These limitations are discussed further in $\S$8.2.

\begin{table}[htp!]
\caption{Post-Adoption Changes in Total, Offline, and Online Purchasing}
\label{tab:post_adoption_total_offline_online}
\centering
\footnotesize{
\begin{tabular}{lccc}
\toprule
Outcome & Total & Offline & Online \\
\midrule
\multicolumn{4}{l}{\emph{Panel A: Store-Visit Frequency}} \\
% (defined as customer-month with a purchase in any category)
PostOnlineAdoption 
& \makecell{0.0905$^{***}$ \\ (0.0057)}
& \makecell{-0.1105$^{***}$ \\ (0.0057)}
& \makecell{0.3160$^{***}$ \\ (0.0053)} \\
\addlinespace

\multicolumn{4}{l}{\emph{Panel B: Dry Dog Food Purchase Incidence}} \\
PostOnlineAdoption 
& \makecell{0.0920$^{***}$ \\ (0.0057)}
& \makecell{-0.1145$^{***}$ \\ (0.0057)}
& \makecell{0.2348$^{***}$ \\ (0.0050)} \\
\addlinespace

\multicolumn{4}{l}{\emph{Panel C: Dry Dog Food Quantity}} \\
PostOnlineAdoption 
& \makecell{0.3936$^{***}$ \\ (0.0596)}
& \makecell{-0.1588$^{**}$ \\ (0.0488)}
& \makecell{0.5524$^{***}$ \\ (0.0329)} \\
\addlinespace

\multicolumn{4}{l}{\emph{Panel D: Dry Dog Food Spend}} \\
PostOnlineAdoption 
& \makecell{22.05$^{***}$ \\ (1.563)}
& \makecell{-27.16$^{***}$ \\ (1.620)}
& \makecell{49.22$^{***}$ \\ (1.493)} \\

\midrule
Customer fixed effects & Yes & Yes & Yes \\
Month fixed effects & Yes & Yes & Yes \\
Observations & 278,966 & 278,966 & 278,966 \\
\bottomrule
%\multicolumn{4}{l}{\footnotesize Notes: This table reports two-way fixed effects estimates at the customer-month level.} \\
\multicolumn{4}{l}{\footnotesize Standard errors clustered at the customer level are reported in parentheses.} \\
%\multicolumn{4}{l}{\footnotesize We define store-visit as long as a customer has made purchase in any category that the retailer offers during a month.} \\
\multicolumn{4}{l}{\footnotesize $^{*}p<0.05$; $^{**}p<0.01$; $^{***}p<0.001$.} \\
\end{tabular}
}
\end{table}

\clearpage

\subsection{Detailed Results for Alternative Demand Model Specifications}
\label{appsec:alternative.demand.model}

\subsubsection{Results for Alternate Demand Specification with Binary Adoption Variable}\label{appsec:alternative.demand.binary}

The alternative demand model specification is given by: 
\begin{equation}
    \begin{aligned}
    \psi_{ijt} & = \alpha_0 \textrm{Price}_{jt} +
    \alpha_{1} \textrm{Post\_OnlineAdoption}_{it} + \alpha_2 \textrm{Price}_{jt} \times \textrm{Post\_OnlineAdoption}_{it} \\
    & +  \alpha_{3} \mathcal{I}\{\textrm{BuyPrevOcca}_{ijt}\}   + \beta_{1}\textrm{CF\_Price}_{jt} +\beta_{2}\textrm{CF\_Price\_Post\_OnlineAdoption}_{ijt} + \delta_{ij} ~~ \forall j \in \{1, \ldots, J\}\\
    \end{aligned}
\label{eq:util_coef_M7.2.1}
\end{equation}

To address the endogeneity of $\textrm{Price}_{jt}$ and its interaction with $\textrm{Post\_OnlineAdoption}_{it}$, we use the control function approach with instruments $\textrm{Cost}_{jt}$ and $\textrm{Cost}_{jt} \times \textrm{Post\_OnlineAdoption}_{it}$, and construct two control function terms, $\textrm{CF\_Price}_{jt}$ and $\textrm{CF\_Price\_Post\_OnlineAdoption}_{ijt}$. 
We then estimate the above demand model and compute the average elasticity at the customer-month level for each product category following the same steps as before. The results are shown below.

Table \ref{tab:append.7.2.1.demand.model} presents the estimates of the demand model. The standard errors for the demand parameters are based on the customer-level block bootstrap with 500 replications. 

\begin{table}[htp!]
\caption{Coefficient Estimates of Demand Model in Equation \eqref{eq:util_coef_M7.2.1} - Binary Online Adoption Variable}
\label{tab:append.7.2.1.demand.model}
\centering
\footnotesize{
\begin{tabular}{lcccc}
\toprule
 & Dry Dog Food & Dry Cat Food & Dog Hygiene & Cat Hygiene \\
\textit{Variables} &  &  &  &  \\
\midrule
Price & \makecell{-0.3031$^{***}$\\(0.0171)} & \makecell{-0.1577$^{***}$\\(0.0202)} & \makecell{-1.0463$^{***}$\\(0.1123)} & \makecell{-0.2038$^{***}$\\(0.0199)} \\
Post\_OnlineAdoption & \makecell{-0.0834\\(0.1341)} & \makecell{-0.2483\\(0.398)} & \makecell{2.1377$^{***}$\\(0.3641)} & \makecell{1.8521$^{***}$\\(0.3083)} \\
Price * Post\_OnlineAdoption & \makecell{-0.034$^{**}$\\(0.0123)} & \makecell{-0.004\\(0.0337)} & \makecell{-1.0684$^{***}$\\(0.159)} & \makecell{-0.1432$^{***}$\\(0.0218)} \\
BuyPrevOcca & \makecell{0.1561$^{***}$\\(0.0273)} & \makecell{-0.4977$^{***}$\\(0.0451)} & \makecell{-0.1802$^{***}$\\(0.0283)} & \makecell{0.3216$^{***}$\\(0.0295)} \\
CF\_Price & \makecell{0.042\\(0.0237)} & \makecell{0.039\\(0.0268)} & \makecell{0.0572\\(0.1622)} & \makecell{0.2797$^{***}$\\(0.0261)} \\
CF\_Price\_Post\_OnlineAdoption & \makecell{0.0521$^{**}$\\(0.0175)} & \makecell{0.038\\(0.0343)} & \makecell{0.6873$^{**}$\\(0.2196)} & \makecell{0.0609$^{*}$\\(0.0256)} \\
Customer\_Brand\_FE & Yes & Yes & Yes & Yes\\ \midrule
Observations       & 136,411         & 45,443          & 78,253         & 55,410\\ 
Log-likelihood & -84085.187 & -30700.989 &  -55426.656 & -37292.32\\
\bottomrule
       \multicolumn{5}{l}{\emph{Bootstrapped standard-errors in parentheses via 500 replications}}\\
      \multicolumn{5}{l}{\emph{Signif. Codes: ***: 0.001, **: 0.01, *: 0.05}}\\
\end{tabular}
}
\end{table}

\begin{figure}[htp!]
    \centering
    \caption{Bootstrapped distribution of Overall ATT Estimates based on 500 replications with the point estimate from Table \ref{tab:7.2.1.CS_Result}.}
    \label{fig:7.2.1.CS.IT.Boot_Est}
    \begin{subfigure}{0.32\textwidth}
        \centering
        \caption{Dry Dog Food}
        \includegraphics[width=\linewidth]{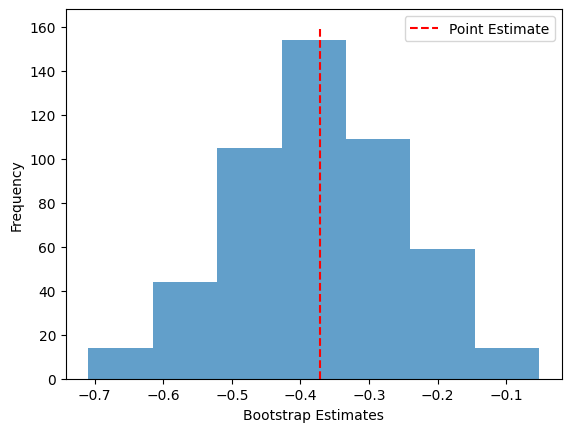}
    \end{subfigure}%
    \hfill
    \begin{subfigure}{0.32\textwidth}
        \centering
        \caption{Dog Hygiene}
        \includegraphics[width=\linewidth]{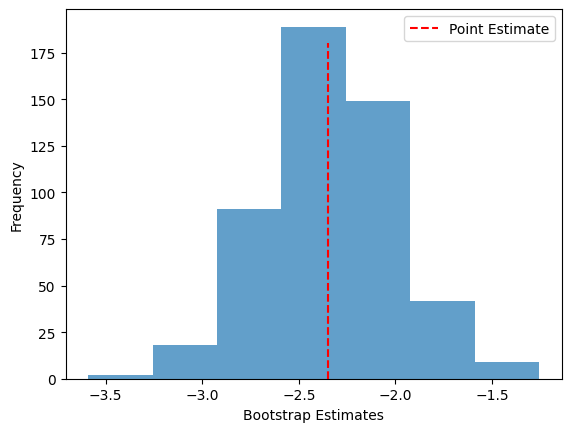}
    \end{subfigure}%
    \hfill
    \begin{subfigure}{0.32\textwidth}
        \centering
        \caption{Cat Hygiene}
        \includegraphics[width=\linewidth]{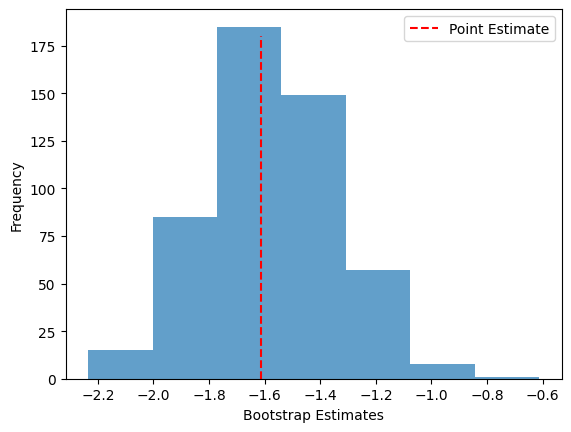}
    \end{subfigure}
\end{figure}

Table \ref{tab:7.2.1.CS_Result} presents the ATT estimates of offline price elasticity from CS Estimator based on the demand model in Equation \eqref{eq:util_coef_M7.2.1}. The standard errors for the ATT estimates and the demand model parameters are based on the customer-level block bootstrap with 500 replications; See Figure \ref{fig:7.2.1.CS.IT.Boot_Est}.

\begin{table}[htp!]
   \caption{ATT of Online Adoption on Offline Price Elasticity (CS Estimator) - Binary Adoption Variable}
   \centering
   \label{tab:7.2.1.CS_Result}
   \footnotesize{
   \begin{tabular}{lccc}
      \tabularnewline \midrule \midrule
      Dependent Variable: & \multicolumn{3}{c}{Offline Price Elasticity}\\
       & \multicolumn{1}{c}{\makecell{Dry Dog Food}}  
       & \multicolumn{1}{c}{\makecell{Dog Hygiene}}
       & \multicolumn{1}{c}{\makecell{Cat Hygiene}}  \\ 
      Model:             & (1)             & (2)             & (3)                   \\  
      \midrule
%      \emph{Variables}\\
      CS               & -0.3711$^{**}$  & -2.3513$^{***}$ & -1.614$^{***}$\\   
                         & (0.1224)        & (0.3407)        & (0.2353)       \\  \midrule
 %     \emph{Fit statistics}\\
     Observations       & 136,411          & 78,253         & 55,410\\  
      \midrule %\midrule
       \multicolumn{4}{l}{\emph{Bootstrapped standard-errors in parentheses via 500 replications}}\\
      \multicolumn{4}{l}{\emph{Signif. Codes: ***: 0.001, **: 0.01, *: 0.05}}\\
   \end{tabular}
   }
\end{table}

%% Post OnlineAdoption_it: 4 categories boot dist; 
\begin{comment}
\begin{figure}[htp!]
    \centering
    \caption{Bootstrapped distribution of Overall ATT Estimates based on 500 replications with the point estimate from Table \ref{tab:7.2.1.CS_Result}.}
    \label{fig:7.2.1.CS.IT.Boot_Est}
    \begin{subfigure}{0.49\textwidth}
        \centering
        \caption{Dry Dog Food}
        \includegraphics[scale=0.5]{0_Table_Figure_Summary/6_Bootstrap/5_DogSP_Boot_Dist500_CS_AdoptIT.png}
    \end{subfigure}%
    ~ %add desired spacing between the subfigures, if needed
    \begin{subfigure}{0.49\textwidth}
        \centering
        \caption{Dry Cat Food}
        \includegraphics[scale=0.5]{0_Table_Figure_Summary/6_Bootstrap/5_CatSP_Boot_Dist500_CS_AdoptIT.png}
    \end{subfigure}

    \vspace{1ex} % add some vertical space between the rows

    \begin{subfigure}{0.49\textwidth}
        \centering
        \caption{Dog Hygiene}
        \includegraphics[scale=0.5]{0_Table_Figure_Summary/6_Bootstrap/5_HM_Boot_Dist500_CS_AdoptIT.png}
    \end{subfigure}%
    ~ %add desired spacing between the subfigures, if needed
    \begin{subfigure}{0.49\textwidth}
        \centering
        \caption{Cat Hygiene}
        \includegraphics[scale=0.5]{0_Table_Figure_Summary/6_Bootstrap/5_Sand_Boot_Dist500_CS_AdoptIT.png}
    \end{subfigure}
\end{figure}
\end{comment}

\newpage

\subsubsection{Results for Alternate Demand Model Controlling for State Dependence due to Online Adoption}\label{appsec:alternative.demand.state}

The alternative utility specification is given by:
\begin{equation}
    \begin{aligned}
    \psi_{ijt} & = \alpha_0 \textrm{Price}_{jt} +
    \alpha_{1} \textrm{cumPercOnlineSpend}_{it} + \alpha_2 \textrm{Price}_{jt} \times \textrm{cumPercOnlineSpend}_{it} \\
    & +  \alpha_{3} \mathcal{I}\{\textrm{BuyPrevOcca}_{ijt}\} + \alpha_{4} \mathcal{I}\{\textrm{BuyPrevOcca}_{ijt}\} \times \textrm{cumPercOnlineSpend}_{it} \\
   & + \beta_{1}\textrm{CF\_Price}_{jt} 
+\beta_{2}\textrm{CF\_Price\_cumPercOnlineSpend}_{ijt} + \delta_{ij} ~~ \forall j \in \{1, \ldots, J\}\\
    \end{aligned}
\label{eq:util_coef_M7.2.2}
\end{equation}

We estimate this model and again obtain the customer-month level average elasticity for each category as before. Table \ref{tab:append.7.2.2.demand.model} presents the estimates of the demand model. The standard errors for the demand model parameters are based on the customer-level block bootstrap with 500 replications. 

\begin{table}[htp!]
\caption{Coefficient Estimates of Demand Model in Equation \eqref{eq:util_coef_M7.2.2} - Controlling for Interaction between Online Spend and Brand Loyalty}
\centering
\label{tab:append.7.2.2.demand.model}
\footnotesize{
\begin{tabular}{lcccc}
\toprule
 & Dry Dog Food & Dry Cat Food & Dog Hygiene & Cat Hygiene \\
\textit{Variables} &  &  &  &  \\
\midrule
Price & \makecell{-0.3157$^{***}$\\(0.0159)} & \makecell{-0.1618$^{***}$\\(0.0172)} & \makecell{-1.2497$^{***}$\\(0.1142)} & \makecell{-0.2209$^{***}$\\(0.0186)} \\
cumPercOnlineSpend & \makecell{-0.3383\\(0.4039)} & \makecell{-1.5509\\(1.2012)} & \makecell{8.6901$^{***}$\\(1.8059)} & \makecell{3.4632$^{***}$\\(0.9725)} \\
Price * cumPercOnlineSpend & \makecell{-0.1092$^{**}$\\(0.0366)} & \makecell{0.0227\\(0.0994)} & \makecell{-4.1435$^{***}$\\(0.7895)} & \makecell{-0.284$^{***}$\\(0.0661)} \\
BuyPrevOcca & \makecell{0.1592$^{***}$\\(0.0278)} & \makecell{-0.5249$^{***}$\\(0.047)} & \makecell{-0.1894$^{***}$\\(0.0294)} & \makecell{0.3242$^{***}$\\(0.0306)} \\
CF\_Price & \makecell{0.0521$^{*}$\\(0.0219)} & \makecell{0.048\\(0.0245)} & \makecell{0.1936\\(0.1673)} & \makecell{0.2803$^{***}$\\(0.0247)} \\
CF\_Price\_cumPercOnlineSpend & \makecell{0.0212\\(0.0719)} & \makecell{-0.0595\\(0.1169)} & \makecell{3.6167$^{***}$\\(0.9742)} & \makecell{0.3009$^{**}$\\(0.0915)} \\
BuyPrevOcca * cumPercOnlineSpend & \makecell{-0.3059\\(0.2918)} & \makecell{1.0106$^{**}$\\(0.3853)} & \makecell{0.2991\\(0.3022)} & \makecell{0.0133\\(0.2768)} \\
Customer\_Brand\_FE & Yes & Yes & Yes & Yes\\ \midrule
Observations       & 136,411         & 45,443          & 78,253         & 55,410\\ 
Log-likelihood & -84209.589& -30693.216& -55438.072& -37327.394 \\ \bottomrule
 \multicolumn{5}{l}{\emph{Bootstrapped standard-errors in parentheses via 500 replications}}\\
\multicolumn{5}{l}{\emph{Signif. Codes: ***: 0.001, **: 0.01, *: 0.05}}\\
\end{tabular}
}
\end{table}

\begin{figure}[htp!]
    \centering
    \caption{Bootstrapped distribution of Overall ATT Estimates based on 500 replications with the point estimate from Table \ref{tab:7.2.2.CS.BD.Result}.}
    \label{fig:7.2.2.CS.BD.Boot_Est}
    \begin{subfigure}{0.32\textwidth}
        \centering
        \caption{Dry Dog Food}
        \includegraphics[width=\linewidth]{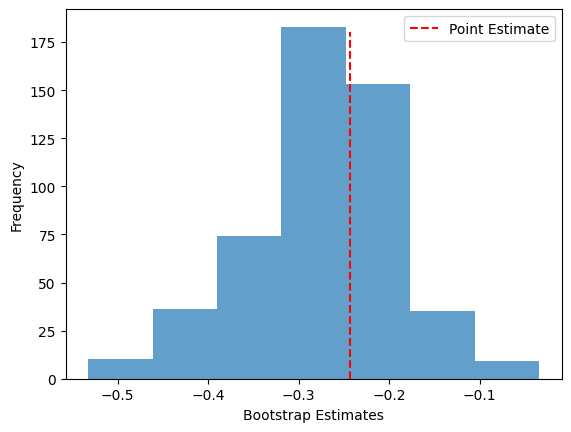}
    \end{subfigure}%
    \hfill
    \begin{subfigure}{0.32\textwidth}
        \centering
        \caption{Dog Hygiene}
        \includegraphics[width=\linewidth]{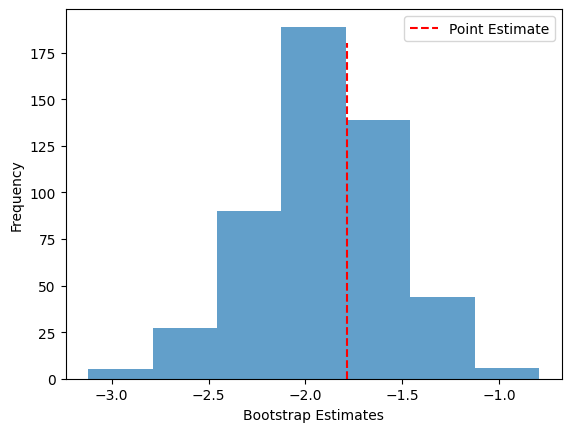}
    \end{subfigure}%
    \hfill
    \begin{subfigure}{0.32\textwidth}
        \centering
        \caption{Cat Hygiene}
        \includegraphics[width=\linewidth]{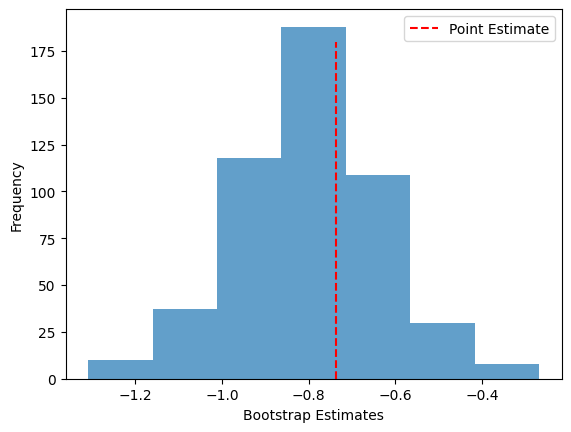}
    \end{subfigure}
\end{figure}

Table \ref{tab:7.2.2.CS.BD.Result} presents the ATT estimates of offline price elasticity from the CS estimator based on the demand model in Equation \eqref{eq:util_coef_M7.2.2}. The standard errors for the ATT estimates are based on the customer-level block bootstrap with 500 replications; See Figure \ref{fig:7.2.2.CS.BD.Boot_Est}.

\begin{table}[htp!]
   \caption{ATT of Online Adoption on Offline Price Elasticity (CS Estimator) - Controlling for State Dependence due to Online Adoption}
   \centering
   \label{tab:7.2.2.CS.BD.Result}
   \footnotesize{
   \begin{tabular}{lccc}
      \tabularnewline \midrule \midrule
      Dependent Variable: & \multicolumn{3}{c}{Offline Price Elasticity}\\
       & \multicolumn{1}{c}{\makecell{Dry Dog Food}}  
       & \multicolumn{1}{c}{\makecell{Dog Hygiene}}
       & \multicolumn{1}{c}{\makecell{Cat Hygiene}}  \\ 
      Model:             & (1)             & (2)             & (3)                \\  
      \midrule
%      \emph{Variables}\\
      CS               & -0.2429$^{**}$  & -1.7868$^{***}$ & -0.7382$^{***}$\\   
                         & (0.0797)            & (0.3544)        & (0.1668)       \\   
  
      \midrule
 %     \emph{Fit statistics}\\
     Observations       & 136,411         & 78,253         & 55,410\\  
      \midrule %\midrule
        \multicolumn{4}{l}{\emph{Bootstrapped standard-errors in parentheses via 500 replications}}\\
      \multicolumn{4}{l}{\emph{Signif. Codes: ***: 0.001, **: 0.01, *: 0.05}}\\
   \end{tabular}
   }
\end{table}

%% State Dependence * online share: all four bootstrap figure

\clearpage

\subsubsection{Results for Alternate Demand Model Controlling for Recency of Channel Effects with Lagged Online Channel Choice}\label{appsec:alternative.demand.recent}

Our main specification relates offline price sensitivity to a customer's cumulative online spending share, which captures the persistent effect of online adoption and subsequent online engagement. One potential concern is that the estimated relationship may instead reflect a short-lived effect of having recently shopped online. For example, prices observed during a recent online purchase may remain salient and temporarily affect the customer's price sensitivity during a subsequent offline purchase occasion.

To distinguish this recent channel-use effect from the cumulative effect of online adoption, we augment the demand model with a lagged online channel-choice indicator. Specifically, $\mathcal{I}{\textrm{lastMonthOnline}{it}}$ equals one if customer $i$ made at least one online purchase in any product category during the month immediately preceding offline purchase occasion $t$, and zero otherwise.
We also interact this indicator with price to allow recent online shopping to affect the customer's offline price sensitivity. The resulting utility specification is

%We run a demand model including the lagged channel choice variable in Equation \ref{eq:lagged_channel_choice}: $ \mathcal{I}\{\textrm{lastMonthOnline}_{it}\}$ is defined as whether customer $i$ has an online purchase (in any category) during his/her previous (lag 1 month) of the focal purchase occasion. 

\begin{equation}
\label{eq:lagged_channel_choice}
    \begin{aligned}
    U_{ijt} & = \alpha_{0} \textrm{Price}_{jt} +
    \alpha_{1} \textrm{cumPercOnlineSpend}_{it} + 
    \alpha_2  \textrm{Price}_{jt}\times \textrm{cumPercOnlineSpend}_{it} \\
& + \alpha_{3} \mathcal{I}\{\textrm{BuyPrevOcca}_{ijt}\} + \delta_{ij}  + \alpha_{4} \mathcal{I}\{\textrm{lastMonthOnline}_{it}\} \\
& + \alpha_{5}  \textrm{Price}_{jt}\times \mathcal{I}\{\textrm{lastMonthOnline}_{it}\}  + \lambda_1 \textrm{CF\_Price}_{jt} \\
& + \lambda_2 \textrm{CF\_Price\_cumPercOnlineSpend}_{ijt} + \lambda_3 \textrm{CF\_Price\_lastMonthOnline}_{ijt} + \epsilon_{ijt} \\
& ~~ \forall j \in \{1, \ldots, J\} \\
\end{aligned}
\end{equation}
The coefficient $\alpha_{5}$ captures the incremental change in offline price sensitivity associated with having purchased online in the preceding month, holding constant the customer's cumulative online spending share. We include the corresponding control function, $\textrm{CF\_Price\_lastMonthOnline}_{ijt}$, to account for the endogeneity of the interaction between price and recent online channel use.

\begin{figure}[htp!]
    \centering
    \caption{Average price elasticity between adopters (shown in twelve dotted lines) and non-adopters (shown in one dashed line) for each category.}
    \label{fig:r2_include_lagged_channel_choice}

    \begin{subfigure}{0.49\textwidth}
        \centering
        \caption{Dry Dog Food}
        \includegraphics[scale=0.26]{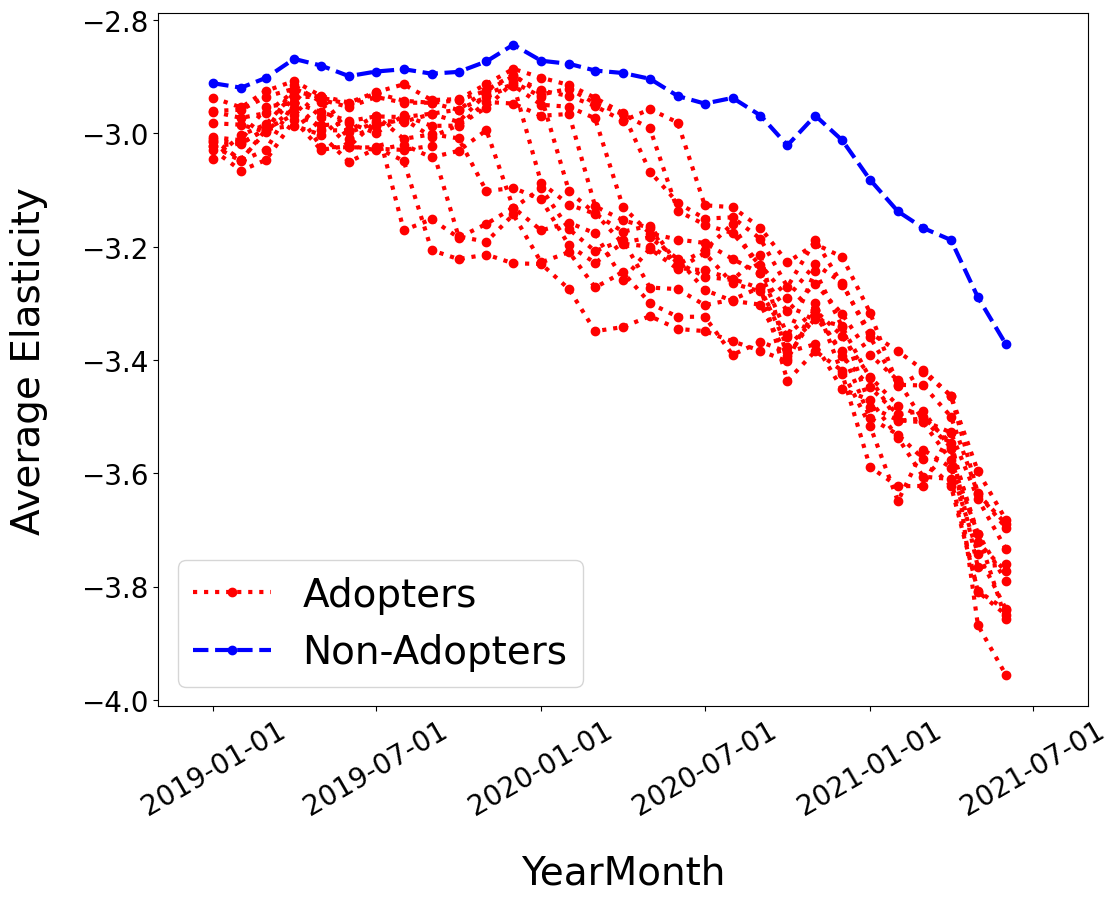}
    \end{subfigure}%
    ~ %add desired spacing between the subfigures, if needed
    \begin{subfigure}{0.49\textwidth}
        \centering
        \caption{Dry Cat Food}
        \includegraphics[scale=0.26]{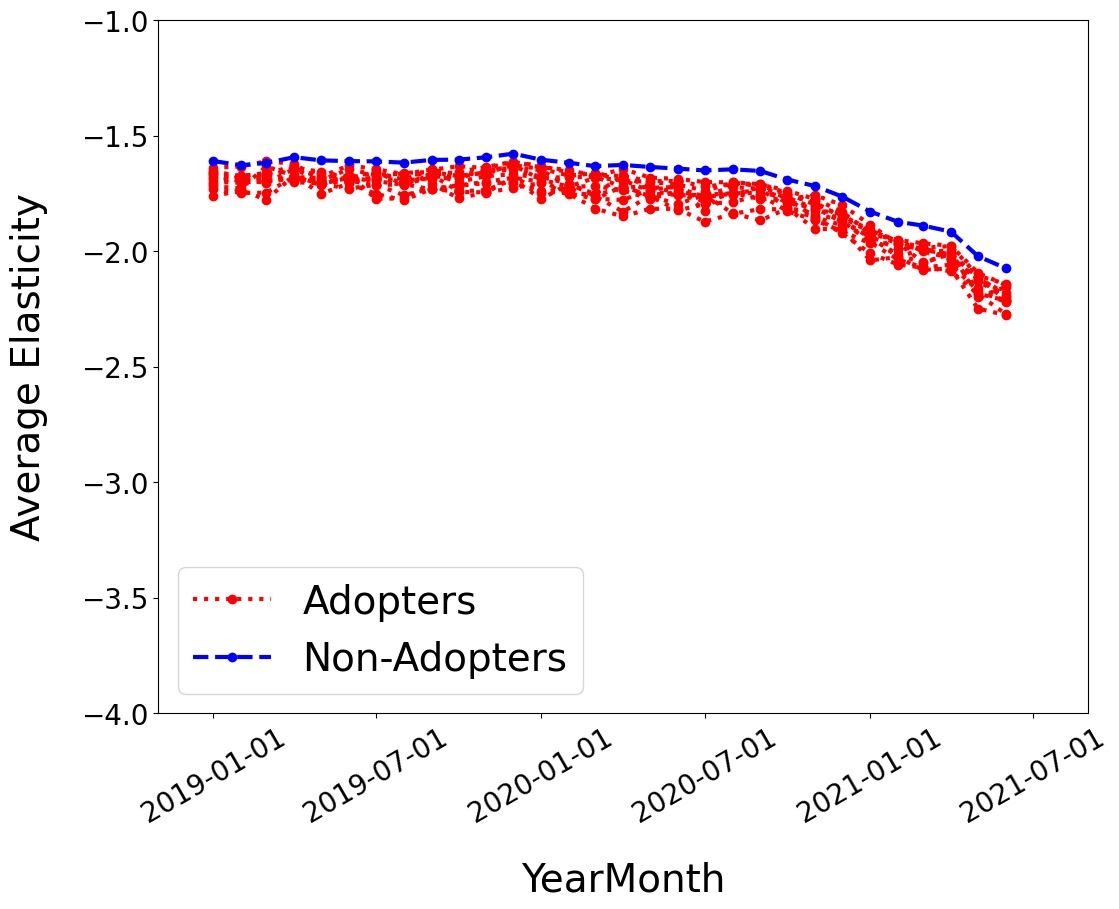}
    \end{subfigure}

    \vspace{1ex} % add some vertical space between the rows

    \begin{subfigure}{0.49\textwidth}
        \centering
        \caption{Dog Hygiene}
        \includegraphics[scale=0.26]{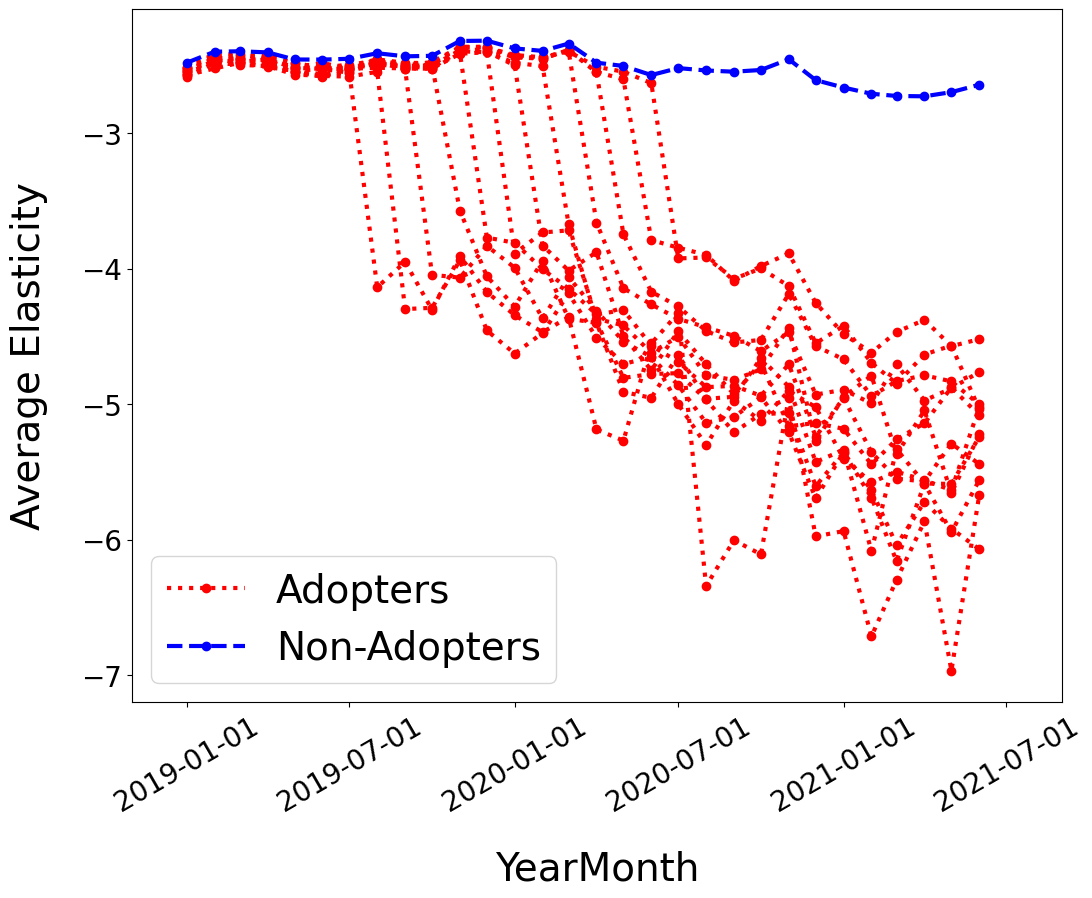}
    \end{subfigure}%
    ~ %add desired spacing between the subfigures, if needed
    \begin{subfigure}{0.49\textwidth}
        \centering
        \caption{Cat Hygiene}
        \includegraphics[scale=0.26]{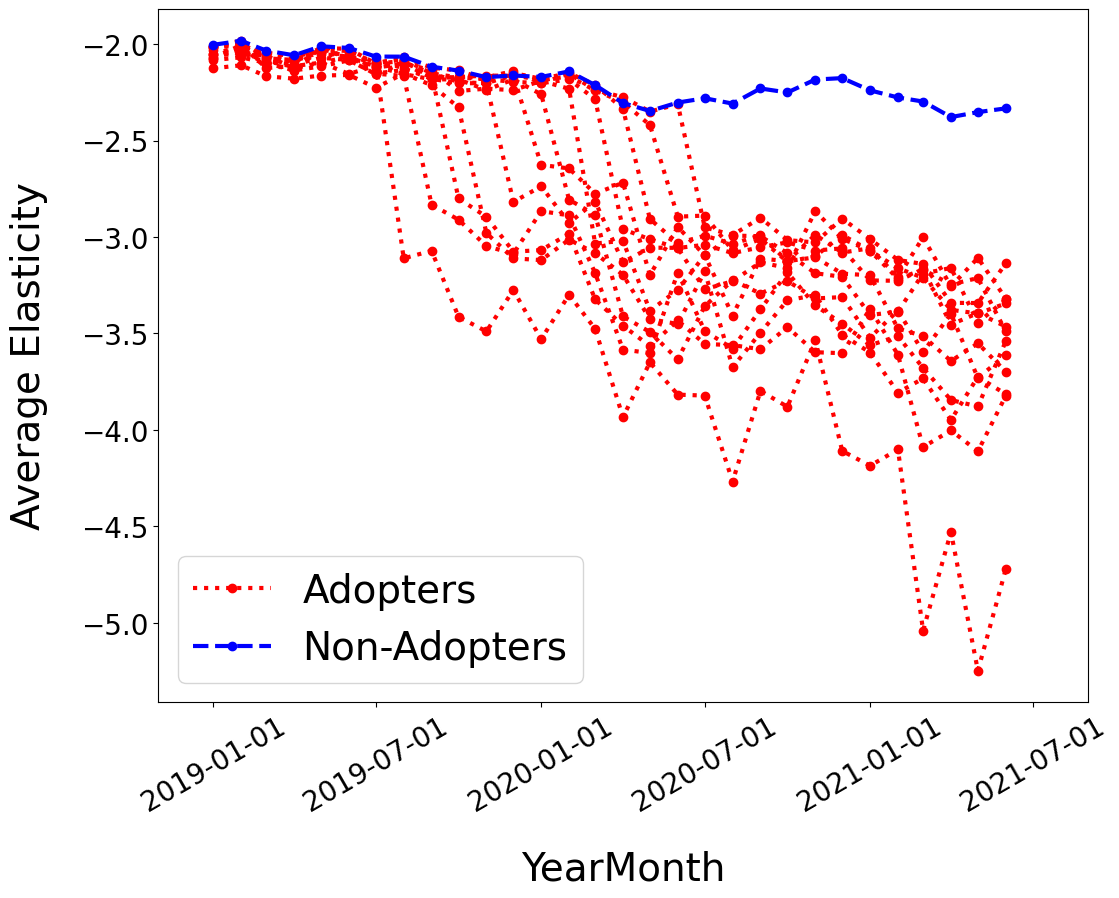}
    \end{subfigure}
\end{figure}

Table~\ref{tab:append.demand.model.lag.online.choice} presents the demand model estimates. The interaction between price and recent online channel use, $\textrm{Price}{jt}\times\mathcal{I}\{\textrm{lastMonthOnline}_{it}\}$, is not statistically significant in any of the four categories. In contrast, the coefficient on $\textrm{Price}_{jt}\times\textrm{cumPercOnlineSpend}_{it}$ remains negative and statistically significant for dry dog food, dog hygiene, and cat hygiene, while remaining statistically insignificant for dry cat food. Figure~\ref{fig:r2_include_lagged_channel_choice} plots the average elasticities for the twelve adopter cohorts and the non-adopter group. Table~\ref{tab:CS_Result_Lagged_Channel_Choice} reports the aggregate ATT estimates over all post-adoption periods. The results remain qualitatively the same as the main findings.

\begin{table}[htp!]
\caption{Coefficients of Demand Model Estimate - Including Lagged Channel Choice}
\label{tab:append.demand.model.lag.online.choice}
\centering
\footnotesize{
\begin{tabular}{lcccc}
\toprule
 & Dry Dog Food & Dry Cat Food & Dog Hygiene & Cat Hygiene \\
VarName &  &  &  &  \\
\midrule
Price & \makecell{-0.3185$^{***}$\\(0.016)} & \makecell{-0.1652$^{***}$\\(0.0177)} & \makecell{-1.2498$^{***}$\\(0.1136)} & \makecell{-0.2213$^{***}$\\(0.0182)} \\
cumPercOnlineSpend & \makecell{-0.2318\\(0.4447)} & \makecell{-1.7559\\(1.3252)} & \makecell{9.2839$^{***}$\\(1.6934)} & \makecell{4.4637$^{***}$\\(1.0803)} \\
Price * cumPercOnlineSpend & \makecell{-0.0952$^{*}$\\(0.0395)} & \makecell{0.0848\\(0.1112)} & \makecell{-4.3426$^{***}$\\(0.7352)} & \makecell{-0.3318$^{***}$\\(0.0743)} \\
BuyPrevOcca & \makecell{0.1475$^{***}$\\(0.0268)} & \makecell{-0.5121$^{***}$\\(0.0436)} & \makecell{-0.1847$^{***}$\\(0.0265)} & \makecell{0.3215$^{***}$\\(0.0298)} \\
CF\_Price & \makecell{0.0537$^{*}$\\(0.0228)} & \makecell{0.0507$^{*}$\\(0.0243)} & \makecell{0.1981\\(0.1696)} & \makecell{0.2807$^{***}$\\(0.0251)} \\
CF\_Price\_cumPercOnlineSpend & \makecell{0.0395\\(0.0769)} & \makecell{-0.1487\\(0.1204)} & \makecell{4.2198$^{***}$\\(0.9732)} & \makecell{0.3555$^{***}$\\(0.1028)} \\
lastMonthOnline & \makecell{-0.2752\\(0.3039)} & \makecell{0.1587\\(0.7859)} & \makecell{-1.1265\\(1.1549)} & \makecell{-1.5009$^{*}$\\(0.6109)} \\
Price * lastMonthOnline & \makecell{-0.0129\\(0.0264)} & \makecell{-0.0786\\(0.0679)} & \makecell{0.3696\\(0.5059)} & \makecell{0.0715\\(0.0432)} \\
CF\_Price\_lastMonthOnline & \makecell{-0.0415\\(0.0435)} & \makecell{0.0655\\(0.0773)} & \makecell{-1.5179$^{*}$\\(0.6741)} & \makecell{-0.1024\\(0.0572)} \\
Customer\_Brand\_FE & Yes & Yes & Yes & Yes\\ \midrule
Observations       & 136,411         & 45,443          & 78,253         & 55,410\\ 
Log-likelihood & -84085.721 & -30670.478 & -55433.895 & -37307.734 \\
\bottomrule
\multicolumn{5}{l}{\emph{Bootstrapped standard-errors in parentheses via 500 replications}}\\
\multicolumn{5}{l}{\emph{Signif. Codes: ***: 0.001, **: 0.01, *: 0.05}}\\
\end{tabular}
}
\end{table}

\begin{table}[htp!]
\caption{ATT of Online Adoption on Offline Price Elasticity Over All Post-Adoption Period (CS Estimator) -- Including Lagged Channel Choice}
\centering
\label{tab:CS_Result_Lagged_Channel_Choice}
\footnotesize{
  \begin{tabular}{lccc}
  \tabularnewline \midrule \midrule
  Dependent Variable: & \multicolumn{3}{c}{Offline Price Elasticity}\\
  & \multicolumn{1}{c}{\makecell{Dry Dog Food}}  
  & \multicolumn{1}{c}{\makecell{Dog Hygiene}}
  & \multicolumn{1}{c}{\makecell{Cat Hygiene}}  \\ 
  Model:             & (1)             & (2)             & (3)                      \\  
  \midrule
  %      \emph{Variables}\\
  ATT               &   -0.2122$^{*}$  & -1.8688$^{***}$ & -0.8522$^{***}$\\   
      & (0.0935)               & (0.3016)   & (0.1771)    \\   \midrule
  \midrule
  %     \emph{Fit statistics}\\
  Observations       &  136,411  & 78,253 & 55,410 \\   
  \midrule %\midrule
  \multicolumn{4}{l}{\emph{Bootstrapped standard-errors in parentheses via 500 replications}}\\
  \multicolumn{4}{l}{\emph{Signif. Codes: ***: 0.001, **: 0.01, *: 0.05}}\\
  \end{tabular}
}
\end{table}

\clearpage

\subsection{Detailed results for testing the Parallel Pre-Trend Assumption}
\label{appsec:parallel_pretrend}

\begin{figure}[htp!]
    \centering
    \caption{Parallel Pre-Trends Evaluation and Dynamic Effect during the Pre- and Post-Adoption Periods}
    \label{fig:7_Lead_Lag_Model_CS}

    \begin{subfigure}{0.7\textwidth}
        \centering
        \caption{Dry Dog Food}
        \includegraphics[width=0.7\linewidth]{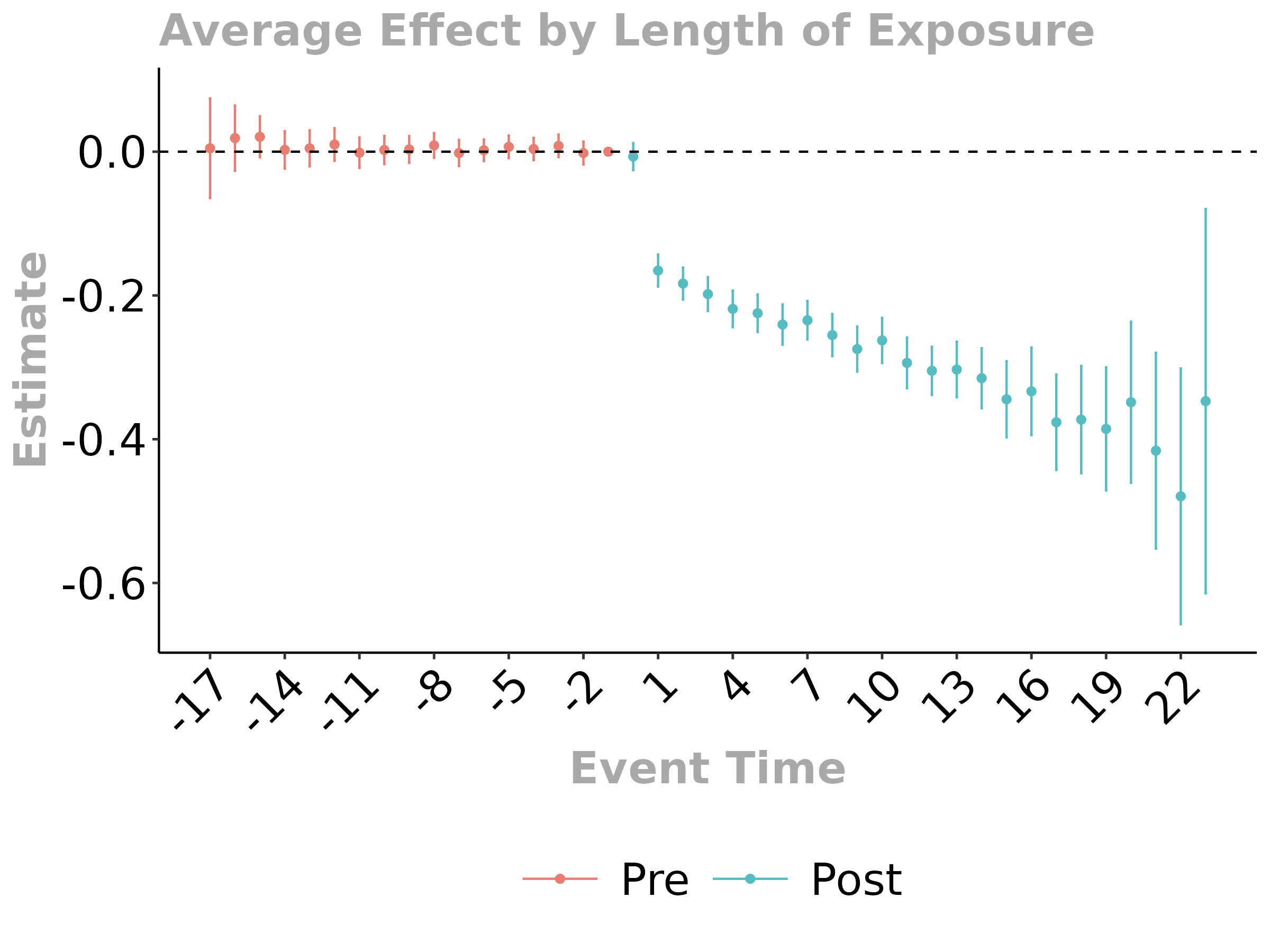}
    \end{subfigure}

    \vspace{1ex}

    \begin{subfigure}{0.7\textwidth}
        \centering
        \caption{Dog Hygiene}
        \includegraphics[width=0.7\linewidth]{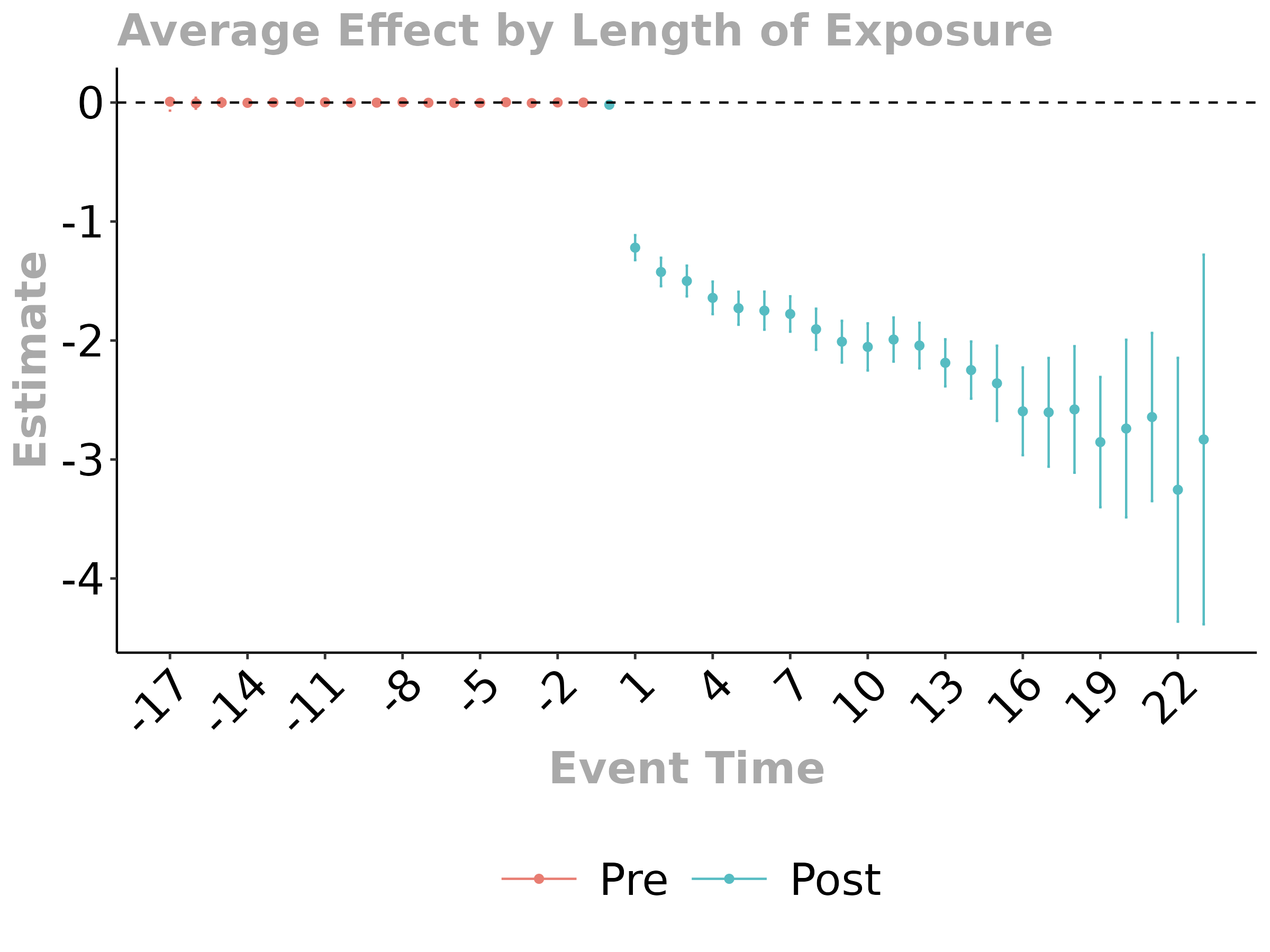}
    \end{subfigure}

    \vspace{1ex}
    \begin{subfigure}{0.7\textwidth}
        \centering
        \caption{Cat Hygiene}
        \includegraphics[width=0.7\linewidth]{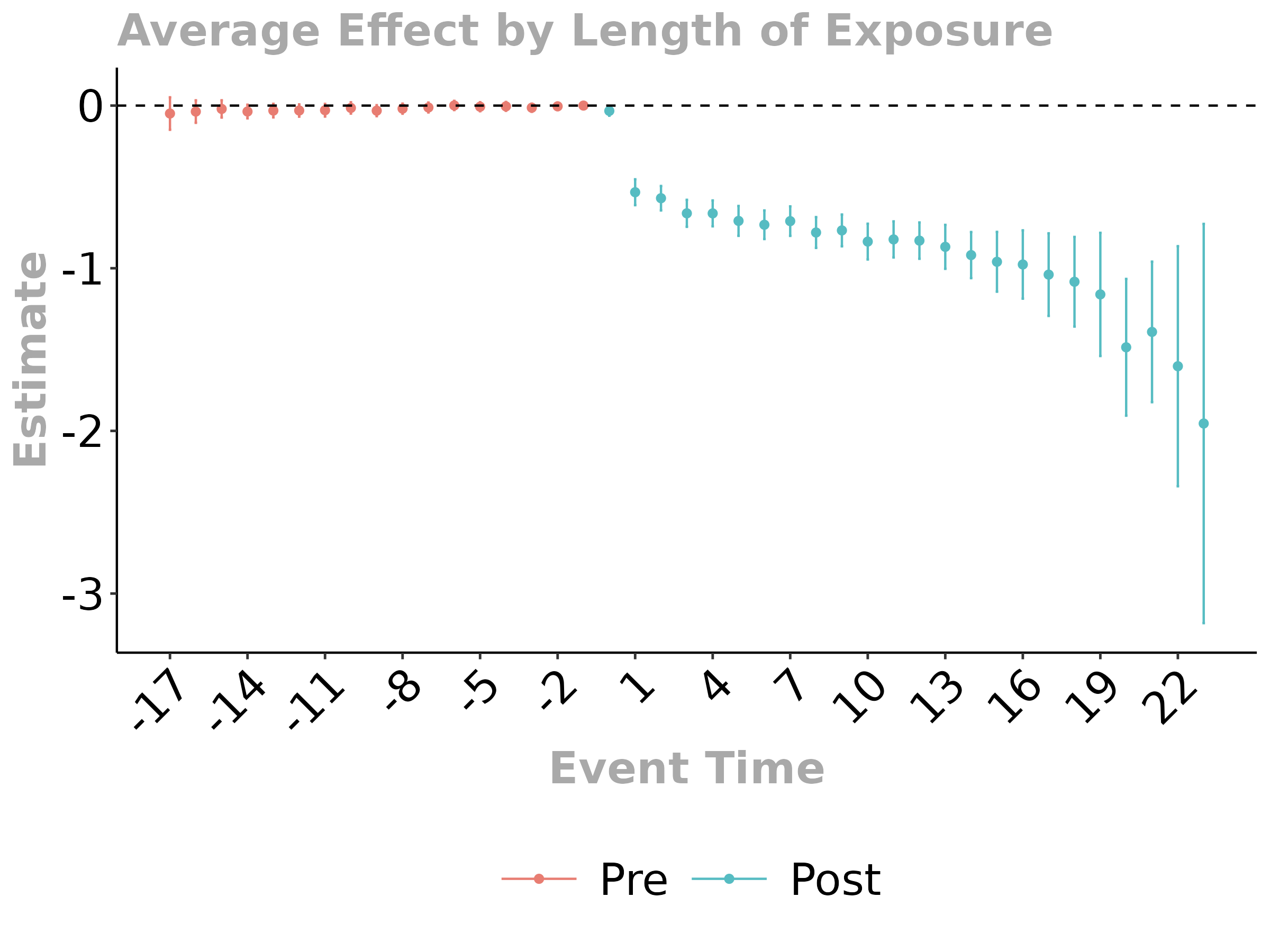}
    \end{subfigure}
\end{figure}

A fundamental condition for the credibility of the estimates presented in $\S$\ref{ssec:did} is the parallel trends assumption. We assess it with an event-study figure following \cite{callaway2021difference}. Figure \ref{fig:7_Lead_Lag_Model_CS} plots the point estimates and 95\% bootstrapped confidence interval for each pre- and post-treatment period, and shows that all pre-treatment coefficients are insignificant from zero. We do not exhibit discernible pre-trends. In addition, the p-values from the Wald tests of the pre-trend assumption are greater than 5\% level of significance in each category. We therefore conclude that the parallel trend assumption is reasonably satisfied throughout the analysis.

\clearpage

\subsection{Detailed Results for Selection into Adoption Timing}
\label{appsec:adoption_timing_parallel_pretrend}

\paragraph{Cohort-specific elasticity estimation.} We first examine whether customers who adopt the online channel at different times exhibit systematically different offline price sensitivity before adoption. For each of the twelve monthly adoption cohorts, we compare adopters with customers who had not yet adopted by the focal cohort's adoption month. We estimate the demand model separately for each cohort and its corresponding control group:
\begin{equation}
\begin{aligned}
\label{eq:se2_lag_model_CF_V1}
U_{ijt}&=
\alpha_{-1} \times \textrm{Price}_{jt} + \sum_{l=-6}^{-2} \kappa_{l} \times\mathbb{I}\{\textrm{Adopter}_{i}\} \times \mathbb{I}\{t - T_{i}^{Adopt} = l\} \times \textrm{Price}_{jt} \\
&+ \kappa_{A} \times  \mathbb{I}\{\textrm{Adopter}_{i}\} \times \mathbb{I}\{t - T_{i}^{Adopt} \geq 0\} \times \textrm{Price}_{jt} + \beta_{2} \mathcal{I}\{\textrm{BuyPrevOcca}_{ijt}\}  +
\delta_{ij} \\
& + \eta_{-1}\times \text{CF\_Price}_{jt} + \sum_{l=-6}^{-2}\eta_{l}\times\text{CF\_Price\_Adopter\_Lagl}_{ijt} \\
& + \eta_{A} \times \text{CF\_Price\_Adopter\_PostAdoption}_{ijt} +  \varepsilon_{ijt},
\end{aligned}
\end{equation}

Event month $-1$ is the omitted reference period. Thus, $\kappa_{\ell}$ captures the difference in the price coefficient between adopters and their corresponding controls in pre-adoption month $\ell$, relative to the difference in month $-1$. The coefficient $\kappa_A$ captures the corresponding average post-adoption difference. The control-function terms account for the endogeneity of each price component. Using the resulting estimates, we calculate customer-month-level offline price elasticities following the procedure described in $\S$\ref{ssec:elasticity_est} of the main paper. 

\begin{figure}[htp!]
    \centering
    \caption{Dry Dog Food: Average price elasticity between each adopter cohort (shown in dotted lines) and non-adopters (shown in one dashed line)}
    \label{fig:3.21F_BootBonf_bonf_dog_Q1_CF}

    \begin{subfigure}{0.49\textwidth}
        \centering
        \caption{2019-07 Adoption}
        \includegraphics[scale=0.22]{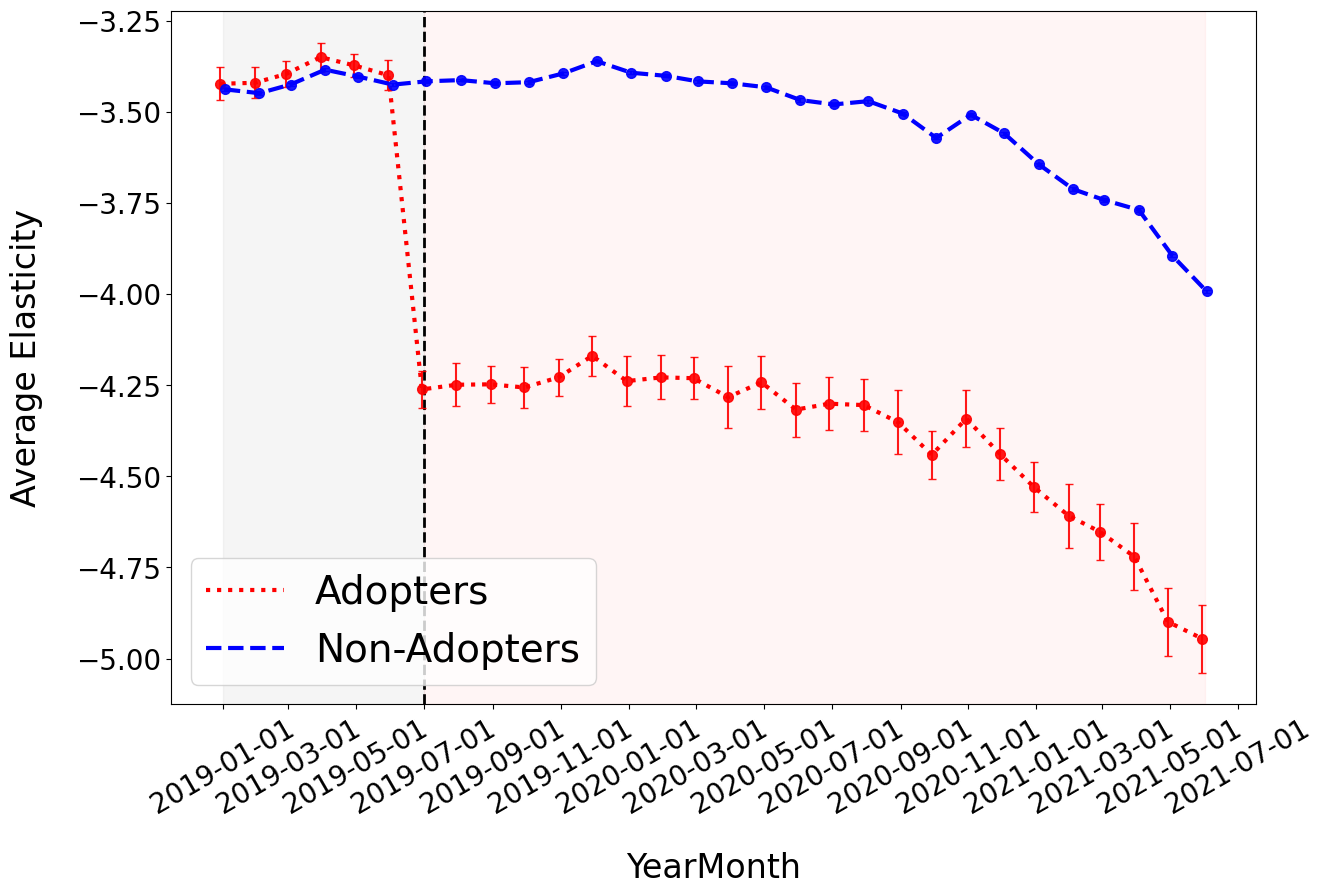}
    \end{subfigure}%
    ~ %add desired spacing between the subfigures, if needed
    \begin{subfigure}{0.49\textwidth}
        \centering
        \caption{2019-08 Adoption}
        \includegraphics[scale=0.22]{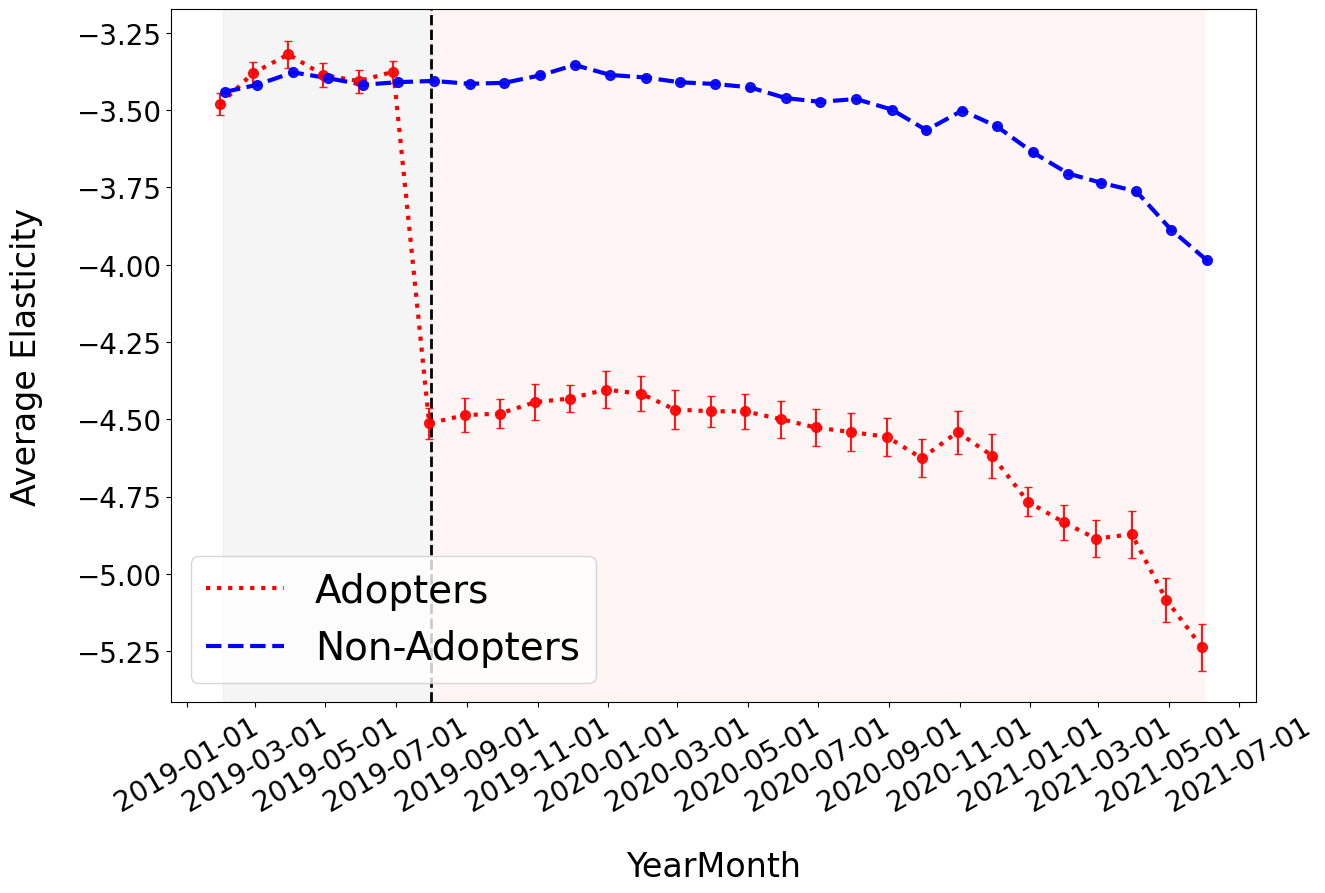}
    \end{subfigure}

    \vspace{0.5ex} % add some vertical space between the rows

    \begin{subfigure}{0.49\textwidth}
        \centering
        \caption{2019-09 Adoption}
        \includegraphics[scale=0.22]{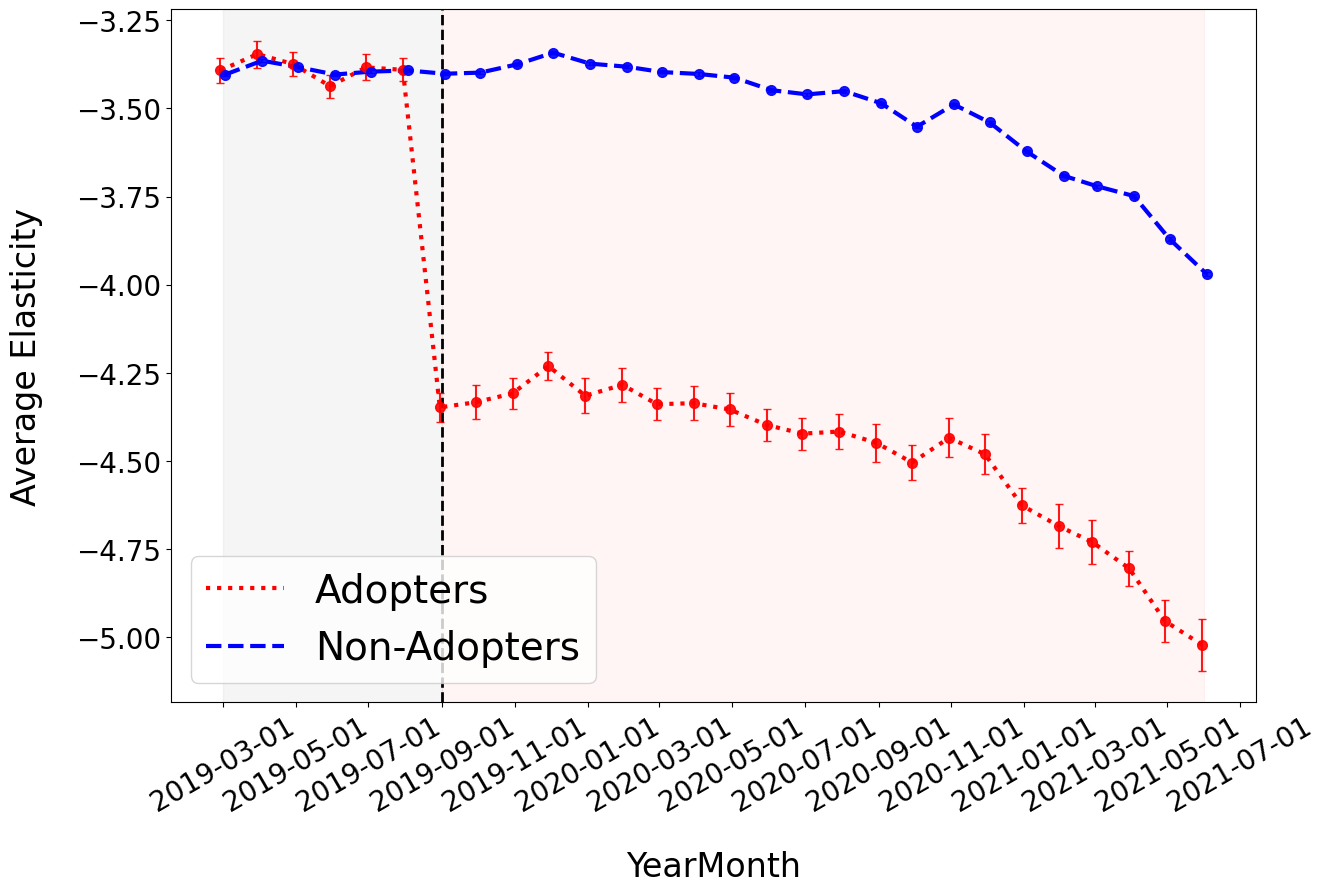}
    \end{subfigure}%
    ~ %add desired spacing between the subfigures, if needed
    \begin{subfigure}{0.49\textwidth}
        \centering
        \caption{2019-10 Adoption}
        \includegraphics[scale=0.22]{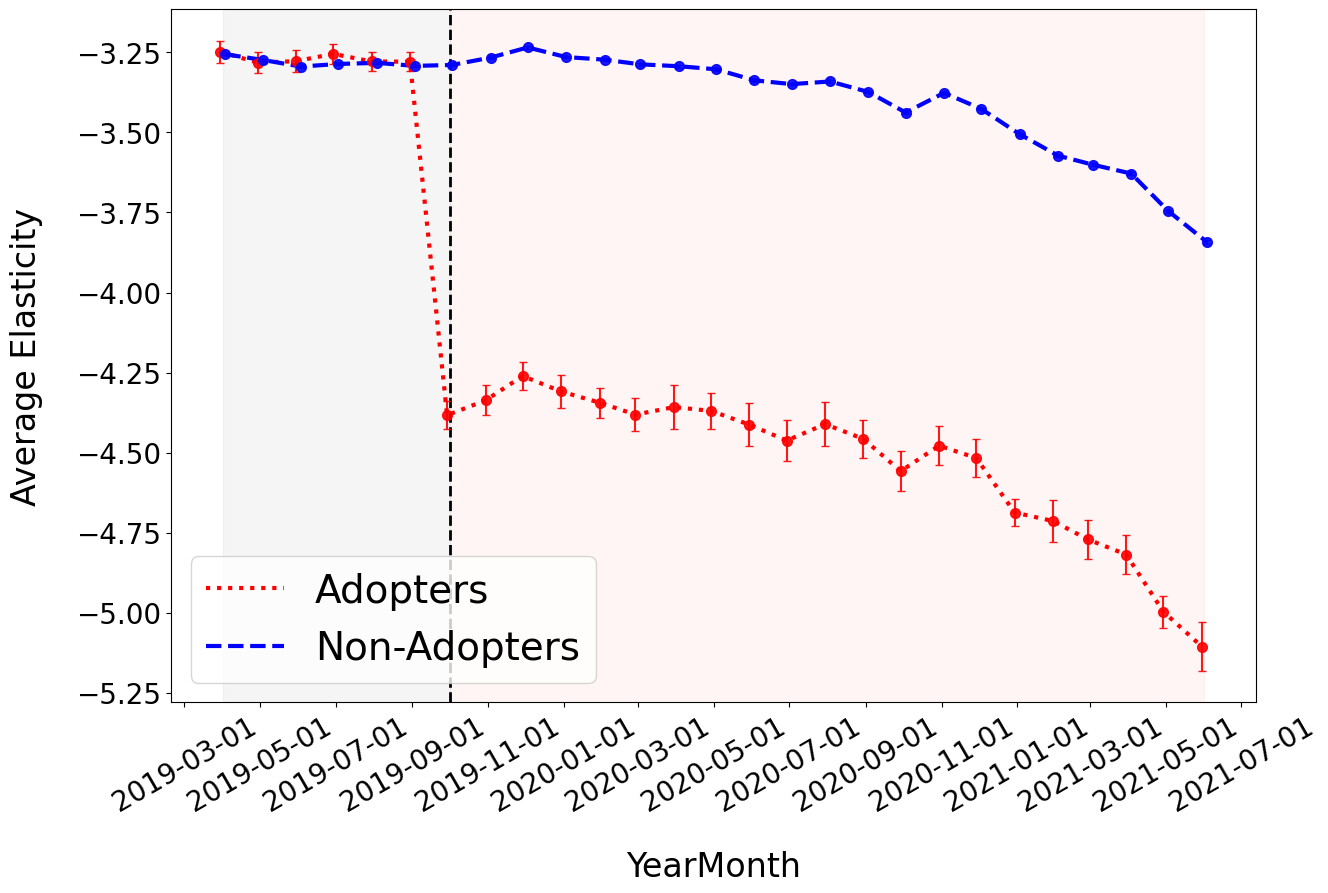}
    \end{subfigure}

     \vspace{0.5ex} % add some vertical space between the rows

    \begin{subfigure}{0.49\textwidth}
        \centering
        \caption{2019-11 Adoption}
        \includegraphics[scale=0.22]{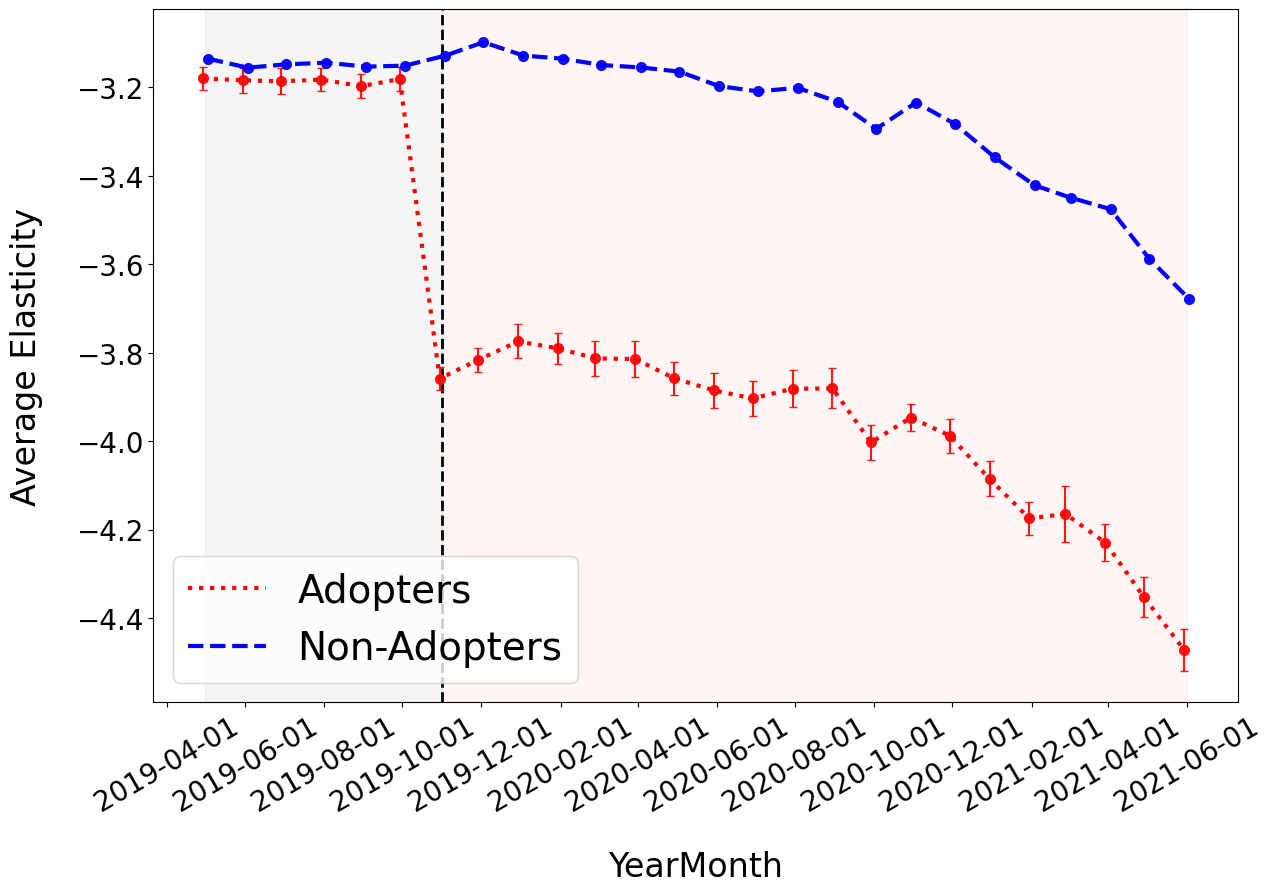}
    \end{subfigure}%
    ~ %add desired spacing between the subfigures, if needed
    \begin{subfigure}{0.49\textwidth}
        \centering
        \caption{2019-12 Adoption}
        \includegraphics[scale=0.22]{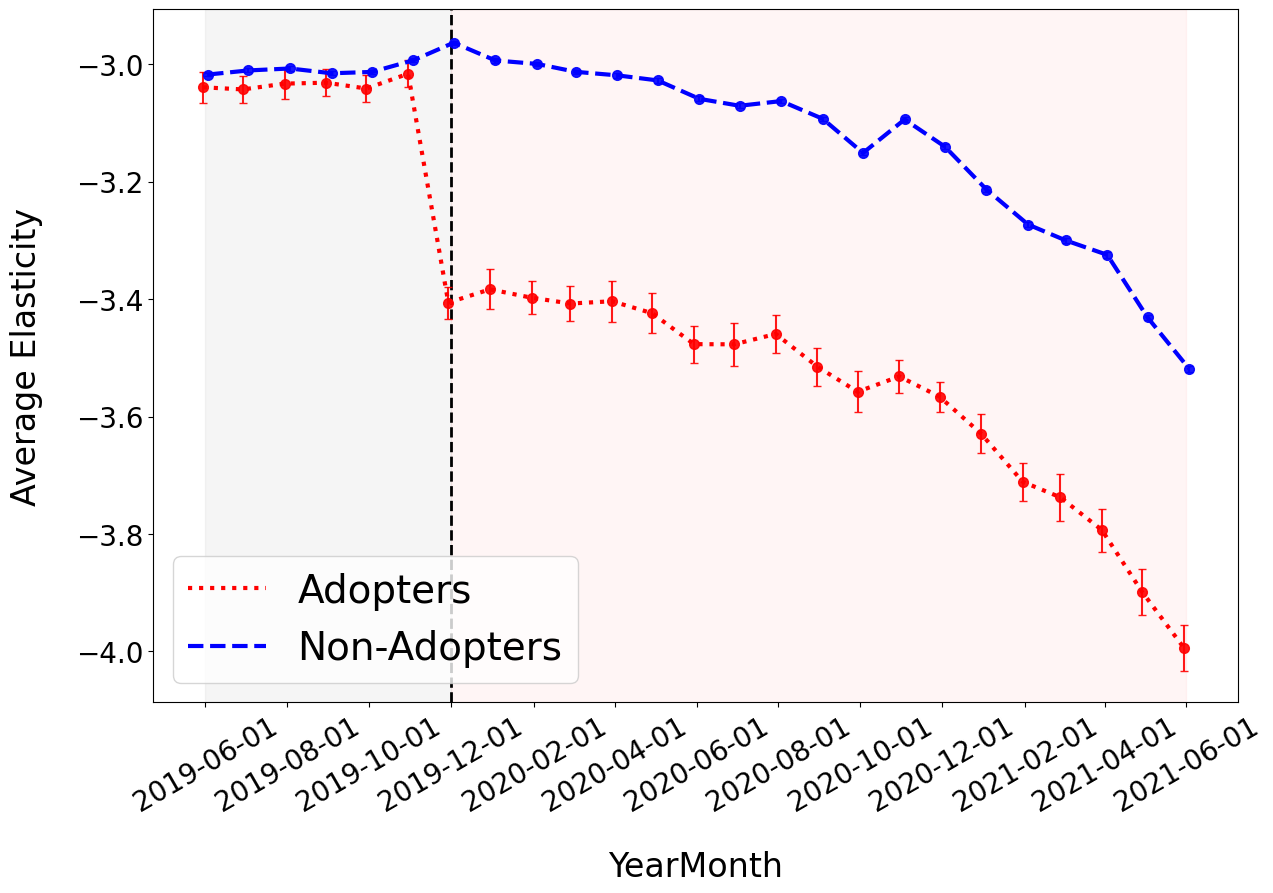}
    \end{subfigure}
\end{figure}

\begin{figure}[htp!]
    \centering
      \caption{Dry Dog Food: Average price elasticity between each adopter cohort (shown in dotted lines) and non-adopters (shown in one dashed line)}
     \label{fig:3.21F_BootBonf_bonf_dog_Q2_CF}
    \begin{subfigure}{0.49\textwidth}
        \centering
        \caption{2020-01 Adoption}
        \includegraphics[scale=0.22]{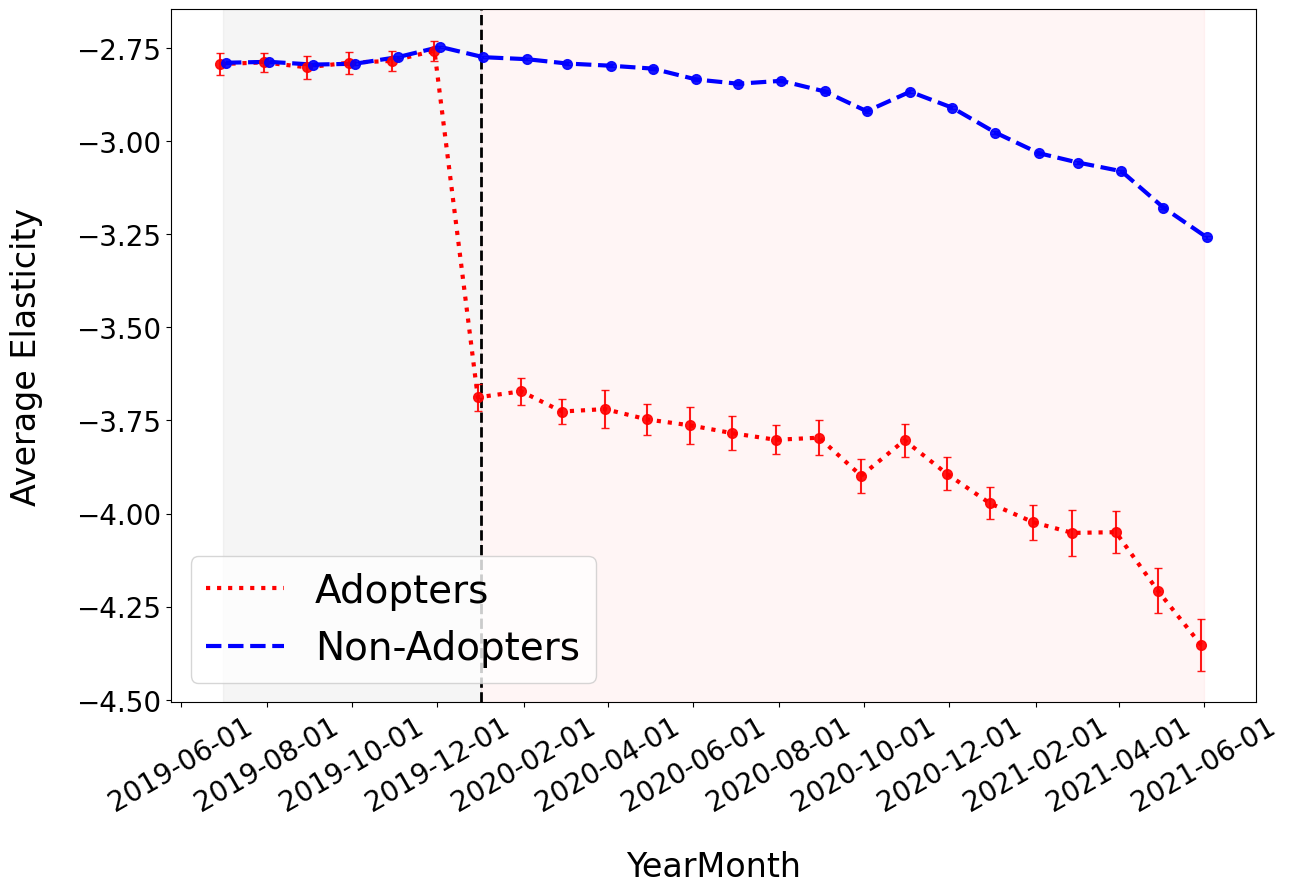}
    \end{subfigure}%
    ~ %add desired spacing between the subfigures, if needed
    \begin{subfigure}{0.49\textwidth}
        \centering
        \caption{2020-02 Adoption}
        \includegraphics[scale=0.22]{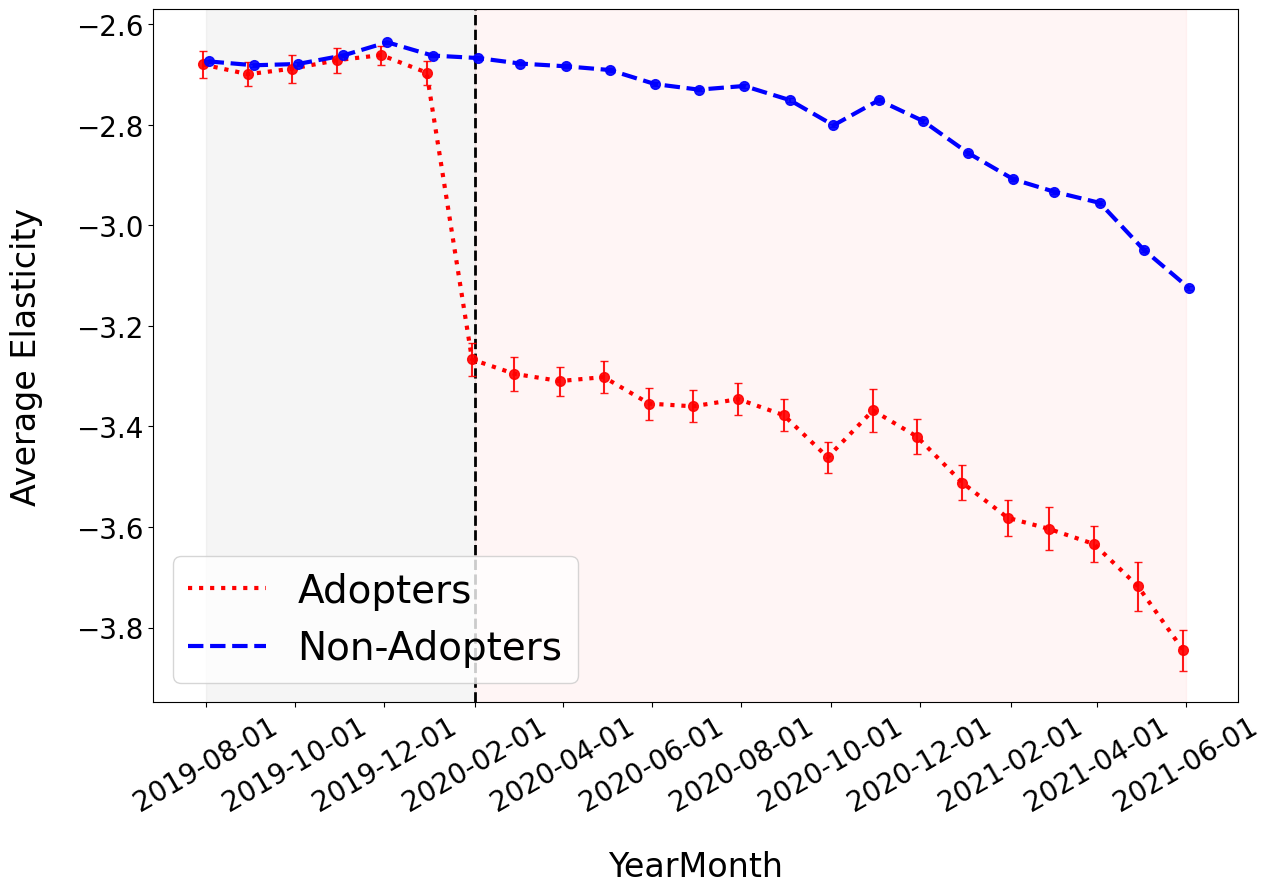}
    \end{subfigure}

\vspace{1ex} % add some vertical space between the rows
    \begin{subfigure}{0.49\textwidth}
        \centering
        \caption{2020-03 Adoption}
        \includegraphics[scale=0.22]{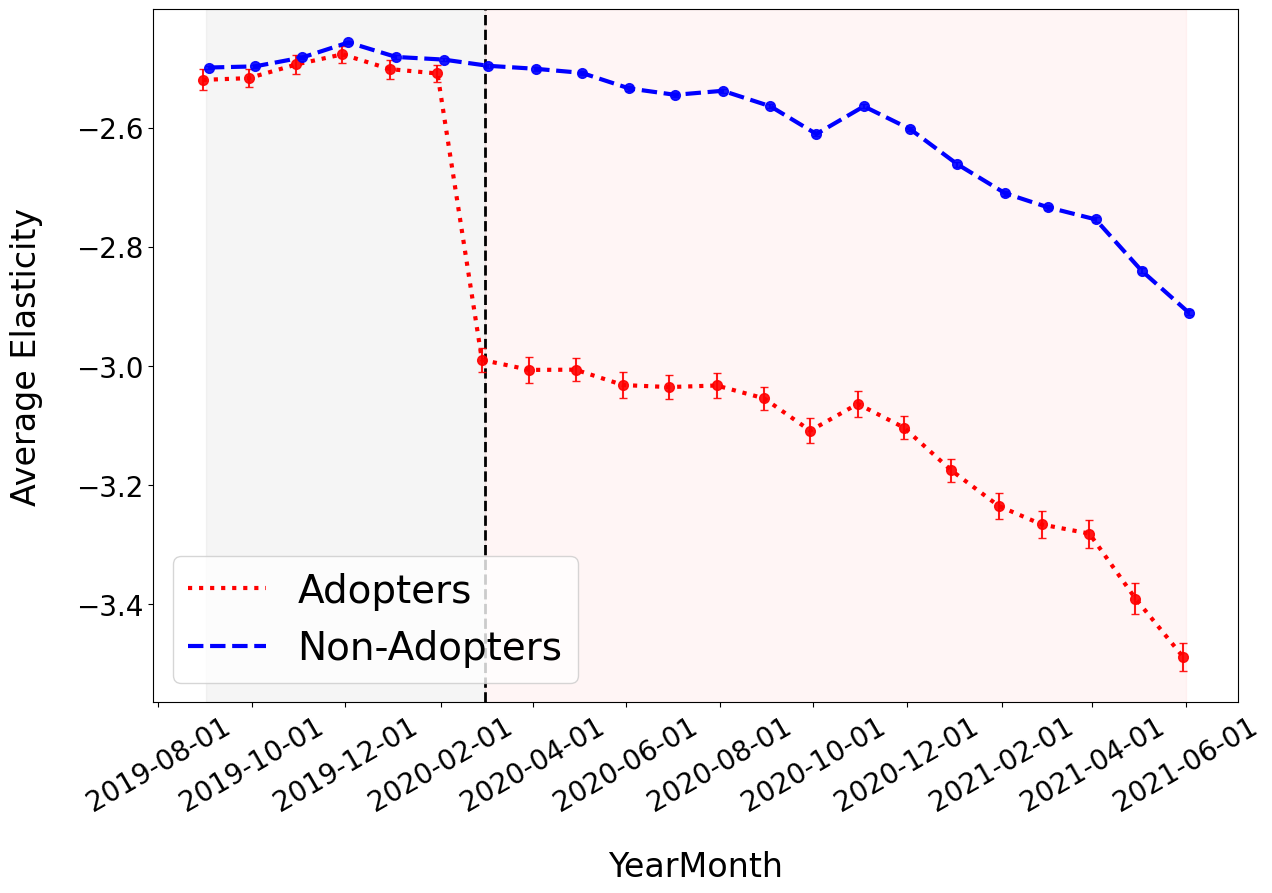}
    \end{subfigure}%
    ~ %add desired spacing between the subfigures, if needed
    \begin{subfigure}{0.49\textwidth}
        \centering
        \caption{2020-04 Adoption}
        \includegraphics[scale=0.22]{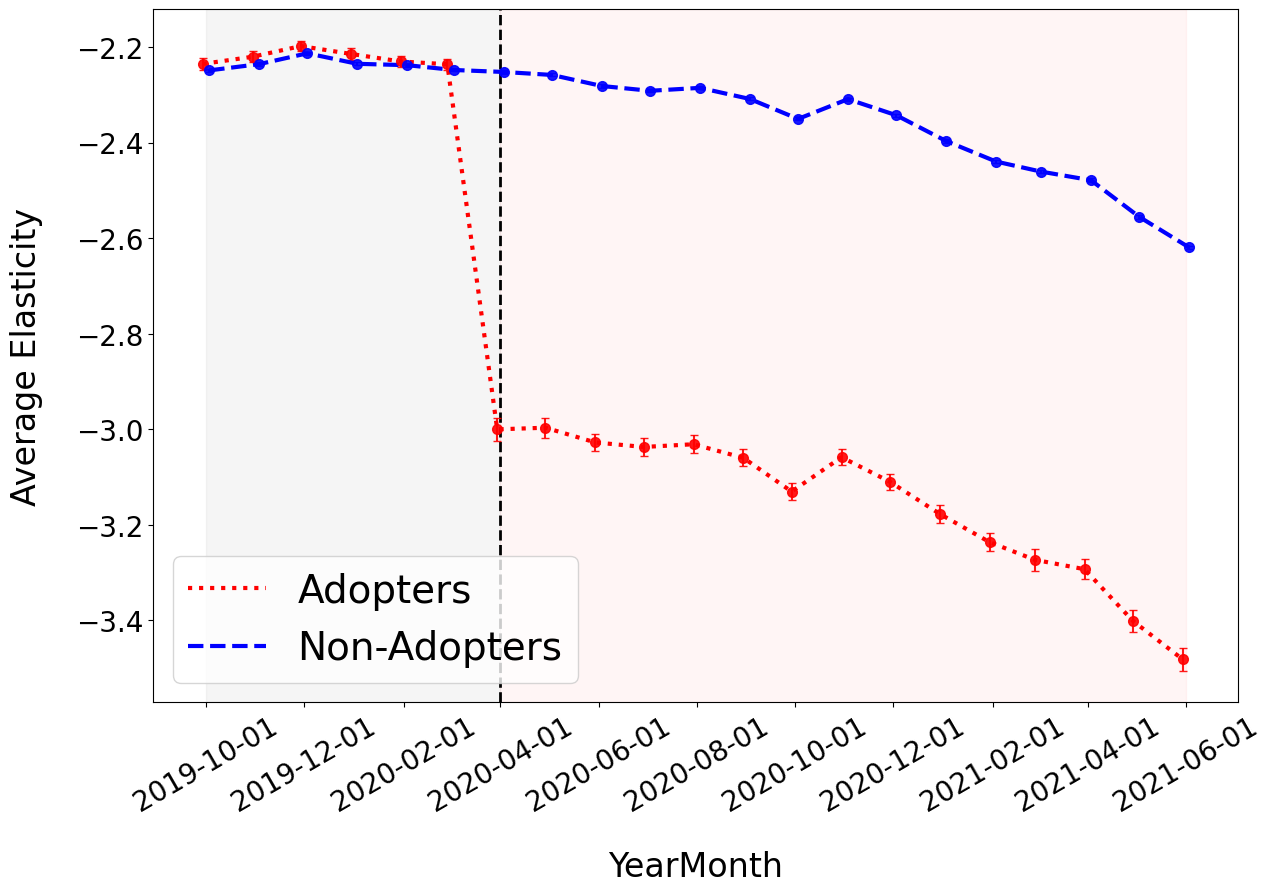}
    \end{subfigure}

    \vspace{1ex} % add some vertical space between the rows

    \begin{subfigure}{0.49\textwidth}
        \centering
        \caption{2020-05 Adoption}
        \includegraphics[scale=0.22]{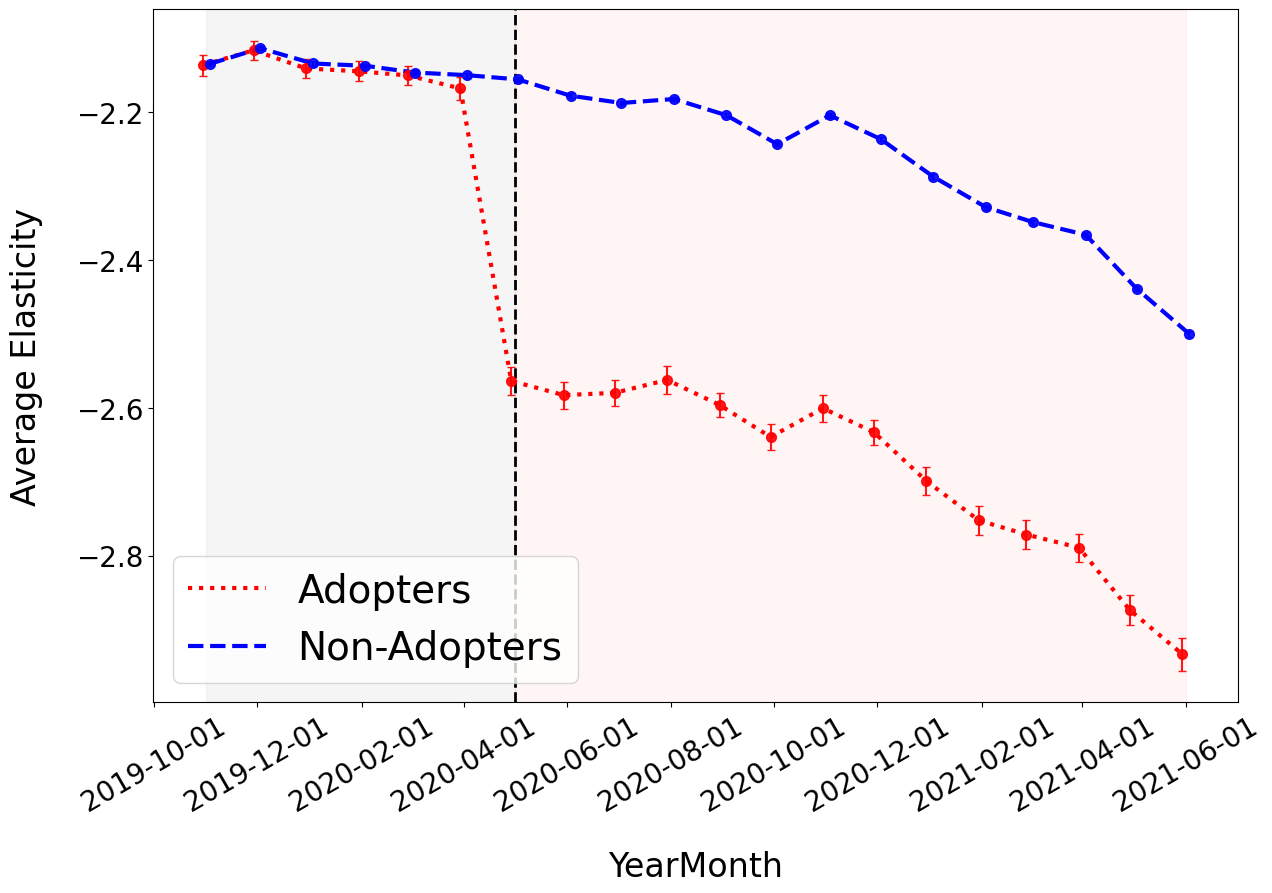}
    \end{subfigure}%
    ~ %add desired spacing between the subfigures, if needed
    \begin{subfigure}{0.49\textwidth}
        \centering
        \caption{2020-06 Adoption}
        \includegraphics[scale=0.22]{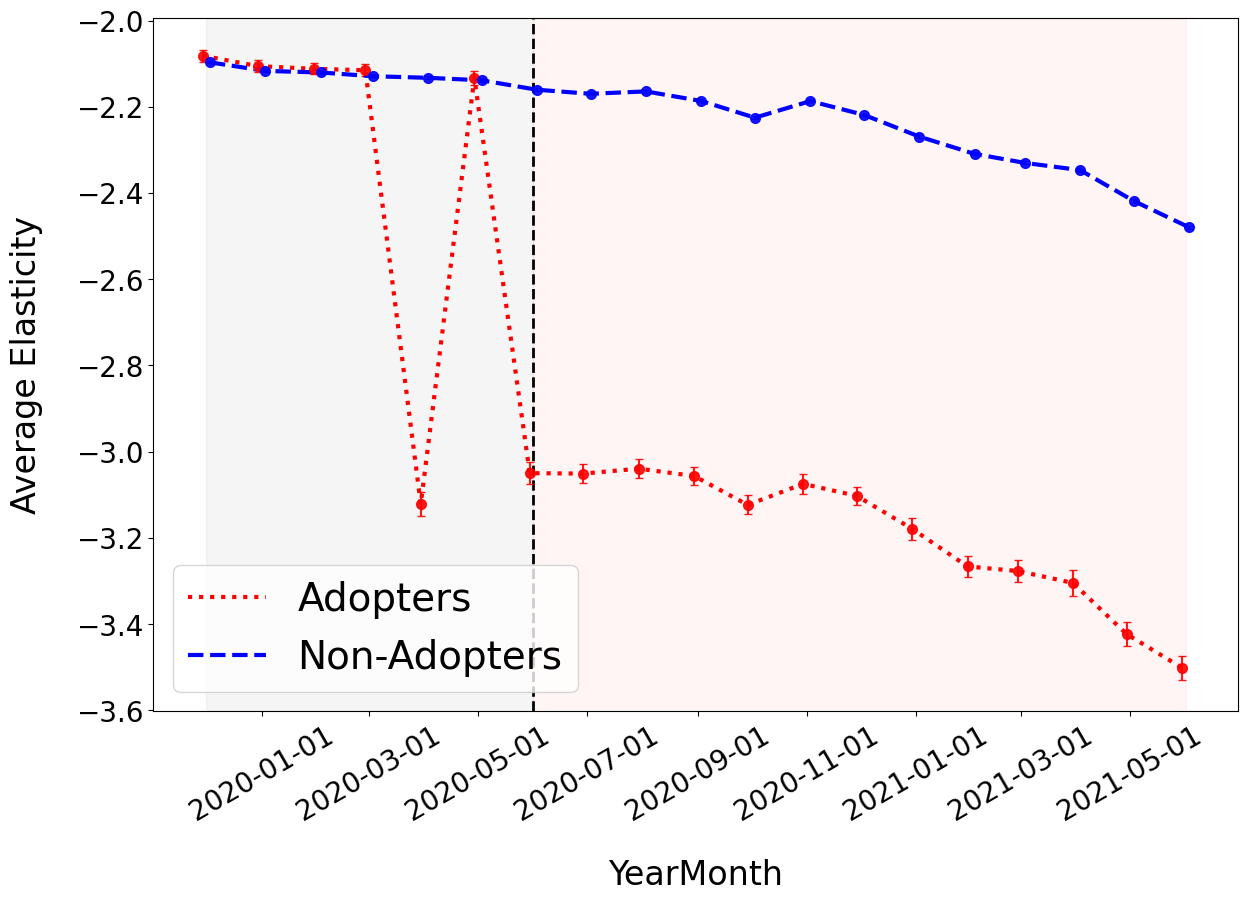}
    \end{subfigure}
\end{figure}

Figures~\ref{fig:3.21F_BootBonf_bonf_dog_Q1_CF} and \ref{fig:3.21F_BootBonf_bonf_dog_Q2_CF} report the cohort-specific elasticity paths for dry dog food. Each panel compares adopters in the indicated cohort with their corresponding not-yet-adopted control group and provides a visual assessment of differential pre-adoption trends. The average elasticity curves for the two groups exhibit parallel trends, with no discernible differences in their trend prior to adoption. We observe similar patterns across the other product categories, including dry cat food, dog hygiene, and cat hygiene.

\clearpage

\paragraph{Stacked pre-trend tests.} We next formally test for differential pre-adoption trends by stacking the cohort-specific adopter and control panels over the six months preceding adoption. We estimate the following specification across all twelve cohorts:

\begin{equation}
\label{eq:stacked_linear_trend_model}
    y_{igt} = \text{Customer}_{ig} + \text{YearMonth}_{gt} + \beta \times \text{Treatment}_{ig} \times \textrm{MonthNumber}_{gt} + e_{igt}
\end{equation}
Here, $y_{igt}$ denotes the outcome of interest (offline price elasticity or other purchase measures) for customer $i$ in cohort $g$ at event time $t$. The term $\text{Customer}_{ig}$ captures customer-by-cohort fixed effects, absorbing time-invariant heterogeneity across customers and cohorts, while $\text{YearMonth}_{gt}$ denotes calendar month fixed effects for cohorts. $\text{Treatment}_{ig}$ is an indicator equal to one for adopters in cohort $g$, and $\textrm{MonthNumber}_{gt}$ indexes the month number of year-month $t$ relative to adoption, which can take values from -6 for the lag 6 to -1 for the lag 1 period before the adoption time. The coefficient $\beta$ therefore captures the average difference in linear pre-trends between adopters and non-adopters before online adoption. Table~\ref{tab:stacked_pretrend_elasticity} reports the results for offline price elasticity. The estimated differential slopes are small and are not statistically distinguishable from zero for the three product categories with significant effects. These estimates provide no evidence of systematic linear differences in pre-adoption elasticity trends for those categories. 

\begin{table}[htp!]
   \caption{Stacked Linear Pre-Trend Test -- Offline Price Elasticity}
\label{tab:stacked_pretrend_elasticity}
   \centering
   \footnotesize{
   \begin{tabular}{lcccc}
      \tabularnewline \midrule \midrule
      Dependent Variable: & \multicolumn{3}{c}{Offline Price Elasticity}\\
      Model:                  & Dry Dog Food                                    & Dog Hygiene         & Cat Hygiene\\  
      \midrule
      \emph{Variables}\\
      Treatment $\times$ MonthNumber  & -0.0009                & 0.0042   & 0.0018\\     
     & (0.0009)              & (0.0031) & (0.0021)\\   
      \midrule
      \emph{Fixed-effects}\\
      $\text{Customer}_{ig}$    & Yes                                   & Yes          & Yes\\  
      $\text{YearMonth}_{gt}$      & Yes                                     & Yes          & Yes\\  
      \midrule
      \emph{Fit statistics}\\
      Observations            & 331,041                             & 173,015      & 130,157\\  
      R$^2$                  & 0.99917               & 0.99822  & 0.99504\\  
      Within R$^2$            & 0.00129   & 0.01017  & 0.00069\\  
      \midrule \midrule
    %  \multicolumn{5}{l}{\emph{Clustered (Customer \& YearMonth) standard-errors in parentheses}}\\
    \multicolumn{4}{l}{\emph{Bootstrapped standard-errors in parentheses via 500 replications}}\\
      \multicolumn{4}{l}{\emph{Signif. Codes: ***: 0.001, **: 0.01, *: 0.05}}\\
   \end{tabular}
   }
\end{table}

\paragraph{Additional purchase outcomes: Adopters vs. Non-Adopters}

In addition to offline price elasticity, we examine whether selection into online adoption is reflected in customers' broader pre-adoption purchasing behavior. For each of the twelve adoption cohorts, we construct a cohort-specific panel containing adopters in the focal cohort and non-adopters. We retain the six months preceding the focal cohort's adoption month and stack the resulting cohort-specific panels. We then estimate the linear pre-trend specification in Equation~\eqref{eq:stacked_linear_trend_model} for monthly spending, profit, quantity, number of orders, and measures of product variety. As shown in Table~\ref{tab:stacked_pretrend_purchase}, we do not find a significant linear differential trend between adopters and non-adopters in these monthly purchase metrics. 

\begin{table}[htp!]
   \caption{Stacked Linear Trend Estimates for Monthly Purchase Metrics - Adopters vs. Non-Adopters}
   \label{tab:stacked_pretrend_purchase}
   \centering
      \resizebox{\textwidth}{!}{
   \footnotesize{
   \begin{tabular}{lcccccccc}
      \tabularnewline \midrule \midrule
      DVs:    & Spend                 & Profit                & Quantity              & Unique Items          & Orders                & Unique Brands         & Subcategories         & Categories\\  
      Model:                  & (1)                   & (2)                   & (3)                   & (4)                   & (5)                   & (6)                   & (7)                   & (8)\\  
      \midrule
      \emph{Variables}\\
      \makecell[l]{Treatment $\times$ \\
      MonthNumber}  & -3.514                & -1.785                & -0.0440               & -0.0307               & -0.0030               & -0.0258               & -0.0257               & -0.0199\\   
                              & (3.790)               & (1.455)               & (0.0306)              & (0.0172)              & (0.0052)              & (0.0137)              & (0.0121)              & (0.0097)\\   
      \midrule
      \emph{Fixed-effects}\\
      $\text{Customer}_{ig}$    & Yes                   & Yes                   & Yes                   & Yes                   & Yes                   & Yes                   & Yes                   & Yes\\  
      $\text{YearMonth}_{gt}$      & Yes                   & Yes                   & Yes                   & Yes                   & Yes                   & Yes                   & Yes                   & Yes\\  
      \midrule
      \emph{Fit statistics}\\
      Observations            & 1,169,836             & 1,169,836             & 1,169,836             & 1,169,836             & 1,169,836             & 1,169,836             & 1,169,836             & 1,169,836\\  
      R$^2$                   & 0.53227               & 0.45620               & 0.66825               & 0.53379               & 0.54840               & 0.49563               & 0.47413               & 0.45513\\  
      Within R$^2$            & $5.36\times 10^{-6}$  & $1.03\times 10^{-5}$  & $6.42\times 10^{-6}$  & $1.76\times 10^{-5}$  & $1.61\times 10^{-6}$  & $2.18\times 10^{-5}$  & $2.86\times 10^{-5}$  & $2.56\times 10^{-5}$\\   
      \midrule \midrule
      \multicolumn{9}{l}{\emph{Clustered (Customer \& YearMonth) standard-errors in parentheses}}\\
      \multicolumn{9}{l}{\emph{Signif. Codes: ***: 0.001, **: 0.01, *: 0.05}}\\
   \end{tabular}
   }
   }
\end{table}

\begin{figure}[htp!]
    \centering
    \caption{Monthly Average Metrics -- \textit{Adopters (2020-03)} vs. \textit{Non-Adopters}}
\label{fig:se2_linear_trend_plot_2020_03}

\begin{subfigure}{0.49\textwidth}
        \centering
        \caption{Spend}
        \includegraphics[scale=0.20]{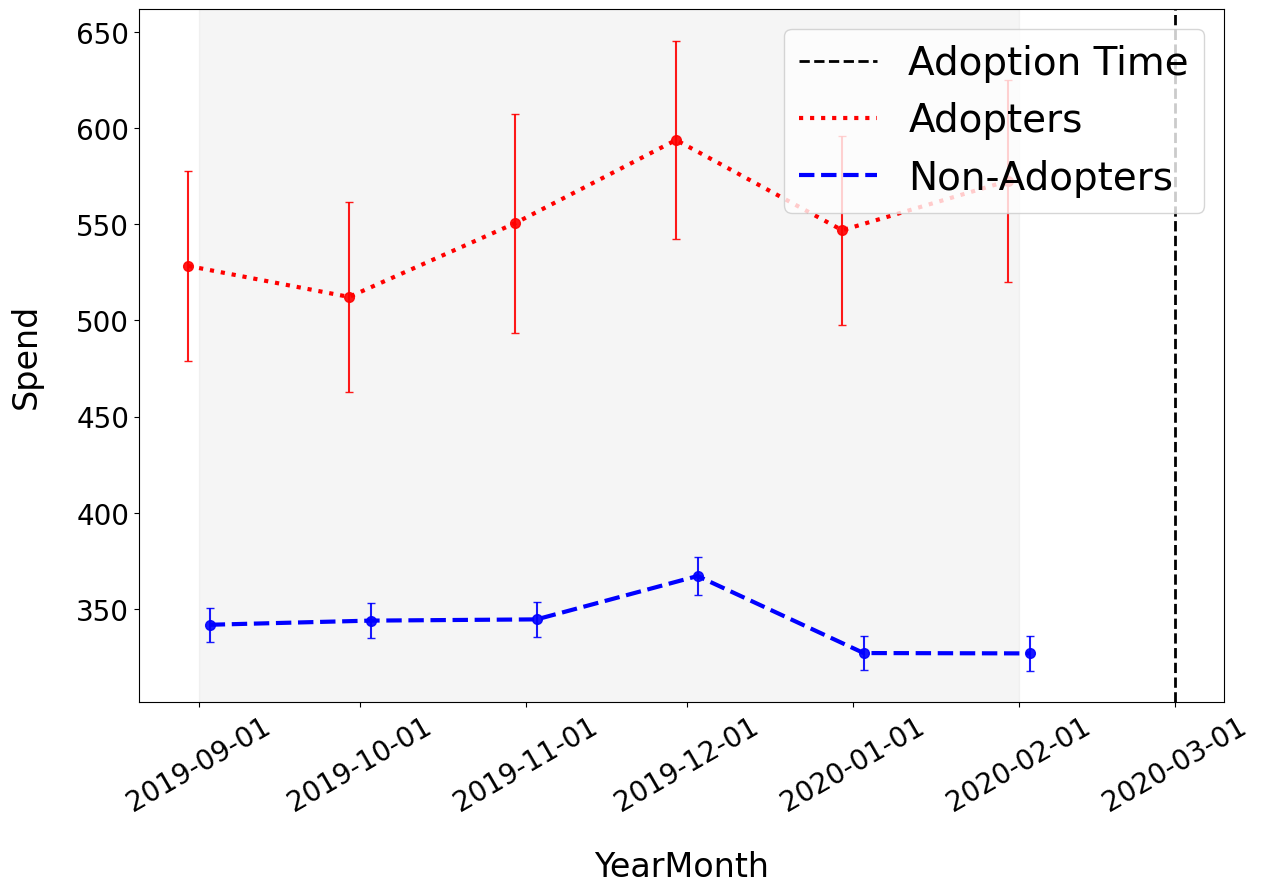}
    \end{subfigure}%
    ~ %add desired spacing between the subfigures, if needed
    \begin{subfigure}{0.49\textwidth}
        \centering
        \caption{Profit}
        \includegraphics[scale=0.20]{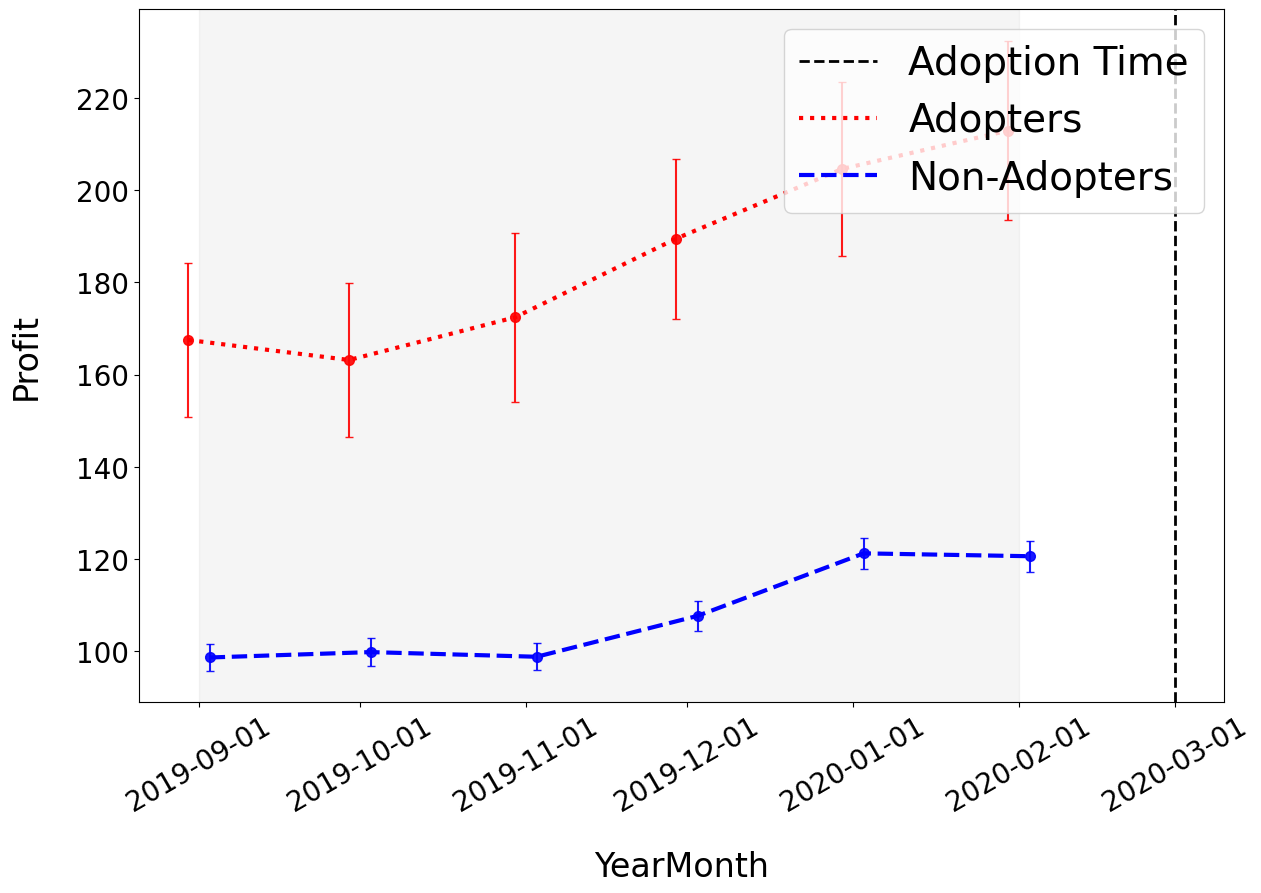}
    \end{subfigure}

    \begin{subfigure}{0.49\textwidth}
        \centering
        \caption{Quantity}
        \includegraphics[scale=0.20]{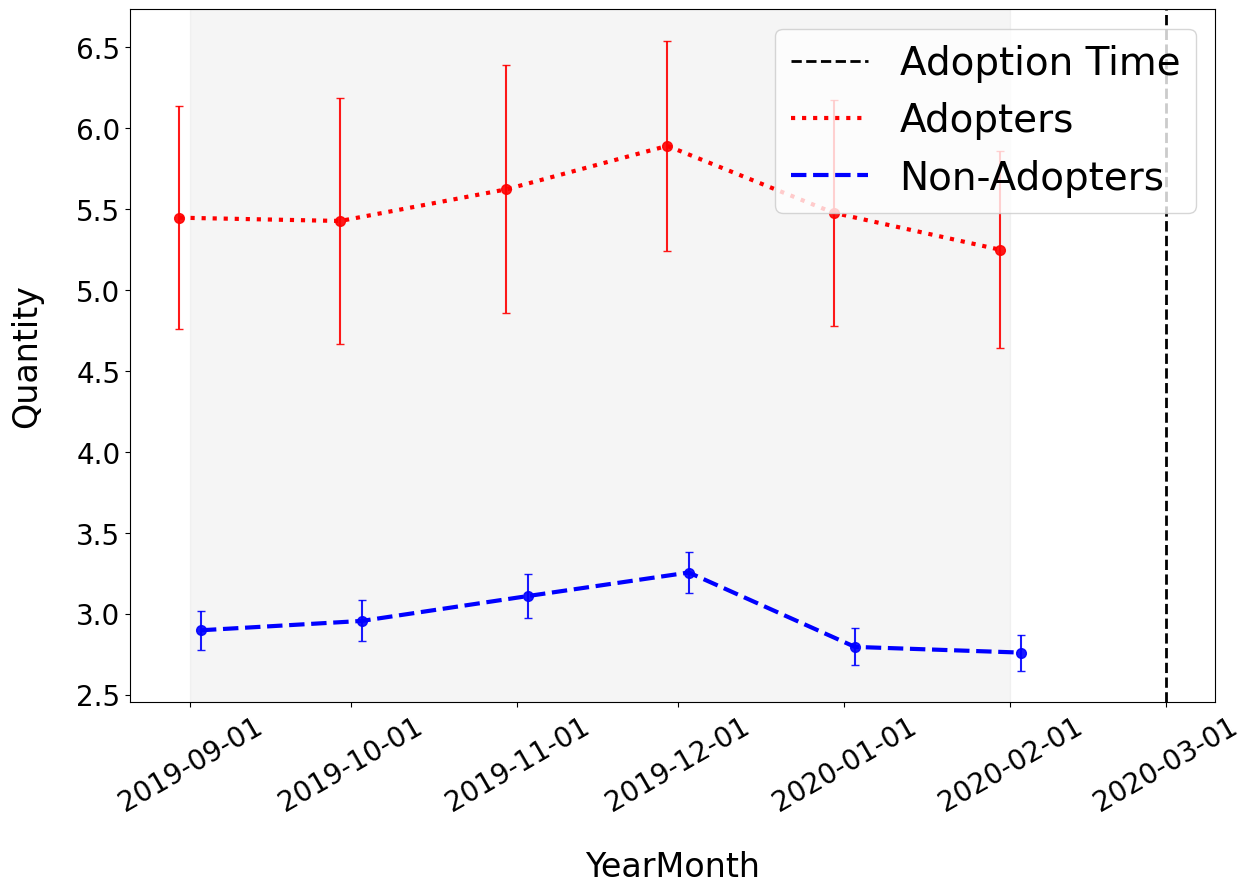}
    \end{subfigure}%
    ~ %add desired spacing between the subfigures, if needed
    \begin{subfigure}{0.49\textwidth}
        \centering
        \caption{Orders}
        \includegraphics[scale=0.20]{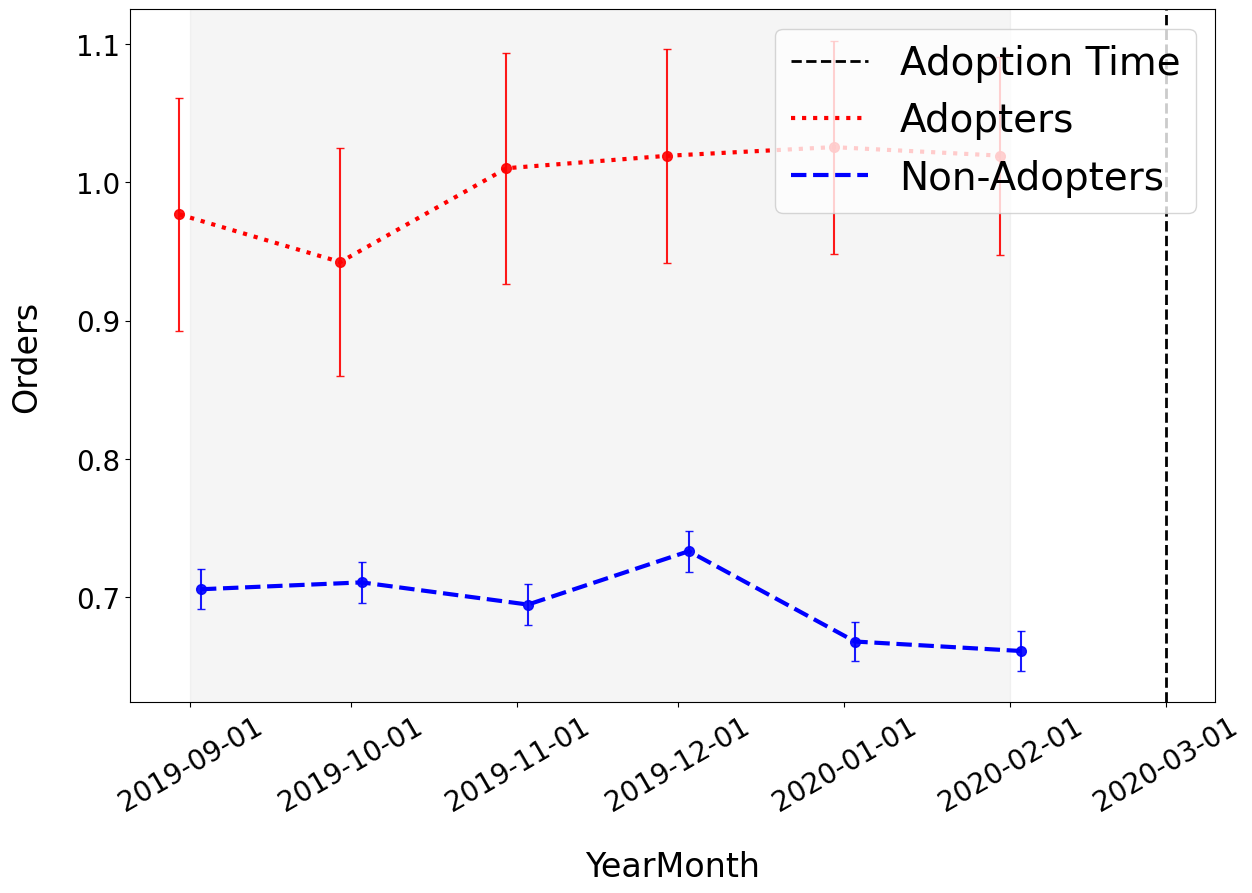}
    \end{subfigure}

    \vspace{1ex} % add some vertical space between the rows

    \begin{subfigure}{0.49\textwidth}
        \centering
        \caption{Unique Items}
        \includegraphics[scale=0.20]{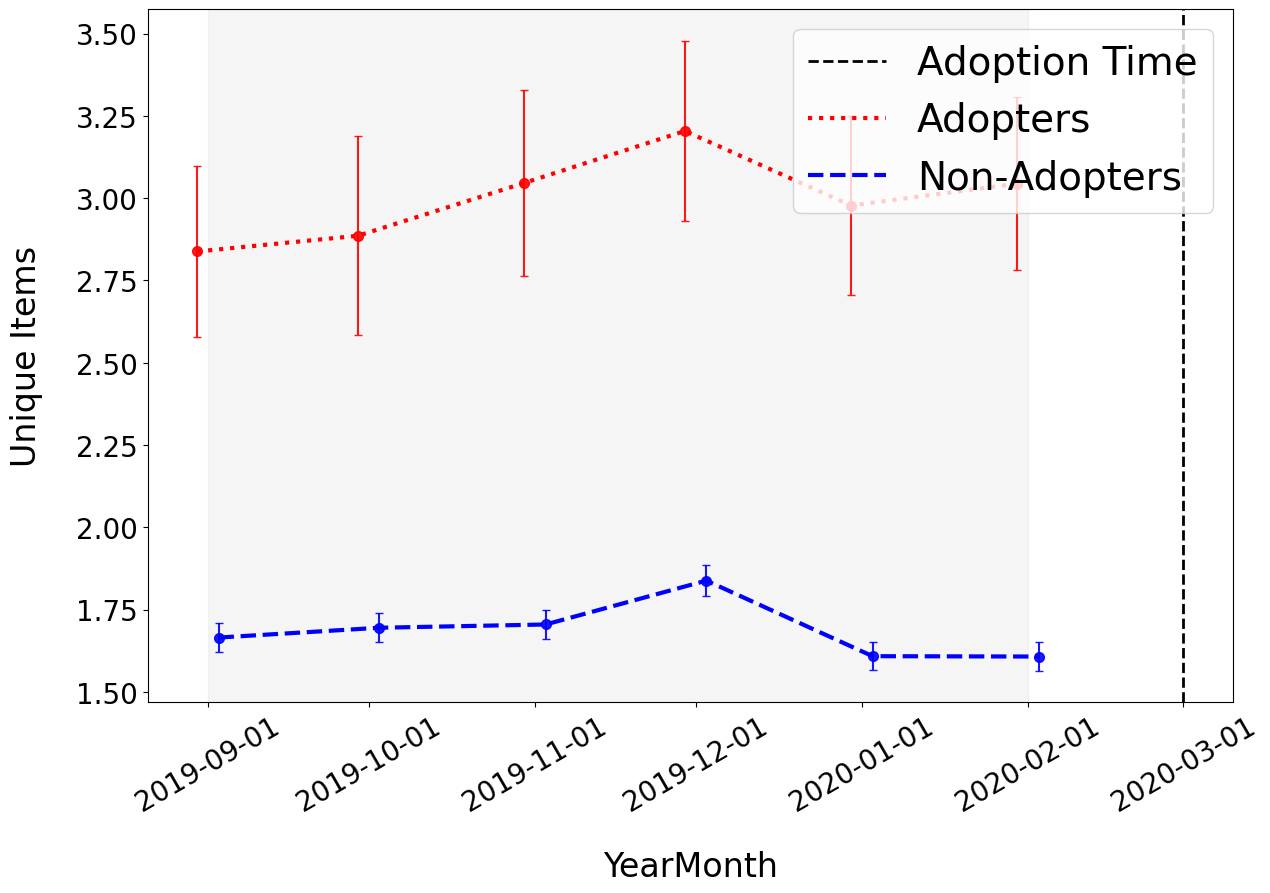}
    \end{subfigure}%
    ~ %add desired spacing between the subfigures, if needed
    \begin{subfigure}{0.49\textwidth}
        \centering
        \caption{Unique Brands}
        \includegraphics[scale=0.20]{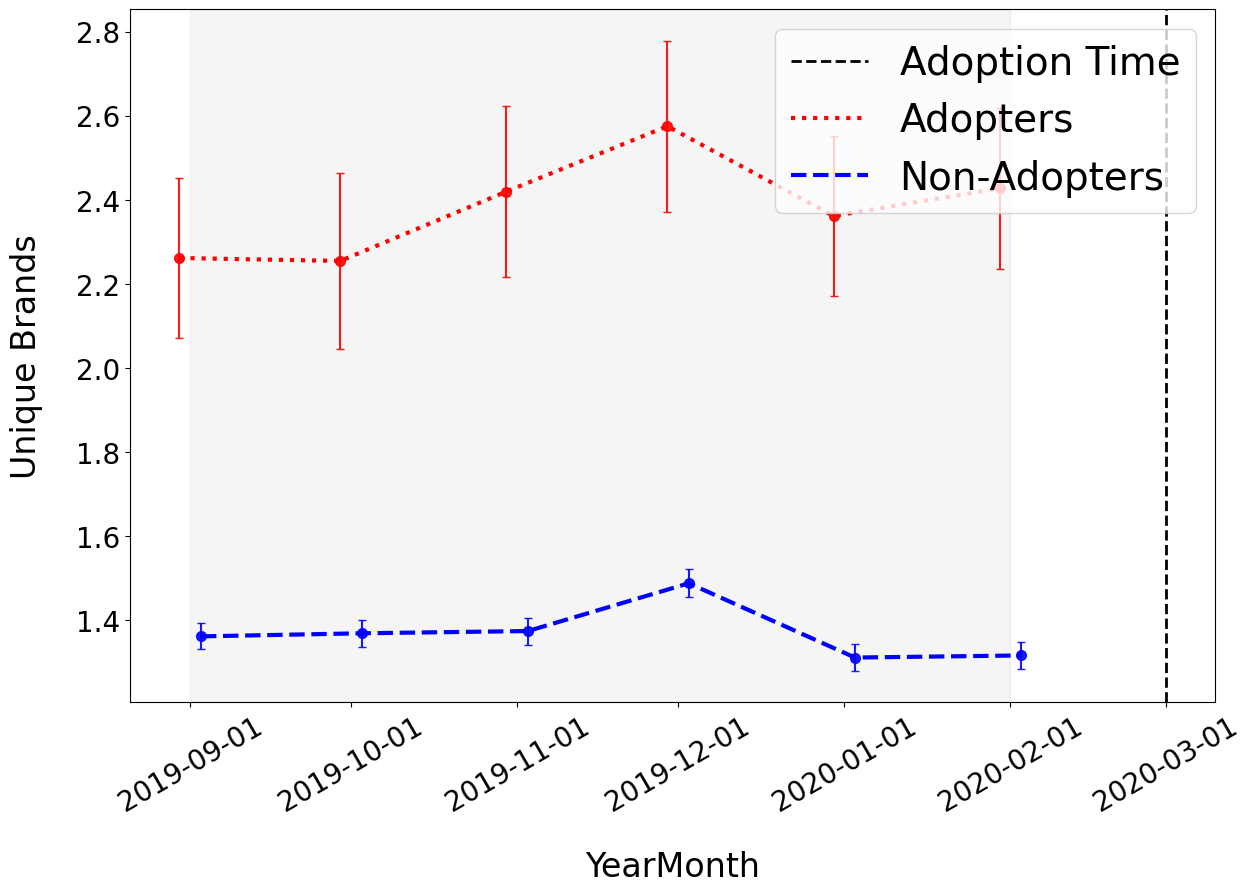}
    \end{subfigure}

    \vspace{1ex} % add some vertical space between the rows

    \begin{subfigure}{0.49\textwidth}
        \centering
        \caption{Unique Subcategories}
        \includegraphics[scale=0.20]{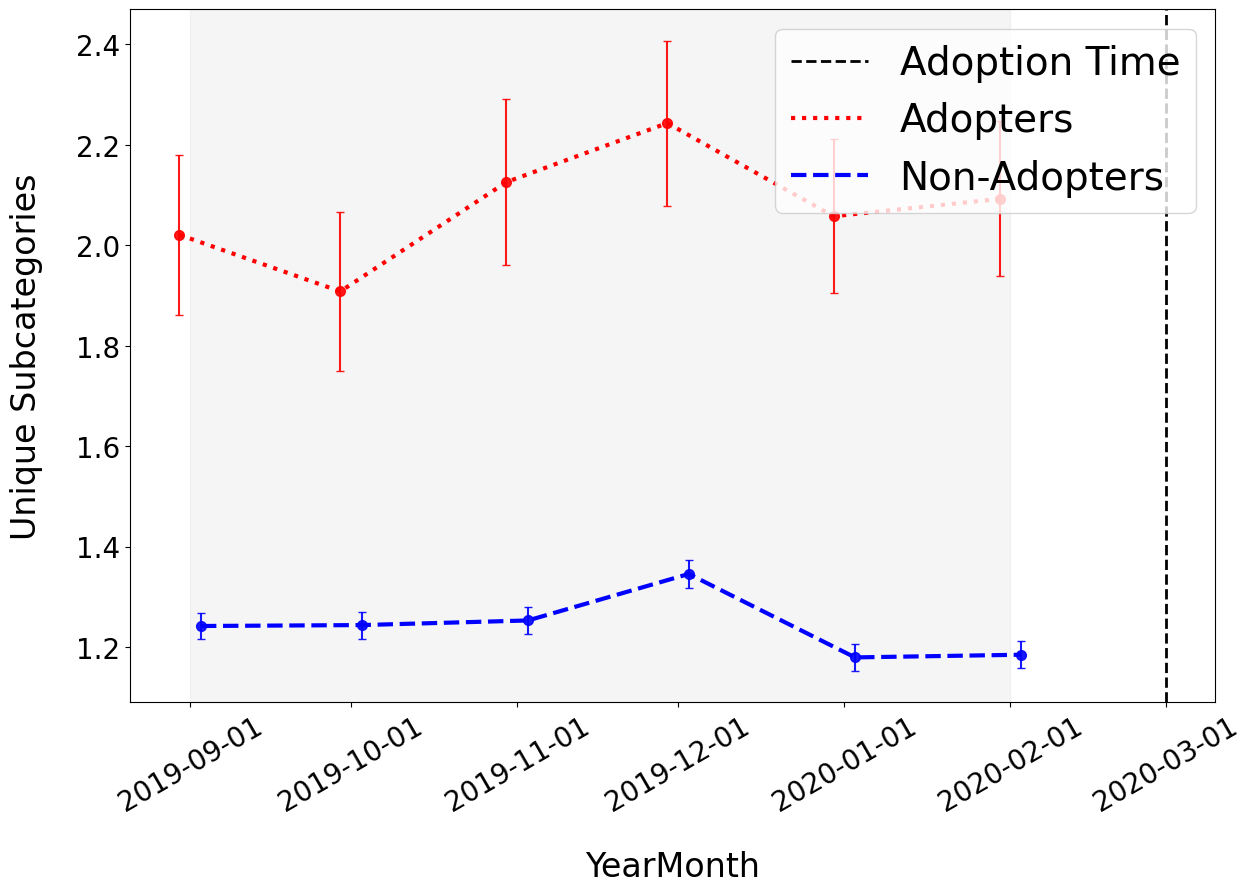}
    \end{subfigure}%
    ~ %add desired spacing between the subfigures, if needed
    \begin{subfigure}{0.49\textwidth}
        \centering
        \caption{Unique Categories}
        \includegraphics[scale=0.20]{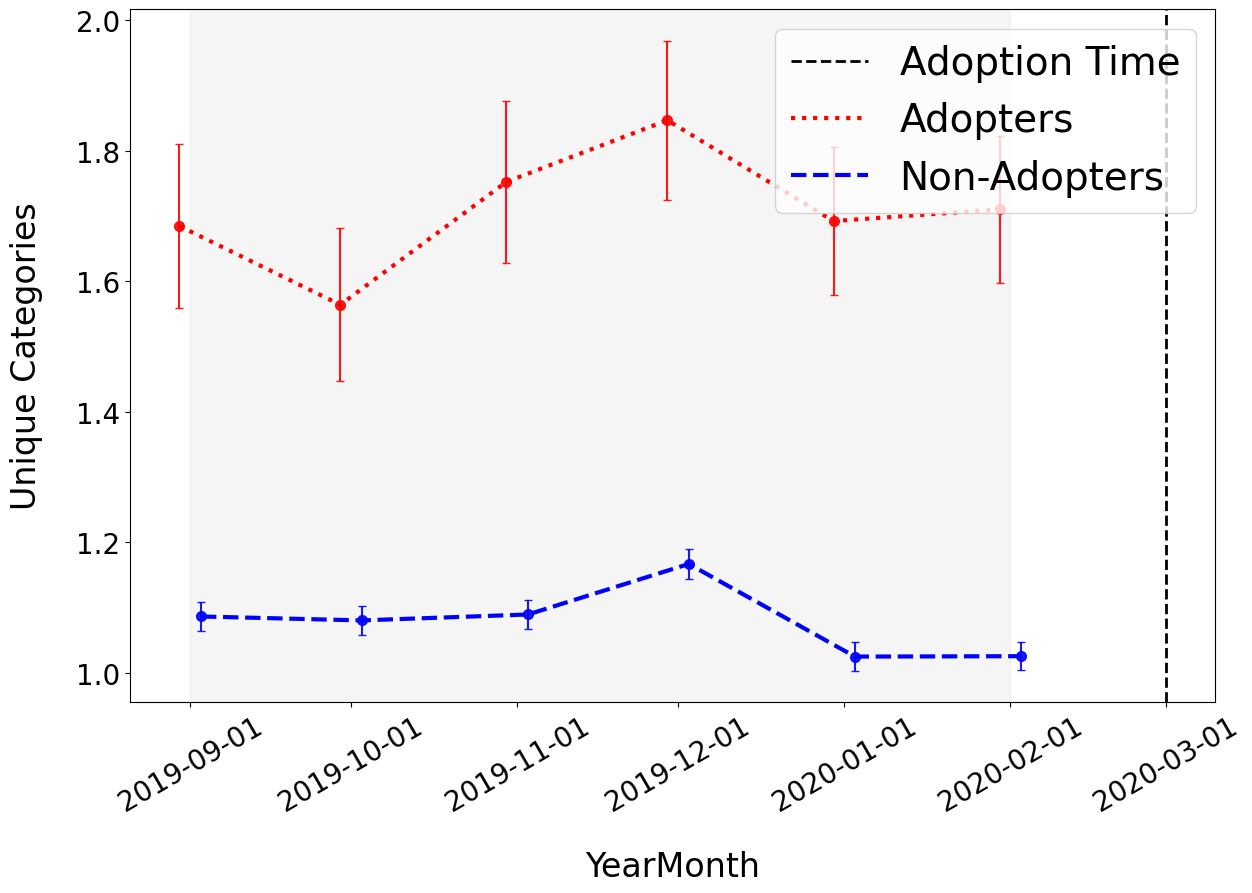}
    \end{subfigure}

\end{figure}

To complement the stacked estimates, Figure~\ref{fig:se2_linear_trend_plot_2020_03} illustrates the corresponding pre-adoption trajectories for the March 2020 adoption cohort and the non-adopter control group. The two groups exhibit broadly similar pre-adoption trends across these purchase outcomes. We observe comparable patterns for the other eleven adoption cohorts.

\clearpage

\subsection{Detailed Results for Matching, IPW Approaches to Control for Selection on Observables}
\label{appsec:selection.psm}

For each of the adoption cohorts under a product category, we estimate the following propensity score model: 
\begin{equation}
\label{eq:psm.logit}
    \textrm{PropensityScore}_{i} = f(X_{i}^{T}\theta),
\end{equation}
where $f(\cdot)$ is the logit link function where $f(X_{i}^{T}\theta) = \frac{1}{1+\exp(-X_{i}^{T}\theta)}$. $\theta\in \mathcal{R}^{M\times 1}$ is a column vector of coefficients. $X_{i}\in\mathcal{R} ^ {M \times 1}$ is a column vector with $M$ covariates of customer $i$, including demographics (e.g., gender, age and income) and the pre-adoption monthly purchase average measured using data up to the adoption period of the adoption cohort. For example, for customers who adopted in July 2019, we measure the pre-adoption variables for both adopters and non-adopters using the data up to June 30, 2019, and estimate the propensity score model for the adoption cohort in July 2019. 

Tables \ref{tab:psm.cohort.match.dog.2019} and \ref{tab:psm.cohort.match.dog.2020} present the propensity score model estimates for the twelve adoption cohorts in the dry dog food, dog hygiene and cat hygiene categories, respectively. Similarly, Tables \ref{tab:psm.cohort.match.hm.2019} and \ref{tab:psm.cohort.match.hm.2020} provide results for dog hygiene and Tables \ref{tab:psm.cohort.match.sand.2019} and \ref{tab:psm.cohort.match.sand.2020} provide results for cat hygiene. We find that certain observables, such as being female, having a high household income, and having a high pre-adoption monthly spend, increase the likelihood of adopting online shopping. 

Using the predicted propensity score, we apply the R package \textit{MatchIt} to conduct one-to-one nearest neighbor matching with replacement for each adoption cohort \citep{stuart2011matchit}. For each product category, we first estimate the ATT using the CS based on the matched sample, with results shown in Web Appendix $\S$\ref{appsec:selection.psm} Table \ref{tab:7.2.2.cohort.based.matching}. Second, we estimate the ATT using the CS estimator with the inverse propensity score weighting (IPW) method, with results shown in Table \ref{tab:7.3.IPTW}. These findings are consistent with our main results, demonstrating that consumers become more price sensitive offline after adopting online shopping. Detailed results are below.

\begin{table}[!htp] \centering 
  \caption{Propensity Score Logit Model for Dry Dog Food -  Adopter Cohorts between 2019-07 and 2019-12 vs. Non-Adopters} 
  \label{tab:psm.cohort.match.dog.2019}
  \footnotesize{
\begin{tabular}{@{\extracolsep{5pt}}lcccccc} 
\\[-1.8ex]\hline 
\hline \\[-1.8ex] 
 & \multicolumn{6}{c}{\textit{Dependent variable:}} \\ 
\cline{2-7} 
\\[-1.8ex] & \multicolumn{6}{c}{Adopter} \\ 
\\[-1.8ex] & 2019-07 & 2019-08 & 2019-09 & 2019-10 & 2019-11 & 2019-12\\ 
\hline \\[-1.8ex] 
 Gender\_Female & 0.288$^{***}$ & 0.382$^{***}$ & 0.671$^{***}$ & 1.030$^{***}$ & 0.655$^{***}$ & 0.283$^{***}$ \\ 
  & (0.078) & (0.072) & (0.078) & (0.073) & (0.062) & (0.064) \\ 

 Gender\_NoInformation & $-$0.944$^{***}$ & $-$0.177 & 0.354$^{***}$ & $-$0.114 & 0.367$^{***}$ & $-$0.180 \\ 
  & (0.155) & (0.111) & (0.104) & (0.122) & (0.083) & (0.095) \\ 

 Age$\geq$50 & $-$0.599$^{***}$ & $-$0.874$^{***}$ & $-$0.354$^{*}$ & $-$0.777$^{***}$ & $-$1.654$^{***}$ & $-$1.523$^{***}$ \\ 
  & (0.166) & (0.175) & (0.160) & (0.153) & (0.111) & (0.112) \\ 

 Age30\_40 & 0.140 & 0.667$^{***}$ & 0.410$^{**}$ & $-$0.047 & $-$0.269$^{**}$ & $-$0.677$^{***}$ \\ 
  & (0.161) & (0.163) & (0.156) & (0.149) & (0.100) & (0.106) \\ 

 Age40\_50 & $-$0.077 & 0.134 & $-$0.143 & 0.204 & $-$1.012$^{***}$ & $-$0.891$^{***}$ \\ 
  & (0.163) & (0.165) & (0.161) & (0.145) & (0.106) & (0.107) \\ 

 Age\_NoInformation & $-$1.798$^{***}$ & $-$1.188$^{***}$ & $-$1.496$^{***}$ & $-$2.105$^{***}$ & $-$2.036$^{***}$ & $-$2.120$^{***}$ \\ 
  & (0.272) & (0.226) & (0.245) & (0.267) & (0.163) & (0.174) \\ 

 Income$<$1000 & $-$0.417$^{***}$ & $-$0.302$^{**}$ & 0.038 & $-$0.047 & $-$0.256$^{***}$ & $-$0.314$^{***}$ \\ 
  & (0.105) & (0.102) & (0.093) & (0.088) & (0.075) & (0.081) \\ 

 Income$\geq$2000 & 0.223$^{*}$ & 0.762$^{***}$ & 0.149 & 0.043 & 0.197$^{**}$ & 0.181$^{*}$ \\ 
  & (0.095) & (0.088) & (0.093) & (0.088) & (0.073) & (0.077) \\ 

 Income\_NoInformation & $-$0.063 & 0.060 & $-$0.272$^{*}$ & $-$0.651$^{***}$ & $-$0.363$^{***}$ & $-$0.075 \\ 
  & (0.120) & (0.116) & (0.125) & (0.129) & (0.097) & (0.097) \\ 

 AvgSpendPerMonth & 0.001$^{***}$ & 0.001$^{***}$ & 0.001$^{***}$ & 0.001$^{***}$ & 0.0002$^{**}$ & 0.0001 \\ 
  & (0.0001) & (0.0001) & (0.0001) & (0.0001) & (0.0001) & (0.0001) \\ 

 AvgQuantitiesPerMonth & $-$0.095$^{***}$ & $-$0.108$^{***}$ & $-$0.058$^{**}$ & 0.010$^{*}$ & $-$0.008 & 0.010 \\ 
  & (0.025) & (0.024) & (0.018) & (0.005) & (0.008) & (0.009) \\ 

 AvgUniqueItemsPerMonth & $-$0.221$^{*}$ & 0.231$^{*}$ & $-$0.507$^{***}$ & $-$0.444$^{***}$ & $-$0.235$^{**}$ & $-$0.166$^{*}$ \\ 
  & (0.105) & (0.090) & (0.105) & (0.090) & (0.074) & (0.079) \\ 

 AvgUniqueOrdersPerMonth & 0.200$^{**}$ & $-$0.653$^{***}$ & $-$0.195$^{*}$ & $-$0.555$^{***}$ & 0.048 & $-$0.027 \\ 
  & (0.075) & (0.102) & (0.084) & (0.102) & (0.062) & (0.071) \\ 

 AvgUniqueBrandsPerMonth & 0.298$^{*}$ & 0.148 & 0.662$^{***}$ & 0.778$^{***}$ & 0.188 & 0.365$^{**}$ \\ 
  & (0.145) & (0.133) & (0.151) & (0.150) & (0.121) & (0.126) \\ 

 AvgUniqueSubcategoriesPerMonth & 0.132 & $-$0.079 & 1.189$^{***}$ & $-$0.129 & 0.243 & $-$0.306 \\ 
  & (0.164) & (0.175) & (0.181) & (0.183) & (0.164) & (0.162) \\ 

 AvgUniqueCategoriesPerMonth & 0.046 & 0.019 & $-$1.069$^{***}$ & 0.172 & 0.222 & 0.522$^{***}$ \\ 
  & (0.142) & (0.159) & (0.160) & (0.158) & (0.144) & (0.155) \\ 

 Intercept & $-$2.549$^{***}$ & $-$2.504$^{***}$ & $-$2.851$^{***}$ & $-$2.536$^{***}$ & $-$1.535$^{***}$ & $-$1.311$^{***}$ \\ 
  & (0.178) & (0.180) & (0.177) & (0.165) & (0.117) & (0.118) \\ 

\hline \\[-1.8ex] 
Observations & 8,023 & 8,049 & 8,038 & 8,048 & 8,112 & 8,082 \\ 
Log Likelihood & $-$2,586.499 & $-$2,989.092 & $-$2,841.850 & $-$2,963.057 & $-$4,088.072 & $-$3,736.394 \\ 
Akaike Inf. Crit. & 5,206.999 & 6,012.184 & 5,717.700 & 5,960.113 & 8,210.144 & 7,506.788 \\ 
\hline 
\hline \\[-1.8ex] 
\textit{Note:}  & \multicolumn{6}{r}{$^{*}$p$<$0.05; $^{**}$p$<$0.01; $^{***}$p$<$0.001} \\ 
\end{tabular}
}
\end{table}

\begin{table}[!htp] \centering 
  \caption{Propensity Score Logit Model for Dry Dog Food -  Adopter Cohorts between 2020-01 and 2020-06 vs. Non-Adopters} 
\label{tab:psm.cohort.match.dog.2020}
\footnotesize{
\begin{tabular}{@{\extracolsep{5pt}}lcccccc} 
\\[-1.8ex]\hline 
\hline \\[-1.8ex] 
 & \multicolumn{6}{c}{\textit{Dependent variable:}} \\ 
\cline{2-7} 
\\[-1.8ex] & \multicolumn{6}{c}{Adopter} \\ 
\\[-1.8ex] & 2020-01 & 2020-02 & 2020-03 & 2020-04 & 2020-05 & 2020-06 \\ 
\hline \\[-1.8ex] 
 Gender\_Female & 0.677$^{***}$ & 0.473$^{***}$ & 0.386$^{***}$ & 0.658$^{***}$ & 0.736$^{***}$ & 0.475$^{***}$ \\ 
  & (0.065) & (0.064) & (0.048) & (0.044) & (0.047) & (0.050) \\ 

 Gender\_NoInformation & $-$0.223$^{*}$ & $-$0.331$^{**}$ & $-$0.098 & 0.176$^{**}$ & 0.134$^{*}$ & $-$0.072 \\ 
  & (0.105) & (0.103) & (0.070) & (0.062) & (0.067) & (0.074) \\ 

 Age$\geq$50 & $-$0.957$^{***}$ & $-$0.966$^{***}$ & $-$1.150$^{***}$ & $-$1.143$^{***}$ & $-$0.923$^{***}$ & $-$1.165$^{***}$ \\ 
  & (0.118) & (0.129) & (0.093) & (0.084) & (0.090) & (0.097) \\ 

 Age30\_40 & $-$0.378$^{**}$ & $-$0.105 & $-$0.313$^{***}$ & $-$0.403$^{***}$ & $-$0.060 & $-$0.113 \\ 
  & (0.115) & (0.122) & (0.090) & (0.083) & (0.088) & (0.092) \\ 

 Age40\_50 & $-$0.682$^{***}$ & $-$0.363$^{**}$ & $-$0.698$^{***}$ & $-$0.602$^{***}$ & $-$0.602$^{***}$ & $-$0.742$^{***}$ \\ 
  & (0.118) & (0.124) & (0.091) & (0.083) & (0.090) & (0.095) \\ 

 Age\_NoInformation & $-$2.308$^{***}$ & $-$2.600$^{***}$ & $-$1.399$^{***}$ & $-$2.220$^{***}$ & $-$2.318$^{***}$ & $-$1.690$^{***}$ \\ 
  & (0.215) & (0.256) & (0.119) & (0.127) & (0.151) & (0.136) \\ 

 Income$<$1000 & $-$0.211$^{*}$ & 0.212$^{**}$ & 0.051 & $-$0.223$^{***}$ & $-$0.061 & $-$0.470$^{***}$ \\ 
  & (0.082) & (0.080) & (0.061) & (0.055) & (0.058) & (0.060) \\ 

 Income$\geq$2000 & 0.115 & 0.163$^{*}$ & 0.481$^{***}$ & 0.335$^{***}$ & 0.210$^{***}$ & $-$0.334$^{***}$ \\ 
  & (0.079) & (0.083) & (0.058) & (0.053) & (0.057) & (0.061) \\ 

 Income\_NoInformation & $-$0.093 & $-$0.149 & $-$0.159$^{*}$ & $-$0.450$^{***}$ & $-$0.215$^{**}$ & $-$0.570$^{***}$ \\ 
  & (0.100) & (0.106) & (0.078) & (0.071) & (0.074) & (0.078) \\ 

 AvgSpendPerMonth & 0.0004$^{***}$ & 0.0002 & 0.0005$^{***}$ & 0.0004$^{***}$ & 0.001$^{***}$ & 0.0001 \\ 
  & (0.0001) & (0.0001) & (0.0001) & (0.0001) & (0.0001) & (0.0001) \\ 

 AvgQuantitiesPerMonth & $-$0.051$^{**}$ & $-$0.008 & $-$0.011 & 0.009 & $-$0.013 & $-$0.034$^{**}$ \\ 
  & (0.016) & (0.012) & (0.008) & (0.007) & (0.008) & (0.011) \\ 

 AvgUniqueItemsPerMonth & $-$0.285$^{**}$ & $-$0.557$^{***}$ & $-$0.156$^{*}$ & $-$0.076 & $-$0.028 & $-$0.029 \\ 
  & (0.095) & (0.099) & (0.061) & (0.054) & (0.069) & (0.076) \\ 

 AvgUniqueOrdersPerMonth & $-$0.223$^{**}$ & $-$0.409$^{***}$ & $-$0.558$^{***}$ & $-$0.455$^{***}$ & $-$0.786$^{***}$ & $-$0.267$^{***}$ \\ 
  & (0.085) & (0.084) & (0.066) & (0.062) & (0.075) & (0.065) \\ 

 AvgUniqueBrandsPerMonth & 0.553$^{***}$ & 1.275$^{***}$ & 0.286$^{**}$ & $-$0.134 & 0.413$^{***}$ & 0.547$^{***}$ \\ 
  & (0.143) & (0.150) & (0.103) & (0.086) & (0.113) & (0.119) \\ 

 AvgUniqueSubcategoriesPerMonth & 0.015 & $-$0.410$^{*}$ & 0.467$^{***}$ & 0.088 & $-$0.627$^{***}$ & $-$0.426$^{*}$ \\ 
  & (0.184) & (0.181) & (0.126) & (0.126) & (0.155) & (0.166) \\ 

 AvgUniqueCategoriesPerMonth & 0.243 & 0.279 & 0.025 & 0.836$^{***}$ & 0.907$^{***}$ & 0.374$^{*}$ \\ 
  & (0.165) & (0.178) & (0.121) & (0.122) & (0.139) & (0.149) \\ 

 Intercept & $-$1.852$^{***}$ & $-$2.011$^{***}$ & $-$1.158$^{***}$ & $-$0.870$^{***}$ & $-$1.039$^{***}$ & $-$0.715$^{***}$ \\ 
  & (0.132) & (0.136) & (0.099) & (0.091) & (0.099) & (0.101) \\ 

\hline \\[-1.8ex] 
Observations & 8,076 & 8,081 & 8,272 & 8,376 & 8,295 & 8,214 \\ 
Log Likelihood & $-$3,577.410 & $-$3,603.534 & $-$6,105.810 & $-$7,120.955 & $-$6,365.540 & $-$5,585.395 \\ 
Akaike Inf. Crit. & 7,188.819 & 7,241.067 & 12,245.620 & 14,275.910 & 12,765.080 & 11,204.790 \\ 
\hline 
\hline \\[-1.8ex] 
\textit{Note:}  & \multicolumn{6}{r}{$^{*}$p$<$0.05; $^{**}$p$<$0.01; $^{***}$p$<$0.001} \\ 
\end{tabular}
}
\end{table}

\begin{table}[!htp] \centering 
  \caption{Propensity Score Logit Model for Dog Hygiene -  Adopter Cohorts between 2019-07 and 2019-12 vs. Non-Adopters} 
\label{tab:psm.cohort.match.hm.2019}
\footnotesize{
\begin{tabular}{@{\extracolsep{5pt}}lcccccc} 
\\[-1.8ex]\hline 
\hline \\[-1.8ex] 
 & \multicolumn{6}{c}{\textit{Dependent variable:}} \\ 
\cline{2-7} 
\\[-1.8ex] & \multicolumn{6}{c}{Adopter} \\ 
\\[-1.8ex] & 2019-07 & 2019-08 & 2019-09 & 2019-10 & 2019-11 & 2019-12\\ 
\hline \\[-1.8ex] 
 Gender\_Female & 0.809$^{***}$ & 0.697$^{***}$ & 0.552$^{***}$ & 0.267$^{**}$ & 0.436$^{***}$ & 0.253$^{**}$ \\ 
  & (0.118) & (0.105) & (0.089) & (0.092) & (0.075) & (0.080) \\ 

 Gender\_NoInformation & $-$0.169 & 0.501$^{***}$ & $-$0.695$^{***}$ & $-$0.035 & 0.076 & 0.037 \\ 
  & (0.197) & (0.145) & (0.164) & (0.135) & (0.109) & (0.116) \\ 

 Age$\geq$50 & $-$0.636$^{*}$ & $-$1.062$^{***}$ & $-$1.063$^{***}$ & $-$0.713$^{***}$ & $-$0.771$^{***}$ & $-$1.237$^{***}$ \\ 
  & (0.275) & (0.198) & (0.171) & (0.183) & (0.165) & (0.170) \\ 

 Age30\_40 & 0.610$^{*}$ & $-$0.246 & $-$0.244 & $-$0.264 & 0.077 & $-$0.046 \\ 
  & (0.261) & (0.192) & (0.164) & (0.179) & (0.160) & (0.159) \\ 

 Age40\_50 & 0.148 & $-$0.400$^{*}$ & $-$0.705$^{***}$ & $-$0.442$^{*}$ & $-$0.103 & $-$0.354$^{*}$ \\ 
  & (0.264) & (0.194) & (0.167) & (0.179) & (0.160) & (0.161) \\ 

 Age\_NoInformation & $-$0.059 & $-$1.842$^{***}$ & $-$2.730$^{***}$ & $-$1.562$^{***}$ & $-$0.930$^{***}$ & $-$2.127$^{***}$ \\ 
  & (0.303) & (0.302) & (0.358) & (0.266) & (0.208) & (0.277) \\ 

 Income$<$1000 & $-$0.354$^{*}$ & $-$0.558$^{***}$ & $-$0.183 & 0.186 & 0.167 & $-$0.492$^{***}$ \\ 
  & (0.176) & (0.163) & (0.124) & (0.119) & (0.099) & (0.117) \\ 

 Income$\geq$2000 & 0.307$^{*}$ & 0.345$^{**}$ & $-$0.239$^{*}$ & $-$0.286$^{**}$ & 0.074 & $-$0.043 \\ 
  & (0.127) & (0.107) & (0.101) & (0.104) & (0.084) & (0.087) \\ 

 Income\_NoInformation & 0.535$^{**}$ & $-$0.225 & 0.540$^{***}$ & $-$0.029 & $-$0.043 & 0.009 \\ 
  & (0.165) & (0.169) & (0.123) & (0.143) & (0.119) & (0.122) \\ 

 AvgSpendPerMonth & 0.001$^{***}$ & 0.001$^{***}$ & 0.0004$^{**}$ & 0.001$^{***}$ & 0.0002 & 0.001$^{***}$ \\ 
  & (0.0001) & (0.0001) & (0.0001) & (0.0001) & (0.0001) & (0.0001) \\ 

 AvgQuantitiesPerMonth & $-$0.147$^{***}$ & 0.015 & $-$0.148$^{***}$ & $-$0.029 & $-$0.039$^{**}$ & $-$0.087$^{***}$ \\ 
  & (0.036) & (0.009) & (0.029) & (0.016) & (0.013) & (0.020) \\ 

 AvgUniqueItemsPerMonth & $-$0.043 & $-$0.249$^{***}$ & $-$0.013 & $-$0.145 & $-$0.327$^{***}$ & $-$0.149 \\ 
  & (0.115) & (0.072) & (0.089) & (0.078) & (0.071) & (0.092) \\ 

 AvgUniqueOrdersPerMonth & $-$0.065 & 0.187$^{**}$ & $-$0.684$^{***}$ & $-$0.455$^{***}$ & $-$0.149$^{*}$ & $-$0.103 \\ 
  & (0.091) & (0.069) & (0.095) & (0.091) & (0.061) & (0.092) \\ 

 AvgUniqueBrandsPerMonth & 0.475$^{***}$ & 0.462$^{***}$ & 0.448$^{***}$ & 0.399$^{**}$ & 0.425$^{***}$ & $-$0.253$^{*}$ \\ 
  & (0.141) & (0.123) & (0.136) & (0.123) & (0.114) & (0.129) \\ 

 AvgUniqueSubcategoriesPerMonth & 0.161 & 0.023 & 0.199 & 0.048 & 0.655$^{***}$ & 1.241$^{***}$ \\ 
  & (0.167) & (0.161) & (0.169) & (0.151) & (0.141) & (0.178) \\ 

 AvgUniqueCategoriesPerMonth & $-$0.463$^{**}$ & $-$0.239 & 0.068 & 0.095 & $-$0.305$^{*}$ & $-$0.512$^{***}$ \\ 
  & (0.161) & (0.156) & (0.157) & (0.146) & (0.134) & (0.152) \\ 

 Intercept & $-$3.145$^{***}$ & $-$2.318$^{***}$ & $-$1.544$^{***}$ & $-$1.800$^{***}$ & $-$1.721$^{***}$ & $-$1.348$^{***}$ \\ 
  & (0.296) & (0.223) & (0.196) & (0.205) & (0.186) & (0.182) \\ 

\hline \\[-1.8ex] 
Observations & 3,634 & 3,653 & 3,678 & 3,664 & 3,725 & 3,702 \\ 
Log Likelihood & $-$1,353.099 & $-$1,680.290 & $-$1,967.035 & $-$1,897.595 & $-$2,712.319 & $-$2,353.311 \\ 
Akaike Inf. Crit. & 2,740.197 & 3,394.581 & 3,968.071 & 3,829.189 & 5,458.638 & 4,740.623 \\ 
\hline 
\hline \\[-1.8ex] 
\textit{Note:}  & \multicolumn{6}{r}{$^{*}$p$<$0.05; $^{**}$p$<$0.01; $^{***}$p$<$0.001} \\ 
\end{tabular}
}
\end{table}

\begin{table}[!htp] \centering 
  \caption{Propensity Score Logit Model for Dog Hygiene -  Adopter Cohorts between 2020-01 and 2020-06 vs. Non-Adopters} 
\label{tab:psm.cohort.match.hm.2020}
\footnotesize{
\begin{tabular}{@{\extracolsep{5pt}}lcccccc} 
\\[-1.8ex]\hline 
\hline \\[-1.8ex] 
 & \multicolumn{6}{c}{\textit{Dependent variable:}} \\ 
\cline{2-7} 
\\[-1.8ex] & \multicolumn{6}{c}{Adopter} \\ 
\\[-1.8ex] & 2020-01 & 2020-02 & 2020-03 & 2020-04 & 2020-05 & 2020-06 \\ 
\hline \\[-1.8ex] 
 Gender\_Female & 0.627$^{***}$ & 0.370$^{***}$ & 0.413$^{***}$ & 0.582$^{***}$ & 0.571$^{***}$ & 0.435$^{***}$ \\ 
  & (0.092) & (0.086) & (0.056) & (0.054) & (0.057) & (0.063) \\ 

 Gender\_NoInformation & 0.579$^{***}$ & 0.172 & $-$0.093 & $-$0.333$^{***}$ & 0.053 & $-$0.056 \\ 
  & (0.123) & (0.121) & (0.083) & (0.085) & (0.084) & (0.093) \\ 

 Age$\geq$50 & $-$1.372$^{***}$ & $-$1.130$^{***}$ & $-$1.080$^{***}$ & $-$0.834$^{***}$ & $-$1.256$^{***}$ & $-$1.169$^{***}$ \\ 
  & (0.171) & (0.168) & (0.114) & (0.119) & (0.121) & (0.128) \\ 

 Age30\_40 & $-$0.134 & $-$0.341$^{*}$ & $-$0.483$^{***}$ & $-$0.122 & $-$0.318$^{**}$ & $-$0.274$^{*}$ \\ 
  & (0.157) & (0.162) & (0.114) & (0.118) & (0.118) & (0.123) \\ 

 Age40\_50 & $-$0.802$^{***}$ & $-$0.800$^{***}$ & $-$0.964$^{***}$ & $-$0.261$^{*}$ & $-$0.675$^{***}$ & $-$0.825$^{***}$ \\ 
  & (0.163) & (0.165) & (0.114) & (0.117) & (0.119) & (0.125) \\ 

 Age\_NoInformation & $-$1.945$^{***}$ & $-$2.592$^{***}$ & $-$2.108$^{***}$ & $-$1.115$^{***}$ & $-$1.837$^{***}$ & $-$2.046$^{***}$ \\ 
  & (0.252) & (0.286) & (0.162) & (0.148) & (0.160) & (0.184) \\ 

 Income$<$1000 & 0.277$^{*}$ & $-$0.157 & $-$0.118 & $-$0.144 & $-$0.222$^{**}$ & $-$0.182$^{*}$ \\ 
  & (0.118) & (0.125) & (0.081) & (0.077) & (0.081) & (0.090) \\ 

 Income$\geq$2000 & 0.304$^{**}$ & 0.246$^{**}$ & 0.430$^{***}$ & 0.331$^{***}$ & 0.233$^{***}$ & 0.263$^{***}$ \\ 
  & (0.100) & (0.095) & (0.062) & (0.060) & (0.062) & (0.069) \\ 

 Income\_NoInformation & $-$0.080 & 0.119 & $-$0.263$^{**}$ & $-$0.317$^{***}$ & $-$0.059 & $-$0.137 \\ 
  & (0.144) & (0.134) & (0.096) & (0.092) & (0.093) & (0.105) \\ 

 AvgSpendPerMonth & 0.001$^{***}$ & 0.001$^{***}$ & 0.001$^{***}$ & 0.001$^{***}$ & 0.001$^{***}$ & 0.0005$^{***}$ \\ 
  & (0.0001) & (0.0001) & (0.0001) & (0.0001) & (0.0001) & (0.0001) \\ 

 AvgQuantitiesPerMonth & $-$0.150$^{***}$ & $-$0.007 & 0.011 & $-$0.046$^{***}$ & $-$0.009 & $-$0.014 \\ 
  & (0.024) & (0.007) & (0.008) & (0.012) & (0.009) & (0.009) \\ 

 AvgUniqueItemsPerMonth & 0.140 & $-$0.345$^{***}$ & $-$0.139$^{**}$ & 0.056 & $-$0.185$^{**}$ & 0.138$^{*}$ \\ 
  & (0.085) & (0.075) & (0.054) & (0.065) & (0.058) & (0.070) \\ 

 AvgUniqueOrdersPerMonth & $-$0.124 & $-$0.080 & $-$0.172$^{***}$ & $-$0.498$^{***}$ & $-$0.372$^{***}$ & $-$0.605$^{***}$ \\ 
  & (0.084) & (0.088) & (0.038) & (0.064) & (0.067) & (0.084) \\ 

 AvgUniqueBrandsPerMonth & $-$0.050 & 0.436$^{***}$ & 0.087 & $-$0.224$^{*}$ & 0.242$^{**}$ & $-$0.292$^{*}$ \\ 
  & (0.133) & (0.113) & (0.091) & (0.102) & (0.093) & (0.116) \\ 

 AvgUniqueSubcategoriesPerMonth & 0.951$^{***}$ & 0.664$^{***}$ & 0.487$^{***}$ & 0.922$^{***}$ & 0.502$^{***}$ & 0.919$^{***}$ \\ 
  & (0.171) & (0.156) & (0.116) & (0.126) & (0.131) & (0.147) \\ 

 AvgUniqueCategoriesPerMonth & $-$0.616$^{***}$ & $-$0.769$^{***}$ & $-$0.139 & $-$0.257$^{*}$ & $-$0.264$^{*}$ & $-$0.394$^{**}$ \\ 
  & (0.154) & (0.160) & (0.113) & (0.117) & (0.135) & (0.139) \\ 

 Intercept & $-$2.091$^{***}$ & $-$1.383$^{***}$ & $-$0.282$^{*}$ & $-$0.497$^{***}$ & $-$0.247 & $-$0.426$^{**}$ \\ 
  & (0.196) & (0.190) & (0.125) & (0.134) & (0.132) & (0.141) \\ 

\hline \\[-1.8ex] 
Observations & 3,689 & 3,688 & 3,964 & 4,024 & 3,935 & 3,842 \\ 
Log Likelihood & $-$2,067.730 & $-$2,175.618 & $-$4,499.026 & $-$4,832.667 & $-$4,400.367 & $-$3,656.464 \\ 
Akaike Inf. Crit. & 4,169.460 & 4,385.235 & 9,032.052 & 9,699.334 & 8,834.733 & 7,346.928 \\ 
\hline 
\hline \\[-1.8ex] 
\textit{Note:}  & \multicolumn{6}{r}{$^{*}$p$<$0.05; $^{**}$p$<$0.01; $^{***}$p$<$0.001} \\ 
\end{tabular}
}
\end{table}

\begin{table}[!htp] \centering 
  \caption{Propensity Score Logit Model for Cat Hygiene -  Adopter Cohorts between 2019-07 and 2019-12 vs. Non-Adopters} 
\label{tab:psm.cohort.match.sand.2019}
\footnotesize{
\begin{tabular}{@{\extracolsep{5pt}}lcccccc} 
\\[-1.8ex]\hline 
\hline \\[-1.8ex] 
 & \multicolumn{6}{c}{\textit{Dependent variable:}} \\ 
\cline{2-7} 
\\[-1.8ex] & \multicolumn{6}{c}{Adopter} \\ 
\\[-1.8ex] & 2019-07 & 2019-08 & 2019-09 & 2019-10 & 2019-11 & 2019-12\\ 
\hline \\[-1.8ex] 
 Gender\_Female & 0.835$^{***}$ & 0.450$^{***}$ & 0.986$^{***}$ & 0.823$^{***}$ & 0.526$^{***}$ & 0.364$^{***}$ \\ 
  & (0.141) & (0.129) & (0.130) & (0.131) & (0.107) & (0.107) \\ 

 Gender\_NoInformation & 0.055 & 0.315 & $-$0.655$^{*}$ & 0.164 & 0.034 & $-$0.424$^{*}$ \\ 
  & (0.229) & (0.179) & (0.259) & (0.195) & (0.158) & (0.179) \\ 

 Age$\geq$50 & $-$0.024 & $-$0.217 & $-$0.605$^{*}$ & $-$0.916$^{***}$ & $-$0.503 & $-$0.764$^{**}$ \\ 
  & (0.361) & (0.353) & (0.277) & (0.218) & (0.264) & (0.236) \\ 

 Age30\_40 & 1.468$^{***}$ & 1.077$^{**}$ & 0.730$^{**}$ & $-$0.334 & 1.052$^{***}$ & 0.182 \\ 
  & (0.352) & (0.345) & (0.270) & (0.219) & (0.254) & (0.234) \\ 

 Age40\_50 & 0.302 & 0.325 & 0.057 & $-$0.929$^{***}$ & 0.504$^{*}$ & 0.031 \\ 
  & (0.361) & (0.349) & (0.269) & (0.222) & (0.255) & (0.229) \\ 

 Age\_NoInformation & $-$15.840 & $-$0.103 & $-$16.557 & $-$17.598 & $-$1.475$^{***}$ & $-$1.916$^{***}$ \\ 
  & (380.738) & (0.404) & (382.142) & (380.148) & (0.408) & (0.394) \\ 

 Income$<$1000 & 0.269 & $-$0.407$^{*}$ & $-$0.334$^{*}$ & $-$0.417$^{*}$ & $-$0.544$^{***}$ & $-$0.396$^{*}$ \\ 
  & (0.229) & (0.182) & (0.161) & (0.199) & (0.153) & (0.171) \\ 

 Income$\geq$2000 & 1.356$^{***}$ & $-$0.137 & $-$0.495$^{***}$ & 0.368$^{**}$ & 0.076 & 0.440$^{***}$ \\ 
  & (0.171) & (0.134) & (0.126) & (0.130) & (0.110) & (0.117) \\ 

 Income\_NoInformation & 0.072 & $-$0.148 & $-$1.423$^{***}$ & $-$0.066 & $-$0.364$^{*}$ & $-$0.507$^{*}$ \\ 
  & (0.281) & (0.187) & (0.255) & (0.193) & (0.171) & (0.200) \\ 

 AvgSpendPerMonth & 0.001$^{**}$ & 0.001$^{***}$ & 0.001$^{***}$ & 0.002$^{***}$ & $-$0.0001 & 0.0002 \\ 
  & (0.0002) & (0.0001) & (0.0002) & (0.0002) & (0.0001) & (0.0002) \\ 

 AvgQuantitiesPerMonth & 0.018$^{*}$ & $-$0.003 & 0.005 & $-$0.076$^{***}$ & 0.044$^{***}$ & 0.009 \\ 
  & (0.008) & (0.008) & (0.009) & (0.015) & (0.006) & (0.008) \\ 

 AvgUniqueItemsPerMonth & $-$0.213$^{*}$ & 0.101 & $-$0.104 & 0.241$^{**}$ & $-$0.649$^{***}$ & $-$0.268$^{**}$ \\ 
  & (0.090) & (0.053) & (0.079) & (0.078) & (0.094) & (0.082) \\ 

 AvgUniqueOrdersPerMonth & $-$0.076 & $-$0.176$^{*}$ & $-$0.340$^{**}$ & $-$0.804$^{***}$ & $-$0.962$^{***}$ & $-$0.716$^{***}$ \\ 
  & (0.109) & (0.084) & (0.104) & (0.116) & (0.126) & (0.115) \\ 

 AvgUniqueBrandsPerMonth & 0.116 & $-$0.003 & 0.143 & $-$0.528$^{***}$ & 0.813$^{***}$ & $-$0.002 \\ 
  & (0.186) & (0.117) & (0.138) & (0.157) & (0.172) & (0.159) \\ 

 AvgUniqueSubcategoriesPerMonth & 0.417 & $-$0.514$^{**}$ & $-$0.142 & $-$0.009 & 0.608$^{**}$ & $-$0.093 \\ 
  & (0.225) & (0.160) & (0.206) & (0.206) & (0.207) & (0.193) \\ 

 AvgUniqueCategoriesPerMonth & $-$0.786$^{***}$ & 0.757$^{***}$ & 0.394$^{*}$ & 0.798$^{***}$ & $-$0.209 & 1.401$^{***}$ \\ 
  & (0.228) & (0.167) & (0.198) & (0.211) & (0.196) & (0.194) \\ 

 Intercept & $-$3.263$^{***}$ & $-$3.235$^{***}$ & $-$2.481$^{***}$ & $-$2.102$^{***}$ & $-$2.167$^{***}$ & $-$2.533$^{***}$ \\ 
  & (0.401) & (0.371) & (0.299) & (0.253) & (0.268) & (0.256) \\ 

\hline \\[-1.8ex] 
Observations & 2,761 & 2,766 & 2,769 & 2,772 & 2,792 & 2,783 \\ 
Log Likelihood & $-$908.558 & $-$1,079.992 & $-$1,078.742 & $-$1,114.206 & $-$1,440.825 & $-$1,336.839 \\ 
Akaike Inf. Crit. & 1,851.117 & 2,193.985 & 2,191.483 & 2,262.411 & 2,915.651 & 2,707.678 \\ 
\hline 
\hline \\[-1.8ex] 
\textit{Note:}  & \multicolumn{6}{r}{$^{*}$p$<$0.05; $^{**}$p$<$0.01; $^{***}$p$<$0.001} \\ 
\end{tabular} 
}
\end{table}

\begin{table}[!htp] \centering 
  \caption{Propensity Score Logit Model for Cat Hygiene -  Adopter Cohorts between 2020-01 and 2020-06 vs. Non-Adopters} 
\label{tab:psm.cohort.match.sand.2020}
\footnotesize{
\begin{tabular}{@{\extracolsep{5pt}}lcccccc} 
\\[-1.8ex]\hline 
\hline \\[-1.8ex] 
 & \multicolumn{6}{c}{\textit{Dependent variable:}} \\ 
\cline{2-7} 
\\[-1.8ex] & \multicolumn{6}{c}{Adopter} \\ 
\\[-1.8ex] & 2020-01 & 2020-02 & 2020-03 & 2020-04 & 2020-05 & 2020-06 \\ 
\hline \\[-1.8ex] 
 Gender\_Female & 0.798$^{***}$ & 0.887$^{***}$ & 0.490$^{***}$ & 0.459$^{***}$ & 0.357$^{***}$ & 0.702$^{***}$ \\ 
  & (0.131) & (0.122) & (0.071) & (0.068) & (0.068) & (0.078) \\ 

 Gender\_NoInformation & 0.835$^{***}$ & 0.334 & $-$0.072 & 0.224$^{*}$ & $-$0.370$^{***}$ & 0.232$^{*}$ \\ 
  & (0.164) & (0.176) & (0.108) & (0.094) & (0.106) & (0.111) \\ 

 Age$\geq$50 & $-$0.962$^{***}$ & $-$1.164$^{***}$ & $-$0.881$^{***}$ & $-$0.896$^{***}$ & $-$1.128$^{***}$ & $-$1.857$^{***}$ \\ 
  & (0.253) & (0.202) & (0.141) & (0.130) & (0.132) & (0.133) \\ 

 Age30\_40 & 0.505$^{*}$ & $-$0.194 & $-$0.145 & $-$0.237 & $-$0.319$^{*}$ & $-$0.801$^{***}$ \\ 
  & (0.240) & (0.195) & (0.142) & (0.131) & (0.133) & (0.130) \\ 

 Age40\_50 & $-$0.014 & $-$0.816$^{***}$ & $-$0.727$^{***}$ & $-$0.629$^{***}$ & $-$0.890$^{***}$ & $-$1.147$^{***}$ \\ 
  & (0.240) & (0.199) & (0.141) & (0.130) & (0.133) & (0.128) \\ 

 Age\_NoInformation & $-$0.407 & $-$2.189$^{***}$ & $-$2.352$^{***}$ & $-$2.549$^{***}$ & $-$1.864$^{***}$ & $-$3.821$^{***}$ \\ 
  & (0.298) & (0.370) & (0.237) & (0.231) & (0.190) & (0.344) \\ 

 Income$<$1000 & $-$0.463$^{**}$ & 0.008 & $-$0.241$^{*}$ & 0.499$^{***}$ & $-$0.160 & $-$0.097 \\ 
  & (0.174) & (0.140) & (0.104) & (0.091) & (0.095) & (0.107) \\ 

 Income$\geq$2000 & 0.122 & $-$0.463$^{***}$ & 0.428$^{***}$ & 0.465$^{***}$ & 0.038 & 0.307$^{***}$ \\ 
  & (0.124) & (0.123) & (0.077) & (0.075) & (0.074) & (0.082) \\ 

 Income\_NoInformation & $-$0.508$^{*}$ & $-$0.588$^{**}$ & $-$0.516$^{***}$ & 0.224$^{*}$ & $-$0.367$^{**}$ & $-$0.214 \\ 
  & (0.205) & (0.194) & (0.131) & (0.108) & (0.113) & (0.128) \\ 

 AvgSpendPerMonth & $-$0.002$^{***}$ & 0.0004$^{*}$ & $-$0.00003 & 0.0003$^{**}$ & $-$0.00001 & $-$0.0002 \\ 
  & (0.0003) & (0.0002) & (0.0001) & (0.0001) & (0.0001) & (0.0001) \\ 

 AvgQuantitiesPerMonth & 0.031$^{**}$ & 0.012 & 0.008 & $-$0.001 & 0.008 & 0.012 \\ 
  & (0.012) & (0.010) & (0.006) & (0.006) & (0.006) & (0.007) \\ 

 AvgUniqueItemsPerMonth & $-$0.673$^{***}$ & $-$0.410$^{***}$ & $-$0.090$^{*}$ & $-$0.029 & $-$0.072 & $-$0.146$^{*}$ \\ 
  & (0.108) & (0.091) & (0.040) & (0.038) & (0.045) & (0.058) \\ 

 AvgUniqueOrdersPerMonth & $-$0.330$^{**}$ & $-$0.696$^{***}$ & $-$0.214$^{***}$ & $-$0.482$^{***}$ & $-$0.512$^{***}$ & $-$0.174$^{**}$ \\ 
  & (0.119) & (0.112) & (0.060) & (0.068) & (0.066) & (0.058) \\ 

 AvgUniqueBrandsPerMonth & 0.586$^{**}$ & 0.323$^{*}$ & 0.167 & 0.006 & $-$0.135 & 0.205 \\ 
  & (0.192) & (0.157) & (0.085) & (0.085) & (0.093) & (0.111) \\ 

 AvgUniqueSubcategoriesPerMonth & 0.745$^{**}$ & 0.984$^{***}$ & 0.073 & $-$0.027 & 0.787$^{***}$ & 0.762$^{***}$ \\ 
  & (0.253) & (0.169) & (0.115) & (0.095) & (0.128) & (0.157) \\ 

 AvgUniqueCategoriesPerMonth & 0.231 & $-$0.441$^{**}$ & 0.326$^{**}$ & 0.642$^{***}$ & $-$0.109 & $-$0.721$^{***}$ \\ 
  & (0.240) & (0.161) & (0.117) & (0.111) & (0.127) & (0.162) \\ 

 Intercept & $-$2.797$^{***}$ & $-$1.602$^{***}$ & $-$0.789$^{***}$ & $-$0.740$^{***}$ & $-$0.040 & $-$0.024 \\ 
  & (0.274) & (0.220) & (0.157) & (0.147) & (0.142) & (0.151) \\ 

\hline \\[-1.8ex] 
Observations & 2,772 & 2,780 & 2,915 & 2,966 & 2,934 & 2,882 \\ 
Log Likelihood & $-$1,163.361 & $-$1,274.134 & $-$2,813.474 & $-$3,215.689 & $-$3,002.450 & $-$2,487.669 \\ 
Akaike Inf. Crit. & 2,360.722 & 2,582.269 & 5,660.947 & 6,465.378 & 6,038.899 & 5,009.338 \\ 
\hline 
\hline \\[-1.8ex] 
\textit{Note:}  & \multicolumn{6}{r}{$^{*}$p$<$0.05; $^{**}$p$<$0.01; $^{***}$p$<$0.001} \\ 
\end{tabular}
}
\end{table}

\begin{table}[htp!]
   \caption{ATT of Online Adoption on Offline Price Elasticity (CS Estimator) - Propensity Score Matching based on each adoption cohort's time of adoption}
   \centering
   \label{tab:7.2.2.cohort.based.matching}
   \footnotesize{
   \begin{tabular}{lccc}
      \tabularnewline \midrule \midrule
      Dependent Variable: & \multicolumn{3}{c}{Offline Price Elasticity}\\
       & \multicolumn{1}{c}{\makecell{Dry Dog Food}}  
       & \multicolumn{1}{c}{\makecell{Dog Hygiene}}
       & \multicolumn{1}{c}{\makecell{Cat Hygiene}}  \\ 
      Model:             & (1)             & (2)             & (3)                      \\  
      \midrule
%      \emph{Variables}\\
      CS               & -0.2473$^{**}$  & -1.788$^{***}$ & -0.7409$^{***}$\\   
                         & (0.0945)             & (0.3678)        & (0.1794)       \\  \midrule
 %     \emph{Fit statistics}\\
     Observations       & 58,580               & 50,913       & 29,039\\  
      \midrule %\midrule
       \multicolumn{4}{l}{\emph{Bootstrapped standard-errors in parentheses via 500 replications}}\\
      \multicolumn{4}{l}{\emph{Signif. Codes: ***: 0.001, **: 0.01, *: 0.05}}\\
   \end{tabular}
   }
\end{table}

\begin{table}[htp!]
   \caption{ATT of Online Adoption on Offline Price Elasticity (CS Estimator) - IPW}
   \centering
   \label{tab:7.3.IPTW}
   \footnotesize{
   \begin{tabular}{lccc}
      \tabularnewline \midrule \midrule
      Dependent Variable: & \multicolumn{3}{c}{Offline Price Elasticity}\\
       & \multicolumn{1}{c}{\makecell{Dry Dog Food}}  
       & \multicolumn{1}{c}{\makecell{Dog Hygiene}}
       & \multicolumn{1}{c}{\makecell{Cat Hygiene}}  \\ 
      Model:             & (1)             & (2)             & (3)                      \\  
      \midrule
%      \emph{Variables}\\
      CS               & -0.2479$^{**}$  & -1.7903$^{***}$ & -0.7399$^{***}$\\   
                         & (0.0945)             & (0.3678)        & (0.1793)       \\  \midrule
 %     \emph{Fit statistics}\\
     Observations       & 136,411               & 78,253         & 55,410\\  
      \midrule %\midrule
       \multicolumn{4}{l}{\emph{Bootstrapped standard-errors in parentheses via 500 replications}}\\
      \multicolumn{4}{l}{\emph{Signif. Codes: ***: 0.001, **: 0.01, *: 0.05}}\\
   \end{tabular}
   }
\end{table}

%% compare to later adopter; using later adopters as control, since there are some unobserved factors that drive people to adopt online shopping; 
\clearpage

\subsection{Alternative Control Groups}
\label{appsec:alternative.control.group}

\subsubsection{ATT Estimates - Using Not-Yet-Treated Adopters as Control Groups}
\label{appssec:alternative_control_not_yet_treated}

Table \ref{tab:7.3.Alternative.Control} presents ATT estimates using Not-Yet-Treated adopters as the control group, as discussed in \S\ref{sssec:alternative.control}.

\begin{table}[htp!]
   \caption{ATT of Online Adoption on Offline Price Elasticity (CS Estimator) - Only Use Not-Yet-Treated Adopters as Comparison Units}
   \centering
   \label{tab:7.3.Alternative.Control}
   \footnotesize{
   \begin{tabular}{lccc}
      \tabularnewline \midrule \midrule
      Dependent Variable: & \multicolumn{3}{c}{Offline Price Elasticity}\\
       & \multicolumn{1}{c}{\makecell{Dry Dog Food}}  
       & \multicolumn{1}{c}{\makecell{Dog Hygiene}}
       & \multicolumn{1}{c}{\makecell{Cat Hygiene}}  \\ 
      Model:             & (1)             & (2)             & (3)                 \\  
      \midrule
%      \emph{Variables}\\
      CS               & -0.1685$^{*}$  & -1.1729$^{***}$ & -0.5222$^{***}$\\   
                         & (0.0814)             & (0.3006)        & (0.1533)       \\  \midrule
 %     \emph{Fit statistics}\\
     Observations       & 29,416          & 28,375        & 15,081\\  
      \midrule %\midrule
       \multicolumn{4}{l}{\emph{Bootstrapped standard-errors in parentheses via 500 replications}}\\
      \multicolumn{4}{l}{\emph{Signif. Codes: ***: 0.001, **: 0.01, *: 0.05}}\\
   \end{tabular}
   }
\end{table}

\subsubsection{Detailed Results for Alternative Filter for Control Groups -- Non-Adopters made a purchase during the adoption period}
\label{appssec:alternative.filter.control.group}

\begin{figure}[htp!]
    \centering
    \caption{Average price elasticity between adopters (shown in twelve dotted lines) and non-adopters (shown in one dashed line) for each category.}
    \label{fig:se5_customer_filter_robustness}

    \begin{subfigure}{0.49\textwidth}
        \centering
        \caption{Dry Dog Food}
        \includegraphics[scale=0.26]{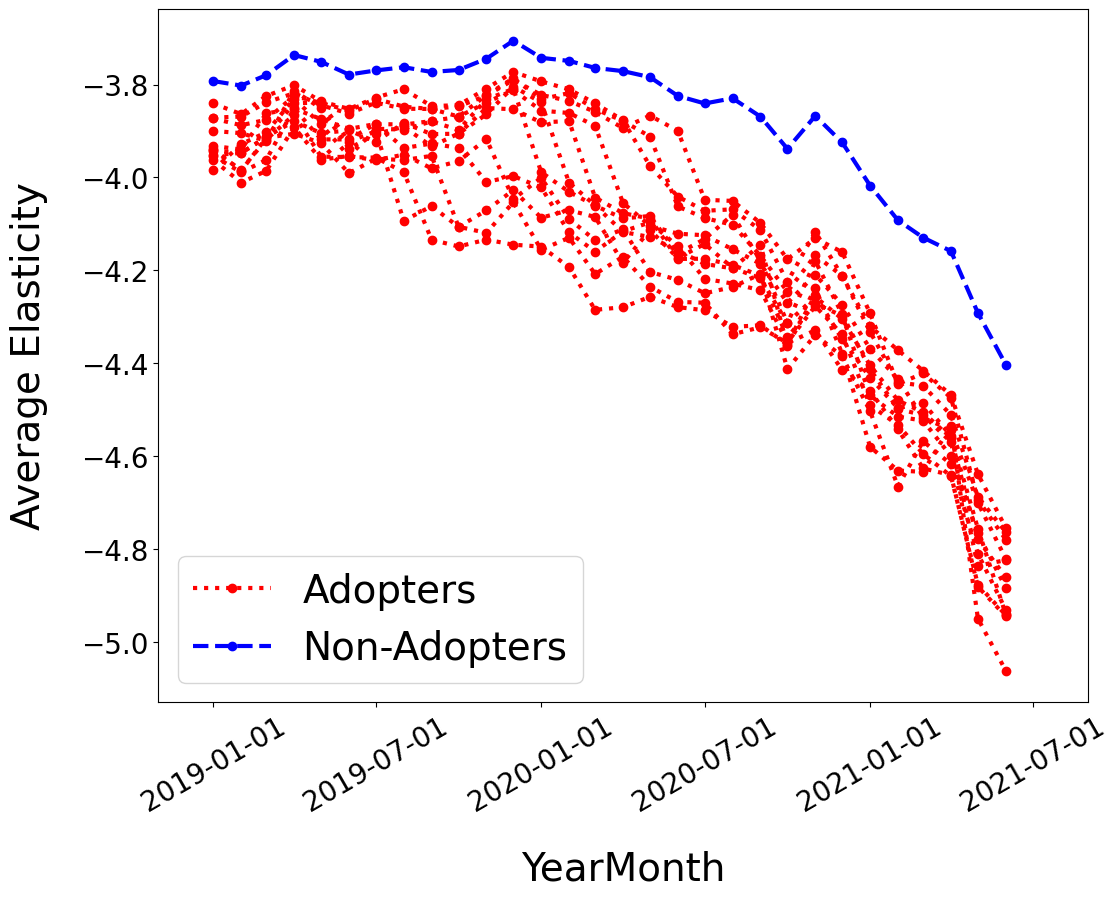}
    \end{subfigure}%
    ~ %add desired spacing between the subfigures, if needed
    \begin{subfigure}{0.49\textwidth}
        \centering
        \caption{Dry Cat Food}
        \includegraphics[scale=0.26]{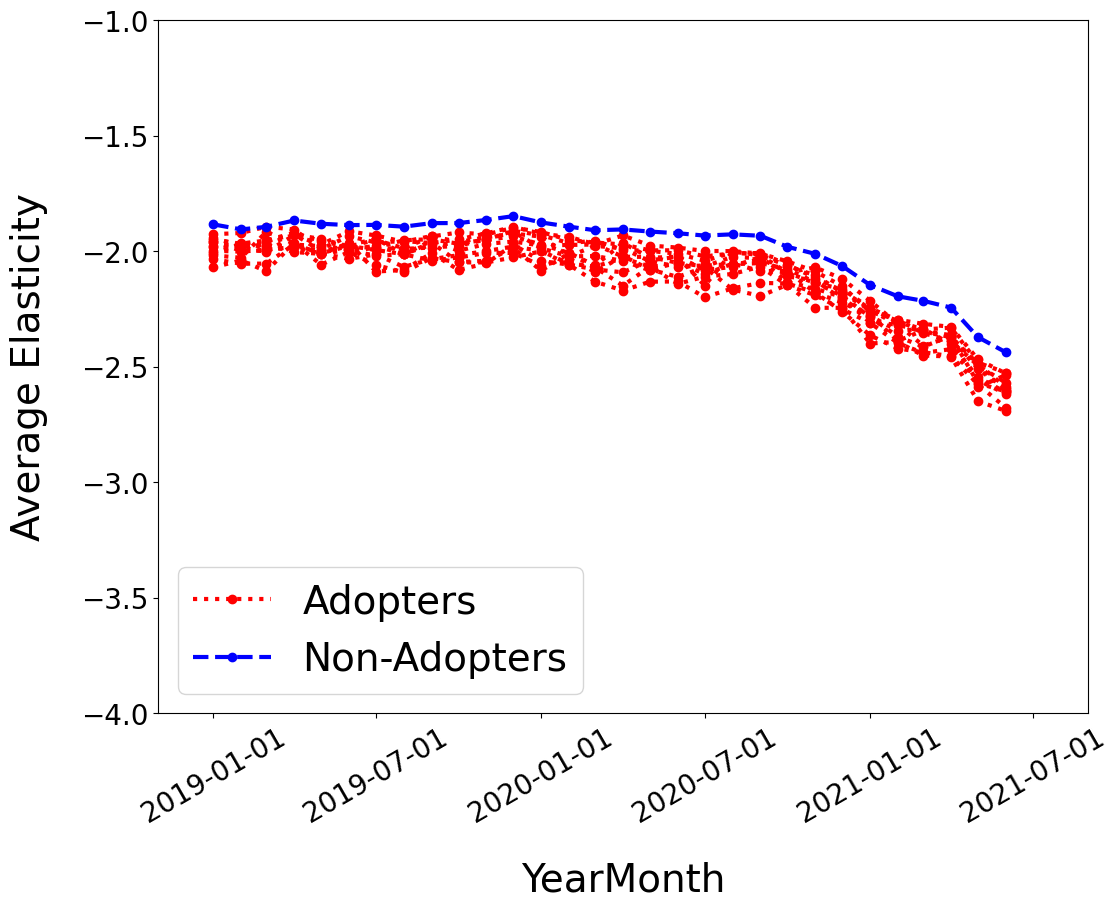}
    \end{subfigure}

    \vspace{1ex} % add some vertical space between the rows

    \begin{subfigure}{0.49\textwidth}
        \centering
        \caption{Dog Hygiene}
        \includegraphics[scale=0.26]{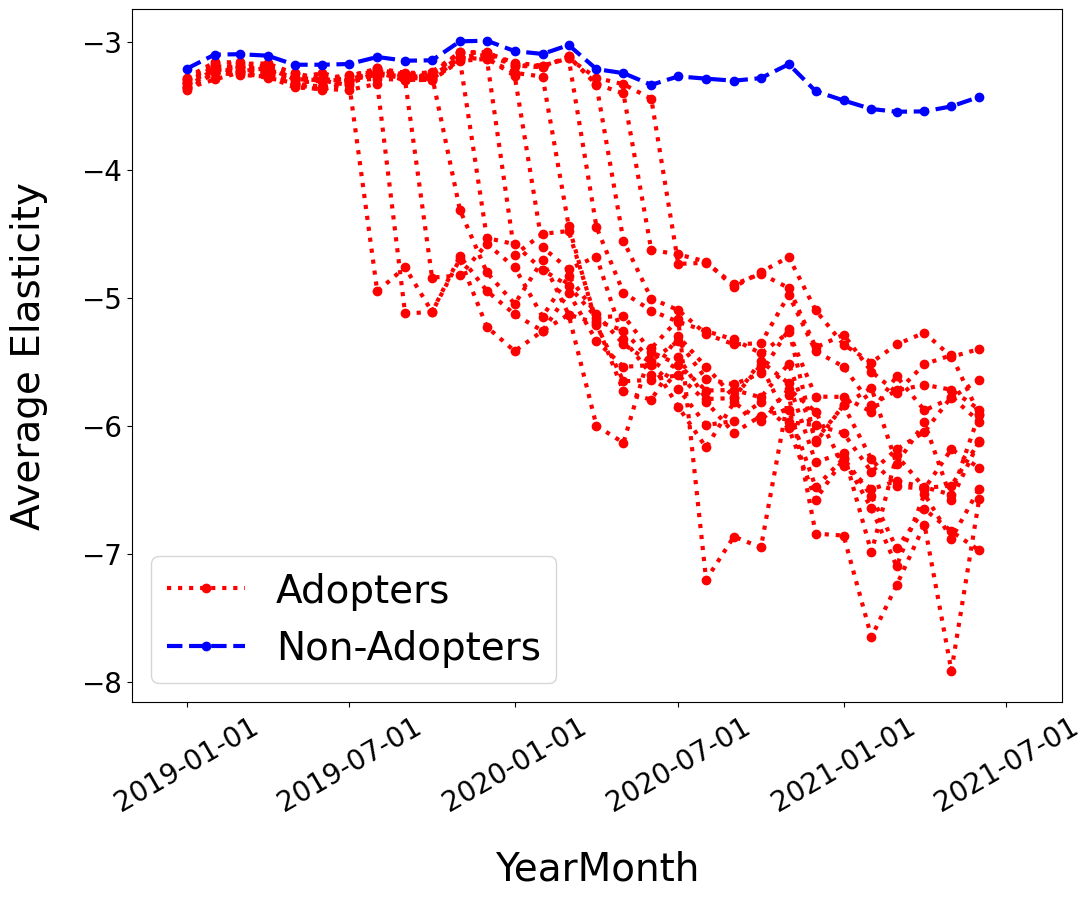}
    \end{subfigure}%
    ~ %add desired spacing between the subfigures, if needed
    \begin{subfigure}{0.49\textwidth}
        \centering
        \caption{Cat Hygiene}
        \includegraphics[scale=0.26]{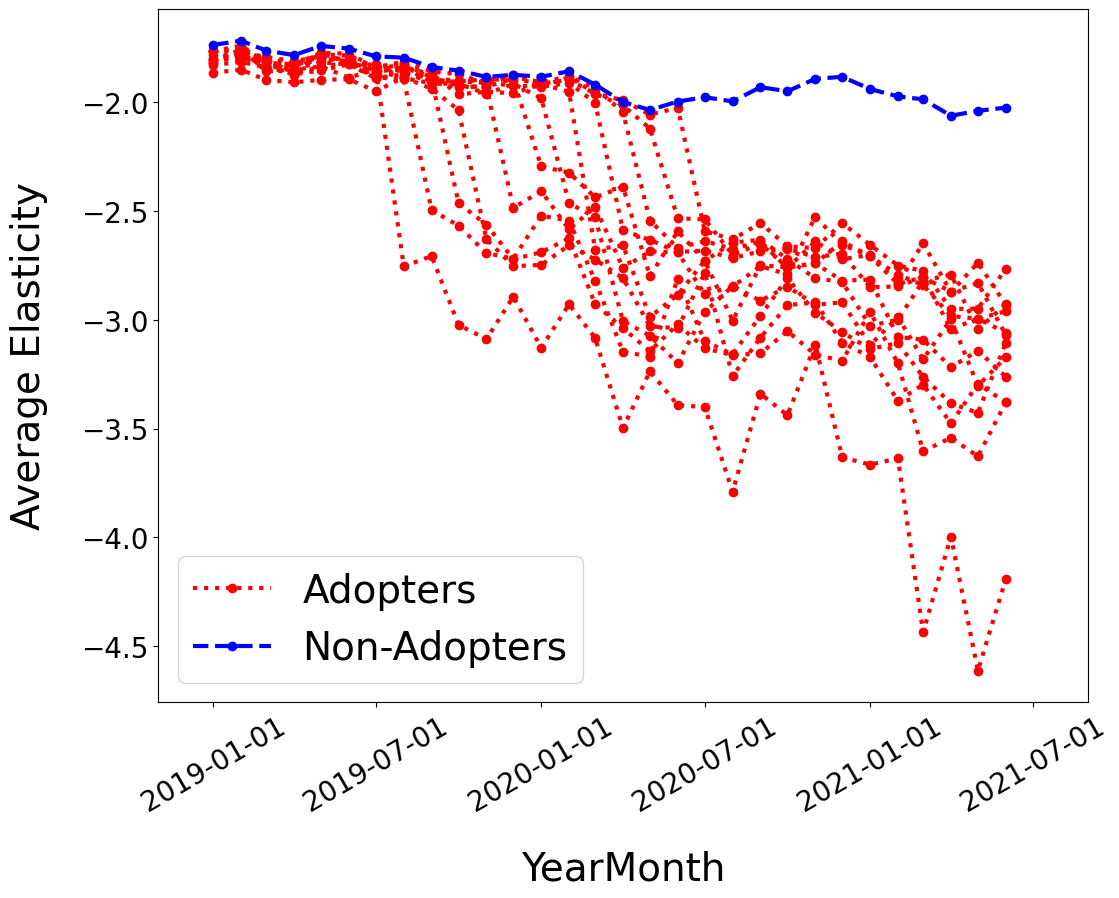}
    \end{subfigure}
\end{figure}

As an additional robustness check, we impose a more symmetric sample-selection criterion by requiring non-adopters to have made at least one purchase during the 12-month adoption window. This restriction alleviates concerns that adopters may be mechanically more active or engaged with the retailer because they are required to transact during the period in which adoption is observed, whereas non-adopters are not subject to the same requirement in the main analysis. Table~\ref{tab:append.demand.model.altDef.nonadopters.filter} reports the demand-model estimates using this alternative control-group definition. Figure~\ref{fig:se5_customer_filter_robustness} presents the corresponding offline price-elasticity trajectories for adopters and non-adopters, which are similar to those in the main analysis. Table~\ref{tab:CS_Result_Alternative_Nonadopter_Filter} reports the ATT estimates, with bootstrap standard errors based on 500 replications. The results remain consistent with our main findings.

\begin{table}[htp!]
\caption{Coefficients of Demand Model Estimate - Alternative Definition of Non-Adopters' Filter}
\label{tab:append.demand.model.altDef.nonadopters.filter}
\centering
\footnotesize{
\begin{tabular}{lcccc}
\toprule
 & Dry Dog Food & Dry Cat Food & Dog Hygiene & Cat Hygiene \\
\textit{Variables} &  &  &  &  \\
\midrule
Price & \makecell{-0.4165$^{***}$\\(0.0176)} & \makecell{-0.1943$^{***}$\\(0.0197)} & \makecell{-1.6336$^{***}$\\(0.1127)} & \makecell{-0.1938$^{***}$\\(0.0207)} \\
cumPercOnlineSpend & \makecell{-0.399\\(0.4712)} & \makecell{-1.7622\\(1.1641)} & \makecell{9.4316$^{***}$\\(1.7703)} & \makecell{3.5865$^{***}$\\(0.9454)} \\
Price * cumPercOnlineSpend & \makecell{-0.0929$^{*}$\\(0.0443)} & \makecell{0.0545\\(0.0983)} & \makecell{-4.387$^{***}$\\(0.7728)} & \makecell{-0.2986$^{***}$\\(0.0668)} \\
BuyPrevOcca & \makecell{0.1345$^{***}$\\(0.0288)} & \makecell{-0.5121$^{***}$\\(0.045)} & \makecell{-0.1867$^{***}$\\(0.03)} & \makecell{0.323$^{***}$\\(0.0301)} \\
CF\_Price & \makecell{0.1573$^{***}$\\(0.0241)} & \makecell{0.0733$^{**}$\\(0.0283)} & \makecell{0.5877$^{***}$\\(0.1731)} & \makecell{0.2242$^{***}$\\(0.0262)} \\
CF\_Price\_cumPercOnlineSpend & \makecell{0.0342\\(0.0734)} & \makecell{-0.049\\(0.1227)} & \makecell{4.4216$^{***}$\\(0.9184)} & \makecell{0.3559$^{***}$\\(0.0893)} \\
Customer\_Brand\_FE & Yes & Yes & Yes & Yes\\ \midrule
Observations       & 127,912         & 42,936          & 71,074         & 51,781\\ 
Log-likelihood & -78572.563& -29105.397& -50968.497& -34895.806 \\ \bottomrule
\multicolumn{5}{l}{\emph{Bootstrapped standard-errors in parentheses via 500 replications}}\\
\multicolumn{5}{l}{\emph{Signif. Codes: ***: 0.001, **: 0.01, *: 0.05}}\\
\bottomrule
\end{tabular}
}
\end{table}

%% Analysis41 ATT with CF
\begin{table}[htp!]
   \caption{ATT of Online Adoption on Offline Price Elasticity Over All Post-Adoption Period (CS Estimator) -- Alternative Definition of Non-Adopters' Filter}
   \centering
\label{tab:CS_Result_Alternative_Nonadopter_Filter}
   \footnotesize{
   \begin{tabular}{lccc}
      \tabularnewline \midrule \midrule
      Dependent Variable: & \multicolumn{3}{c}{Offline Price Elasticity}\\
       & \multicolumn{1}{c}{\makecell{Dry Dog Food}}  
       & \multicolumn{1}{c}{\makecell{Dog Hygiene}}
       & \multicolumn{1}{c}{\makecell{Cat Hygiene}}  \\ 
      Model:             & (1)             & (2)             & (3)                      \\  
      \midrule
%      \emph{Variables}\\
      ATT               & -0.2159$^{*}$  &  -1.8991$^{***}$ & -0.7684$^{***}$\\   
                         & (0.0926)              & (0.3223)               & (0.1571)       \\   \midrule
      \midrule
 %     \emph{Fit statistics}\\
     Observations       & 127,912         & 71,074         & 51,781\\  
      \midrule %\midrule
        \multicolumn{4}{l}{\emph{Bootstrapped standard-errors in parentheses via 500 replications}}\\
      \multicolumn{4}{l}{\emph{Signif. Codes: ***: 0.001, **: 0.01, *: 0.05}}\\
   \end{tabular}
   }
\end{table}

\clearpage

\subsection{Alternative Purchase Occasion Definitions}
\label{appsec:alternative.def.purchase.occasion}

\subsubsection{Detailed Results for Alternative Definition of Purchase Occasions}
\label{appssec:alternative.def.po}

Table \ref{tab:append.demand.model.altDef.PO} presents the estimates of the main demand model based on an alternative definition of the purchase occasion (i.e., customer-month level during which we observe any interaction with the retailer, either offline or online interaction). The standard errors for the demand parameters are based on the customer-level block bootstrap with 500 replications. 

\begin{table}[htp!]
\caption{Coefficients of Demand Model Estimate - Alternative Definition of Purchase Occasions}
\label{tab:append.demand.model.altDef.PO}
\centering
\footnotesize{
\begin{tabular}{lcccc}
\toprule
 & Dry Dog Food & Dry Cat Food & Dog Hygiene & Cat Hygiene \\
\textit{Variables} &  &  &  &  \\
\midrule
Price & \makecell{-0.3171$^{***}$ \\(0.0204)} & \makecell{-0.1637$^{***}$ \\(0.0185)} & \makecell{-1.4982$^{***}$ \\(0.1166)} & \makecell{-0.2933$^{***}$ \\(0.0205)} \\
cumPercOnlineSpend & \makecell{-1.1362$^{**}$ \\(0.365)} & \makecell{-2.8399$^{*}$ \\(1.118)} & \makecell{4.822$^{**}$ \\(1.6143)} & \makecell{0.8407\\(0.9654)} \\
Price * cumPercOnlineSpend & \makecell{-0.1354$^{***}$ \\(0.0333)} & \makecell{0.0413\\(0.0926)} & \makecell{-2.8394$^{***}$ \\(0.7037)} & \makecell{-0.1615$^{*}$ \\(0.0652)} \\
BuyPrevOcca & \makecell{0.173$^{***}$ \\(0.0263)} & \makecell{-0.4511$^{***}$ \\(0.0437)} & \makecell{-0.1633$^{***}$ \\(0.0275)} & \makecell{0.3556$^{***}$ \\(0.0295)} \\
CF\_Price & \makecell{0.0421\\(0.0257)} & \makecell{0.0596$^{*}$ \\(0.0247)} & \makecell{0.4807$^{**}$ \\(0.1714)} & \makecell{0.3616$^{***}$ \\(0.0272)} \\
CF\_Price\_cumPercOnlineSpend & \makecell{-0.01\\(0.0623)} & \makecell{-0.1603\\(0.1016)} & \makecell{2.372$^{**}$ \\(0.881)} & \makecell{0.1731\\(0.0889)} \\
Customer\_Brand\_FE & Yes & Yes & Yes & Yes\\ \midrule
Observations       & 144,094         & 48,737          & 84,689         & 57,344\\ 
Log-likelihood & -88735.39 &  -32496.48 & -58287.349 &  -38561.677 \\ 
\bottomrule
\multicolumn{5}{l}{\emph{Bootstrapped standard-errors in parentheses via 500 replications}}\\
\multicolumn{5}{l}{\emph{Signif. Codes: ***: 0.001, **: 0.01, *: 0.05}}\\
\end{tabular}
}
\end{table}

\begin{figure}[htp!]
    \centering
    \caption{Bootstrapped distribution of Overall ATT Estimates based on 500 replications with the point estimate from Table \ref{tab:CS_Result_AltDef_PurchaseOccasion}.}
\label{fig:tab:CS_Result_AltDef_PurchaseOccasion.Boot_Est}
    \begin{subfigure}{0.32\textwidth}
        \centering
        \caption{Dry Dog Food}
        \includegraphics[width=\linewidth]{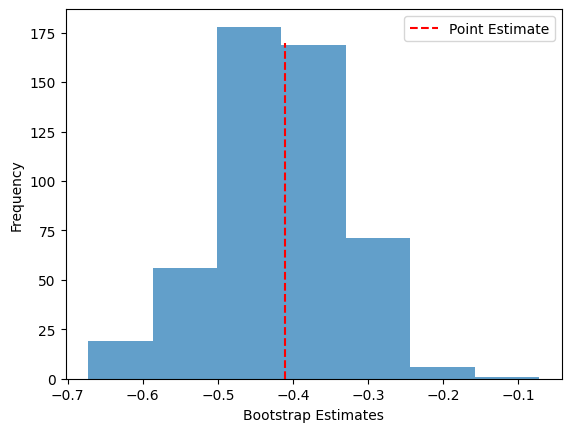}
    \end{subfigure}%
    \hfill
    \begin{subfigure}{0.33\textwidth}
        \centering
        \caption{Dog Hygiene}
        \includegraphics[width=\linewidth]{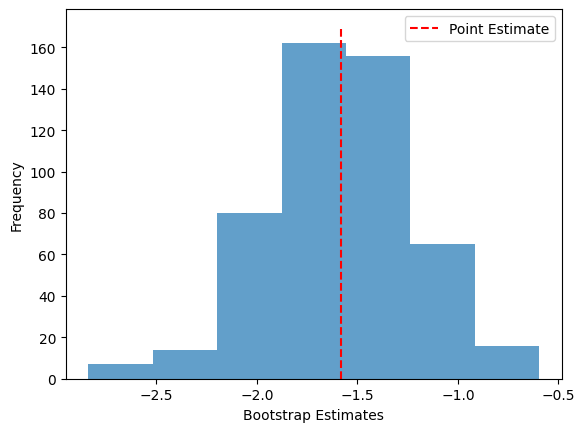}
    \end{subfigure}%
    \hfill
    \begin{subfigure}{0.32\textwidth}
        \centering
        \caption{Cat Hygiene}
        \includegraphics[width=\linewidth]{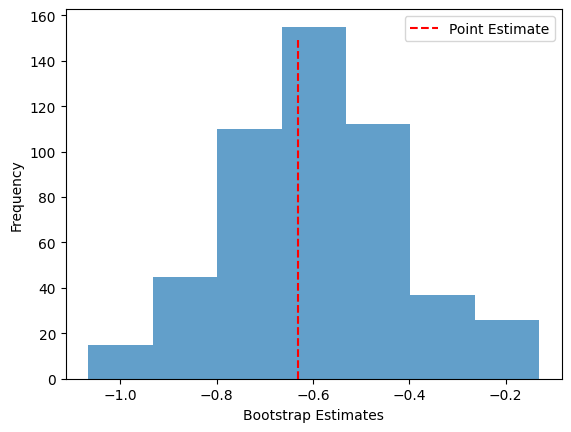}
    \end{subfigure}
\end{figure}

Table \ref{tab:CS_Result_AltDef_PurchaseOccasion} presents the ATT estimates of offline price elasticity from CS Estimator under this alternative definition of purchase occasions and customer-month panel. The standard errors for the ATT estimates and the demand model parameters are based on the customer-level block bootstrap with 500 replications; See Figure \ref{fig:tab:CS_Result_AltDef_PurchaseOccasion.Boot_Est}.

\begin{table}[htp!]
   \caption{ATT of Online Adoption on Offline Price Elasticity (CS Estimator) -- Alternative Definition of Purchase Occasions}
   \centering  \label{tab:CS_Result_AltDef_PurchaseOccasion}
   \footnotesize{
   \begin{tabular}{lccc}
      \tabularnewline \midrule \midrule
      Dependent Variable: & \multicolumn{3}{c}{Offline Price Elasticity}\\
       & \multicolumn{1}{c}{\makecell{Dry Dog Food}}  
       & \multicolumn{1}{c}{\makecell{Dog Hygiene}}
       & \multicolumn{1}{c}{\makecell{Cat Hygiene}}  \\ 
      Model:             & (1)             & (2)             & (3)                      \\  
      \midrule
%      \emph{Variables}\\
      ATT               & -0.4104$^{***}$  &  -1.5819$^{***}$ &  -0.6315$^{***}$\\   
                         & (0.0867)       & (0.3688)        & (0.1799)       \\   \midrule
 %     \emph{Fit statistics}\\
     Observations       & 144,094             & 84,689         & 57,344\\  
      \midrule %\midrule
        \multicolumn{4}{l}{\emph{Bootstrapped standard-errors in parentheses via 500 replications}}\\
      \multicolumn{4}{l}{\emph{Signif. Codes: ***: 0.001, **: 0.01, *: 0.05}}\\
   \end{tabular}
   }
\end{table}

\subsubsection{Detailed Results for Alternative Definition of Purchase Occasions -- Customer-Month Full Panel}
\label{appssec:customer.month.full.panel}

\begin{figure}[htp!]
    \centering
    \caption{Average price elasticity between adopters (shown in twelve dotted lines) and non-adopters (shown in one dashed line) for each category.}
    \label{fig:se6.2_customer_month_full_panel}

    \begin{subfigure}{0.49\textwidth}
        \centering
        \caption{Dry Dog Food}
        \includegraphics[scale=0.26]{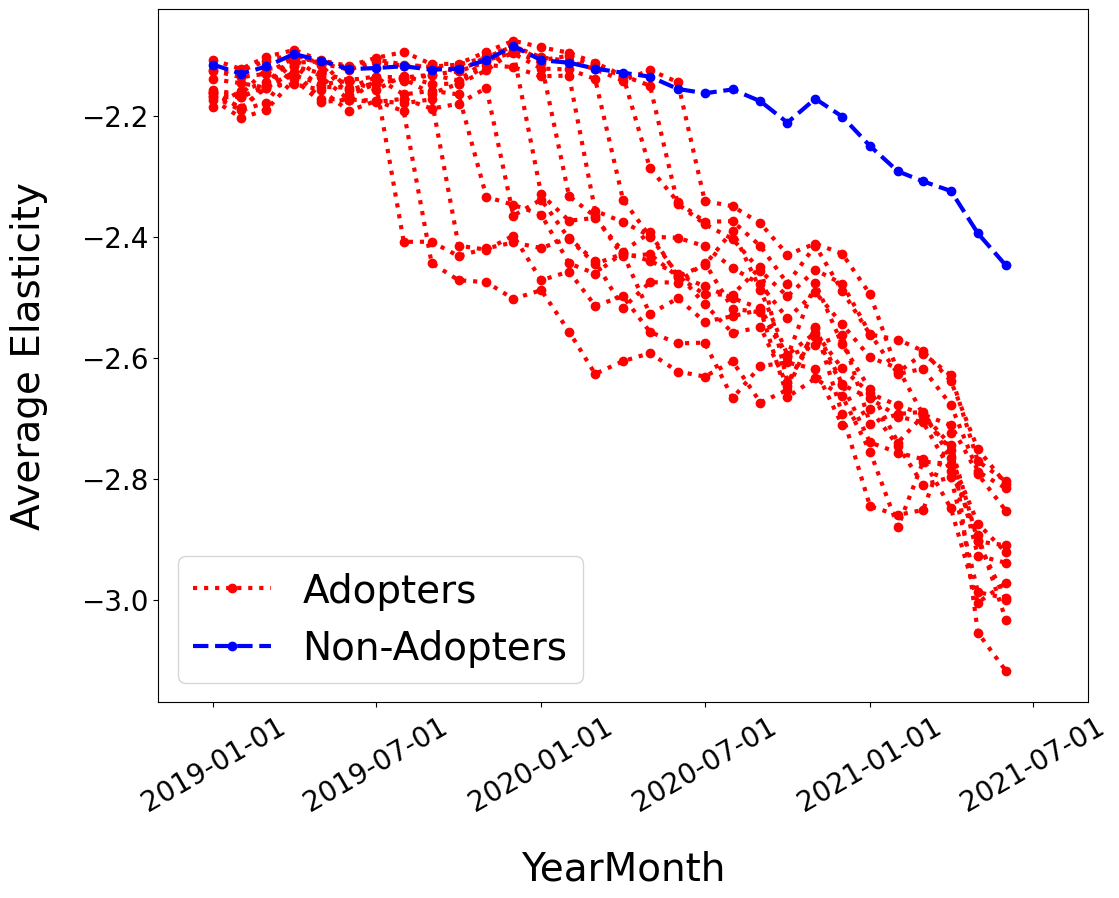}
    \end{subfigure}%
    ~ %add desired spacing between the subfigures, if needed
    \begin{subfigure}{0.49\textwidth}
        \centering
        \caption{Dry Cat Food}
        \includegraphics[scale=0.26]{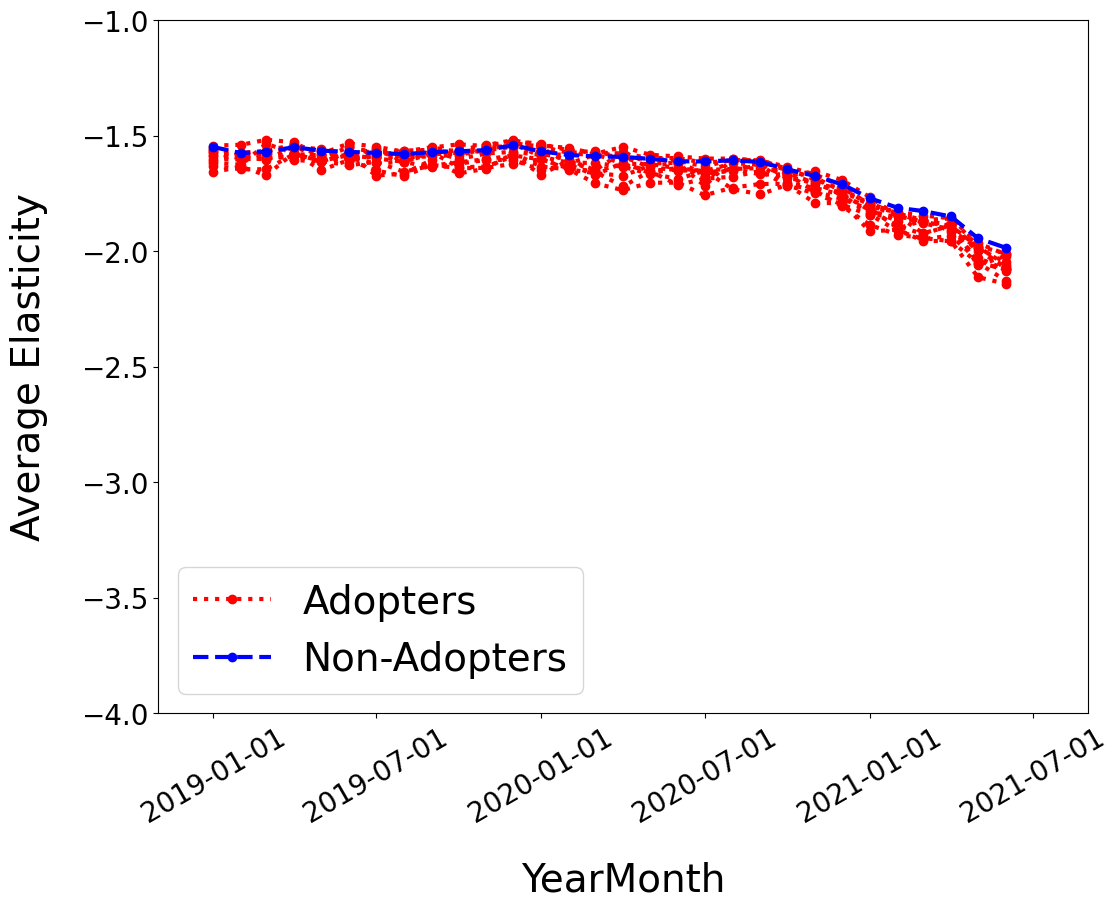}
    \end{subfigure}

    \vspace{1ex} % add some vertical space between the rows

    \begin{subfigure}{0.49\textwidth}
        \centering
        \caption{Dog Hygiene}
        \includegraphics[scale=0.26]{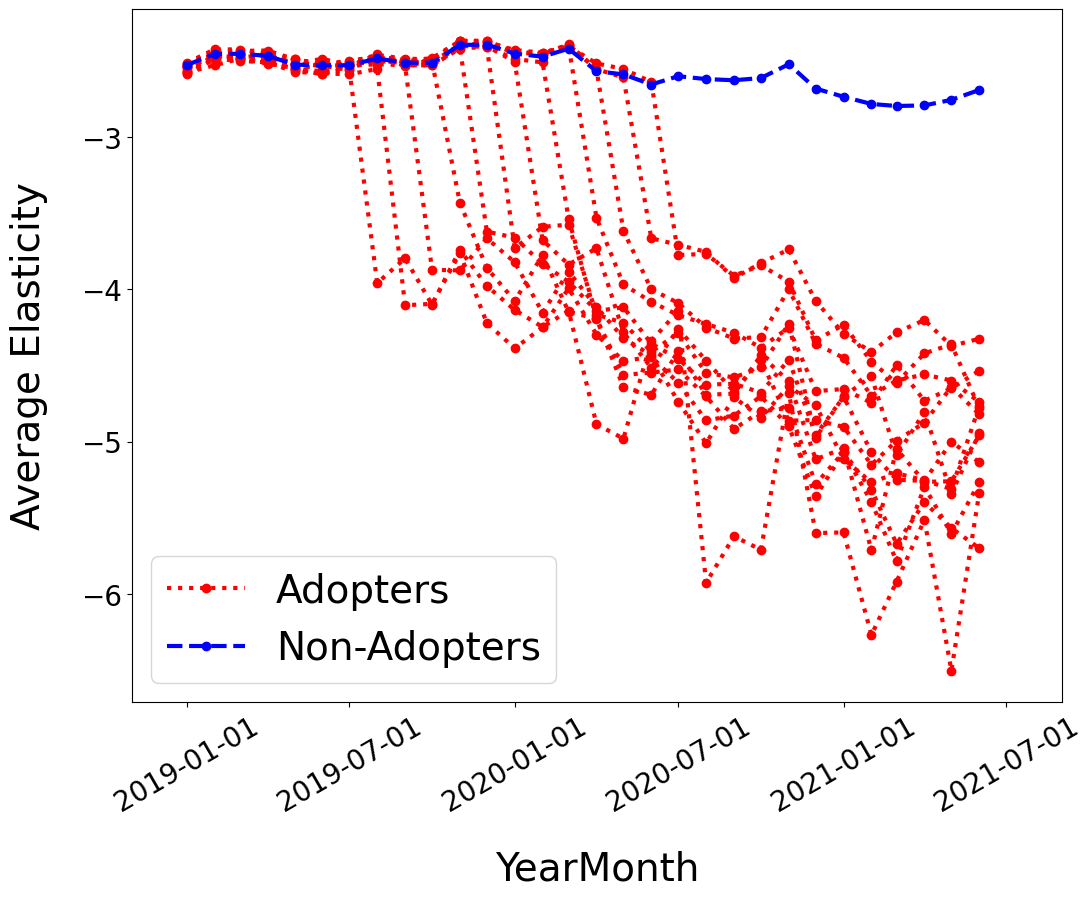}
    \end{subfigure}%
    ~ %add desired spacing between the subfigures, if needed
    \begin{subfigure}{0.49\textwidth}
        \centering
        \caption{Cat Hygiene}
        \includegraphics[scale=0.26]{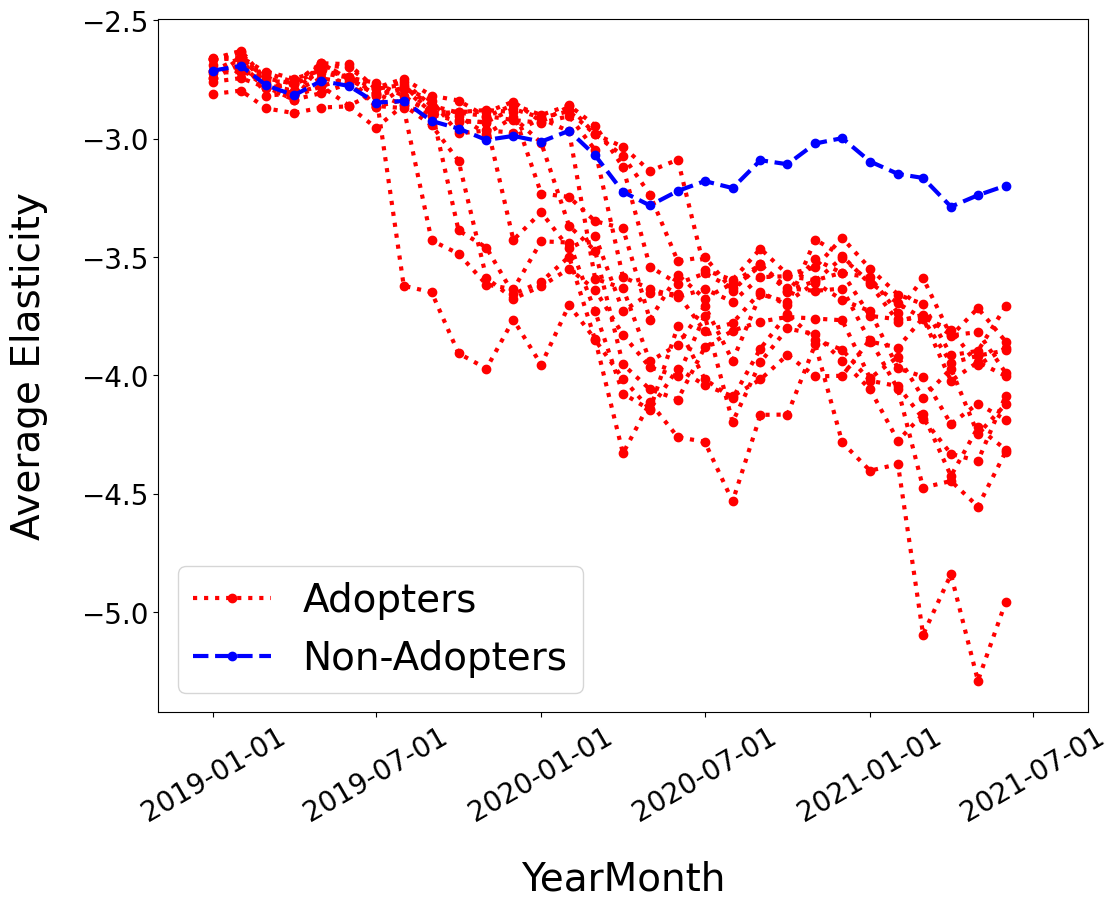}
    \end{subfigure}
\end{figure}

We further provide a robustness check for constructing a customer-month full panel. 
Table \ref{tab:append.demand.model.altDef.PO.full.panel} presents the demand model estimates. Figure \ref{fig:se6.2_customer_month_full_panel} shows that online adoptions are still significant across these product categories when we define every customer-month as a purchase occasion. Table \ref{tab:CS_Result_Customer_Month_Full_Panel} presents the ATT estimates, with bootstrapped standard errors with 500 replications. We obtain consistent findings with the main results.

\begin{table}[htp!]
\caption{Coefficients of Demand Model Estimate - Alternative Definition of Purchase Occasion (Customer-Month Full Panel)}
\label{tab:append.demand.model.altDef.PO.full.panel}
\centering
\footnotesize{
\begin{tabular}{lcccc}
\toprule
 & Dry Dog Food & Dry Cat Food & Dog Hygiene & Cat Hygiene \\
\textit{Variables} &  &  &  &  \\
\midrule
Price & \makecell{-0.2289$^{***}$\\(0.0137)} & \makecell{-0.1556$^{***}$\\(0.0184)} & \makecell{-1.2531$^{***}$\\(0.1129)} & \makecell{-0.2942$^{***}$\\(0.0171)} \\
cumPercOnlineSpend & \makecell{-0.2775\\(0.4073)} & \makecell{-1.5972\\(1.344)} & \makecell{7.9921$^{***}$\\(1.768)} & \makecell{2.9808$^{**}$\\(1.0155)} \\
Price * cumPercOnlineSpend & \makecell{-0.1393$^{***}$\\(0.0367)} & \makecell{0.0307\\(0.1131)} & \makecell{-3.8323$^{***}$\\(0.7692)} & \makecell{-0.2356$^{***}$\\(0.0691)} \\
BuyPrevOcca & \makecell{-0.123$^{***}$\\(0.0248)} & \makecell{-0.5365$^{***}$\\(0.0406)} & \makecell{-0.219$^{***}$\\(0.0258)} & \makecell{0.2492$^{***}$\\(0.0279)} \\
CF\_Price & \makecell{-0.0222\\(0.0191)} & \makecell{0.0332\\(0.0259)} & \makecell{0.2492\\(0.1769)} & \makecell{0.3684$^{***}$\\(0.0233)} \\
CF\_Price\_cumPercOnlineSpend & \makecell{-0.0025\\(0.0708)} & \makecell{-0.1134\\(0.1247)} & \makecell{3.0106$^{**}$\\(0.9993)} & \makecell{0.2586$^{**}$\\(0.0909)} \\
Customer\_Brand\_FE & Yes & Yes & Yes & Yes\\ \midrule
Observations  &     163,326 & 53,672 & 90,254 & 63,024 \\ 
Log-likelihood & -110999.11 & -36618.153 & -62535.493 & -44408.812 \\
\bottomrule
\multicolumn{5}{l}{\emph{Bootstrapped standard-errors in parentheses via 500 replications}}\\
\multicolumn{5}{l}{\emph{Signif. Codes: ***: 0.001, **: 0.01, *: 0.05}}\\
\end{tabular}
}
\end{table}

\begin{table}[htp!]
\caption{ATT of Online Adoption on Offline Price Elasticity Over All Post-Adoption Period (CS Estimator) -- Customer-Month Full Panel}
\centering
\label{tab:CS_Result_Customer_Month_Full_Panel}
\footnotesize{
  \begin{tabular}{lccc}
  \tabularnewline \midrule \midrule
  Dependent Variable: & \multicolumn{3}{c}{Offline Price Elasticity}\\
  & \multicolumn{1}{c}{\makecell{Dry Dog Food}}  
  & \multicolumn{1}{c}{\makecell{Dog Hygiene}}
  & \multicolumn{1}{c}{\makecell{Cat Hygiene}}  \\ 
  Model:             & (1)             & (2)             & (3)                      \\  
  \midrule
  %      \emph{Variables}\\
  ATT               &  -0.3038$^{***}$  & -1.6611$^{***}$ & -0.6452$^{***}$\\   
      & (0.0749)               & (0.3236)   & (0.1677)    \\   \midrule
  \midrule
  %     \emph{Fit statistics}\\
  Observations       &  163,326  & 90,254 & 63,024 \\   
  \midrule %\midrule
  \multicolumn{4}{l}{\emph{Bootstrapped standard-errors in parentheses via 500 replications}}\\
  \multicolumn{4}{l}{\emph{Signif. Codes: ***: 0.001, **: 0.01, *: 0.05}}\\
  \end{tabular}
}
\end{table}

\clearpage

\subsection{Price Endogeneity and Control-Function Specification}
\label{appsec:price.endogeneity.control.function}

\subsubsection{Detailed Results for Alternative Demand Equation without Control Function}
\label{appssec:alternative.imple.no.control.function}

\begin{figure}[htp!]
    \centering
    \caption{Average price elasticity between adopters (shown in twelve dotted lines) and non-adopters (shown in one dashed line) for each category.}
    \label{fig:se1.main.elas.fig.noCF}

    \begin{subfigure}{0.49\textwidth}
        \centering
        \caption{Dry Dog Food}
        \includegraphics[scale=0.26]{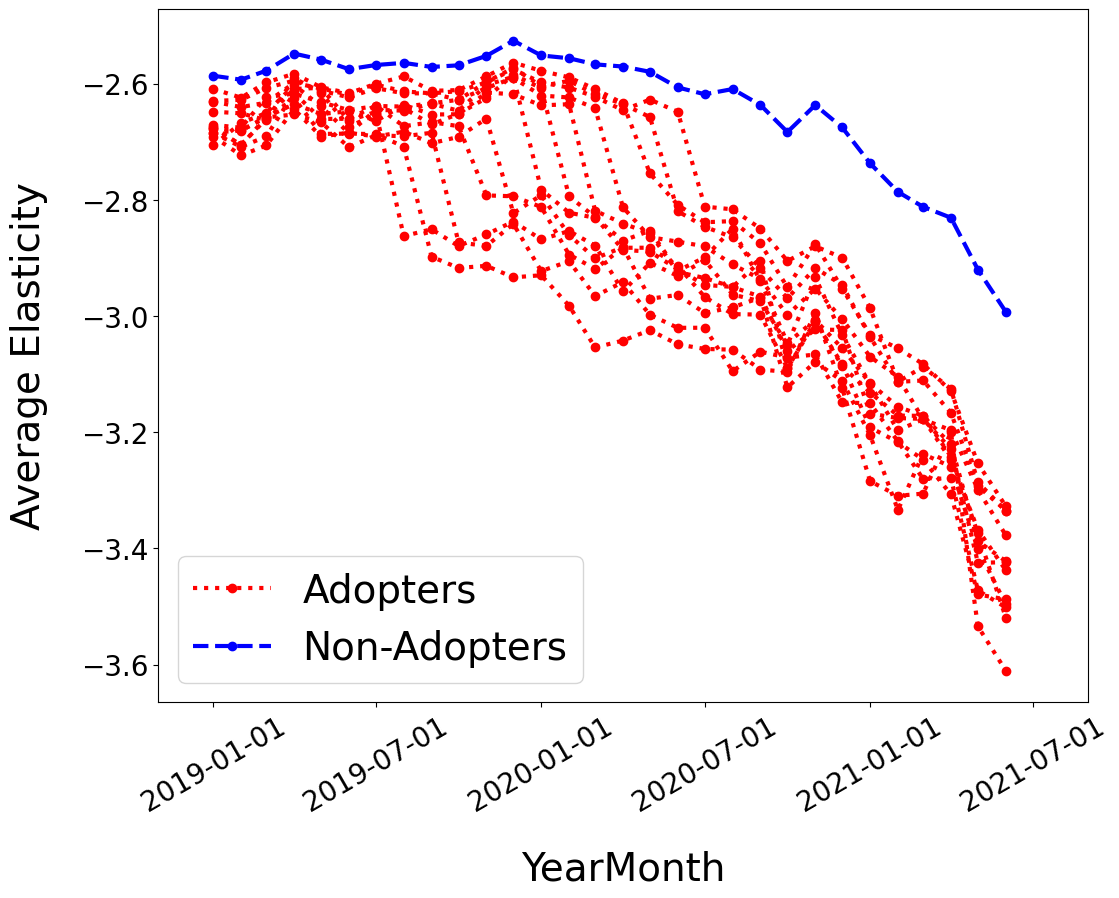}
    \end{subfigure}%
    ~ %add desired spacing between the subfigures, if needed
    \begin{subfigure}{0.49\textwidth}
        \centering
        \caption{Dry Cat Food}
        \includegraphics[scale=0.26]{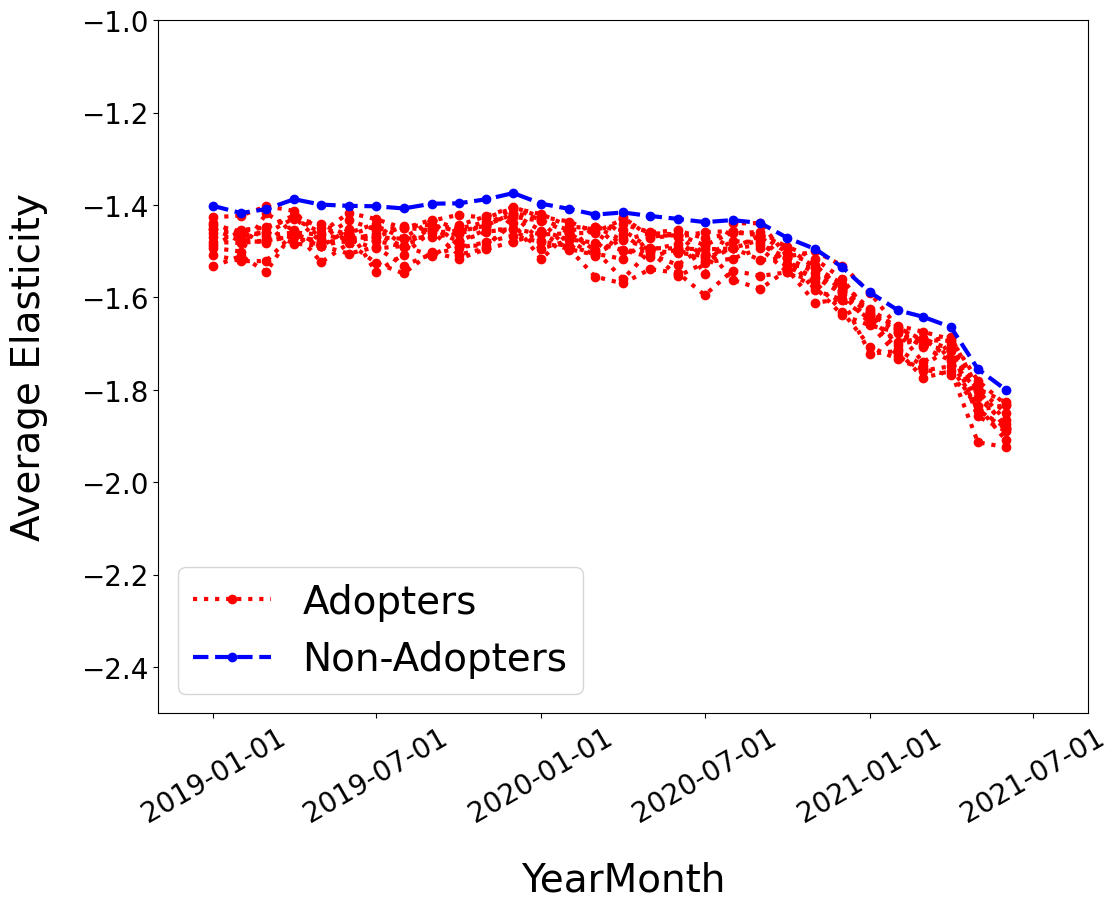}
    \end{subfigure}

    \vspace{1ex} % add some vertical space between the rows

    \begin{subfigure}{0.49\textwidth}
        \centering
        \caption{Dog Hygiene}
        \includegraphics[scale=0.26]{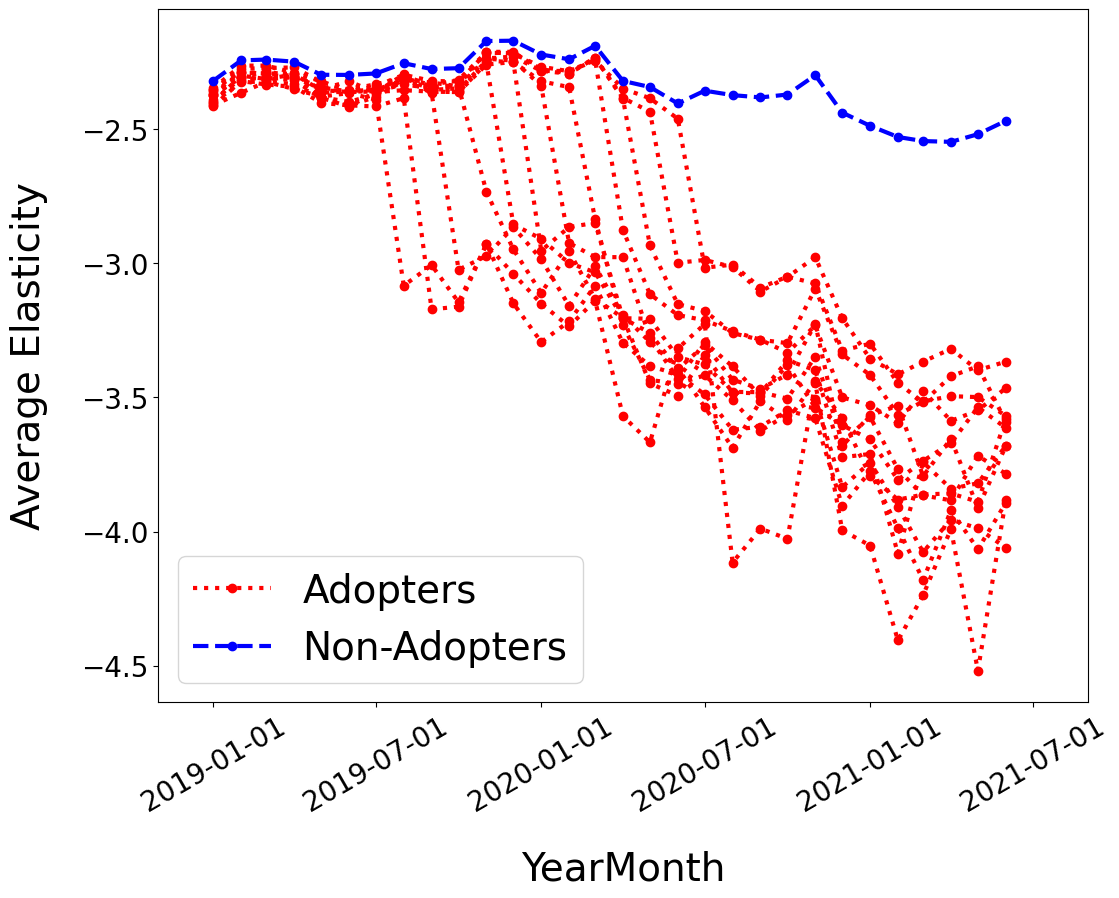}
    \end{subfigure}%
    ~ %add desired spacing between the subfigures, if needed
    \begin{subfigure}{0.49\textwidth}
        \centering
        \caption{Cat Hygiene}
        \includegraphics[scale=0.26]{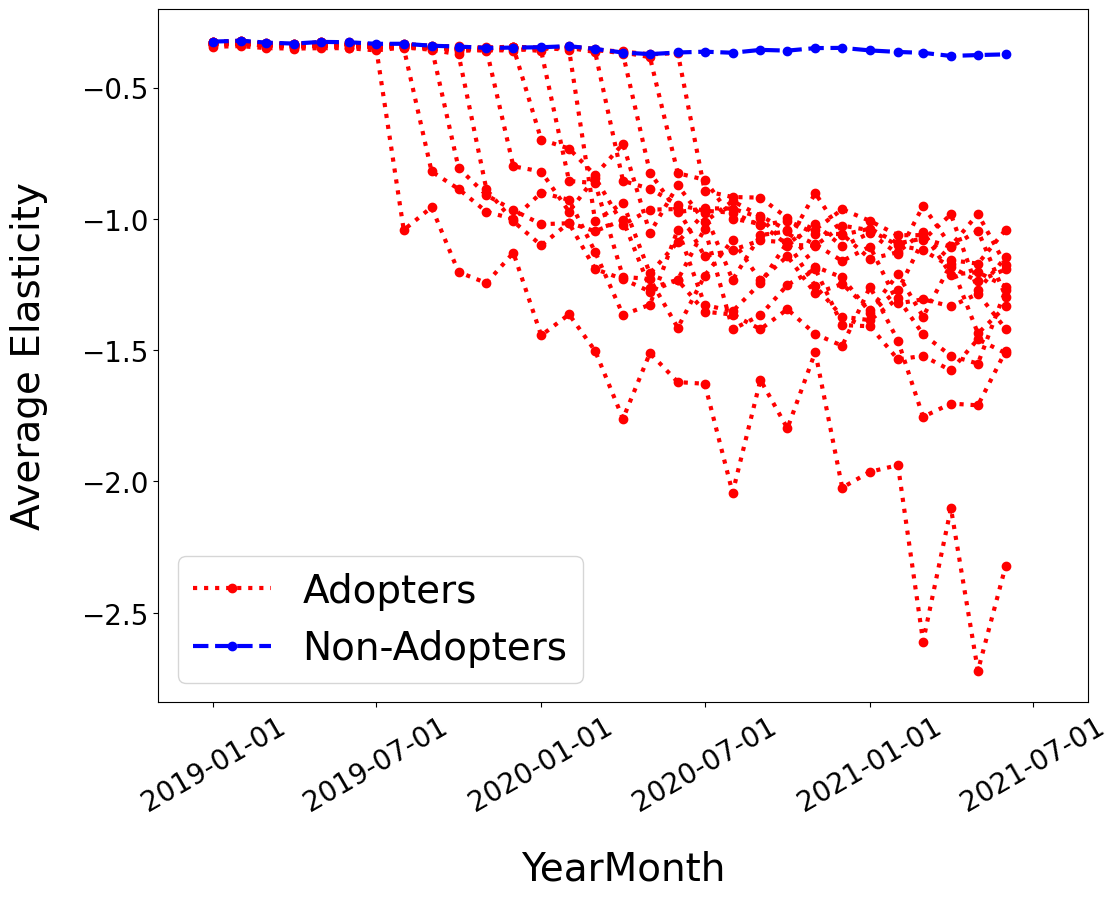}
    \end{subfigure}
\end{figure}

We present the model estimation results without including the control function terms. Table \ref{tab:se1.main.model.noCF} presents demand model estimates for models without control functions. Figure 
\ref{fig:se1.main.elas.fig.noCF} presents elasticity curves based on model estimates without control functions. Table \ref{tab:se1.CS_Result} presents ATT estimates, all consistent with the main results.

\begin{table}[htp!]
\caption{Coefficient Estimates - Main Results without Control Functions}
\centering
\label{tab:se1.main.model.noCF}
\footnotesize{
\begin{tabular}{lcccc}
\toprule
 & Dry Dog Food & Dry Cat Food & Dog Hygiene & Cat Hygiene \\
\textit{Variables} &  &  &  &  \\
\midrule
Price & \makecell{-0.2829$^{***}$\\(0.0136)} & \makecell{-0.1438$^{***}$\\(0.0159)} & \makecell{-1.1687$^{***}$\\(0.0789)} & \makecell{-0.0355$^{***}$\\(0.0094)} \\
cumPercOnlineSpend & \makecell{-0.4024\\(0.3971)} & \makecell{-1.0924\\(1.1567)} & \makecell{3.5462$^{**}$\\(1.139)} & \makecell{2.9824$^{**}$\\(0.9355)} \\
Price * cumPercOnlineSpend & \makecell{-0.1097$^{**}$\\(0.0369)} & \makecell{-0.0102\\(0.0983)} & \makecell{-1.9093$^{***}$\\(0.5018)} & \makecell{-0.2909$^{***}$\\(0.0628)} \\
BuyPrevOcca & \makecell{0.1538$^{***}$\\(0.0272)} & \makecell{-0.4999$^{***}$\\(0.0449)} & \makecell{-0.178$^{***}$\\(0.0283)} & \makecell{0.3215$^{***}$\\(0.0293)} \\
Customer\_Brand\_FE & Yes & Yes & Yes & Yes\\ \midrule
Observations       & 136,411         & 45,443          & 78,253         & 55,410\\ 
Log-likelihood & -84145.962 & -30701.339 & -55460.807 & -37433.254 \\
\bottomrule
\multicolumn{5}{l}{\emph{Bootstrapped standard-errors in parentheses via 500 replications}}\\
\multicolumn{5}{l}{\emph{Signif. Codes: ***: 0.001, **: 0.01, *: 0.05}}\\
\end{tabular}
}
\end{table}

\begin{table}[htp!]
   \caption{ATT of Online Adoption on Offline Price Elasticity Over All Post-Adoption Period (CS Estimator)}
   \centering
   \label{tab:se1.CS_Result}
   \footnotesize{
   \begin{tabular}{lccc}
      \tabularnewline \midrule \midrule
      Dependent Variable: & \multicolumn{3}{c}{Offline Price Elasticity}\\
       & \multicolumn{1}{c}{\makecell{Dry Dog Food}}  
       & \multicolumn{1}{c}{\makecell{Dog Hygiene}}
       & \multicolumn{1}{c}{\makecell{Cat Hygiene}}  \\ 
      Model:             & (1)             & (2)             & (3)                      \\  
      \midrule
%      \emph{Variables}\\
      ATT               & -0.2417$^{**}$  & -0.8444 $^{***}$ & -0.701$^{***}$\\   
                         & (0.0791)              & (0.2044)               & (0.1543)       \\   \midrule
      \midrule
 %     \emph{Fit statistics}\\
     Observations       & 136,411                   & 78,253         & 55,410\\  
      \midrule %\midrule
        \multicolumn{4}{l}{\emph{Bootstrapped standard-errors in parentheses via 500 replications}}\\
      \multicolumn{4}{l}{\emph{Signif. Codes: ***: 0.001, **: 0.01, *: 0.05}}\\
   \end{tabular}
   }
\end{table}

\clearpage

\subsubsection{Detailed Results for Alternative Demand Equation with Control Function for Price Term Only}
\label{appssec:alternative.imple.control.function}

In this section, we follow \cite{ebbes2016dealing} to implement the control function method by running Equation \eqref{eq:Price_FirstStage} and adding the control function term only for price. We show that our main findings are consistent when only adding the control function terms for price. Table \ref{tab:append.main.demand.model.singleCF.price} presents the estimates of the main demand model with control function for price term only. The standard errors for the demand parameters are based on the customer-level block bootstrap with 500 replications.

\begin{table}[htp!]
\caption{Coefficient Estimates - Demand Model with Single Control Function Term for Price}
\centering
\label{tab:append.main.demand.model.singleCF.price}
\footnotesize{
\begin{tabular}{lcccc}
\toprule
 & Dry Dog Food & Dry Cat Food & Dog Hygiene & Cat Hygiene \\
\textit{Variables} &  &  &  &  \\
\midrule
Price & \makecell{-0.3165$^{***}$\\(0.0212)} & \makecell{-0.1588$^{***}$\\(0.018)} & \makecell{-1.3137$^{***}$\\(0.1163)} & \makecell{-0.2401$^{***}$\\(0.0197)} \\
cumPercOnlineSpend & \makecell{-0.3837\\(0.4008)} & \makecell{-1.1528\\(1.1453)} & \makecell{3.7018$^{**}$\\(1.1297)} & \makecell{2.4235$^{*}$\\(0.9528)} \\
Price * cumPercOnlineSpend & \makecell{-0.1076$^{**}$\\(0.0374)} & \makecell{-0.0017\\(0.0974)} & \makecell{-1.9529$^{***}$\\(0.4989)} & \makecell{-0.2061$^{**}$\\(0.0638)} \\
BuyPrevOcca & \makecell{0.151$^{***}$\\(0.0272)} & \makecell{-0.5032$^{***}$\\(0.045)} & \makecell{-0.1805$^{***}$\\(0.0283)} & \makecell{0.326$^{***}$\\(0.0294)} \\
CF\_Price & \makecell{0.0524\\(0.0268)} & \makecell{0.0432\\(0.0247)} & \makecell{0.3133\\(0.1697)} & \makecell{0.3093$^{***}$\\(0.0257)} \\
Customer\_Brand\_FE & Yes & Yes & Yes & Yes\\ \midrule
Observations       & 136,411         & 45,443          & 78,253         & 55,410\\ 
Log-likelihood & -84211.483 & -30698.359 & -55455.669 & -37337.88 \\
\bottomrule
\multicolumn{5}{l}{\emph{Bootstrapped standard-errors in parentheses via 500 replications}}\\
\multicolumn{5}{l}{\emph{Signif. Codes: ***: 0.001, **: 0.01, *: 0.05}}\\
\end{tabular}
}
\end{table}

\begin{figure}[htp!]
    \centering
    \caption{Bootstrapped distribution of Overall ATT Estimates based on 500 replications with the point estimate from Table \ref{tab:CS_Result_CFPriceOnly}.}
\label{fig:tab:CS_Result_CS_Result_CFPriceOnly.Boot_Est}
    \begin{subfigure}{0.32\textwidth}
        \centering
        \caption{Dry Dog Food}
        \includegraphics[width=\linewidth]{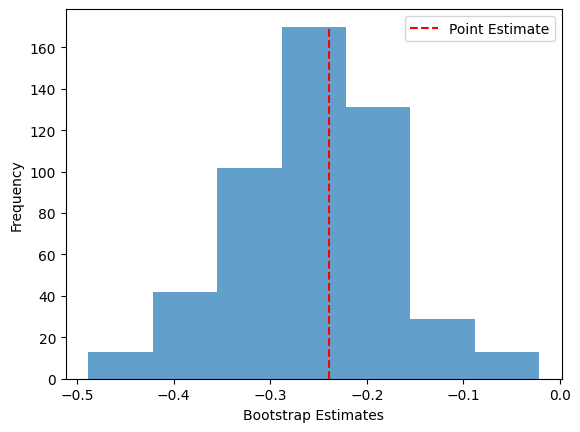}
    \end{subfigure}%
    \hfill
    \begin{subfigure}{0.33\textwidth}
        \centering
        \caption{Dog Hygiene}
        \includegraphics[width=\linewidth]{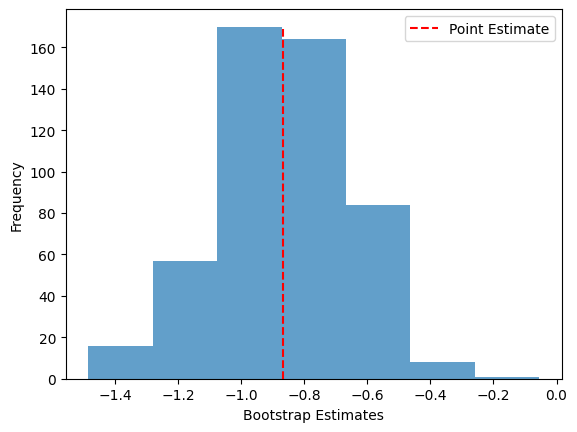}
    \end{subfigure}%
    \hfill
    \begin{subfigure}{0.32\textwidth}
        \centering
        \caption{Cat Hygiene}
        \includegraphics[width=\linewidth]{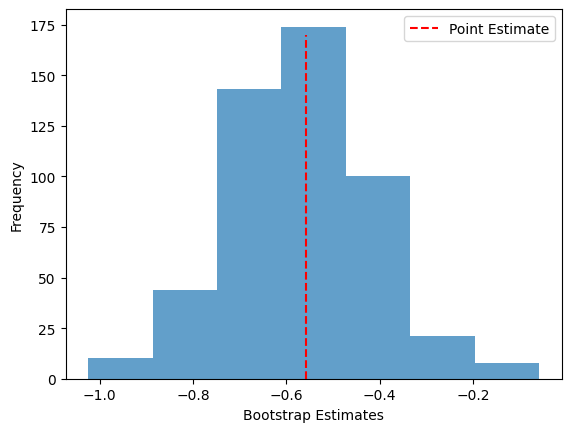}
    \end{subfigure}
\end{figure}

Table \ref{tab:CS_Result_CFPriceOnly} presents the ATT estimates of offline price elasticity when we estimate the demand model in Equation \eqref{eq:util_choice_M2} with a control function term for price only. The standard errors for the ATT estimates and the demand model parameters are based on the customer-level block bootstrap with 500 replications; See Figure \ref{fig:tab:CS_Result_CS_Result_CFPriceOnly.Boot_Est}.

\begin{table}[htp!]
   \caption{ATT of Online Adoption on Offline Price Elasticity (CS Estimator) -- Including Single CF Term for Price Endogeneity}
   \centering
   \label{tab:CS_Result_CFPriceOnly}
   \footnotesize{
   \begin{tabular}{lccc}
      \tabularnewline \midrule \midrule
      Dependent Variable: & \multicolumn{3}{c}{Offline Price Elasticity}\\
       & \multicolumn{1}{c}{\makecell{Dry Dog Food}}  
       & \multicolumn{1}{c}{\makecell{Dog Hygiene}}
       & \multicolumn{1}{c}{\makecell{Cat Hygiene}}  \\ 
      Model:             & (1)                         & (2)             & (3)             \\  
      \midrule
%      \emph{Variables}\\
      ATT               & -0.2394$^{**}$  & -0.867$^{***}$ & -0.5584$^{***}$\\   
                         & (0.0807)       & (0.2119)        & (0.1566)       \\   \midrule
      \midrule
 %     \emph{Fit statistics}\\
     Observations       & 136,411              & 78,253         & 55,410\\  
      \midrule %\midrule
        \multicolumn{4}{l}{\emph{Bootstrapped standard-errors in parentheses via 500 replications}}\\
      \multicolumn{4}{l}{\emph{Signif. Codes: ***: 0.001, **: 0.01, *: 0.05}}\\
   \end{tabular}
   }
\end{table}

\clearpage

\subsubsection{Results for Price Variations and Control Function for Price}
\label{appssec:price_variation_control_function}

We now document the sources of price variation used in the price-elasticity analysis and provide additional evidence supporting the control-function approach for price. This analysis addresses two related questions. First, we examine whether input costs generate meaningful supply-side variation in the offline retail prices used in the demand estimation. Second, we examine whether online prices exhibit large independent channel-specific variation that could mechanically explain the estimated changes in offline price sensitivity after online adoption.

\paragraph{First stage for offline prices.}
Table \ref{tab:first_stage_price_regression} reports the first-stage regressions for the three categories used in the main elasticity analysis. The dependent variable is the offline store-brand-month retail price used in the demand estimation. We regress price on input cost while absorbing store, brand, and year-month fixed effects. Across the three categories, input costs positively and significantly predict offline retail prices after these fixed effects are included. This provides support for the relevance of the cost-based control-function approach: the identifying price variation is not solely driven by persistent differences across stores, brands, or months, but also reflects cost-related supply-side variation.

\begin{table}[htp!]
   \caption{First-Stage Regression for Brand Price}
   \label{tab:first_stage_price_regression}
   \centering
   \footnotesize{
   \begin{tabular}{lccc}
      \tabularnewline \midrule \midrule
      Dependent Variable: & \multicolumn{3}{c}{Price}\\
      Model:     &  Dry Dog Food & Dog Hygiene & Cat Hygiene \\
      \midrule
      \emph{Variables}\\
      Cost         & 1.061$^{**}$ & 0.3898$^{**}$ & 2.403$^{***}$\\   
                   & (0.2401)     & (0.0544)      & (0.1558)\\   
      \midrule
      \emph{Fixed-effects}\\
      Store        & Yes          & Yes           & Yes\\  
      Brand        & Yes          & Yes           & Yes\\  
      YearMonth    & Yes          & Yes           & Yes\\  
      \midrule
      \emph{Fit statistics}\\
      Observations & 31,170       & 13,080        & 15,960\\  
      R$^2$        & 0.93974      & 0.86905       & 0.89381\\  
      Within R$^2$ & 0.22909      & 0.06799       & 0.49604\\  
      \midrule
      Excluded-IV Wald F-statistic & 19.53 & 51.27 & 237.94 \\
      \midrule \midrule
      \multicolumn{4}{l}{\emph{Clustered (Brand \& YearMonth) standard-errors in parentheses}}\\
      \multicolumn{4}{l}{\emph{Signif. Codes: ***: 0.001, **: 0.01, *: 0.05}}\\
   \end{tabular}
   }
\end{table}

Table \ref{tab:offline_brand_price_decomposition} further decomposes offline store-brand-month price variation for Dry Dog Food. Column (1) includes only cost and shows that cost alone explains a substantial share of price variation. Columns (2)--(4) sequentially add brand fixed effects, year-month fixed effects, and store fixed effects. The $R^2$ increases from 0.786 in Column (1) to 0.940 in Column (4), indicating that offline retail prices reflect a combination of cost variation, persistent brand differences, common time shocks, and store-level differences. Importantly, the coefficient on cost remains positive and statistically significant after absorbing brand, year-month, and store fixed effects. Thus, the first-stage residuals used in the control-function specification are constructed from cost-related price variation after accounting for the main sources of persistent price heterogeneity.

\begin{table}[htp!]
   \caption{Store-Brand-Month Price Variation Decomposition - Dry Dog Food}
   \label{tab:offline_brand_price_decomposition}
   \centering
   \footnotesize{
   \begin{tabular}{lcccc}
      \tabularnewline \midrule \midrule
      Dependent Variable: & \multicolumn{4}{c}{Price}\\
      Model:       & No FE           & Brand FE            & Brand + YearMonth FE            & Brand + YearMonth + Store FE \\  
      \midrule
      \emph{Variables}\\
      Constant     & 1.731         &                &                &   \\   
                   & (0.8775)      &                &                &   \\   
      Cost         & 1.505$^{***}$ & 0.9319$^{***}$ & 0.8335$^{***}$ & 1.061$^{**}$\\   
                   & (0.1707)      & (0.1695)       & (0.1512)       & (0.2401)\\   
      \midrule
      \emph{Fixed-effects}\\
      Brand        &               & Yes            & Yes            & Yes\\  
      YearMonth    &               &                & Yes            & Yes\\  
      Store        &               &                &                & Yes\\  
      \midrule
      \emph{Fit statistics}\\
      Observations & 31,170        & 31,170         & 31,170         & 31,170\\  
      R$^2$        & 0.78621       & 0.90973        & 0.91826        & 0.93974\\  
      Within R$^2$ &               & 0.28735        & 0.24973        & 0.22909\\  
      \midrule \midrule
      \multicolumn{5}{l}{\emph{Clustered (Brand \& YearMonth) standard-errors in parentheses}}\\
      \multicolumn{5}{l}{\emph{Signif. Codes: ***: 0.001, **: 0.01, *: 0.05}}\\
   \end{tabular}
   }
\end{table}

\begin{figure}[htp!]
    \centering
        \caption{Store-Level Price Variation Over Time for the Top One Selling Brand}
\label{fig:store_level_price_variation_top1_selling_brand}
    \includegraphics[width=0.8\linewidth]{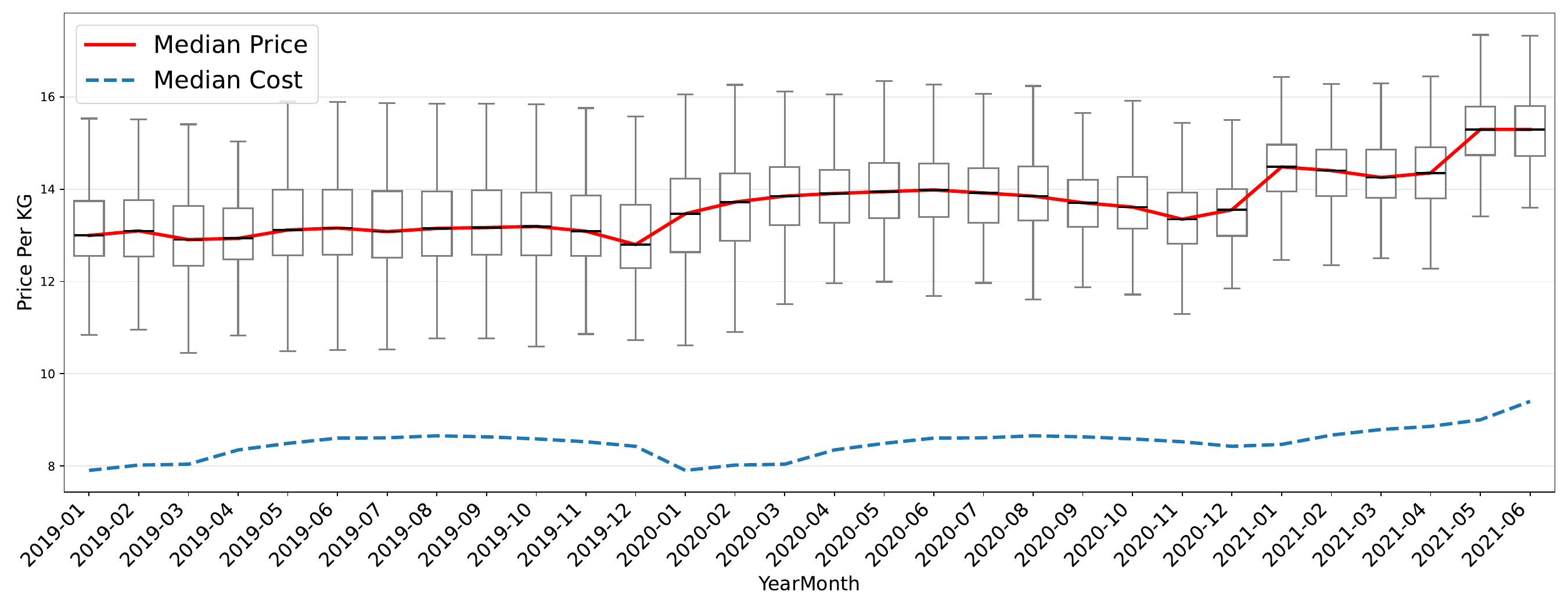}
\end{figure}

\begin{figure}[htp!]
    \centering
        \caption{Store-Level Cost Variation Over Time for the Top One Selling Brand}
\label{fig:store_level_cost_variation_top1_selling_brand}
    \includegraphics[width=0.8\linewidth]{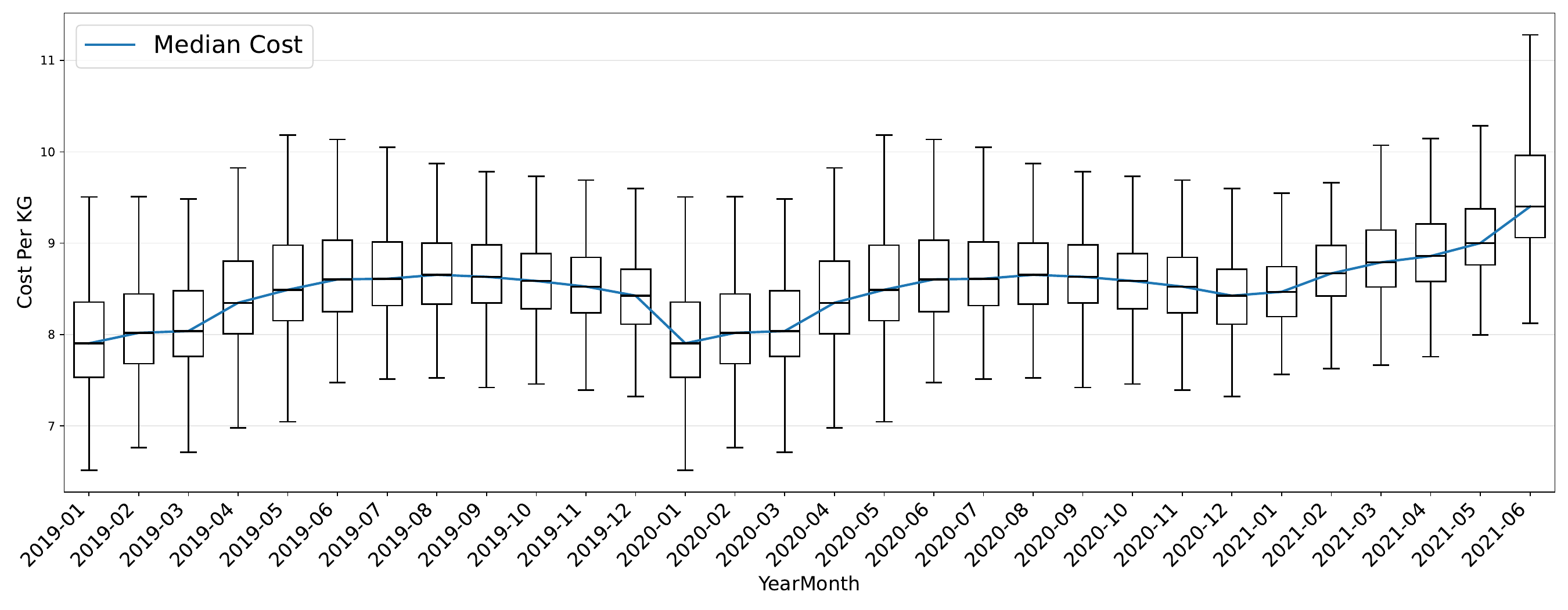}
\end{figure}

Figures \ref{fig:store_level_price_variation_top1_selling_brand} and \ref{fig:store_level_cost_variation_top1_selling_brand} provide complementary visual evidence using the top-selling brand. Figure \ref{fig:store_level_price_variation_top1_selling_brand} shows the distribution of store-level prices by month, together with the median price and median cost. Prices vary substantially across stores within a month and also change over time. Figure \ref{fig:store_level_cost_variation_top1_selling_brand} shows that costs also vary across stores and over time. These patterns illustrate that the data contain both cross-store and time-series variation in prices and costs, which helps separate cost-related price movements from persistent store and brand differences.

\paragraph{First stage for the endogenous interaction term.}
Table \ref{tab:first_stage_price_onlineShare_regression} reports the first-stage regression for $\textrm{Price} \times \textrm{cumPercOnlineSpend}$, where the focal excluded instrument is $\textrm{Cost} \times \textrm{cumPercOnlineSpend}$. The specifications also control for cost, cumulative online spending share, prior brand purchase, and customer, brand, and year-month fixed effects. The coefficient on
$\textrm{Cost}\times\textrm{cumPercOnlineSpend}$ is positive and statistically significant in all three categories. The excluded-IV Wald F-statistics are also large, providing strong evidence that the instrument is relevant for the endogenous price-interaction term.

\begin{table}[htp!]
   \caption{First-Stage Regression for Price * cumPercOnlineSpend}
   \label{tab:first_stage_price_onlineShare_regression}
   \centering
   \footnotesize{
   \begin{tabular}{lccc}
      \tabularnewline \midrule \midrule
      Dependent Variable: & \multicolumn{3}{c}{Price\_cumPercOnlineSpend}\\
      Model:                    & (1)             & (2)                     & (3)\\  
      \midrule
      \emph{Variables}\\
 Cost                      & -0.0054$^{*}$  & -0.0077$^{**}$ & -0.0291$^{***}$\\   
                                & (0.0027)       & (0.0026)       & (0.0038)\\   
      cumPercOnlineSpend        & 2.366$^{***}$  & 1.679$^{***}$  & 1.660$^{**}$\\   
                                & (0.3544)       & (0.0359)       & (0.5211)\\   
      Cost * cumPercOnlineSpend & 1.409$^{***}$  & 0.7774$^{***}$ & 2.510$^{***}$\\   
                                & (0.0639)       & (0.0328)       & (0.1225)\\   
      BuyPrevOcca               & 0.0059$^{***}$ & -0.0004        & 0.0035\\   
                                & (0.0011)       & (0.0002)       & (0.0025)\\   
      \midrule
      \emph{Fixed-effects}\\
      Customer                  & Yes            & Yes            & Yes\\  
      Brand                     & Yes            & Yes            & Yes\\  
      YearMonth                 & Yes            & Yes            & Yes\\  
      \midrule
      \emph{Fit statistics}\\
      Observations              & 1,091,288      & 313,012        & 221,640\\  
      R$^2$                     & 0.98249        & 0.99616        & 0.98954\\  
      Within R$^2$              & 0.96176        & 0.99040        & 0.97622\\  
             \midrule
      Excluded-IV Wald F-statistic & 496.14 & 285.95 & 432.65 \\
 % Sanderson–Windmeijer (SW) conditional F-statistic & 395.65 & 413.23 & 78.44 \\
      \midrule \midrule
      \multicolumn{4}{l}{\emph{Clustered (Customer \& Brand-YearMonth) standard-errors in parentheses}}\\
      \multicolumn{4}{l}{\emph{Signif. Codes: ***: 0.001, **: 0.01, *: 0.05}}\\
   \end{tabular}
       }
\end{table}

\paragraph{Supplementary order-level price decomposition.} 

As an additional check, we decompose transaction-level prices into product and time components. This analysis uses a different unit of observation from the store-brand-month prices used in the demand model and produces the qualitatively similar conclusion in source of price variation. For dry dog food, cost explains 63 percentage points of the total order-level price variation. Adding SKU fixed effects increases the $(R^2)$ by 22.9 percentage points for online orders. Adding year-month fixed effects contributes a further 3.9 and 3.8 percentage points, respectively. Thus, most of the additional explanatory power at the order level comes from persistent differences across products, with a smaller contribution from common time shocks.

%%% comment out the order-level online price variation decomposition; also we do not use online price at all to the research question.

\begin{comment}
\begin{table}[htp!]
   \caption{Order-Level Online Price Variation Decomposition -- Dry Dog Food}
\label{tab:order_level_online_price_variation}
   \centering
   \footnotesize{
   \begin{tabular}{lccc}
      \tabularnewline \midrule \midrule
      Dependent Variable: & \multicolumn{3}{c}{Price\_Paid}\\
      Model:       & (1)           & (2)      & (3)\\  
      \midrule
      \emph{Variables}\\
      Constant     & 47.37$^{***}$ &          &   \\   
                   & (6.879)       &          &   \\   
      Cost         & 1.146$^{***}$ & 0.0331   & -0.0703\\   
                   & (0.0690)      & (0.0696) & (0.0467)\\   
      \midrule
      \emph{Fixed-effects}\\
      SKU          &               & Yes      & Yes\\  
      YearMonth    &               &          & Yes\\  
      \midrule
      \emph{Fit statistics}\\
      Observations & 7,818         & 7,818    & 7,818\\  
      R$^2$        & 0.63133       & 0.86001  & 0.89853\\  
      Within R$^2$ &               & 0.00066  & 0.00394\\  
      \midrule \midrule
      \multicolumn{4}{l}{\emph{Clustered (SKU) standard-errors in parentheses}}\\
      \multicolumn{4}{l}{\emph{Signif. Codes: ***: 0.001, **: 0.01, *: 0.05}}\\
   \end{tabular}
   }
\end{table}
\end{comment}

Table \ref{tab:order_level_offline_price_variation} repeats the decomposition using the broader order-level price data for Dry Dog Food. Cost is strongly associated with raw transaction prices in Column (1). However, the cost coefficient becomes much smaller after adding SKU fixed effects in Column (2), further attenuates after adding year-month fixed effects in Column (3), and becomes close to zero after additionally controlling for store fixed effects in Column (4). The $R^2$ increases from 0.631 in Column (1) to 0.941 in Column (4), suggesting that order-level price variation is primarily explained by SKU, time, and store components.

\begin{table}[htp!]
   \caption{Order-Level Offline Price Variation Decomposition -- Dry Dog Food}
\label{tab:order_level_offline_price_variation}
   \centering
   \footnotesize{
   \begin{tabular}{lcccc}
      \tabularnewline \midrule \midrule
      Dependent Variable: & \multicolumn{4}{c}{Price\_Paid}\\
      Model:       & (1)           & (2)            & (3)          & (4)\\  
      \midrule
      \emph{Variables}\\
      Constant     & 57.95$^{***}$ &                &              &   \\   
                   & (6.898)       &                &              &   \\   
      Cost         & 1.094$^{***}$ & 0.0875$^{***}$ & 0.0379$^{*}$ & 0.0076\\   
                   & (0.0810)      & (0.0219)       & (0.0166)     & (0.0068)\\   
      \midrule
      \emph{Fixed-effects}\\
      SKU          &               & Yes            & Yes          & Yes\\  
      YearMonth    &               &                & Yes          & Yes\\  
      Store        &               &                &              & Yes\\  
      \midrule
      \emph{Fit statistics}\\
      Observations & 99,233     & 99,233 & 99,233     & 99,233\\  
      R$^2$        & 0.63096       & 0.89400        & 0.93170      & 0.94055\\  
      Within R$^2$ &               & 0.00997        & 0.00288      & 0.00013\\  
      \midrule \midrule
      \multicolumn{5}{l}{\emph{Clustered (SKU \& Store) standard-errors in parentheses}}\\
      \multicolumn{5}{l}{\emph{Signif. Codes: ***: 0.001, **: 0.01, *: 0.05}}\\
   \end{tabular}
   }
\end{table}

Taken together, these results clarify the identifying variation behind the price-elasticity estimates. The control-function approach relies on cost-related variation in offline retail prices after absorbing persistent store, brand, and common time components.  The supplementary offline order-level decomposition shows that product, time, and store differences account for much of the variation in offline transaction prices.\footnote{Similarly, for online order-level prices, we find that product accounts for 81.7\% of variation, while time accounts for only about 4.6\%.} Combined with Web Appendix $\S$\ref{appssec:cross.channel.price.differences}, which shows that systematic online-offline price differences are modest relative to SKU-level variation, these results indicate that the estimated changes in offline price sensitivity after online adoption are unlikely to be mechanically driven by large online-offline price gaps. Instead, the estimates capture how consumers respond to offline price variation after online adoption, conditional on the main sources of price heterogeneity.

\clearpage

\subsection{Alternative Product Category Market Definition}
\label{appsec:alternative.product.category.definition}

\begin{figure}[htp!]
    \centering
    \caption{Alternative Category Definition: Average price elasticity between adopters (shown in twelve dotted lines) and non-adopters (shown in one dashed line) for each category.}
    \label{fig:se4_category_definition_robustness}

    \begin{subfigure}{0.32\textwidth}
        \centering
        \caption{Dry Dog Food}
        \includegraphics[width=\linewidth]{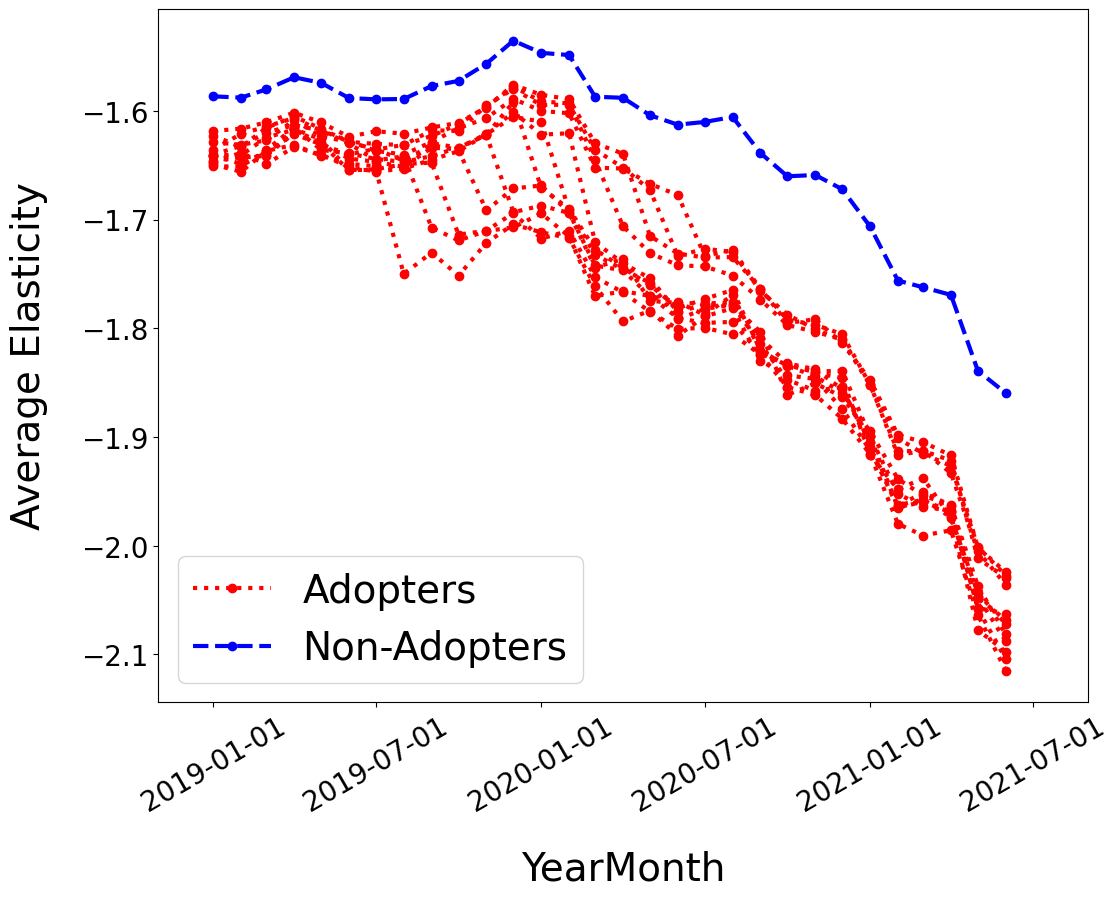}
    \end{subfigure}%
    \hfill
    \begin{subfigure}{0.32\textwidth}
        \centering
        \caption{Dog Hygiene}
        \includegraphics[width=\linewidth]{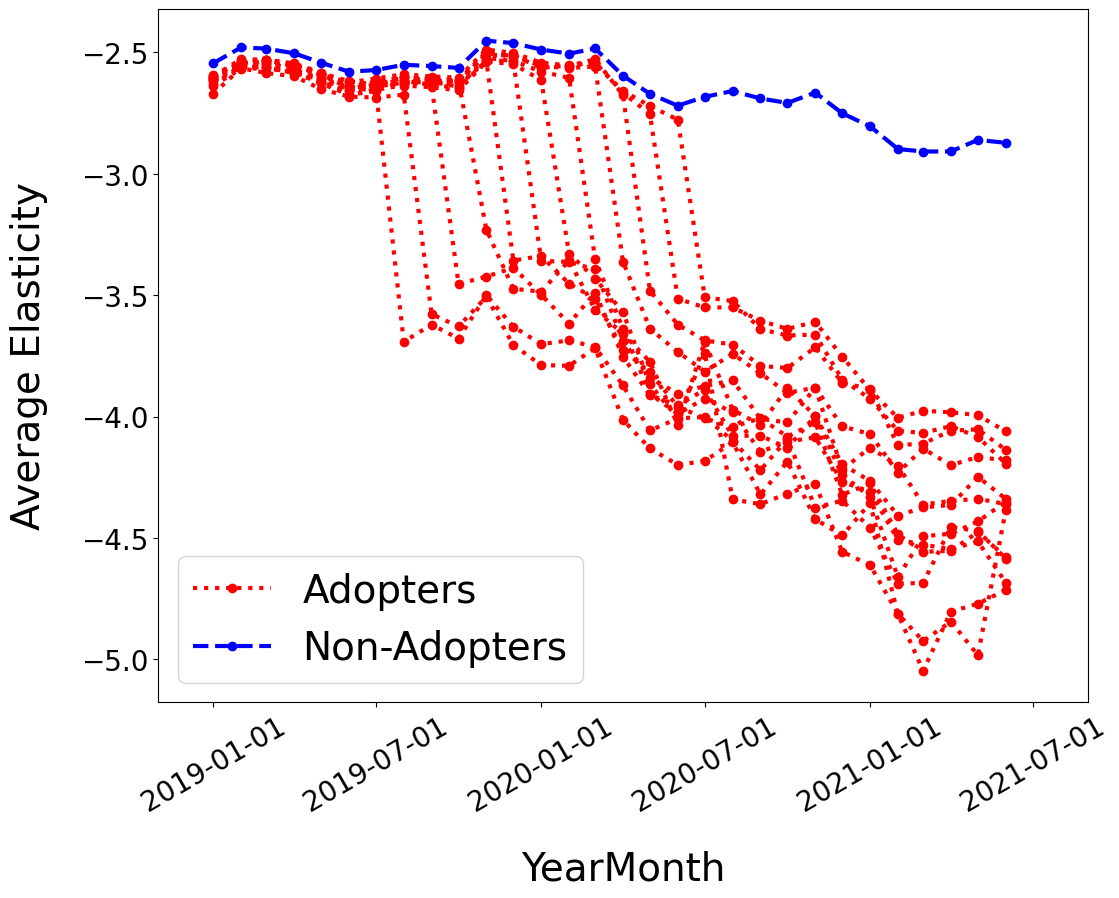}
    \end{subfigure}%
    \hfill
    \begin{subfigure}{0.32\textwidth}
        \centering
        \caption{Cat Hygiene}
        \includegraphics[width=\linewidth]{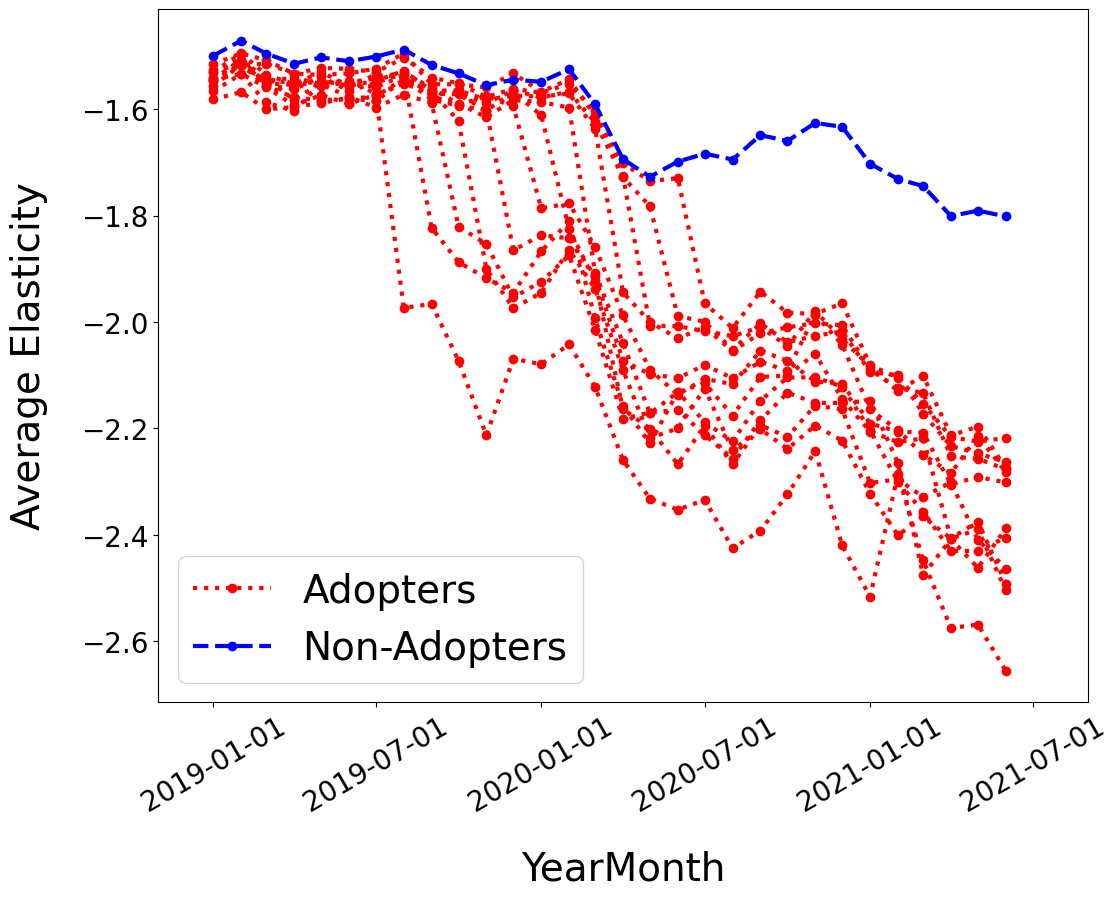}
    \end{subfigure}
\end{figure}

We provide a robustness check by including additional pack sizes that consist of at least 80\% of the total categories' sales. For example, for the dry dog food, besides 15KG (top-selling pack size), we also include other top-selling pack sizes, including 2.5KG, 10KG, 12.5KG, and 20KG, into our product category definition, and correspondingly include customers who shift between these pack sizes. Given this alternative product category definition, we still observe significant online adoption effects across the three categories, consistent with the paper's main findings. See Figure \ref{fig:se4_category_definition_robustness} for elasticity prediction. Table \ref{tab:append.demand.model.altDef.category.morepacksizes} presents the demand model estimates, and Table \ref{tab:CS_Result_Alternative_Definition_Categories} presents the ATT estimates, with bootstrapped standard errors with 500 replications. We obtain consistent findings with the main results.

\begin{table}[htp!]
\caption{Coefficients of Demand Model Estimate - Alternative Definition of Product Categories (More Packsizes)}
\centering
\label{tab:append.demand.model.altDef.category.morepacksizes}
\footnotesize{
\begin{tabular}{lcccc}
\toprule
 & Dry Dog Food & Dry Cat Food & Dog Hygiene & Cat Hygiene \\
\textit{Variables} &  &  &  &  \\
\midrule
Price & \makecell{-0.1183$^{***}$\\(0.0036)} & \makecell{-0.1552$^{***}$\\(0.0081)} & \makecell{-1.3376$^{***}$\\(0.0563)} & \makecell{-0.1728$^{***}$\\(0.0107)} \\
cumPercOnlineSpend & \makecell{-0.7823$^{***}$\\(0.1139)} & \makecell{-3.8597$^{***}$\\(0.9229)} & \makecell{4.9911$^{***}$\\(1.1205)} & \makecell{0.6706\\(0.4006)} \\
Price * cumPercOnlineSpend & \makecell{-0.0298$^{***}$\\(0.0067)} & \makecell{0.107\\(0.0693)} & \makecell{-2.6093$^{***}$\\(0.5007)} & \makecell{-0.1445$^{***}$\\(0.0326)} \\
BuyPrevOcca & \makecell{0.3172$^{***}$\\(0.0122)} & \makecell{-0.1262$^{***}$\\(0.0189)} & \makecell{-0.0488$^{*}$\\(0.0197)} & \makecell{0.3691$^{***}$\\(0.0195)} \\
CF\_Price & \makecell{-0.0049\\(0.005)} & \makecell{0.0063\\(0.0114)} & \makecell{-0.6658$^{***}$\\(0.0775)} & \makecell{0.1469$^{***}$\\(0.0167)} \\
CF\_Price\_cumPercOnlineSpend & \makecell{0.1454$^{***}$\\(0.0232)} & \makecell{-0.1979$^{*}$\\(0.0817)} & \makecell{3.4866$^{***}$\\(0.684)} & \makecell{0.5027$^{***}$\\(0.0694)} \\
Customer\_Brand\_FE & Yes & Yes & Yes & Yes\\ \midrule
Observations       & 547,683         & 186,423          & 179,707         & 110,444\\ 
Log-likelihood & -373370.82 & -125724.78 & -143028.75& -72882.275 \\
\bottomrule
\multicolumn{5}{l}{\emph{Bootstrapped standard-errors in parentheses via 500 replications}}\\
\multicolumn{5}{l}{\emph{Signif. Codes: ***: 0.001, **: 0.01, *: 0.05}}\\
\end{tabular}
}
\end{table}

%%Analysis51 ATT with CF
\begin{table}[t!]
   \caption{ATT of Online Adoption on Offline Price Elasticity Over All Post-Adoption Period (CS Estimator) -- Alternative Definition of Product Categories (More Packsizes)}
   \centering
   \label{tab:CS_Result_Alternative_Definition_Categories}
   \footnotesize{
   \begin{tabular}{lccc}
      \tabularnewline \midrule \midrule
      Dependent Variable: & \multicolumn{3}{c}{Offline Price Elasticity}\\
       & \multicolumn{1}{c}{\makecell{Dry Dog Food}}  
       & \multicolumn{1}{c}{\makecell{Dog Hygiene}}
       & \multicolumn{1}{c}{\makecell{Cat Hygiene}}  \\ 
      Model:             & (1)             & (2)             & (3)                      \\  
      \midrule
%      \emph{Variables}\\
      ATT               & -0.0966$^{***}$  & -1.0663$^{***}$ & -0.3602$^{***}$\\   
                         & (0.0188)              & (0.1976)               & (0.0723)       \\   \midrule
      \midrule
 %     \emph{Fit statistics}\\
     Observations       & 547,683            & 179,707         & 110,444\\  
      \midrule %\midrule
        \multicolumn{4}{l}{\emph{Bootstrapped standard-errors in parentheses via 500 replications}}\\
      \multicolumn{4}{l}{\emph{Signif. Codes: ***: 0.001, **: 0.01, *: 0.05}}\\
   \end{tabular}
   }
\end{table}

\clearpage

\section{Detailed Results for Implications for Retailer Pricing and Profitability}
\label{appsec:counterfactual.implication}

So far, we have established that when consumers begin to shop online, their price sensitivity in the offline channel increases. We now examine how the retailer should modify its offline pricing strategy in response to these customer-level changes. In particular, given the firm's current level of online adoption, how much \textit{offline} profit can be gained by properly accounting for the change in price elasticity relative to a scenario where the firm ignores this effect? To quantify the profit implications, we conduct counterfactual analyses that compare offline profits under pricing policies that incorporate the online adoption effect vs. those that do not. In $\S$\ref{appssec:naive.demand.model}, we first outline how a {\it naive} retailer who ignores the online adoption effect would (incorrectly) estimate the offline price elasticities. Next, in \S\ref{appssec:optimal_prices}, we characterize the profit-maximizing prices that would be set under the true demand model (accounting for the online adoption effect) and under the naive demand model (ignoring the online adoption effect) for two pricing strategies: uniform pricing and personalized pricing. In \S\ref{appssec.counterfactual.scenarios}, we describe how we compute the profits under four potential counterfactual pricing strategies and quantify the profit implications of accounting for the online adoption effect. We also mention some caveats of our counterfactual results. 

\subsection{Naive demand model without adoption effect}
\label{appssec:naive.demand.model}

First, we characterize how a naive retailer would estimate consumers' offline price sensitivities. In this case, we assume that the retailer uses a demand model that does not account for online adoption effects. 

Following notation of the demand model in Equation \eqref{eq:util_choice_M2} in \S\ref{sec:model}, we specify the mean utility for consumer $i$ choosing brand $j$ at purchase occasion $t$ in this case as follows: 
\begin{equation}
\label{eq:naive.model}
    \tilde{\psi}_{ijt} =  \tilde{\alpha}_0 \textrm{Price}_{jt} +  \tilde{\alpha}_{1} \mathcal{I}\{\textrm{BuyPrevOcca}_{ijt}\}   + \tilde{\beta}_{1}\textrm{CF\_Price}_{jt} + \tilde{\delta}_{ij} ~~ \forall j \in \{1, \ldots, J\}\\
\end{equation}
Here, $\tilde{\alpha}_0$ is the \textit{naive} estimate of the price sensitivity parameter. $\tilde{\alpha}_1$ and $\tilde{\beta}_{1}$ are coefficients for the state dependence variable and price control function, respectively, and $\tilde{\delta}_{ij}$ denotes customer-brand fixed effects. The corresponding choice probability for consumer $i$ choosing brand $j$ on purchase occasion $t$ is then given by $\tilde{Pr}_{ijt} = \frac{\exp(\tilde{\psi}_{ijt})}{1 + \sum_{j'=1}^{J}\exp(\tilde{\psi}_{ij't})}$.

We follow the estimation procedure described in \S\ref{ssec:estimation} and report the parameter estimates in Table \ref{tab:naive.demand.model}. 

\begin{table}[!htp]
\caption{Coefficient Estimates of the Naive Demand Model}
\centering
\label{tab:naive.demand.model}
\footnotesize{
\begin{tabular}{llll}
\toprule
 & Dry Dog Food & Dog Hygiene & Cat Hygiene \\
\textit{Variables} &  &  &  \\
\midrule
Price & \makecell{-0.3808$^{***}$ \\(0.0237)} & \makecell{-1.5921$^{***}$ \\(0.1129)} & \makecell{-0.2822$^{***}$ \\(0.0189)} \\
BuyPrevOcca & \makecell{0.1581$^{***}$ \\(0.0316)} & \makecell{-0.1772$^{***}$ \\(0.0299)} & \makecell{0.3306$^{***}$ \\(0.0289)} \\
CF\_Price & \makecell{0.1068$^{***}$ \\(0.0279)} & \makecell{0.5272$^{**}$ \\(0.1847)} & \makecell{0.3522$^{***}$ \\(0.0261)} \\
Customer\_Brand\_FE & Yes & Yes  & Yes\\ \midrule
Observations       & 136,411      & 78,253         & 55,410\\ 
Log-likelihood & -84222.285 & -55490.014 & -37358.772 \\ 
\bottomrule
\multicolumn{4}{l}{\emph{Bootstrapped standard-errors in parentheses via 500 replications}}\\
\multicolumn{4}{l}{\emph{Signif. Codes: ***: 0.001, **: 0.01, *: 0.05}}\\
\end{tabular}
}
\end{table}

Compared to the parameter estimates in Table \ref{append.main.demand.model}, we find that the estimated price coefficient $\tilde{\alpha}_{0}$ from the naive model reflects a weighted average of price sensitivities for both adopters and non-adopters. This naive estimate serves as a key input for computing the counterfactual profit-maximizing prices that the retailer would set if they ignored the effect of online adoption.

\subsection{Optimal Offline Price Characterization Under Counterfactual Scenarios}
\label{appssec:optimal_prices}
We now characterize the profit-maximizing offline prices for a given month, under two pricing strategies: (1) Uniform pricing, where all consumers who visit an offline store during a particular time period see the same price for a given brand, and (2) Personalized pricing, where each consumer may potentially receive a different price vector depending on their demographics, prior purchase behavior, and under our demand model, their online shopping status (whether they have adopted online shopping and how much of their cumulative spend is on the online channel). In practice, the personalized pricing strategy would need to be implemented using targeted coupons \citep{rossi1996value}. Personalized pricing is being increasingly adopted by firms in retail settings and has been the focus of recent research in marketing \citep{dube2023personalized,smith2023optimal,jain2024effective}. Therefore the impact of the online adoption effect on both uniform and personalized pricing is of interest for managers. 

Let $\mathcal{M}_{k}$ be a generic demand model, where the subscript $k$ denotes the model used. Specifically, $\mathcal{M}_{N}$ denotes the {\it N}aive model defined in Equation \eqref{eq:naive.model}, whereas $\mathcal{M}_{O}$ refers to {\it O}ur model in Equation \eqref{eq:util_choice_M2}. We now characterize the optimal offline pricing for a generic demand model $\mathcal{M}_{k}$ and then derive the profit-maximizing price vectors for both the uniform and personalized pricing strategies. 

Let $N$ be the market size, which is the number of offline shoppers who make at least one offline store visit during the month. Since we focus on monthly profit, we drop the month index ($t$) for notational simplicity. Let $\mathbf{p}\in\mathcal{R}^{J\times 1}$ denote the price vector, with $p_j$ being the price of brand $j$. We use the notation $p_j$ for price here (instead of Price$_j$) since price is now a decision variable rather than an observable. Under the naive model $\mathcal{M}_{N}$, the mean utility for consumer $i$ choosing brand $j$ based on Equation \eqref{eq:naive.model} can be written as:
\begin{equation}
\psi_{ij}(\mathbf{p};\mathcal{M}_{N}) = \tilde{\alpha}_0 p_{j} +  \tilde{\alpha}_{1} \mathcal{I}\{\textrm{BuyPrevOcca}_{ij}\}   + \tilde{\beta}_{1}\textrm{CF\_Price}_{j} + \tilde{\delta}_{ij}\nonumber  
\end{equation}
In contrast, the mean utility of our demand model $\mathcal{M}_{O}$ based on Equation \eqref{eq:util_choice_M2} can be written as:
\begin{equation}
\begin{aligned}
    \psi_{ij}(\mathbf{p};\mathcal{M}_{O}) &= \alpha_{0} p_{j} +
    \alpha_{1} \textrm{cumPercOnlineSpend}_{i} + 
    \alpha_2   p_{j}\times \textrm{cumPercOnlineSpend}_{i} \\
    & + \alpha_{3} \mathcal{I}\{\textrm{BuyPrevOcca}_{ij}\}  +\lambda_1 \textrm{CF\_Price}_{j} + \lambda_2 \textrm{CF\_Price\_cumPercOnlineSpend}_{ij} + \delta_{ij} 
    \end{aligned} \nonumber
\end{equation}

\subsubsection{Uniform Pricing}
\label{appsssec.optimal.offline.price}

Under uniform pricing, the retailer maximizes the expected profit per customer, defined as: 
\begin{equation}
\begin{aligned}
\label{eq:uniform.profit.function}
R(\mathbf{p}; \mathcal{M}_{k}) & = \frac{1}{N}\sum_{i=1}^{N} \sum_{j=1}^{J}(p_{j} - m_{j})\times Pr_{ij}(\mathbf{p}; \mathcal{M}_{k}) = \frac{1}{N}\sum_{i=1}^{N} R_{i}(\mathbf{p}; \mathcal{M}_{k})
    \end{aligned}
\end{equation}
where $m_j$ is the cost of brand $j$.
$Pr_{ij}(\mathbf{p};\mathcal{M}_{k}) = \frac{\exp(\psi_{ij}(\mathbf{p}; {\mathcal{M}_{k}})}{1 + \sum_{j'=1}^{J}\exp(\psi_{ij'}(\mathbf{p};{\mathcal{M}_{k}}))}$ refers to the choice probability of consumer $i$ for brand $j$ given price vector $\mathbf{p}$ and demand model $\mathcal{M}_{k}$, and $R_{i}(\mathbf{p}; \mathcal{M}_{k})$ represents customer $i$'s profit contribution. By taking the derivative of $R(\mathbf{p};\mathcal{M}_{k})$ with respect to each brand's price $p_{j}$, we characterize the optimal uniform pricing condition under a demand model $\mathcal{M}_{k}$: 

\begin{defnitalic}
\textit{Profit-Maximizing Uniform Price under Model $\mathcal{M}_{k}$:} Let $\alpha_{i}^{\mathcal{M}_{k}}$ denote the price sensitivity estimate for each customer $i$ under model $\mathcal{M}_{k}$. Then the optimal uniform price vector $\mathbf{p}_{*} = (p_{*,1},\ldots, p_{*,J})'$ satisfies the following system of equations:
\begin{equation}
\label{eq:j_th_FOC}
    p_{*,j} - m_j = c_{j}(\mathbf{p}_{*}; \mathcal{M}_{k}) + \sum_{i=1}^{N} w_{ij}(\mathbf{p}_{*};\mathcal{M}_{k})R_{i}(\mathbf{p}_{*};\mathcal{M}_{k}),
\end{equation}
where $c_{j}(\mathbf{p}_{*};\mathcal{M}_{k}) = -\sum_{i=1}^{N} \frac{Pr_{ij}(\mathbf{p}_{*};\mathcal{M}_{k})}{\sum_{i'=1}^{N}\alpha_{i'}^{\mathcal{M}_{k}}  Pr_{i'j}(\mathbf{p}_{*};\mathcal{M}_{k})}$ and $w_{ij}(\mathbf{p}_{*};\mathcal{M}_{k}) = \frac{\alpha_i^{\mathcal{M}_{k}} Pr_{ij}(\mathbf{p}_{*};\mathcal{M}_{k})}{\sum_{i'=1}^{N}\alpha_{i'}^{\mathcal{M}_{k}} Pr_{i'j}(\mathbf{p}_{*}; \mathcal{M}_{k})}$. Further, when we use the naive model, $\alpha_{i}^{\mathcal{M}_{N}} = \tilde{\alpha}_{0}$ and when we use our model, the $\alpha_{i}^{\mathcal{M}_{O}} = \alpha_{0} + \alpha_{2} \times \textrm{cumPercOnlineSpend}_{i}$. 
\end{defnitalic}

\noindent \textbf{Proof:} To find the profit-maximizing price vector $\mathbf{p}_{*}$, we take the first derivative of $R(\mathbf{p};\mathcal{M}_{k})$ with respect to each brand $j$'s price $p_j$. The resulting first-order condition (FOC) for the price of brand $j$, $p_{j}$, is:
\begin{equation}
    \frac{\partial R(\mathbf{p};\mathcal{M}_{k})}{\partial p_{j}} = \frac{1}{N}\sum_{i=1}^{N}  \frac{\alpha_i e_{ij}}{1 + \Lambda_{i}(e)} \left[\frac{1}{\alpha_{i}} + (p_{j} - m_j) - \frac{\sum_{l=1}^{J}(p_{l} - m_l)e_{il}}{1 + \Lambda_{i}(e)}\right]
\end{equation}

The profit-maximizing price vector $\mathbf{p}_{*}$ satisfies the following condition for each brand $j$: 
\begin{equation}
  \frac{1}{N}\sum_{i=1}^{N}  \frac{\alpha_i e_{ij}}{1 + \Lambda_{i}(e)} \left[\frac{1}{\alpha_{i}} + (p_{*,j} - m_j) - \frac{\sum_{l=1}^{J}(p_{*,l} - m_l)e_{il}}{1 + \Lambda_{i}(e)}\right] = 0
\end{equation}

Let $Pr_{ij}(\mathbf{p};\mathcal{M}_{k}) = \frac{e_{ij}}{1 + \Lambda_{i}(e)}.$ be the choice probability of consumer $i$ and brand $j$. Substituting this into the FOC for each brand $j$ yields:
\begin{equation}
\begin{aligned}
\label{append.eq:j_th_FOC}
    &(p_{*,j} - m_j)\sum_{i=1}^{N}\alpha_i Pr_{ij}(\mathbf{p}_{*};\mathcal{M}_{k})  = -\sum_{i=1}^{N} Pr_{ij}(\mathbf{p}_{*};\mathcal{M}_{k}) + \sum_{i=1}^{N}\alpha_i Pr_{ij}(\mathbf{p}_{*};\mathcal{M}_{k}) R_{i}(\mathbf{p}_{*};\mathcal{M}_{k}) \\
    & p_{*,j} - m_j = -\sum_{i=1}^{N} \frac{Pr_{ij}(\mathbf{p}_{*};\mathcal{M}_{k})}{\sum_{i'=1}^{N}\alpha_{i'}  Pr_{ij'}(\mathbf{p}_{*};\mathcal{M}_{k})} + \sum_{i=1}^{N} \frac{\alpha_i Pr_{ij}(\mathbf{p}_{*};\mathcal{M}_{k})}{\sum_{i'=1}^{N} \alpha_{i'} Pr_{i'j}(\mathbf{p}_{*};\mathcal{M}_{k})}\times R_{i}(\mathbf{p}_{*};\mathcal{M}_{k}) \\
    & p_{*,j} - m_j = c_{j}(\mathbf{p}_{*};\mathcal{M}_{k}) + \sum_{i=1}^{N} w_{ij}(\mathbf{p}_{*};\mathcal{M}_{k})R_{i}(\mathbf{p}_{*};\mathcal{M}_{k}) \\
\end{aligned}
\end{equation}
where $c_{j}(\mathbf{p}_{*};\mathcal{M}_{k}) = -\sum_{i=1}^{N} \frac{Pr_{ij}(\mathbf{p}_{*};\mathcal{M}_{k})}{\sum_{i'=1}^{N}\alpha_{i'} Pr_{i'j}(\mathbf{p}_{*};\mathcal{M}_{k})}$ and $w_{ij}(\mathbf{p}_{*};\mathcal{M}_{k}) = \frac{\alpha_i Pr_{ij}(\mathbf{p}_{*};\mathcal{M}_{k})}{\sum_{i'=1}^{N}\alpha_{i'} Pr_{i'j}(\mathbf{p}_{*};\mathcal{M}_{k})}$.

Equation \eqref{eq:j_th_FOC} defines a system of $J$ first-order conditions in $J$ unknown prices, which we solve numerically to obtain the optimal uniform price vector under each demand model $\mathcal{M}_{k}$. 

We denote the profit-maximizing uniform prices solution for a given demand model $\mathcal{M}_{k}$ as $\mathbf{p}_{*}^{U, \mathcal{M}_k}$, and $k\in\{N, O\}$. Specifically, $\mathbf{p}_{*}^{U, \mathcal{M}_N}$ denotes the profit-maximizing uniform prices under the naive model $\mathcal{M}_{N}$ and $\mathbf{p}_{*}^{U, \mathcal{M}_O}$ denotes the profit-maximizing uniform prices under our model $\mathcal{M}_{O}$.

\subsubsection{Personalized Pricing}
\label{appsssec.personalized.offline.price}

We now consider a personalized pricing strategy where the firm can tailor prices to each customer based on their price elasticity through targeted coupons and email campaigns. Personalized pricing is now commonly used by firms in retail settings and has been the focus of recent research in marketing \citep{rossi1996value,smith2023optimal,jain2024effective}. 

Here, the retailer's objective is to maximize the profit for each customer $i$:
\begin{equation}
\label{eq:personalized.profit.function}
    R_{i}(\mathbf{p}^{i}; \mathcal{M}_{k}) =  \sum_{j=1}^{J}(p^{i}_{j} - m_{j})\times Pr_{ij}(\mathbf{p}^{i}; \mathcal{M}_{k}),
\end{equation}
where $\mathbf{p}^{i}\in\mathcal{R}^{J\times 1}$ is the personalized price vector offered to customer $i$. As before, let $Pr_{ij}(\mathbf{p}^{i}; \mathcal{M}_{k}) = \frac{\exp(\psi_{ij}(\mathbf{p}^{i};\mathcal{M}_{k})}{1 + \sum_{j'=1}^{J} \exp(\psi_{ij'}(\mathbf{p}^{i};{\mathcal{M}}_{k}))}$ denote the probability that customer $i$ chooses brand $j$ under model $\mathcal{M}_{k}$ given price $\mathbf{p}^{i}$. Following \cite{jain2024effective}, we show that there exists a unique globally optimal revenue and price vector characterized by a fixed-point equation and derive the optimal price vector as follows.

\begin{defnitalic} \textit{Profit-Maximizing Personalized Price under Model $\mathcal{M}_{k}$:} The optimal personalized price vector under model $\mathcal{M}_{k}$ for any customer $i$, $\mathbf{p}^{i, \mathcal{M}_{k}}_{*}:= \arg \max_{\mathbf{p}^{i}\in \mathcal{R}^{J\times 1}_{\geq 0}} R_{i}(\mathbf{p}^{i}; \mathcal{M}_{k})$ satisfies:
\begin{equation}
\label{eq:optimal.personalized.price.condition}
    p^{i, \mathcal{M}_{k}}_{*,j} - m_j = - \frac{1}{\alpha_i^{\mathcal{M}_{k}}} + R_{i}^{*},
\end{equation}
Here, $\alpha_i^{\mathcal{M}_{k}}$ is the price sensitivity estimate of customer $i$ under model $\mathcal{M}_{k}$, and $R_{i}^{*} = R_{i}(\mathbf{p}^{i, \mathcal{M}_{k}}_{*}; \mathcal{M}_{k})$ denotes the corresponding optimal expected profit. The latter $R_{i}^{*}$ is the solution of the following fixed point equation: 
\begin{equation}
\label{eq:8_Optimal_Personalized_Price}
    R_{i}^{*} = -\frac{1}{\alpha_{i}^{\mathcal{M}_{k}}}\sum_{j=1}^{J}\exp(-1 + \alpha_{i}^{\mathcal{M}_{k}}  m_j + v_{ij}({\mathcal{M}_{k}}))\exp(\alpha_i^{^{\mathcal{M}_{k}}}  R_i ^{*}),
\end{equation}
where $v_{ij}({\mathcal{M}_{k}})$ represents the sum of all remaining terms in the utility for customer $i$ and brand $j$ under model $\mathcal{M}_{k}$ that do not include price.
%Specifically, under our model $\mathcal{M}_{O}$, $v_{ij}(\mathcal{M}_{O}) = \alpha_1 \textrm{cumPercOnlineSpend}_{i} + \alpha_{3}\mathcal{I}\{\textrm{BuyPrevOcca}_{ij}\} +\lambda_1 \textrm{CF\_Price}_{j} + \lambda_2 \textrm{CF\_Price\_cumPercOnlineSpend}_{ij} + \delta_{ij}$. Under the naive model $\mathcal{M}_{N}$, $v_{ij}(\mathcal{M}_{N}) =  \tilde{\alpha}_{1}\mathcal{I}\{\textrm{BuyPrevOcca}_{ij}\} + \tilde{\beta}_1 \textrm{CF\_Price}_{j} + \tilde{\delta}_{ij}$.
\end{defnitalic}

\noindent \textbf{Proof:} The first-order condition for maximizing consumer i's profit with respect to brand $j$'s price is given by:

\begin{equation}
\label{eq:foc_personalized_price}
\begin{aligned}
    p^{i, \mathcal{M}_{k}}_{*,j} - m_j & = - \frac{1}{\alpha_{i}^{\mathcal{M}_{k}}} + \frac{\sum_{l=1}^{J}(p^{i, \mathcal{M}_{k}}_{*,l} - m_l) e_{il}}{1 + \Lambda_{i}(e)} \\
    & =- \frac{1}{\alpha_{i}^{\mathcal{M}_{k}}} + R_{i}(\mathbf{p}^{i, \mathcal{M}_{k}}_{*}; \mathcal{M}_{k})
\end{aligned}
\end{equation}

%Denote $R_{i} = R_{i}(\mathbf{p}^{i, \mathcal{M}_{k}}_{*};\mathcal{M}_{k})$ as customer $i$'s optimal profit evaluated at the profit maximizing price vector of customer $i$. 

To derive the fixed-point equation for $R_{i}^{*}$, we multiply both side of Equation \eqref{eq:foc_personalized_price}  by $e_{ij}$ and sum over all brands $j\in[J]$:

\begin{equation}
\begin{aligned}
\label{append.eq:8_Optimal_Personalized_Price}
    \sum_{j=1}^{J} (p^{i}_{*,j} -m_j)e_{ij} & = -\sum_{j=1}^{J} \frac{e_{ij}}{\alpha_i} + \frac{\sum_{j=1}^{J}e_{ij} \left(\sum_{l=1}^{J} (p^{i}_{*,l} - m_l)e_{il}\right)}{1 + \Lambda_{i}(e)} \\
    [1 + \Lambda_{i}(e)]R_{i} & = -\sum_{j=1}^{J} \frac{e_{ij}}{\alpha_i} +R_{i}\sum_{j=1}^{J}e_{ij}\\
    R_{i} & = -\sum_{j=1}^{J} \frac{e_{ij}}{\alpha_i} \\
    & = -\sum_{j=1}^{J} \frac{\exp(\alpha_i p^{i}_{*,j} + v_{ij})}{\alpha_i} \\
    & = -\frac{1}{\alpha_i}\sum_{j=1}^{J}\exp(-1 + \alpha_i m_j + v_{ij})\exp(\alpha_i R_i)
\end{aligned}
\end{equation}
where $v_{ij}$ refers to the sum of all remaining terms that does not include price but affect the utility. Specifically, for our model $\mathcal{M}_{O}$, $v_{ij}(\mathcal{M}_{O}) = \alpha_1 \textrm{cumPercOnlineSpend}_{i} + \alpha_{3}\mathcal{I}\{\textrm{BuyPrevOcca}_{ij}\} +\lambda_1 \textrm{CF\_Price}_{j} + \lambda_2 \textrm{CF\_Price\_cumPercOnlineSpend}_{ij} + \delta_{ij}$.
For the naive model $\mathcal{M}_{N}$, $v_{ij}(\mathcal{M}_{N}) =  \tilde{\alpha}_{1}\mathcal{I}\{\textrm{BuyPrevOcca}_{ij}\} +\tilde{\beta}_1 \textrm{CF\_Price}_{j} + \tilde{\delta}_{ij}$.

Notice that Equation \eqref{eq:8_Optimal_Personalized_Price} has a unique solution since the left-hand side is increasing and starts from zero, and the right-hand side is positive at $R_i(\mathbf{p}^{i};\mathcal{M}_{k}) = 0$ and decreasing. Following \citet{jain2024effective}, we use a binary search algorithm to obtain the optimal $R_{i}(\mathbf{p}^{i}_{*}; \mathcal{M}_{k})$ for each individual consumer $i$ and then calculate her optimal price vector $\mathbf{p}_{*}^{i,\mathcal{M}_{k}}$. Finally, we denote the profit-maximizing personalized prices for each customer $i$ under the naive and our model as $\mathbf{p}_{*}^{i,\mathcal{M}_{N}}$ and $\mathbf{p}_{*}^{i,\mathcal{M}_{O}}$, respectively. 

\subsection{Profit Evaluation under Counterfactual Scenarios}
\label{appssec.counterfactual.scenarios}

We now consider four counterfactual scenarios that allow for different pricing strategies (uniform vs. personalized pricing) and whether the retailer accounts for the online adoption effect (naive vs. our model). Each scenario is characterized by its input parameters, including the individual price sensitivity parameter $\alpha_{i}$ and the mean utility function $\psi_{ij}(\mathbf{p};\mathcal{M}_{k})$ as defined in $\S$\ref{appssec:optimal_prices}. Specifically, we consider four counterfactual scenarios, where the firm sets profit-maximizing prices: 
\squishlist
\item Uniform Price using the Naive Model, denoted by $\mathbf{p}_{*}^{U, \mathcal{M}_{N}}$. 

%This is characterized by using inputs  $\alpha_{i} = \tilde{\alpha}_{0}$ from Equation \eqref{eq:naive.model}, and $\psi_{ij}(\mathbf{p}_{*}^{U, \mathcal{M}_{N}};\mathcal{M}_{N})$. %from the naive model in Equation \eqref{eq:naive.model}. 
\item Uniform Price using Our Model, denoted by $\mathbf{p}_{*}^{U, \mathcal{M}_{O}}$. 

%This is characterized by using inputs $\alpha_{i} = \alpha_{0} + \alpha_{2} \times \textrm{cumPercOnlineSpend}_{i}$ from Equation \eqref{eq:util_choice_M2}, and $\psi_{ij}(\mathbf{p}_{*}^{U, \mathcal{M}_{O}};\mathcal{M}_{O})$. %from our model in Equation \eqref{eq:util_choice_M2}. 

\item Personalized Price under the Naive Model, denoted by $\mathbf{p}_{*}^{i,\mathcal{M}_{N}}$ for customer $i$. 

%This is characterized by using inputs $\alpha_{i} = \tilde{\alpha}_{0}$ from Equation \eqref{eq:naive.model}, and $\psi_{ij}(\mathbf{p}_{*}^{i,\mathcal{M}_{N}};\mathcal{M}_{N})$. %from naive model in Equation \eqref{eq:naive.model}. 

\item Personalized Price under Our Model, denoted by $\mathbf{p}_{*}^{i,\mathcal{M}_{O}}$ for customer $i$. 
%This is characterized by using inputs  $\alpha_{i} = \alpha_{0} + \alpha_{2} \times \textrm{cumPercOnlineSpend}_{i}$ from Equation \eqref{eq:util_choice_M2}, and $\psi_{ij}(\mathbf{p}_{*}^{i,\mathcal{M}_{O}};\mathcal{M}_{O})$. %from our model in Equation \eqref{eq:util_choice_M2}. 
\squishend

The counterfactual analysis consists of two steps: (1) Optimal price calculation and (2) Profit evaluation. In the optimal price calculation step, we numerically solve for the optimal uniform and personalized price vectors under both our model and the naive model, as described in $\S$\ref{appsssec.optimal.offline.price} and $\S$\ref{appsssec.personalized.offline.price}. Next, in the profit evaluation step, we take the optimal price from Step 1 and evaluate the counterfactual profit using Our model $\mathcal{M}_{O}$ (which we treat as the true model that accounts for the online adoption effect). This gives us the counterfactual profits for all four scenarios described above. 

Next, we define how much the firm would gain if it uses our model instead of the naive model and engages in uniform pricing, as follows:
\begin{equation}
    \begin{aligned}
\label{eq:lift_profit_nonPersonalized}
       \Delta \textrm{Gain from Uniform Pricing (\%)} = \frac{R(\mathbf{p}_{*}^{\mathcal{U, M}_{O}};\mathcal{M}_{O}) - R(\mathbf{p}_{*}^{U,\mathcal{M}_{N}};\mathcal{M}_{O})}{R(\mathbf{p}_{*}^{U, \mathcal{M}_{N}};\mathcal{M}_{O})}
    \end{aligned}
\end{equation}
Similarly, we can define the gains from using our model instead of the naive model for the personalized pricing case:
\begin{equation}
    \begin{aligned}
\label{eq:lift_profit_Personalized}
       \Delta \textrm{Gain from Personalized Pricing (\%)} = \frac{\Sigma_{i=1}^N R_i(\mathbf{p}_{*}^{i, \mathcal{M}_{O}};\mathcal{M}_{O}) - \Sigma_{i=1}^{N} R_i(\mathbf{p}_{*}^{i,\mathcal{M}_{N}};\mathcal{M}_{O})}{\Sigma_{i=1}^N R_i(\mathbf{p}_{*}^{i, \mathcal{M}_{N}};\mathcal{M}_{O})}
    \end{aligned}
\end{equation}

Finally, we simulate the four counterfactual scenarios using data from June 2021. We focus on the last month of the observational period to provide a clear interpretation of the firm's \textit{monthly} incremental profit and to avoid any time-specific variations across months that could complicate the interpretation. Table \ref{tab:8_Sample_Size_Comparison} summarizes the sample sizes for each category included in this counterfactual study (i.e., customers defined in $\S$\ref{ssec:construction} with a purchase occasion in June 2021).

\begin{table}[htp!]
\centering
\caption{Sample Sizes for the Counterfactual Study - By Category}
\label{tab:8_Sample_Size_Comparison}
\footnotesize{
\begin{tabular}{lrrrr}
\toprule
 & Dry Dog Food  & Dog Hygiene & Cat Hygiene & Total \\
\midrule
Adopters & 765  & 838 & 450 & 2425\\ 
Non-Adopters & 2879  & 1400 & 1165 & 6314  \\ \midrule
Total & 3644  &  2238 & 1615 & 8739 \\ \midrule
\bottomrule
\end{tabular}
}
\end{table}

We now discuss the results of our counterfactual analysis. Under uniform pricing, we find that using our model (accounts for the online adoption effect)  to optimize prices, increases monthly profit by 1.17\% compared to pricing using the naive model. Further, under personalized pricing, using our model to optimize prices increases monthly profit by 4.1\% compared to pricing using a naive model. Overall, these results suggest that the retailer can benefit from accounting for the effect of online adoption on consumers' offline price sensitivity when optimizing their offline pricing decisions. Further, the magnitude of these gains is non-trivial, economically meaningful and consistent with the earlier literature on personalization \citep{rossi1996value, yoganarasimhan_etal_2023, rafieian_yoganarasimhan_2023}. 

Overall, our findings highlight an important yet overlooked managerial implication for multichannel retailers: retailers need to incorporate online adoption effects into offline pricing strategies, and doing so can improve retailers' profitability in the offline channel, regardless of whether pricing is uniform or personalized. Further, the magnitude of these gains is economically meaningful. As retailers continue to expand their multichannel operations, adjusting for this effect in offline pricing becomes increasingly important. 

Nevertheless, our counterfactual results should be interpreted carefully, with the appropriate caveats.  First, it is possible that consumers' likelihood of shopping offline changes under the counterfactual pricing strategy. However, consistent with the earlier literature studying brand choice within channels, \citep{danaher2003comparison,chu2008research}, we do not model the channel choice decision \citep{valentini2011decision}, and this can potentially bias our counterfactual results. Second, we abstract away from retailer-level competition, and we do not model consumers' choices between different retailers and how changes to offline prices may affect these choices\citep{bell1998determining}. Again, this is outside of the scope of our data. Finally, our counterfactual policy estimates are based on model-based estimates rather than based on off-policy IPW evaluation (such as \cite{smith2023optimal}). This is because there is insufficient overlap in price distributions between monopolistic profit-maximizing prices and observed prices in the data. Nevertheless, in line with model-based evaluation similar to \cite{rossi1996value}, these counterfactual analyses provide important insights into the potential gains from accounting for the effect of online adoption on offline pricing. We hope that future research can build on our findings and make further progress on these issues.

%\putbib

%\end{bibunit}
%\end{bibunit}

\end{appendices}

\end{document}